\documentstyle[epsfig,graphics]{mn2e}

\begin{document}

\newcommand{\Zsolar}{\mbox{\,$\rm Z_{\odot}$}}
\newcommand{\etal}{{et al.}\ }
\newcommand{\ang}{\mbox{$\rm \AA$}}
\newcommand{\xs}{$\chi^{2}$}
\newcommand{\be}{\begin{equation}}   
\newcommand{\ee}{\end{equation}}     

\title[A combined re-analysis of existing blank-field SCUBA surveys]{A 
combined re-analysis of existing blank-field SCUBA surveys:
comparative $\bf 850-\mu m$ source lists, combined number counts, and
evidence for strong clustering of the bright sub-mm galaxy population on
arcminute scales.}

\author[S.E. Scott et al.]
{S.E. Scott$^{1}$, J.S. Dunlop$^{1}$, S. Serjeant$^{2,3}$\\
$^{1}$SUPA\thanks{Scottish Universities Physics Alliance}, 
Institute for Astronomy, University of Edinburgh, Royal Observatory, 
Edinburgh, EH9 3HJ, UK\\
$^{2}$
Centre for Astrophysics and Planetary Science, 
School of Physical Sciences, University of Kent, Canterbury, Kent CT2 7NZ, UK\\
$^{3}$Astrophysics Group, Department of Physics, The Open University,
Milton Keynes, MK7 6AA, UK 
}

\date{Accepted for publication in MNRAS}

\maketitle
  
\begin{abstract}
Since the advent of SCUBA on the JCMT, a series of complementary
surveys has resolved the bulk of the far-infrared extragalactic  
background into
discrete sources. This has revealed a population of heavily
dust-obscured sources at high redshift ($z>1$) undergoing an intense
period of massive star-forming activity with inferred star formation 
rates of several hundred to several thousand solar masses per year.
Taken together, these existing surveys cover a total area of 460
sq. arcmin to a range of depths, but combining the results has
hitherto been complicated by the fact that different survey groups
have used different methods of data reduction and source extraction. 
In this paper we re-reduce and analyse all of
the blank field surveys to date in an almost
identical manner to that employed in the ``SCUBA 8\,mJy Survey''. 
Comparative source catalogues are given which
include a number of new significant source detections as well as
failing to confirm 
some of those objects previously published. These new source
catalogues have been combined to produce the most accurate number
counts to date from 2 to 12.5\,mJy. We find $\rm N(>4\,mJy)
=620^{+110}_{-190}\, deg^{-2}$, $ \rm N(>6\,mJy)=310^{+\phantom{0}60}_{-100}\, deg^{-2}$, $
\rm N(>8\,mJy)=150^{+\phantom{0}40}_{-\phantom{0}60}\, deg^{-2}$ and $\rm
N(>10\,mJy)=\phantom{0}40^{+\phantom{0}20}_{-\phantom{0}20}\, deg^{-2}$ after
correcting for the effects of incompleteness, flux-density boosting
and contamination from spurious / confused detections. Furthermore
the cumulative number counts appear to steepen beyond $\rm S_{850} >
8\, mJy$, which could indicate an intrinsic turn-over in the
underlying luminosity function placing an upper limit on the
luminosity (and hence mass) of a high redshift galaxy. We have also 
investigated
the clustering properties of the bright $\rm S_{850} >5 \,mJy$ SCUBA
population by means of 2-point angular correlation functions. We find a
$\rm \simeq 3.5 \sigma$ excess of pairs within the first 100
arcsec over that expected from a Poisson distribution. Fits of a
standard power-law of the form $\rm w(\theta)=A \theta^{-\delta}$ to
the angular correlation functions for $\rm S_{850} >5 \,mJy$ are
limited in accuracy by the small number of source detections but
appear to be broadly consistent with that measured for
EROs. Nearest-neighbour analyses further
support strong clustering on arcmin scales, rejecting the null
hypothesis that the
distribution of the submm sources is random at the 95\% confidence level for
$\rm S_{850} >5\,mJy$, and at the 99\% confidence level for $\rm
S_{850} >7\,mJy$. 

\end{abstract}


\newpage

\vspace*{1cm}

\section{Introduction}
Over the past seven years, a series of complementary deep 
$850\,{\rm \mu m}$ surveys (eg. Smail et al. 1997, Hughes et al. 1998, Blain et
al. 1999, Barger et al. 1998, Barger et al. 1999, Eales et al. 2000,
Scott et al. 2002, Cowie et al. 2002, Borys et al. 2003, Webb et
al. 2003a) carried out using SCUBA (Holland et al. 1999) on the
JCMT has resolved the bulk of the far-infrared (FIR)
extragalactic background into discrete sources. These surveys vary
in size and depth from ultra-deep surveys exploiting gravitational
lensing from intervening clusters to study the very faintest submm
sources (Smail et al. 1997, Cowie et al. 2002), small and 
deep blank field surveys such as the HDF (6 sq. arcmin to a
uniform $1\sigma_{\rm rms}\simeq 0.5$ mJy/beam; Hughes et
al. 1998, Serjeant et al. 2003), through to moderate area and comparatively
shallower blank field surveys such as the ``SCUBA 8-mJy Survey'' (a
total of 250 sq. arcmin to a uniform $1\sigma_{\rm rms}\simeq
2.5$\,mJy/beam; Scott et al. 2002). 
These surveys have revealed a population of heavily
dust-enshrouded high-redshift ($z>1$) morphologically-irregular
galaxies. Although there
is still considerable uncertainty in the fraction of SCUBA sources 
hosting either low luminosity or Compton-thick AGNs (Smail et
al. 2002, Ivison et al. 2002), deep X-ray observations with the
Chandra and XMM-Newton telescopes have suggested that even when an 
AGN is present, it rarely
dominates the far-infrared / submillimetre emission from the galaxy 
(Frayer et al. 1998,
Alexander et al. 2003). Hence, if the thermal dust emission is
dominated by reprocessed starlight,  the inferred 
star formation rates (SFRs) in the very brightest submillimetre
galaxies may be as high as several thousand solar masses per
year. This is
sufficient to form the most massive elliptical galaxies observed in
the present-day Universe on timescales
of $\sim 1$\,Gyr, implying that the submillimetre sources may be the
progenitors of todays massive ellipticals, and that the
star-formation rate density in the early Universe was at least a factor
of 2 larger than that derived from optical / UV observations (Steidel
et al. 1999). 

A key test of whether the bright submillimetre sources really are the
progenitors of present-day massive spheroids is to measure their
clustering properties. If they are indeed proto-massive ellipticals
then they should be strongly clustered. This is an inevitable result
of gravitational collapse  from Gaussian initial density fluctuations:
the rare high-mass peaks are strongly biased with respect to the mass
(eg. Benson et al. 2001).
Each of the submillimetre survey consortia have performed their own
reduction, source extraction and simulations on their individual
datasets, in order to study the nature of the SCUBA population in
general. However, each survey alone images only $\sim 100 - 200$
sq. arcmin of sky spread over several fields, resulting in
discrepancies in the cumulative number counts at $\rm S_{850} >
8\,mJy$ by over a factor of 5 due to cosmic variance, and also
potentially the effects of clustering. Furthermore, no individual
survey has identified enough sources to make a significant measurement
of the clustering properties, although tentative
evidence was obtained for strong clustering on scales of 1-2 arcmin from the
``SCUBA 8\,mJy Survey'', the largest of these surveys undertaken to
date (Scott et al. 2002).

In this paper we combine the
data from all of the blank field surveys completed
up to the completion date of the ``SCUBA 8\,mJy Survey''. 
In order
to do this we downloaded the raw data for the ``Canada UK Deep
Submillimetre Survey (CUDSS)'', the ``Hawaii Flanking Fields Survey''
and the ``Hubble Deep Field (HDF) North Pencil Beam Survey'' from the
Canadian Astronomy Data Centre (CADC) archive and re-reduced it in an
identical manner to that employed in the ``SCUBA 8\,mJy Survey''. This
is decribed in Section 2. The source extraction algorithm 
developed by Scott et al. (2002) was used to identify significant
submillimetre sources in each of the fields, and this method is
described briefly in Section 3. In Section 4 
we present and discuss the results of simulations carried out both in
conjunction with the real data, and through the production of
fully-simulated survey areas, for all of the survey
fields. Comparative reduction methods and source lists down to a 
signal-to-noise ratio of 3.00 are given in Section 5. These lists include
some previously unidentified significant sources but also cast
doubt on the reality of other published detections. Section 6 combines
these new catalogues into a master source list, from which the most
accurate cumulative number counts to date in the flux density range 
$2-12.5$\,mJy have been calculated. A number of models are tested
against these data points. In Section 7, the new master catalogue has
been used in an attempt to measure the clustering properties of the
bright ($>5$\,mJy) submillimetre sources, through 2-point angular
correlation functions and nearest-neighbour analyses. Section 8 outlines the
resulting motivation and aims of the
current blank field survey work being undertaken by the JCMT + SCUBA,
and the conclusions
from this work are presented in Section 9.

\section{Data Reduction}

The raw data for the ``Canada UK Deep Submillimetre Survey (CUDSS)'',
the ``Hawaii Flanking Fields Survey'', and the ``Hubble Deep Field
Survey'' (pencil beam only) were downloaded from the Canadian
Astronomy Data Centre (CADC) archive. The data were then fully reduced
using the same Interactive Data Language (IDL)-based reduction routines employed in the ``SCUBA
8\,mJy Survey''. This procedure is fully described in Scott et
al. (2002), but is summarised briefly below.

Firstly, in order to take nodding into consideration, the off-position
was subtracted from the on-position in the raw beam-switched data. The
relative sensitivities of the bolometers, with respect to a
reference bolometer (the central bolometers H7 and C14, for the $\rm
850 \, \mu m$ and $\rm 450 \, \mu m$ arrays respectively) were
accounted for by multiplying by the standard flatfield values. The
atmospheric opacity was measured wherever possible using a 6th order
polynomial fit to the 225\,GHz measurements from the Caltech
Submillimetre Observatory (CSO), followed by a linear interpolation to
850 and $\rm 450 \, \mu m$ using the relations given in Archibald et
al. (2002a). On the occasions when the CSO opacity meter was out of
service, or less than 7 reliable CSO observations were available
within $\pm 1$\,hour of each observation, a linear interpolation
between successive skydip values was used to determine $\rm
\tau_{850\mu m}$ and  $\rm \tau_{450\mu m}$.

The next stage in the data reduction process was to remove spikes in
the data resulting from cosmic-rays and bad bolometers, and to remove
any residual sky emission. In this IDL reduction, deglitching and 
residual sky subtraction were
undertaken by an iterative process, each iteration making a temporal
noise estimate and deglitching, followed by a spatial sky
subtraction. There were no bright sources in any of submm survey
fields that would have been significantly detected in any single
jigglemap, let alone in sub-dividing the data-stream into shorter
timescale chunks and so the procedure was as follows:

\noindent$\bullet$ For each bolometer, noise estimates were made by
fitting a Gaussian to the data-stream in chunks of 128 readout groups.

\noindent$\bullet$ These time dependent noise estimates were then used
to remove any spikes by performing a $\rm 3 \sigma$ clip on the data.

\noindent$\bullet$ Using the fits to all of the bolometers in the
array, a modal residual sky level was determined for each of these 128
readout groups, and subtracted from these data.

With each consecutive iteration, the deglitching process makes a
harder cut. Noisy bolometers were assigned a low inverse variance
weight in this way. In the case of the calibration data, however, the
presence of a bright source would likely lead to over-enthusiastic
clipping of the data, and so in this case a timeline without object
signal was constructed. This was created by calculating the mean of
the timestream data points recorded immediately before and after the
readout being considered, and subtracting this from the readout value.

Flux conversion factors (FCFs) were determined by dividing the flux
density within the JCMT beam of a known calibration source by the measured 
peak voltage. 
Each of the individual jiggle-maps comprising the submm survey data were
calibrated prior to producing the final coadded images using the
gain value from whichever calibration source was taken closest in
time.

The final images were produced using an optimal noise-weighted drizzling
algorithm (Fruchter \& Hook, 2002) with a pixel size of 1 sq.
arcsec. This is the same method as that employed in the `SCUBA 8-mJy
Survey' (Scott et al. 2002) and `Hubble Deep Field North' (Serjeant et al.
2003) data reductions. Both output signal and noise maps were created,
the signal in any one sq. arcsec pixel given by the noise-weighted
average of the bolometer readouts at that position, and the corresponding
noise value given by a noise-weighted average of the Gaussian fits to the
readout histograms. Unlike a standard shift-and-add technique which takes
the flux density in each detector pixel and places it into the final map
over an area equivalent to one detector pixel projected on the sky,
drizzling takes the flux density and places it into a smaller area in the
final map. Although this significantly reduces the signal-to-noise ratio
in each pixel, this approach helps preserve information on small angular
scales, provided that there are enough observations to fill in the
resulting gaps. The area in the coadded map receiving the flux from one
detector pixel is termed the `footprint'. This method is an extreme
example of drizzling in which the `footprint' is selected to be as small
as is practicable given the pointing errors invloved (termed the
`zero-footprint'), selected here to be one sq. arcsec.
Unlike in the standard SURF reduction, there is no
intrinsic smoothing or interpolation between neighbouring pixels in this rebin
procedure. Although there is some degree of correlation between pixels
in the output zero-footprint \emph{signal} maps in terms of the beam
pattern, the corresponding pixel \emph{noise} values represent
individual measurements of the temporally varying sky noise averaged
over the dataset integration time, at a specific point on the sky, and
are hence statistically independent from their neighbours. In essence
this method produces a very oversampled image with
statistically independent pixels.
A final $4\sigma $-clip on the signal-to-noise was carried out to
remove any remaining `hot pixels'. A noise-weighted convolution with a
beam-sized Gaussian point spread function (PSF) was used to produce realistic
smoothed maps of the survey areas whilst accounting for variable
signal-to-noise between individual pixels.

\section{Source Extraction}

\begin{table*}

\begin{tabular}{|l|c|c|c|} \hline
Survey & Total area & Uniform noise & $\rm 1 \sigma$ rms noise\\
Field  & /sq. arcmins& area /sq. arcmins & level mJy/beam \\ \hline	
ELAIS N2 & $137$ & $113$ & $2.2 \pm 0.7$ \\
Lockman Hole wide area & $\phantom{0}141$* & $\phantom{0}88$ & $2.7 \pm 0.7$ \\
Lockman Hole deep strip & & $\phantom{0}21$ & $1.8 \pm 0.2$ \\
CUDSS 03h wide area & $\phantom{00}69$* & $\phantom{0}55$ & $1.8 \pm 0.5$ \\
CUDSS 03h deep area &  & $\phantom{00}8$ & $1.1 \pm 0.2$ \\
CUDSS 10h           & $\phantom{0}10$ & $\phantom{00}8$ & $1.3 \pm 0.2$ \\
CUDSS 14h & $\phantom{0}61$ & $\phantom{0}57$ & $1.5 \pm 0.3$ \\
CUDSS 22h & $\phantom{00}7$ & $\phantom{00}5$ & $1.5 \pm 0.3$ \\
Hubble Deep Field & $\phantom{0}10$ & $\phantom{00}6$ & $0.6 \pm 0.1$ \\
SSA13 wide area & $\phantom{00}72$* &$\phantom{0}45$ & $2.5 \pm 0.6$ \\ 
SSA13 deep area & & $\phantom{00}8$ & $0.7 \pm 0.1$ \\
SSA17 & $\phantom{0}24$ & $\phantom{0}21$ & $1.6 \pm 0.5$ \\
SSA22 & $\phantom{0}26$ & $\phantom{0}21$ & $0.9 \pm 0.2$ \\
Lockman Hole deep area & $\phantom{0}11$ & $\phantom{00}8$ & $0.8 \pm
0.1$ \\ \hline
\end{tabular}
\label{table:noise_areas}\caption{\small Survey field areas and $\rm
1\sigma$ rms noise levels in the regions of uniform noise, as given by
the mean and standard deviation measured directly from the 14.5'' FWHM
Gaussian convolved noise maps. Total areas marked with * refer to
fields composed of a small deep region within a wider shallower survey
area and correspond to the full area of that entire field (ie. both
shallow and deep).}
\end{table*}

The chopping-nodding mechanism of the telescope provides a valuable
method of discriminating between real detections and spurious noise
spikes in the data. With the exception of
the Hubble Deep Field (HDF), each of the surveys used a single chop
throw fixed in right ascension (RA), thus creating negative sidelobes, half
the depth of the peak flux density, on either side of a real
source. In the case of the HDF, this strategy was modified to use two
chop throws fixed in RA, each chop throw used for approximately half
of the total integration time. This side-lobe signal can
be recovered to boost the overall signal-to-noise ratio of a detection.

For well-separated sources, convolving
the images with the beam is formally the best method of source
extraction (Eales et al. 1999, 2000, Serjeant et al. 2003). However,
following a careful examination of the reduced ``8\,mJy Survey'' data
(Scott et al. 2002), it became clear
that some of the potential sources were partially confused. This was
particularly prominent in the map of ELAIS N2
where the negative sidelobes of individual sources have overlapped and
are therefore somewhat deepened relative to both source
peaks. Consequently, in order to decouple any confused sidelobes
a source-extraction algorithm was devised based on a simultaneous 
maximum-likelihood fit to the flux densities of all potentially 
significant peaks in the maps. This is  made feasible by the
independent data-points and errors yielded by the zero-footprint
IDL-reduced maps. These peaks were identified as any
positive peak in the noise-weighted Gaussian convolved signal maps.
Using a peak-normalised beam-map as a source template (generated by
binning together all of the observations of Uranus or CRL618
taken with the relevant chop throw), a basic model was constructed by
centring a beam-map at the positions of every peak in the maps. The
normalisation coefficients of each of the positioned beam-maps were then
calculated simultaneously such that the final multi-source model
provided the best description of the submm sky, as judged by a
minimum $\chi^{2}$ fit.

The fitting process is as follows. Suppose one considers a normalised beam-map $B(x,y)$ as a source template
and that at position $(i,j)$ in the unconvolved image the signal is
$S(i,j)$ and the noise is $N(i,j)$. If n peaks above a specified flux
threshold are located in the Gaussian-convolved image, one may construct a
model to the unconvolved zero-footprint image such that beam-maps centred on
each peak position are simultaneously scaled to give an overall best
fit to the entire image. Using a minimum $\chi^{2}$ fit as the
maximum likelihood estimator then \be
\chi^{2} = \left( \sum_{i,j} \frac{S(i,j) - \sum_{k=1}^{n} a_{k} B_{k}(x-i,y-j)}{N(i,j)^{2}} \right)^{2}
\ee where $a_{k}$ is the best fit flux to the $k$th peak.

Defining $\alpha_{mk}$, an $n \times n$ matrix, as
\be
\alpha_{mk}= \sum_{i,j} \frac{B_{k}(x-i,y-j)
B_{m}(x-i,y-j)}{N(i,j)^{2}}
\ee
and $\beta_{m}$, a vector of length $n$ as
\be
\beta_{m} = \sum_{i,j} \frac{S(i,j)
B_{m}(x-i,y-j)}{N(i,j)^{2}}
\ee
then the best fit values of $a_{k}$ are given by 
\be
a_{k} = \sum_{m=1}^{n} [\alpha]_{km}^{-1} \beta_{m}
\ee
and the significances $\sigma(a_{k})$ of the peak detections are given by
\be
\sigma(a_{k}) = \frac{a_{k}}{\sqrt{[\alpha]_{kk}^{-1}}}
\ee

Furthermore, this method can be modified to deal with surveys which have used
more than one chop throw or position angle. The peaks are found in the
same way as before, by regridding all of the individual observations together
(regardless of the chop throw or position angle used) and carrying out
a noise-weighted smoothing with a beam-sized Gaussian. When conducting
the $\chi^{2}$ fit, however, each particular combination of chop throw and
position angle is binned separately. If r different chop
configurations have been used then the expression for $\chi^{2}$ becomes
\be
\chi^{2} = \sum_{p=1}^{r} \left( \sum_{i,j} \frac{S_{p}(i,j) - \sum_{k=1}^{n} a_{k} B_{p,k}(x_{p}-i,y_{p}-j)}{N_{p}(i,j)^{2}} \right)^{2}
\ee
and 
\be
\alpha_{mk}= \sum_{p=1}^{r} \left(\sum_{i,j} \frac{B_{p,k}(x_{p}-i,y_{p}-j)
B_{m,k}(x_{p}-i,y_{p}-j)}{N_{p}(i,j)^{2}} \right)
\ee
and
\be
\beta_{m} = \sum_{p=1}^{r} \left( \sum_{i,j} \frac{S_{p}(i,j)
B_{p,m}(x_{p}-i,y_{p}-j)}{N_{p}(i,j)^{2}} \right)
\ee
where the expressions for the best fit values of $a_{k}$ and
significances $\sigma(a_{k})$ are the same as before.

For a full mathematical description of this source extraction
algorithm, the reader is referred to Scott et al. (2002) and Mortier
et al. (2005). 

\section{Simulations}

In order to assess the effects of confusion and noise on the
reliability of the source-extraction algorithm, Monte Carlo simulations
were carried out on all of the survey fields. The individual fields
vary widely in size and depth from small, deep surveys covering a few
sq. arcmin of sky down to the confusion limit (eg. the Hubble deep
field), to wider, shallower surveys aimed at studying the most
luminous sub-millimetre sources on scales of $\sim 100$ sq.
arcmin (eg. the wide area Lockman Hole field from the ``SCUBA
8-mJy Survey''). The typical noise levels and areas of each of
the fields are given in Table 1. The dependences of positional
error, completeness and error in reclaimed flux density, on input source
flux density and noise in the maps, were determined by planting 
individual sources of known flux density into the real SCUBA
maps. This has the advantage of testing the 
source-reclamation process against the real noise and
confusion properties of the images, accounting for any clustering in
the faint background source population, for example. However, these simulations do not allow assessment of the
level to which false or confused sources can contaminate an extracted
source list. Therefore a number of fully-simulated 
images of the survey areas have also been created by assuming a reasonable 
850~${\rm \mu m}$ source-counts model, derived from a best-fit
power-law to the source counts given later in this paper.
The results of analyzing these sets of simulations are discussed in the
following two subsections.
 
\subsection{Simulations building on the real survey data}

\begin{table*}

\begin{tabular}{|l|c|c|c|c|} \hline
Survey Field & Noise Region & a & b & $\rm \chi^{2}$ \\ \hline
ELAIS N2 & uniform & 0.17043      & 2.3338 &   0.37425 \\
Lockman Hole wide area & uniform & 0.14800  & 3.2690 & 0.28300 \\
Lockman Hole deep strip & uniform & 0.20896 & 2.1064 & 0.22527 \\
CUDSS 03h wide area & uniform & 0.20081    &   2.0013  &  0.21037 \\
CUDSS 03h deep area & uniform & 0.28937      & 1.6500  &  0.39179 \\
CUDSS 10h           & uniform & 0.22587    &  1.3248 &   0.42789 \\
CUDSS 14h & uniform & 0.21016   &    1.5467 &   0.31834 \\
CUDSS 22h & uniform & 0.26450    &   1.9359 &   0.12883 \\
Hubble Deep Field & uniform & 0.39711      & 0.6258  &  0.15015 \\
SSA13 wide area & uniform & 0.14627     &  2.6614 &   0.19068 \\
SSA13 deep area & uniform & 0.22037	& 0.8738 &  0.13963 \\
SSA17 & uniform & 0.24248 &  1.9077  &  0.19140  \\
SSA22 & uniform & 0.21362 &  0.4445  &  0.37697 \\
Lockman Hole deep area & uniform & 0.24862    &   0.8042 & 0.12062\\ \hline
ELAIS N2 & non-uni & 0.00588    &   0.0030 &  0.04114 \\ 
Lockman Hole wide area & non-uni & 0.01388	& 4.6093 &  0.05206
\\ \hline
\end{tabular}
\label{table:completeness}\caption{\small Best fit values determined
for a and b in equation 9, describing the percentage differential
completeness against input source flux density for each of the survey
fields and noise regions.}
\end{table*}

\begin{figure}
 \centering
   \vspace*{3.8cm}
   \leavevmode
   \includegraphics{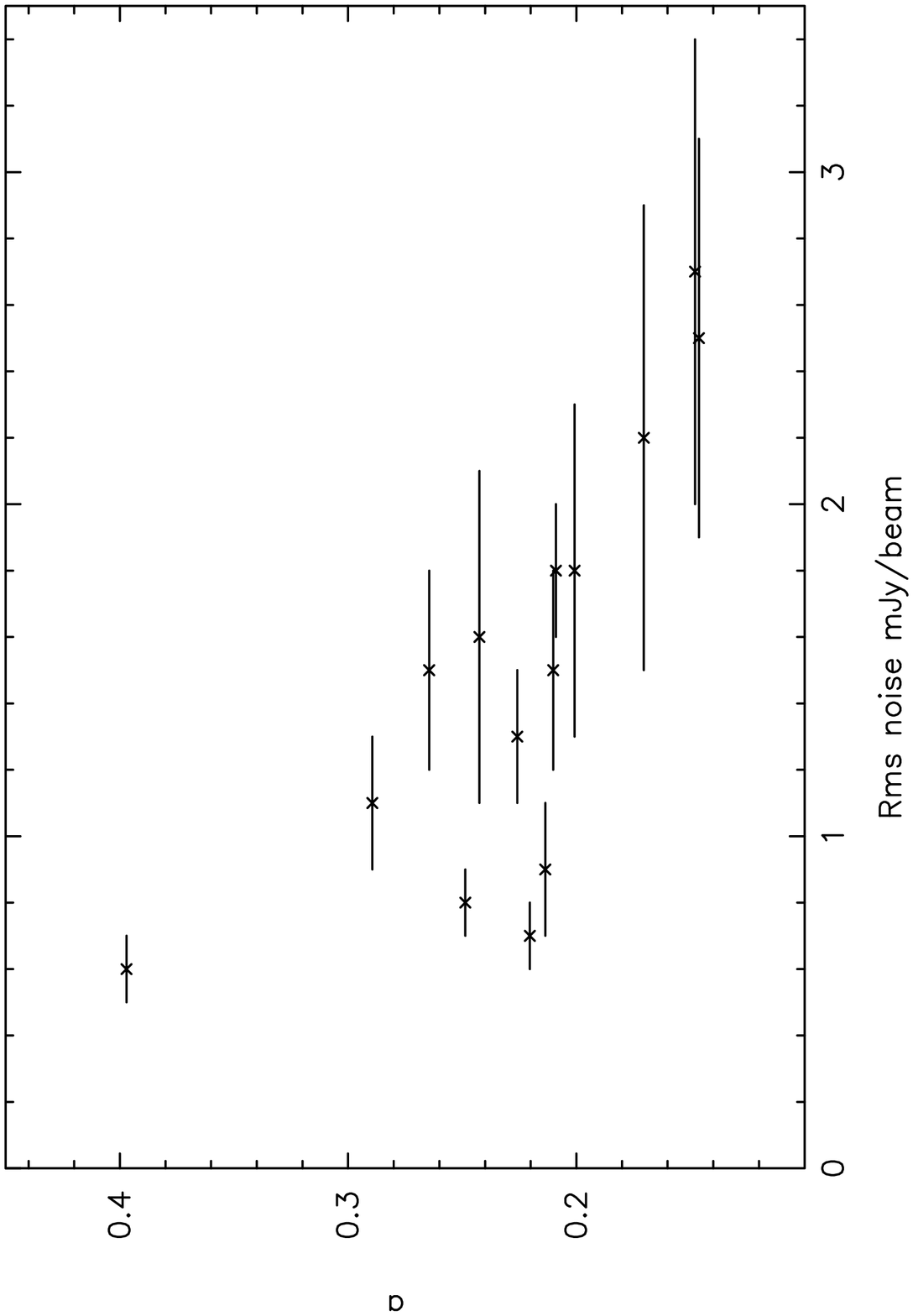}
\label{fig:param_a} 
\caption{\small Best-fit values for the parameter ``a'' as given in
   Table 2, plotted against the $\rm 1\sigma$ rms noise levels as
   determined from the uniform regions of the 14.5'' Gaussian
   convolved noise images (values given in Table 1). The horizontal 
   error bars show the standard deviation of the noise about the mean level.}

 \centering
   \vspace*{3.8cm}
   \leavevmode
   \includegraphics{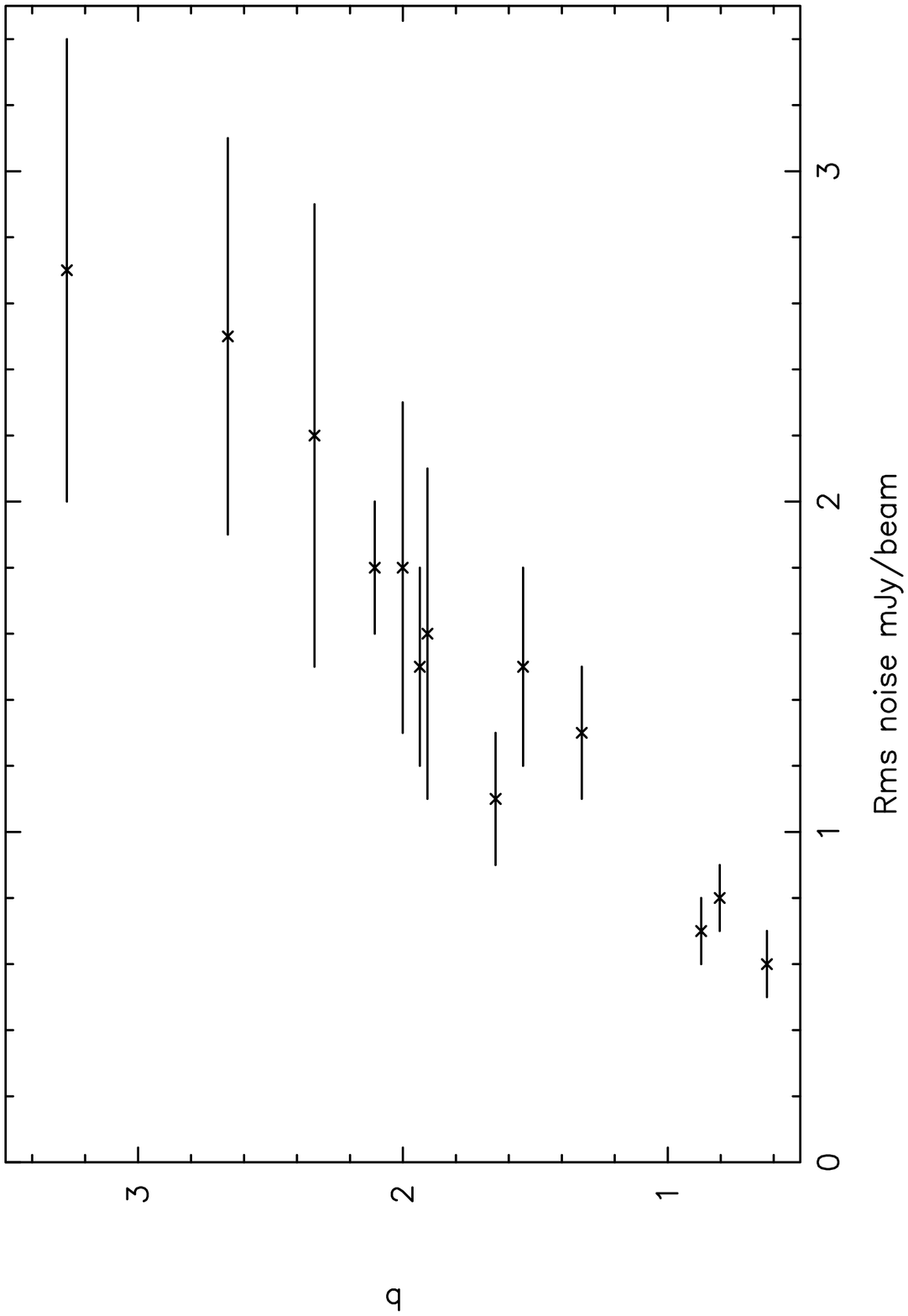}
\label{fig:param_b} 
\caption{\small Best-fit values for the parameter ``b'' as given in
   Table 2, plotted against the $\rm 1\sigma$ rms noise levels as
   determined from the uniform regions of the 14.5'' Gaussian
   convolved noise images (values given in Table 1). The horizontal 
   error bars show the standard deviation of the noise about the mean level.}
\end{figure}

A normalised beam-map, with the same chop throw and position angle as
that used in the real observations, was used as a source
template. At flux density intervals of 0.5\,mJy, spanning the entire
range of flux densities for which real sources were recovered, fake sources
were added into the unconvolved zero-footprint signal maps. This was
done one fake source at a time, so as not to enhance significantly
any existing real confusion noise within the image. The
source-extraction algorithm was then re-run. This exercise was repeated for 100
different randomly-selected positions on each image, at each flux density
level, so that source reclamation could be monitored as a function of
input flux density and position/noise-level within the maps. The source
reclamation was deemed to have been successful if the 
source-extraction algorithm returned the fake source with
signal-to-noise $> 3.50$ (a level selected as a compromise between
recovering a reasonable number of sources and contamination with
spurious / confused sources - see Subsection 4.2)
within less than half a beam-width of the input position, but
excluding from the analysis any fake sources which had fallen upon a position
within half a beam-width of a brighter $\rm >3.00\sigma$ peak already detected
in the map. This is because the flux
densities of the recovered sources within the real data span a broad
range ($\rm \simeq 2-12 \, mJy$), and it is not possible to resolve two
separate sources placed closer together than this - they would appear
as one peak in the Gaussian-smoothed image. It is not realistic to consider, for
example, a fake 2\,mJy source to have been successfully recovered if
it lies almost on top of an 8\,mJy source already detected
significantly in the map. Under this situation it is really the 8\,mJy
source already present in the image which is being recovered. Reversing the situation, however, the
successful reclamation of a fake 8\,mJy source planted into the map in
the near vicinity of an already significantly detected 2\,mJy source
(a possible scenario in the very deep images such as that of the HDF)
would be included in the analysis because the fake source is making the
dominant contribution to the combined flux density. 

It is possible that future interferometers may resolve some of our
point sources into multiple components; this is an inevitable caveat
to any source count analysis. In the meantime however, the JCMT
resolution provides an effective working definition of point source
for the current work. 

The Lockman Hole East field from the ``SCUBA 8\,mJy
Survey'', the 03h field from the ``Canada UK Deep Submillimetre
Survey (CUDSS)'', and the SSA13 field from the ``Hawaii Submillimetre Survey'',
contain sections of map which are markedly deeper than the rest of
the data. In the case of the Lockman Hole this was due to an early
change in survey mapping strategy, and in the 03h and SSA13 fields
this resulted from a deep pencil beam survey being incorporated into
the wider-area images. In each of these cases, separate sets of
simulations were run on the deep and shallower sections of the
fields. 

\begin{table*}
\begin{tabular}{|l|c|c|c|c|c|} \hline
Survey Field & Noise Region & C & d & f & $\rm \chi^{2}$ \\ \hline
ELAIS N2 & uniform & $\phantom{00}6.350$      & 0.4813 &   1.0150 & 1.11683 \\
Lockman Hole wide area & uniform & $\phantom{00}5.555$  & 0.3591 & 1.0246 & 1.08116 \\
Lockman Hole deep strip & uniform & $\phantom{00}6.543$ & 0.5313 & 1.0154 & 1.34420 \\
03h wide area & uniform & $\phantom{00}6.587$   &   0.5667 &  1.0603 & 1.05373 \\
03h deep area & uniform & $\phantom{0}13.575$     & 1.1842  &  1.0537 & 1.67181 \\
10h           & uniform & $\phantom{00}8.786$    &  0.9443 &   1.0439 & 0.76535 \\
14h & uniform & $\phantom{00}8.086$   &   0.7701 &  1.0325 & 1.71510 \\
22h & uniform & $\phantom{00}9.815$   &   0.8236 &  1.1301 & 1.51353 \\
Hubble Deep Field & uniform & $\phantom{0}12.106$      & 1.6358  &  1.0513 & 1.80567 \\
SSA13 wide area & uniform & $\phantom{00}4.606$     & 0.4312 & 1.0430 & 1.52413   \\
SSA13 deep area & uniform & $\phantom{00}2.726$	& 0.8049 &  1.0516 & 1.22079 \\
SSA17 & uniform & $\phantom{00}6.769$ &  0.8624  & 1.0965 & 1.60692  \\
SSA22 & uniform & $\phantom{00}8.262$ &  1.2241  & 1.0168 & 1.23946 \\
Lockman Hole deep area & uniform & $\phantom{00}9.712$    & 1.5104 & 1.0422 &
0.68671 \\ \hline
ELAIS N2 & non-uni & $\phantom{00}3.686$    &   0.2283 &  1.1805 & 1.84850 \\ 
Lockman Hole wide area & non-uni & $112.821$ & 0.6582 &  1.2593 & 3.39015
\\ \hline
\end{tabular}
\label{table:boosting}\caption{\small Best fit values determined
for C, d and f in equation 10, describing the output to input flux
density ratio against input flux density for each of the survey
fields and noise regions.}
\end{table*}

Additionally, regions of uniform and non-uniform noise were
defined for each field (again treating the deep parts of the Lockman
Hole, 03h and SSA13 fields as separate fields from the wider-area
shallower part), using the ``GAIA'' tool to manually cut out a template of
the uniform noise area using the Gaussian-smoothed noise maps. The
deep pencil beam surveys, such as the HDF and CUDSS 22h field,
are comprised of a stack of jiggle-map observations centred on one or
two positions only, and so the non-uniform edge regions in these images
are largely the result of undersampling from the bolometers on the
outer ring of the array. The wider-area images, however, were built up
from a series of jiggle-pointings, offset from each other by some
fraction of an array width. Hence, the pointings forming the outer-most
regions of the survey field lack the next consecutive
set of integrations from what would have been the neighbouring pointing, resulting in a border of shallower (and hence
noisier) observations. In Section 5, which discusses the various
survey fields in detail, any sources recovered in these poorer noise
regions have been marked with the term ``edge'' in the source list
tables.

\subsubsection{Completeness}

The differential completeness is given by the percentage of sources
recovered with signal-to-noise ratio $>3.50$ at each input flux
density level, and was found to be described well by the functional
form
\be \rm differential \phantom{0} completeness = 100(1-e^{-a(x-b)}) \ee
where x is the input flux density, and the values of a and b were
determined by a minimised $\rm \chi^{2}$ fit to the simulation results 
for each $\rm 850\, \mu m$ survey field. The values of a and b
determined from these fits are given in Table 2. The plots for the
percentage of sources recovered against input flux density for the
individual survey fields may be found in Appendix A1.

The primary goal in allowing a fit of this nature was to obtain a
best-fit description of the overall shape of the curve, rather than
a detailed analysis of possible combinations of free-parameters 
`a' and `b' in $\rm
\chi^{2}$ space. However, even simple plots of the best fit values of
a and b against the $\rm 1 \sigma$ rms noise levels as measured from
the beam-sized Gaussian convolved noise images (Figs. 1 and 2
respectively), show clear noise-dependent trends. The horizontal
error bars reflect the standard deviation of the noise values about
the mean, in the uniform regions of the map. The best-fit values of
parameter `a' show a general decrease with increasing rms noise levels,
albeit with a fairly broad dispersion, particularly between the deep
pencil beam surveys such as the Hubble deep field and the SSA13 deep area
field. This is likely a combination of being at the confusion limit
(generally high source density) and the variation in the number
density of sources between these small area fields (cosmic variance
and perhaps clustering effects also). The parameter `b' defines a
lower flux density cut-off below which no sources are successfully
recovered, and shows a much tighter correlation, increasing roughly 
linearly with the rms noise as $\rm b \simeq 1.25\times
noise$. 

Unfortunately, the
scatter in parameter `a' with rms noise is too large to allow for
a general differential completeness formula applicable to any survey 
field to be developed, based on these data.
 
Figures A1, A2 and A3 show the completeness analysis for the
uniform noise regions of the ``SCUBA 8\,mJy Survey'' fields
(ELAIS N2, Lockman Hole wide area and Lockman Hole deep strip). The
error bars are given by the Poisson error on the number of sources
planted into the field in each noise region, and at each flux density. These
fields have the largest shallow border regions of all the
survey fields discussed here, due to the survey strategy
adopted to even out the noise (see Section 5.1). The corresponding completeness
plots for the non-uniform regions of ELAIS N2 and the Lockman Hole
wide area are shown in Figs. A4 and A5 respectively. It is
immediately obvious in comparing plots of uniform and non-uniform
noise that source recovery in the non-uniform edge regions is very
much worse than in the fully-observed central areas, reaching at best
$\sim 10\%$ at $\rm 15\,mJy$ as opposed to the $\sim90\%$ in the
uniform noise regions. The simulations
carried out on the remaining smaller fields did not yield sufficiently
good statistics in the non-uniform noise regions to allow any
meaningful fit to be made, hence only plots for the uniform noise
regions of the remaining fields have been presented (Figs A6 to A16).

\subsubsection{Output versus Input Flux Density}

\begin{figure}
 \centering
   \vspace*{3.8cm}
   \leavevmode
   \includegraphics{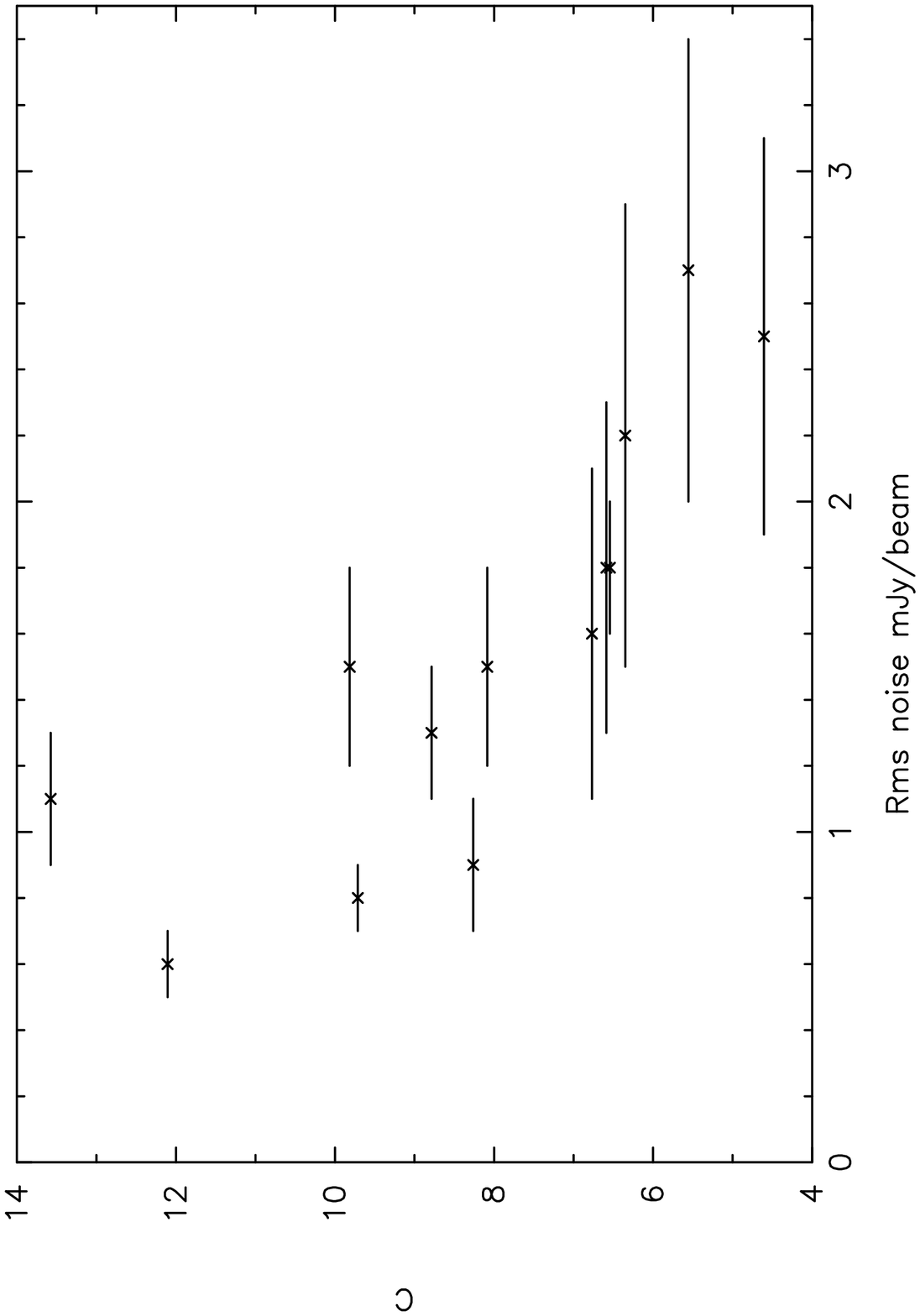}
\label{fig:param_C} 
\caption{\small Best-fit values for the parameter ``C'' as given in
   Table 3, plotted against the $\rm 1\sigma$ rms noise levels as
   determined from the uniform regions of the 14.5'' Gaussian
   convolved noise images (values given in Table 1). The horizontal 
   error bars show the standard deviation of the noise about the mean level.}
 \centering
   \vspace*{3.8cm}
   \leavevmode
   \includegraphics{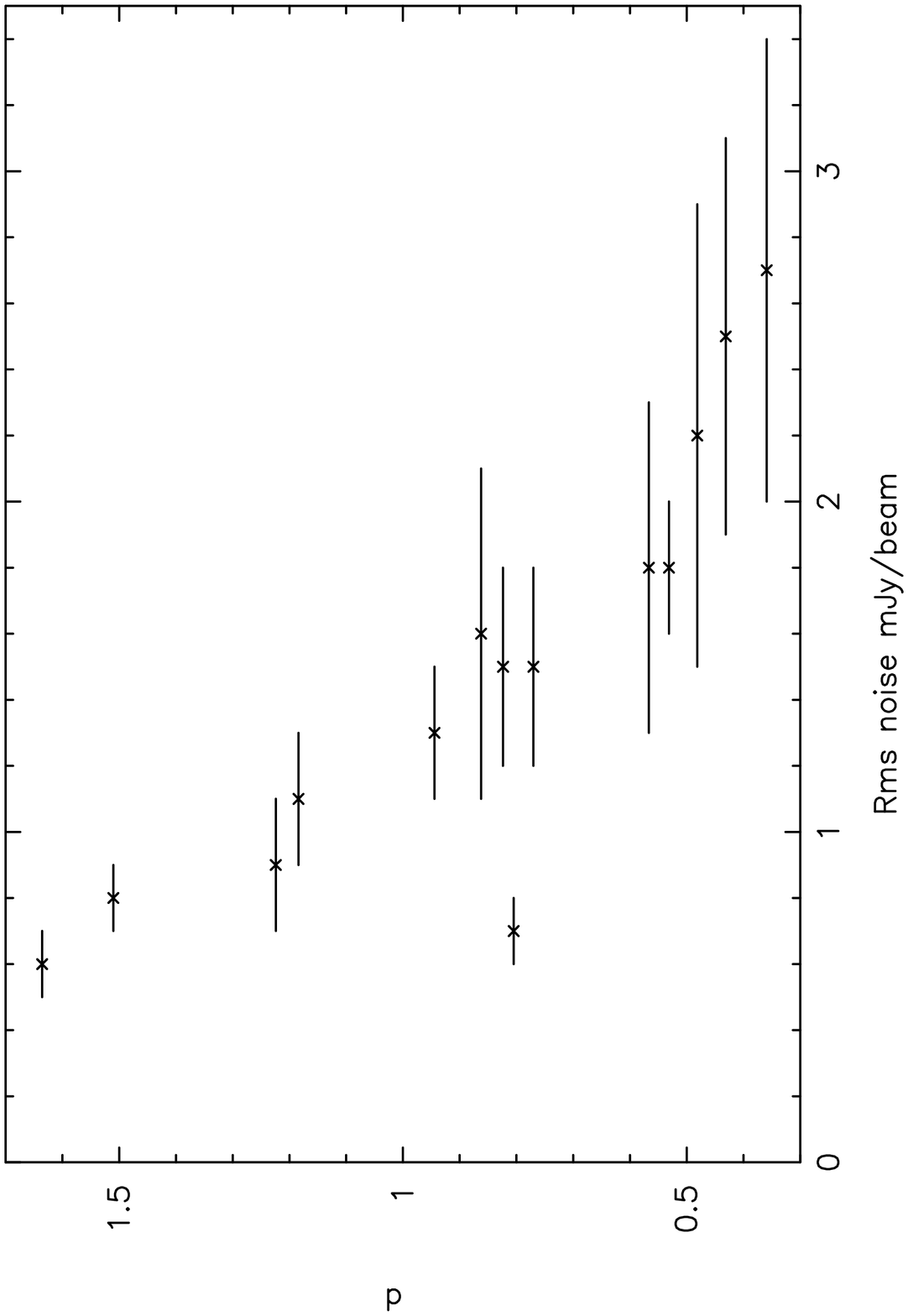}
\label{fig:param_d} 
\caption{\small Best-fit values for the parameter ``d'' as given in
   Table 3, plotted against the $\rm 1\sigma$ rms noise levels as
   determined from the uniform regions of the 14.5'' Gaussian
   convolved noise images (values given in Table 1). The horizontal 
   error bars show the standard deviation of the noise about the mean level.}
 \centering
   \vspace*{3.8cm}
   \leavevmode
   \includegraphics{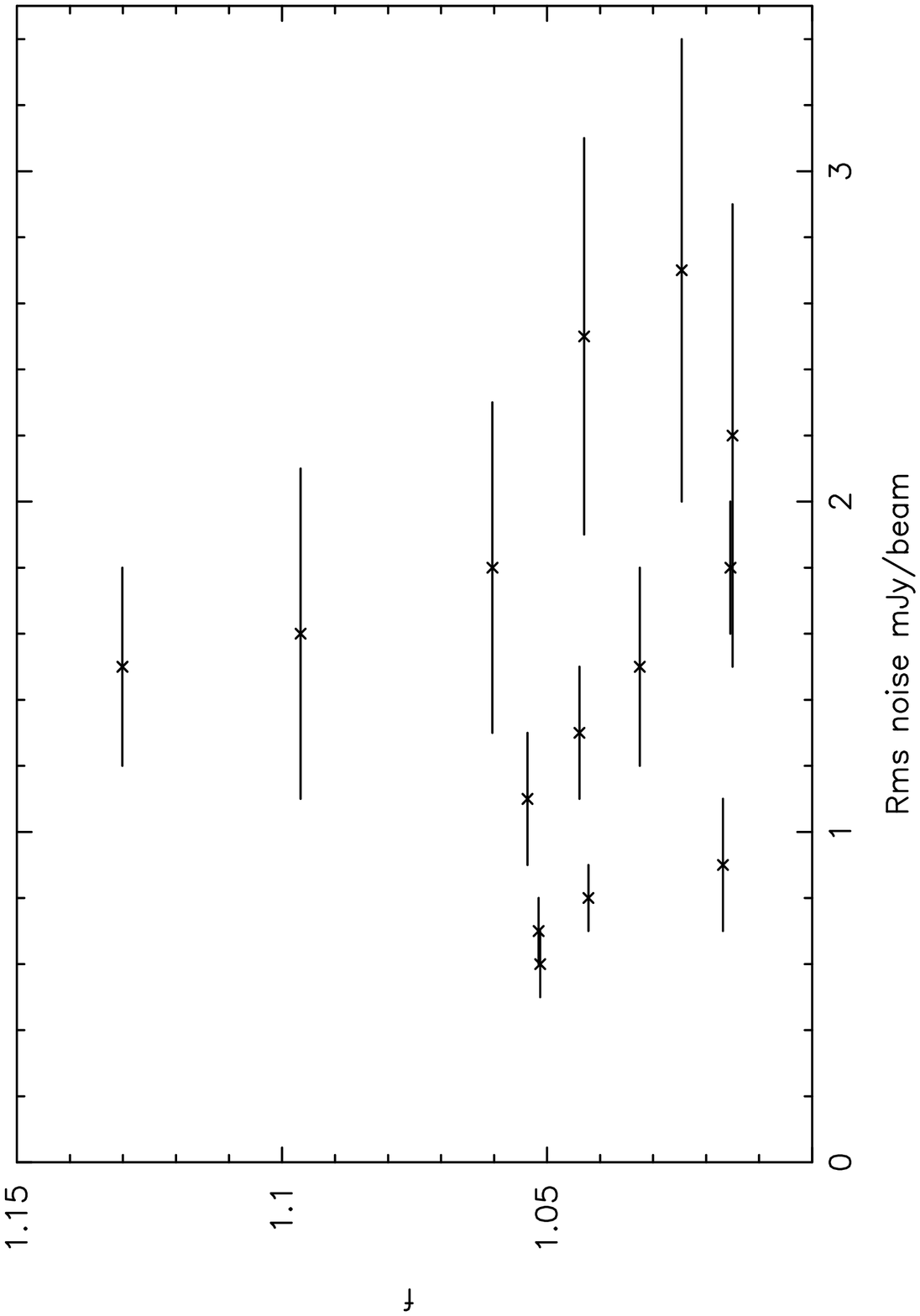}
\label{fig:param_f} 
\caption{\small Best-fit values for the parameter ``f'' as given in
   Table 3, plotted against the $\rm 1\sigma$ rms noise levels as
   determined from the uniform regions of the 14.5'' Gaussian
   convolved noise images (values given in Table 1). The horizontal 
   error bars show the standard deviation of the noise about the mean level.}
\end{figure}

Using these simulations, it is also possible to determine the
dependence of the mean output-to-input flux density ratio as a
function of the
input flux density, for those sources identified with signal-to-noise
ratio $>3.50$. This relation was found to be well described by the expression
\be \rm \frac{output \phantom{0} flux \phantom{0} density}{input
\phantom{0} flux \phantom{0} density} = C e^{-dx} + f \ee
where x is the input flux density, and the values of C, d and f were
determined by a minimised $\rm \chi^{2}$ fit to the simulation results 
for each $\rm 850\, \mu m$ survey field and are given in Table 3. The
plots of the ratio of output-to-input flux density against input flux
density for the individual survey fields may be found in Appendix A2. 

The plots of mean output/input flux density ratio against input flux
density are shown in Figs. A17 to A32. The error bars are the
standard error on the mean.
One of the first things to notice about the subsequent ratio plots, 
is that the effect of noise and confusion is
to produce systematic `flux-boosting', the mean
retrieved flux density always being greater than the input value. This
effect is known as Eddington bias (Eddington 1913) and is apparent in any flux limited
survey where a specific signal-to-noise threshold is employed. The presence of noise
and confusion from the faint background population will vary the flux
densities with which a source of specified input flux density is
retrieved. If, for example, one considers a very simple case of pure
Gaussian noise on a fake source, the measured flux
densities would be expected to have a symmetric distribution about the actual source flux
density, the exact characteristics of the distribution dependent on
the level of noise applied. However, if a fixed
signal-to-noise ratio is applied to the source extraction procedure, one will
preferentially select those sources which have been retrieved with a brighter
flux density, as some of the fainter measured values will fail to
make the signal-to-noise cutoff. Consequently, the mean retrieved flux
density will always be larger than the input flux density. Applying
the same noise characteristics to input sources of increasing
brightness, the mean boosting ratio is reduced,
because only the larger negative fluctuations on the tail of the
Gaussian noise distribution will allow the brighter sources to fall below
the signal-to-noise threshold. For very bright sources the
output-to-input flux density ratio approaches unity. Non-Gaussian noise
and confusion will of course affect the distribution of the retrieved
flux densities - in particular confusion of faint background sources
may lead to a more asymmetric distribution, especially if the SCUBA
population is found to strongly cluster. Simulations such as
these, however, allow for an empirical numerical description on a
field by field basis. 

Trends in the properties of
parameters `C', `d' and `f' with rms noise are shown in Figs. 3, 4 and 5
respectively. Both parameters `C' and `d' decrease with increasing
rms noise. The decline is steep at low rms noise levels, but becomes more
shallow above $\rm 1\sigma_{rms}\simeq 1.5-2\,mJy$. Parameter `d'
shows a fairly tight correlation, however `C' shows too great a level
of scatter to allow a general formula, based solely on rms noise, to
be developed for the output-to-input flux density ratio. The parameter `f'
represents the ratio of output-to-input flux density for very bright
sources with a constant value of $\sim 1$ expected for all fields,
regardless of noise level (the median value is in fact 1.04).

\begin{table*}

\begin{tabular}{|l|c|c|c|c|} \hline
Survey Field & Noise Region & g & h & $\rm \chi^{2}$ \\
 & & arcsec/mJy & arcsec &  \\ \hline
ELAIS N2 & uniform & 0.17364      & 4.6693 &   1.45445 \\
Lockman Hole wide area & uniform & 0.12235  & 4.5791 & 0.90109 \\
Lockman Hole deep strip & uniform & 0.16592 & 4.2357 & 1.65490 \\
03h wide area & uniform & 0.13772   &   4.2672  &  1.05487 \\
03h deep area & uniform & 0.46026      & 5.2313  &  2.22813 \\
10h           & uniform & 0.36036    &  4.8814 &   2.04834 \\
14h & uniform & 0.20604  &    4.2997 &   0.52830 \\
22h & uniform & 0.33656    &   5.6634 &   0.76516 \\
Hubble Deep Field & uniform & 0.42609      & 5.0274  &  1.11186 \\
SSA13 wide area & uniform & 0.11505    &  4.2934 &   0.76635 \\
SSA13 deep area & uniform & 0.43070	& 5.1689 &  0.36683 \\
SSA17 & uniform & 0.37809 &  5.1600  &  0.65348  \\
SSA22 & uniform & 0.29406 &  4.3191  &  1.79829 \\
Lockman Hole deep area & uniform & 0.33253    &   4.2634 & 0.54652\\ \hline
ELAIS N2 & non-uni & 0.51966   &   9.4885 &  4.09028 \\ 
Lockman Hole wide area & non-uni & 0.00000	& 4.1870 &  1.88165
\\ \hline

\end{tabular}
\label{table:poserr}\caption{\small Best fit values determined
for g and h in equation 11, describing the mean positional error
against input flux density for each of the survey fields and noise regions.}
\end{table*}

It can also be readily seen from comparing Figs. A17 and A18, with
A20 and A21, that the level of flux-boosting is much
greater in the non-uniform noise regions and with a much larger degree
of scatter in the data points. For example, a source input to the
ELAIS N2 or Lockman Hole fields (``from the SCUBA 8-mJy Survey'') would appear boosted on average by a factor
of $1.2-1.3$ if extracted from the uniform noise regions. In the
non-uniform regions though, the mean level of boosting is by a factor
2. Due to the combination of a poor level
of retrieval and large flux boosting factors, any sources 
recovered in the non-uniform noise regions (marked as ``edge'' in
Section 5) have been excluded from statistical analyses such
as source counts, and clustering measures etc.

\subsubsection{Positional Uncertainty}
The mean positional uncertainty in retrieving the fake sources was
found to be well approximated by a linear dependence on the input flux
density such that
  
\be \rm positional \phantom{0} error = -gx + h \ee
where x is the input flux density, the values of g and h for each 
$\rm 850\, \mu m$ survey field were determined by a minimised $\rm
\chi^{2}$ fit to the simulation results (given in Table 4), and the
positional uncertainty is given in arcseconds. The plots of the mean
positional uncertainty against input flux density for each of the
individual fields are given in Appendix A3.

Figs. 6 and 7
show the dependence of parameters `g' and `h' on rms noise. One might
expect a general formula for positional error to depend on the ratio
of input flux density to rms noise such that $\rm g_{general} \propto
1\sigma_{rms}^{-1}$ in this straight line desription. The data points are consistent with $\rm g_{general}\times
1\sigma_{rms} \sim 0.35$\,arcsec, but the large scatter means that a simple
straight line with negative gradient provides a similarly good
description. The values of parameter `h' have a median of 
$\sim 4.65$ between
all the survey fields, and there is no obvious trend with rms noise.

\begin{figure}
 \centering
   \vspace*{3.8cm}
   \leavevmode
   \includegraphics{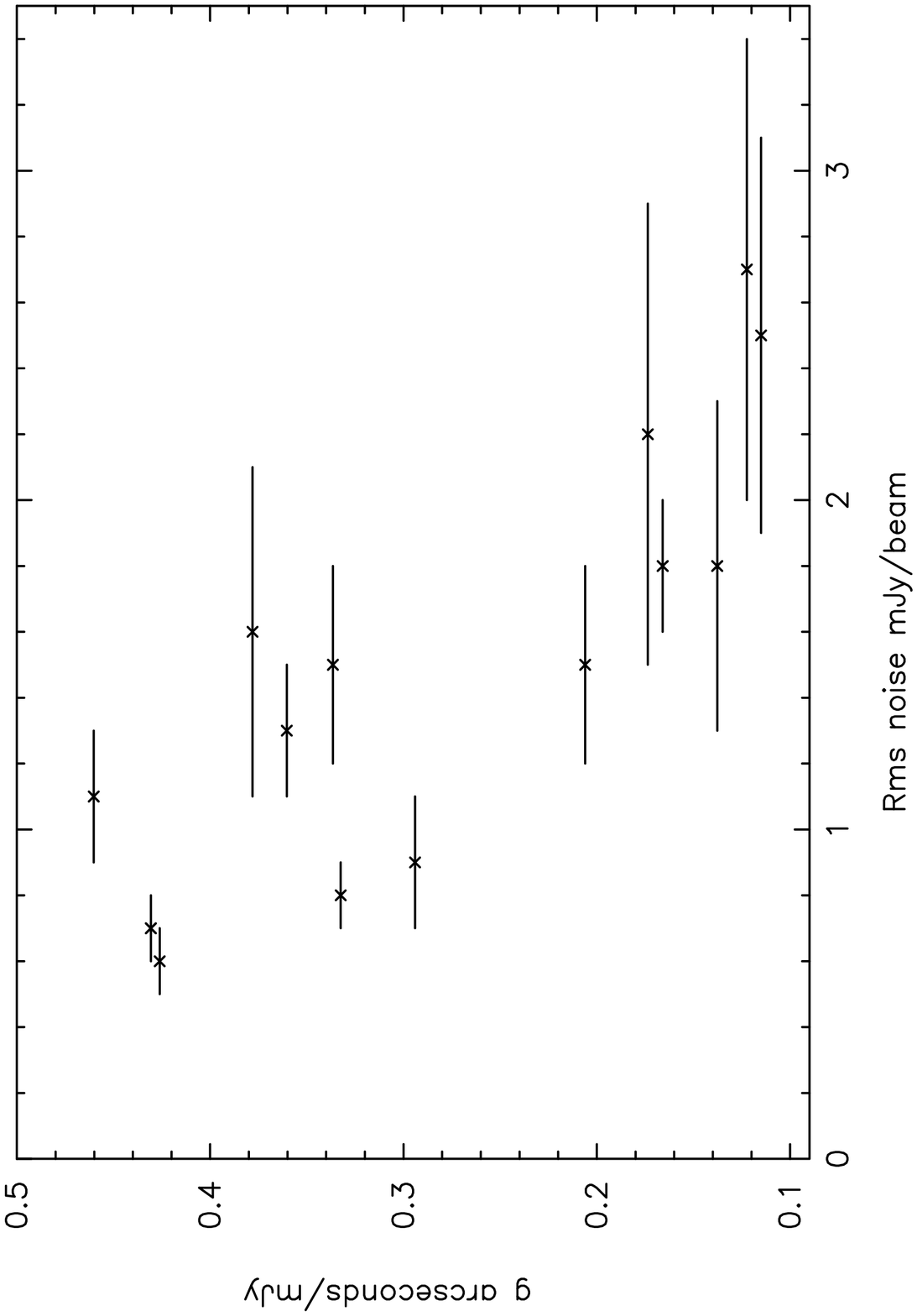}
\label{fig:param_g} 
\caption{\small Best-fit values for the parameter ``g'' as given in
   Table 4, plotted against the $\rm 1\sigma$ rms noise levels as
   determined from the uniform regions of the 14.5'' Gaussian
   convolved noise images (values given in Table 1). The horizontal 
   error bars show the standard deviation of the noise about the mean level.}
 \centering
   \vspace*{3.8cm}
   \leavevmode
   \includegraphics{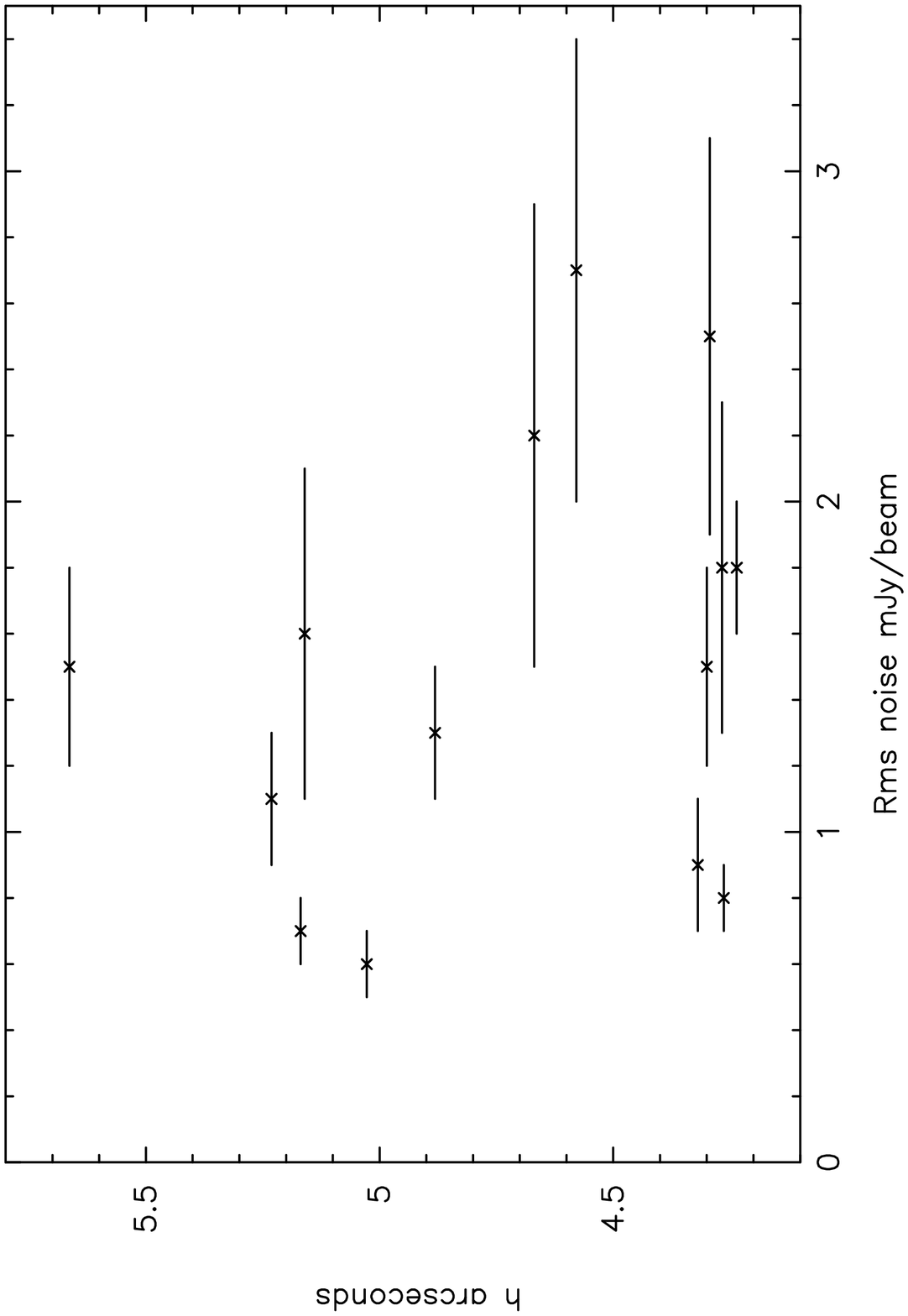}
\label{fig:param_h} 
\caption{\small Best-fit values for the parameter ``h'' as given in
   Table 4, plotted against the $\rm 1\sigma$ rms noise levels as
   determined from the uniform regions of the 14.5'' Gaussian
   convolved noise images (values given in Table 1). The horizontal 
   error bars show the standard deviation of the noise about the mean level.}
\end{figure}

Figures A33 through to A48 show the mean positional error of the
retrieved fake sources against input flux density. The error bars are the
standard error on the mean.
Again, one can see that the data from the ``SCUBA 8\,mJy Survey'' fields
(Figs. A33 to A37) show a greater scatter in the non-uniform noise regions than the
central uniform noise parts, with a generally greater positional
uncertainty in these border areas $4-5$\,arcsec, as compared with
$2-4$\,arcsec in the uniform noise regions). The positional accuracy improves with
higher flux density sources (and hence better signal-to-noise). Such
estimates of positional error do not include any pointing errors
arising whilst the data are being taken at the telescope. 
The pointing of the JCMT is known to be very accurate for such a large
dish; typical pointing errors are less than 3 arcsec, much
less than the $\rm 850 \, \mu m$ beam size of 14.5 arcsec FWHM. Several
pointing problems have been discovered during the period in which
these deep submillimetre surveys were undertaken. However, the fact
that each map is built up from many shorter integration datasets
limits the impact which these pointing inaccuracies can have on the
final image. Overall, the uniform noise regions of all the survey
areas suggest likely positional errors of $2-4$\,arcsec in
source retrieval, arising from the effects of noise and confusion. 

Note that the positional uncertainty is known to be well-described by
a Gaussian distribution in this data reduction methodology
(e.g. Ivison et al. 2002, Serjeant et al. 2003, Coppin et al. 2006).

\subsection{Completely simulated maps}

In order to obtain constraints on the fraction of recovered $\rm 850\,
\mu m$ sources arising from confusion and noise, 100 simulated images of
each of the survey fields were generated. In addition, these
simulations were also used to estimate the integrated
completeness and count correction factors. The assumed source counts
were taken from the best fit of a simple power-law
model (in the format first employed by Barger et al., 1999) to the
differential counts given later in this paper, which were corrected for
completeness and flux-boosting at the $\rm > 3.50 \sigma$ level
using the simulation results of Section 4.1. Specifically,
the differential counts are given by:
\be \rm \frac{dN(S)}{dS} = \frac{N_{0}}{(a+ S^{\alpha})} \ee
\noindent where $\rm N_{0} = 2.67 \times 10^{4}$, $\rm a=0.49$ and
$\rm \alpha=3.14$, which predicts a total $\rm 850 \, \mu m$
background of $\rm 3.8 \times 10^{4}\,mJy\,deg^{-2}$, consistent with
the value of $\rm 4.4 \times 10^{4}\,mJy\,deg^{-2}$ measured by
Fixsen et al. (1998). A realistic model of the background
counts was produced , by randomising the number of sources 
placed into each simulated field at
0.1~mJy intervals, from $0.1 - 14.0$\,mJy, according
to a normal distribution about the number expected. 
Each source was then allocated a random position and the
whole image was convolved with the beam. The simulated field is
initially created to be larger than the actual field,
allowing for the negative sidelobes of sources centred off-field in
the final image to appear in it. The clustering properties of
the SCUBA population are, at present, not well characterised. Results
presented in Section 7 suggest that at least the very brightest SCUBA
sources ($\rm > 5\, mJy$) are strongly clustered on arcminute scales,
consistent with the idea that these objects are progenitors of
present-day 
massive ellipticals, but there is insufficient blank field survey
data available to allow a realistic clustering component to be added
into the selection of positions within the simulations. This means
that these simulations which make the
assumption of a random distribution (ie. no clustering) can only be
used as a first approximation in determining the level of
spurious / confused source contamination.

Noise overlays were constructed by subtracting the full minimised $\rm
\chi^{2}$ fit model (i.e. the model representation of the full sky region
comprised of the best-fit beam profiles to all of the 
peaks in the
image) from the zero-footprint signal maps of the actual
survey data. 

The subsequent simulation analyses were conducted at
various signal-to-noise levels down to very low significance levels
($\rm >1.50\sigma$), even though the source catalogues presented in
Section 5 only reach $\rm >3.00\sigma$. This was to allow an
assessment of whether sources recovered at $\rm > 3\sigma$ by
other submillimetre groups using different reduction and extraction
methods, but which were recovered at lower signal-to-noise in the
analysis presented in Section 5, were likely to be real.

At first glance one might think that this process would
remove a significant amount of real Gaussian noise as well as faint sources
from the signal image. However, one is not simply
fitting a Gaussian to positive flux density peaks (which would indeed
remove a significant amount of the real noise), but
instead fitting the \emph{full beam profile} which has the two 
negative sidelobes, each half the depth of the peak, to the
signal image. Consequently, if a peak arose due to Gaussian noise
rather than a real source, one would not find the accompanying
sidelobes at the relevant position and of the right depth and so the
best-fit normalisation of the beam profile at this position would be
very close to zero, thus resulting in very little of the real
noise being removed.

The source-extraction algorithm was then re-run on these
residual images to determine the number of sources which could be
recovered from the noise overlays
alone, at signal-to-noise thresholds of $\rm 1.50 - 4.00 \sigma$,
spaced regularly at $\rm 0.5 \sigma$ levels. Gaussian statistics predict that given the number of beams in
the 464 sq. arcmin of uniform noise, there are likely to be 
$\sim14$ 

noise peaks recovered at $\rm >3.00 \sigma$ and 
$\sim 0.32$ 
noise peaks 
recovered at $\rm >4.00 \sigma$ (in fact slightly less than this, as
this calculation does not account for the recovery of the negative
sidelobes). The actual numbers recovered are 24 and 1 at $\rm >3.00
\sigma$ and $\rm > 4.00\sigma$ respectively, comparing reasonably well
with the Gaussian estimates. Additional sources of high-significance
($\rm > 3.00 \sigma$) peaks in the residuals might be:\\ \\
\noindent 1) A non-Gaussian component in the noise eg. microphonics.\\
\noindent 2) A poor fit of the model to the data in a small sub-region of
the full dataset.\\
\noindent 3) Incomplete source removal, for example a faint source
confused with a bright source such that only the brighter of the two
sources could be identified by the presence of a peak.\\

Future improvements to the source extraction algorithm
will address points (2) and (3). Currently, however, a poor model-fit
to sub-sections of the original map or the presence of
any remaining real sources in the residual images will lead to an
over-estimate of the level of spurious/confused source contamination,
and so the results presented in subsequent tables may be considered
an upper limit. 

The final signal images were constructed by adding the unsmoothed
noise-overlay to the simulated background counts, and trimming to the
correct size and shape. The original zero-footprint noise maps were
used as noise maps for the simulated images, and any
``hot'' pixels identified with signal-to-noise ratio
$>4.00$ were re-assigned large noise levels, as was done with the real
data. The source extraction algorithm was then applied to each
simulated image in an identical way to the actual survey maps. 

Simulated images of the 03h (CUDSS), SSA13 (Hawaii Survey) and Lockman Hole
wide area (8\,mJy Survey) fields were created with the small deep regions
combined into the wider area surveys, however, in the same way as the
adding of one source into the real data and attempting to retrieve it
(Section 4.1) the results for the deep and shallower areas were
treated separately. The regions of uniform and non-uniform noise were
also treated individually, as before.

These simulations differ slightly to those presented in Scott et
al. (2002), and this is reflected in the results presented for the 
``SCUBA 8\,mJy Survey'' fields (ELAIS N2 and the Lockman Hole East wide
area field). Combining the data from the various surveys improves the
constraints on the $\rm 850\, \mu m$ source counts (as discussed in
Section 6), and a steeper source counts model, fit to the combined 
counts, has been used to create
these mock images. This leads to higher densities of fainter sources,
and hence increases the fraction of significant detections arising
from confusion. The second difference is that a lower flux density cut-off
of 14\,mJy has been applied, corresponding to the retrieved flux
density of the brightest source detected in the uniform regions
of any of the survey fields, as opposed to the 20\,mJy cut-off
employed in the earlier simulations. The presence of sources with
artificially high input flux densities increases the fraction of
objects retrieved with high signal-to-noise ratios, which in turn
overestimates the integral completeness at a given signal-to-noise
threshold, particularly in the
non-uniform areas where the noise levels are higher. The third difference is
the noise overlay added on to the background sources. In previous
simulations this was created by rebinning the individual datasets with
randomised bolometer astrometry so as to smear out any sources
present. This approach was found to have problems in regions with
several significant bright sources, which would become smeared together on
scrambling, creating a patch of
excessive noise. For this reason, and in order to preserve the noise
properties of the real data as far as possible, the residual signal
maps were adopted as the overlaid noise. These residual maps are
the difference between the pixel values of the actual unconvolved
signal maps, and the best-fit model of the full sky region as constructed
from a series of idealised beam-profiles centred on every peak in the
convolved signal image. Hence, the residual maps represent the excess
noise levels superimposed on top of the real data. The final difference is in
the flux densities of peaks in the convolved maps, considered as
potential sources. In Scott et al. (2002), peaks identified at
$>3$\,mJy were considered as possible sources and included in the
source extraction matrix, whereas in these simulations all positive
peaks were included in the maximum likelihood fit.

The extracted sources were each identified with the brightest input
source, located within 8 arcsec of the retrieved peak
position. Regions of uniform or non-uniform noise were assigned
according to whether the input position lay within the uniform noise
cutouts. This raises the possibility of a source located very close to
the uniform/non-uniform boundary being input and assigned one noise area, but
extracted a few arcseconds away under a different noise classification.
In these circumstances, both locations were allocated the input
position classification so as not to underestimate the
completeness. 

The tabulated
results also reflect a broad range of flux density thresholds, some of
which are not of particular interest to every field, for example one would not
expect to recover a 2\,mJy source in a field where the rms noise levels
are $\rm >2\,mJy$. These values were included to allow trends with flux
density in the various quantified properties to be identified.
The tabulated results of these simulations may be found in Appendix B.

The results of the integral completeness analyses are given in Tables
B1 to B16, for the uniform noise regions of each individual field,
and the non-uniform noise regions of the ``8\,mJy Survey''
fields. Flux density thresholds of $2-10$\,mJy at 2\,mJy intervals,
and signal-to-noise thresholds of $\rm 1.50-4.00\sigma$ at $\rm
0.50\sigma$ intervals were considered. The quoted errors are the $\rm 1 \sigma$
Poisson error on the number of sources input to the fields. 
Comparing the values given in
Tables B1 and B3 with B2 and B4, there is a marked contrast in
the fraction of sources successfully retrieved in the uniform and
non-uniform noise regions. For example, at $\rm S_{850}>8$\,mJy and a
significance of $\rm 3.50 \sigma$ the uniform noise regions of the
Lockman Hole East and ELAIS N2 are $70-75$\% complete, whereas the
non-uniform noise regions are only $10-15$\% complete. As the
flux density threshold drops to reach the faint limit at which significant
($\rm >3.00 \sigma$) sources can still be detected in the uniform
noise regions of the respective
surveys (corresponding to an integral completeness of $65-75\%$), there
is a drop of $5-10\%$ in the fraction of sources recovered for every
increase of $0.50 \sigma$ in the signal-to-noise threshold. The 
estimated completeness
percentages in the small deep surveys (such as the 10h and SSA13 deep fields) should be considered as lower limits due to their
small area in relation to the beam size. The undersampling in the
jiggle pattern affects the data taken up to a beam-width into the
field. These elevated noise levels in turn can mask the identification
of a potential source located up to a further half a beam-width into
the map, despite that region being fully sampled and hence uniform
noise. A greater proportion of the field area in a small SCUBA map is
affected by this problem than in a wider-area field and may
artificially increase the number of bright sources which fail to be
recovered. The error bars on the integral completeness measurements at
bright flux densities (8 or 10\,mJy) are also generally large ($\pm
20\%$) in the smaller fields due to a low source density and hence
fewer bright sources being entered into the pencil-beam maps. 

Tables B17 - B32 give the percentage count correction factors, above a
specific flux density level, and at a given 
signal-to-noise ratio threshold,
for the uniform noise regions of each individual field,
and the non-uniform noise regions of the ``8\,mJy Survey''
fields. Count correction factors at 2\,mJy intervals from $\rm
S_{850}>2$\,mJy to $\rm S_{850}>10$\,mJy,
and significances of better than $\rm 1.50-4.00\sigma$ at $\rm
0.50\sigma$ intervals were considered. The first point of note is that
the percentage count corrections in the non-uniform noise regions of
the ``SCUBA 8\,mJy Survey'' fields (Tables B18 and B20) are
in most cases greater than $100\%$ for the $3.50$ and $4.00$ significance
levels at all flux densities, reflecting the poor levels of
completeness in these regions. The converse, however, is true in the
regions of uniform noise, implying that the effects of flux-boosting
(discussed in Subsection 4.1.2) have a stronger effect on the
source counts than incompleteness. In the Lockman Hole East and ELAIS N2
fields, the correction factors applied to the bright counts (8 or
10\,mJy) is $30-40\%$ at $\rm S/N>3.00$, $50-60\%$ at $\rm S/N>3.50$,
and $70-90\%$ at $\rm S/N>4.00$, the trend in signal-to-noise ratio
indicative of an increase in the contamination of spurious/confused
sources in the raw catalogues as the significance threshold is
lowered. The count corrections become less severe with decreasing
noise levels. The intermediate depth and sized areas such as the 03h
wide area field (``CUDSS''), Lockman Hole deep strip (``8\,mJy Survey''), and
SSA17 and SSA22 (``Hawaii Survey'') require a $70-80\%$ correction to
the raw counts at $\rm S/N>3.50$, and $80-90\%$ at $\rm S/N>4.00$. The
deep single-pointing SCUBA maps such as the 10h and 22h fields
(``CUDSS'') and the Lockman Hole deep field and SSA13 deep field
(``Hawaii Survey'') require no count-correction in the $8-10$\,mJy 
range, and a $70-90\%$ correction at
$\rm S_{850}>4$\,mJy to $\rm S_{850}>6$\,mJy,
for applied signal-to-noise thresholds of 3.00 or higher. At 2 or
3\,mJy, the very faintest source flux density levels accessible in
a blank field survey due to the confusion limit being reached, the
count correction factor again exceeds $100\%$. This is because the
extraction of sources becomes less complete as confusion worsens, and
begins to offset and even exceed the effect of flux-boosting which
dominated at the brighter end of the counts. These simulations show
that the raw source counts will be an overestimate of the true source
counts
at the faint end, 
and this will be readdressed in Section 6 where the $\rm 850\,
\mu m$ source counts are considered in greater detail.

The final property of the survey data investigated by these
simulations is the relationship of the output-to-input flux densities
of the sources, with signal-to-noise ratio. The results for the
uniform noise regions of each individual field,
and the non-uniform noise regions of the ``8\,mJy Survey''
fields are given in Tables B33 to B48, for significances in the
range $\rm >1.50\sigma$ to $\rm >4.00\sigma$ at $\rm 0.50\sigma$
intervals. The first quantity to be considered was the fraction of
``sources'' recovered above a specific signal-to-noise threshold which
could be attributed to noise only, by running the source extraction
algorithm purely on the residual signal maps with no background
counts added in (Column 2 in the tables). Each source recovered from
the simulated images was then classified according to the relation
between the output and identified input flux density. The classes
were:\\

\noindent 1) Fainter. The retrieved flux density was fainter than the input
source with which it had been identified ($\rm S_{in}>S_{out}$).\\
\noindent 2) Within error bars. The input flux density lay within the $\rm 1
\sigma_{rms}$ error bars of the retrieved value ($\rm
S_{out}-err_{out} < S_{in} < S_{out}+err_{out}$).\\
\noindent 3) Boosted. The input flux density was less than the lower error
boundary on the output value, but was still within a factor of 2 of
the measured flux density ($\rm S_{out}/2 < S_{in} <
S_{out}-err_{out}$).\\
\noindent 4) Spurious / confused. The fitted flux density to the peak
could not be identified with a source in the input catalogue, located
within 8 arcsec and a factor 2 in brightness ($\rm S_{in} < S_{out}/2$).\\
The percentage of sources classified as (1), (2), (3) and (4) are
given in Columns 3, 4, 5 and 6 respectively. As a point of note, the
peaks identified in the residual signal image do not just contribute
to the confused / spurious fraction, but may affect any of the
classifications of source to some extent. In all cases, the fraction
of sources which are recovered at a fainter flux density than they
were input is $<10\%$. In the uniform noise regions, $\sim 65-70\%$ of
the sources recovered with $\rm S/N>4.00$ may be identified as boosted
or within the error bars. In the wider-area shallower surveys, this
number drops to $\sim 55-65\%$ for $\rm S/N >3.50$ and $\sim 40-55\%$
for $\rm S/N >3.00$. The decline with signal-to-noise ratio is less marked
in the fields with lower rms noise levels, and in fact remains
approximately constant at the $65-70\%$ level in the deep
single-pointing SCUBA fields. The number of sources categorised as
boosted is approximately the same as the number for which the
identified input object fell within the extracted $\rm 1 \sigma_{rms}$
error bars.
The confused / spurious
fraction of sources in the non-uniform regions of the ``8\,mJy
Survey'' fields are markedly higher than their uniform region
counterparts, even cutting at high signal-to-noise levels. 
The Lockman Hole East is the worst of the two, with
$\sim 90\%$ of the recovered ``sources'' unidentified with an input
source at least 1/2 as bright, even at $>4.00\sigma$. The ELAIS N2
field is not quite as severe, but still $\sim 40\%$ of the recovered
$\rm >4.00\sigma$ peaks fall into the spurious / confused category,
rising to $\sim 90\%$ at $\rm >3.00\sigma$. This affect arises due to
redundancy i.e. the number of times a region of sky has been
observed. The central regions of the `8\,mJy Survey' fields have a
higher redundancy and hence according to the ``Central Limit Theorem''
the noise in those areas which have received the full integration time
will be more Gaussian. The noise in the undersampled perimeter regions 
of the wide area maps which have not received the full integration
time, however, will be less Gaussian and lead to a lower source
detection reliability for a given signal-to-noise threshold in these
areas. This casts severe doubt
on the reality of any $>3.00\sigma$ objects identified in the high
noise regions near the edge of these maps. 

The simulations imply that up
to $20-30\%$ of the $\rm >4.00\sigma$ peaks may the result of
confusion or noise, increasing to $30-40\%$ at $\rm >3.50\sigma$.
and $\sim 30-60\%$ at $\rm >3.00\sigma$. The fields with the highest noise
levels (Lockman Hole wide area field, ELAIS N2, and the SSA13 wide
area field) show the highest levels of contamination and the
deepest surveys the least at both $\rm S/N >3.00$ and $\rm S/N>3.50$,
however, there is no obvious trend of spurious / confused fraction
with noise at $\rm >4.00\sigma$. We reiterate that without
particularly tight constraints on the number density of the faint
SCUBA population, and without knowing the clustering properties, these
quantities should be considered a rough guideline only.

The opposing effects of increasing the completeness of a catalogue by
dropping to lower signal-to-noise thresholds, while at the same time
also introducing a larger fraction of spurious / confused sources
suggests that setting a cut-off of $\rm >3.50\sigma$ is a good
compromise for selecting SCUBA sources to follow-up. Unless otherwise
stated, subsequent analyses in Sections 6 and 7 are based on the $\rm
S/N>3.50$ lists.

Simulations of a similar nature to these have been carried out by
Eales et al. 
(2000), who used their raw 14h field counts to produce 5
simulations of this field. They reported an integral completeness of
$\sim 90\%$ at the $\rm S_{850} > 3$\,mJy based on a $\rm 3 \sigma$
catalogue, which is rather higher than that implied from this analysis
(integral completeness $\sim 60\%$ at $\rm S_{850}>4$\,mJy based on
objects retrieved at $\rm >3.00\sigma$). This discrepancy may be
explained as a combination of 2 effects. The first is in the source
counts model used. Eales et al. (2000) used the raw $\rm 850 \, \mu m$
counts from the 14h field to create the simulated images. They also,
however, reported the flux-boosting effect when comparing their output
and input catalogues, quoting a median factor of 1.44, albeit with a
large scatter about this value. This means that the input source
counts will on average have overestimated the real input counts by a
factor of 1.44, and this may in turn have affected the
completeness estimate. The second, and probably dominant point, is that the two
source extraction mechanisms differ, leading to different source lists
and different significances for those sources common to
both. Appoximately 50\% of the sources reported as $\rm >3\sigma$ by
Eales et al., were recovered at $\rm < 3.00 \sigma$ using the
simultaneous maximum-likelihood fit described in Section 3. 
If one instead compares with the completeness values at $\rm
>2.00 \sigma$ or $\rm >2.50\sigma$ from these simulations, the completenss level is closer to
$70\%$, which is slightly more inkeeping with that of Eales et al. (2000).

Hughes \& Gazta\~naga (2000) have also performed more sophisticated 
Monte Carlo simulations
of SCUBA surveys, as part of a broader investigation into the
effects of confusion and noise on submillimetre surveys in general,
for both existing and planned instruments and telescope
facilities. They incorporated a clustering component into the
simulations by employing an N-body simulation with $200^{3}$ particles
in a $\rm 600 h^{-1}$\,Mpc box, produced with the same matter power spectrum as
that measured for the APM Survey galaxies (Gazta\~naga \& Baugh, 1998,
and references therein), which was then replicated to cover the total 
extent of the survey. The local IRAS $\rm 60 \,\mu m$ luminosity
function (Saunders et al. 1990) was interpolated to longer
wavelengths, and a model of pure luminosity evolution of the form
$(1+z)^{3}$ for $0<z<2.2$, $(1+2.2)^{3}$ for $2.2<z<6$ and an
exponential cutoff for $z>6$ was assumed to account for the star
formation histories. An Arp 220 SED was assumed throughout. The
simulations of Hughes \& Gazta\~naga (2000) find the same effects of
confusion and noise on the expected number counts as those simulations
presented here. Their mock images of comparable size and
depth to the ``SCUBA 8\,mJy Survey'' fields (Lockman Hole East wide
area and ELAIS N2) imply that a count-correction factor of $\sim 50\%$
for a $\rm S/N>3$ catalogue at $\rm S_{850}>8$\,mJy is required, which
is in keeping with the quantities presented in Tables B17 and
B19, again implying a large quantity of either boosted or spurious /
confused sources. This affects the brighter flux-limited surveys more,
due to the steep decline in galaxy counts and therefore increased
contribution of the measured counts due to noise.

\section{Comparison with existing source lists}
\subsection{The SCUBA 8\,mJy Survey}

\begin{figure}
 \centering
   \vspace*{10cm}
   \leavevmode
   \includegraphics{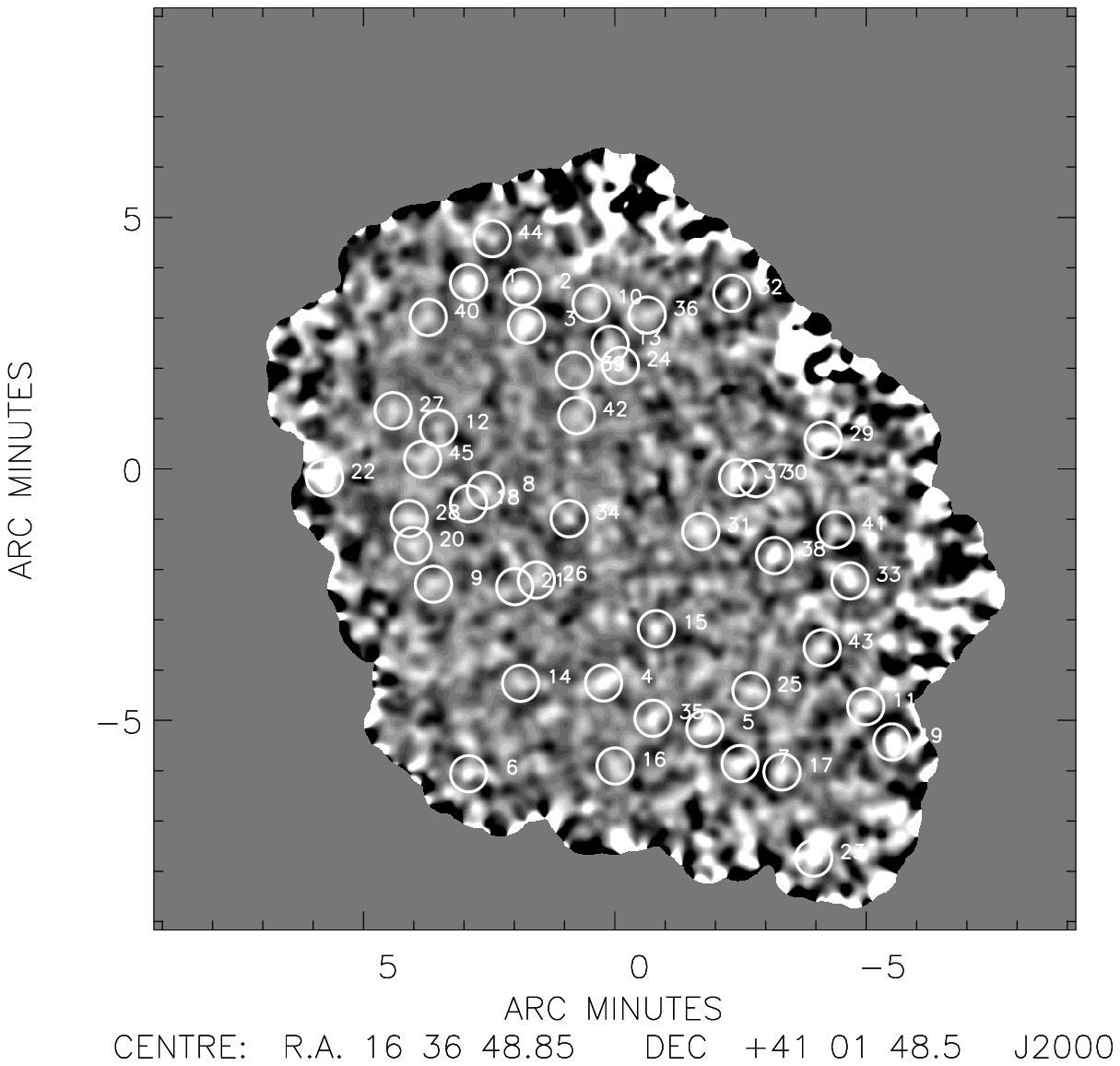}
\label{fig:n2_850} 
\caption{\small{The $\rm 850 \, \mu m$ image of the ELAIS N2 field,
smoothed with a beam-size Gaussian (14.5 arcsec FWHM). The
numbered circles highlight those sources found at a significance of
$>3.00$. The labelling corresponds to the numbers in Table 5.}}

   \vspace*{9.5cm}
   \leavevmode
   \includegraphics{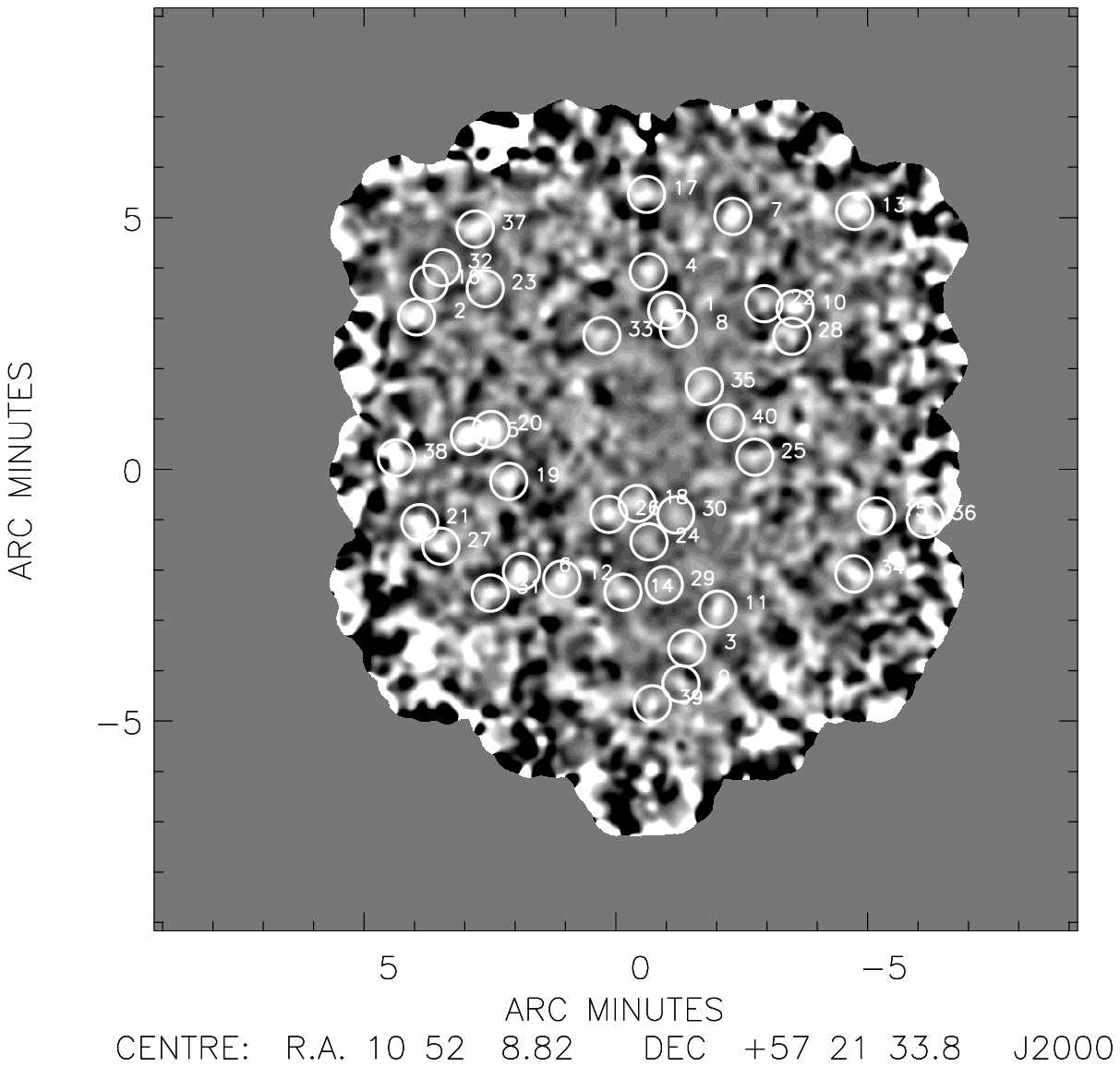}
\label{fig:lhwide_850} 
\caption{\small{The $\rm 850 \, \mu m$ image of the Lockman Hole East field,
smoothed with a beam-size Gaussian (14.5 arcsec FWHM). The
numbered circles highlight those sources found at a significance of
$>3.00$. The labelling corresponds to the numbers in Table 6.}}
 \end{figure}

The ``SCUBA 8\,mJy Survey'' (Lutz et al. 2001, Scott et al. 2002, 
Fox et al. 2002, Ivison et al. 2002, Almaini et al. 2003) is divided 
between two fields; the Lockman Hole East
(centred at RA 10:52:08.82, DEC +57:21:33.8) and
ELAIS N2 (centred at RA 16:36:48.85, DEC +41:01:48.5). These fields
both lie in regions of low galactic cirrus emission (Schlegel,
Finkbeiner \& Davis 1998), and have a vast
quantity of multi-wavelength data available for follow-up
studies. They were also selected to
coincide with deep Infrared Space Observatory (ISO)
surveys at 6.7, 15, 90 and 175~${\rm \mu m}$ (Lockman Hole East -- Elbaz et
al. 1999, Kawara et al. 1998; ELAIS N2 -- Oliver et al. 2000, Serjeant
et al. 2000, Efstathiou et al. 2000, Rowan-Robinson et al. 2004).
This section describes the final results of the ``SCUBA 8\,mJy Survey'',
currently the largest of the deep $\rm 850\,\mu m$ blank field surveys to be
completed. The Lockman Hole East map covers $\simeq 140$ sq. arcmin
of sky of which $\simeq 120$ sq. arcmin of the map have an rms noise
level of $\sigma_{850} \simeq 2.7$~mJy/beam and the remaining 20
sq. arcmin form a deeper strip in the centre of the image to an rms
noise level of $\sigma_{850} \simeq 1.8$~mJy/beam. The ELAIS N2 field
is also $\simeq 140$ sq. arcmin in size and reaches an rms depth of
$\sigma_{850} \simeq 2.2$~mJy/beam. 
\begin{table*}
\begin{tabular}{|l|c|c|r|c|c|l|c|c|} \hline
 & RA    & DEC   & $\rm S_{850}\phantom{000}$ & S/N & Noise & Previous &
       Prev. & Sep. \\
       &(J2000)&(J2000)&     /mJy\phantom{00}  &     & Region& Reference& S/N      & /arcsec \\ \hline
01     & 16:37:04.29 & +41:05:30.9 & $11.1 \pm 1.7\phantom{0}$ & 8.54 & central & S02 (N2.01) & $\phantom{0}8.59$ & $\phantom{0}0.9$ \\
02     & 16:36:58.62 & +41:05:24.9 & $11.0 \pm 1.9\phantom{0}$ & 6.92 & central & S02 (N2.02) & $\phantom{0}6.27$ & $\phantom{0}1.3$ \\
03     & 16:36:58.18 & +41:04:39.9 & $ 9.5 \pm 1.8\phantom{0}$ & 6.02 & central & S02 (N2.03) & $\phantom{0}5.86$ & $\phantom{0}2.1$ \\
04     & 16:36:50.04 & +40:57:33.0 & $ 8.3 \pm 1.8\phantom{0}$ & 5.20 & central & S02 (N2.04) & $\phantom{0}5.18$ & $\phantom{0}0.5$ \\
05     & 16:36:39.36 & +40:56:38.9 & $ 9.1 \pm 2.4\phantom{0}$ & 4.14 & central & S02 (N2.07) & $\phantom{0}4.07$ & $\phantom{0}1.0$ \\
06     & 16:37:04.25 & +40:55:44.9 & $ 9.2 \pm 2.4\phantom{0}$ & 4.12 & central & S02 (N2.06) & $\phantom{0}4.13$ & $\phantom{0}0.6$ \\
07     & 16:36:35.66 & +40:55:56.9 & $ 8.4 \pm 2.2\phantom{0}$ & 4.10 & central & S02 (N2.05) & $\phantom{0}4.16$ & $\phantom{0}1.3$ \\
08     & 16:37:02.50 & +41:01:22.9 & $ 6.1 \pm 1.7\phantom{0}$ & 4.00 & central & S02 (N2.12) & $\phantom{0}3.65$ & $\phantom{0}0.1$ \\ \hline
09     & 16:37:07.97 & +40:59:30.9 & $ 6.0 \pm 1.6\phantom{0}$ & 3.94 & central &                 &                   &                  \\
10     & 16:36:51.37 & +41:05:06.0 & $ 6.5 \pm 1.8\phantom{0}$ & 3.90 & central & S02 (N2.17) & $\phantom{0}3.50$ & $\phantom{0}0.3$ \\
11     & 16:36:22.41 & +40:57:04.8 & $ 9.2 \pm 2.6\phantom{0}$ & 3.84 & edge   & S02 (N2.09) & $\phantom{0}3.76$ & $\phantom{0}0.2$ \\
12     & 16:37:07.46 & +41:02:36.9 & $ 6.2 \pm 1.7\phantom{0}$ & 3.84 & central & S02 (N2.30) & $\phantom{0}3.13$ & $\phantom{0}0.5$ \\
13     & 16:36:49.34 & +41:04:17.0 & $ 7.7 \pm 2.2\phantom{0}$ & 3.73 & central & S02 (N2.20) & $\phantom{0}3.48$ & $\phantom{0}0.7$ \\
14     & 16:36:58.78 & +40:57:32.9 & $ 5.0 \pm 1.4\phantom{0}$ & 3.71 & central & S02 (N2.08) & $\phantom{0}3.82$ & $\phantom{0}0.2$ \\
15     & 16:36:44.48 & +40:58:38.0 & $ 7.3 \pm 2.1\phantom{0}$ & 3.66 & central & S02 (N2.11) & $\phantom{0}3.67$ & $\phantom{0}0.2$ \\
16     & 16:36:48.81 & +40:55:54.0 & $ 5.5 \pm 1.6\phantom{0}$ & 3.65 & central & S02 (N2.10) & $\phantom{0}3.69$ & $\phantom{0}0.1$ \\
17     & 16:36:31.25 & +40:55:46.9 & $ 6.4 \pm 1.9\phantom{0}$ & 3.64 & central & S02 (N2.13) & $\phantom{0}3.56$ & $\phantom{0}0.6$ \\
18     & 16:37:04.27 & +41:01:06.9 & $ 6.4 \pm 1.9\phantom{0}$ & 3.62 & central &                 &                   &                  \\
19     & 16:36:19.68 & +40:56:22.7 & $11.2 \pm 3.3\phantom{0}$ & 3.55 & edge    & S02 (N2.14) & $\phantom{0}3.55$ & $\phantom{0}0.4$ \\
20     & 16:37:10.10 & +41:00:16.8 & $ 5.1 \pm 1.5\phantom{0}$ & 3.54 & central & S02 (N2.15) & $\phantom{0}3.52$ & $\phantom{0}1.1$ \\
21     & 16:36:59.41 & +40:59:57.9 & $ 8.1 \pm 2.5\phantom{0}$ & 3.50 & central &                 &                   &                  \\ \hline
22     & 16:37:19.47 & +41:01:37.7 & $12.4 \pm 3.8\phantom{0}$ & 3.46 & edge    & S02 (N2.23) & $\phantom{0}3.45$ & $\phantom{0}0.5$ \\
23     & 16:36:27.90 & +40:54:03.9 & $13.2 \pm 4.1$\phantom{0} & 3.42 & edge    & S02 (N2.22) & $\phantom{0}3.46$ & $\phantom{0}0.1$ \\
24     & 16:36:48.27 & +41:03:52.0 & $ 7.0 \pm 2.2\phantom{0}$ & 3.25 & central & S02 (N2.27) & $\phantom{0}3.25$ & $\phantom{0}0.3$ \\
25     & 16:36:34.50 & +40:57:23.9 & $ 6.4 \pm 2.0\phantom{0}$ & 3.30 & central &                 &                   &                  \\
26     & 16:36:57.11 & +40:59:36.0 & $ 7.6 \pm 2.4\phantom{0}$ & 3.28 & central &                 &                   &                  \\
27     & 16:37:12.23 & +41:02:57.8 & $ 5.1 \pm 1.6\phantom{0}$ & 3.27 & central &                 &                   &                  \\
28     & 16:37:10.54 & +41:00:48.8 & $ 4.6 \pm 1.5\phantom{0}$ & 3.26 & central & S02 (N2.21) & $\phantom{0}3.47$ & $\phantom{0}0.5$ \\
29     & 16:36:26.89 & +41:02:22.8 & $10.9 \pm 3.5\phantom{0}$ & 3.24 & edge    & S02 (N2.24) & $\phantom{0}3.43$ & $\phantom{0}0.2$ \\
30     & 16:36:33.96 & +41:01:36.9 & $11.8 \pm 3.8\phantom{0}$ & 3.22 & edge    &                 &                   &                  \\
31     & 16:36:39.79 & +41:00:33.9 & $ 5.9 \pm 2.0\phantom{0}$ & 3.14 & central & S02 (N2.32) & $\phantom{0}3.07$ & $\phantom{0}0.2$ \\
32     & 16:36:36.51 & +41:05:17.9 & $11.8 \pm 4.0\phantom{0}$ & 3.13 & edge    &                 &                   &                  \\
33     & 16:36:24.07 & +40:59:34.8 & $10.4 \pm 3.5\phantom{0}$ & 3.12 & central & S02 (N2.29) & $\phantom{0}3.14$ & $\phantom{0}0.4$ \\
34     & 16:36:53.66 & +41:00:49.0 & $ 7.2 \pm 2.4\phantom{0}$ & 3.11 & central &                 &                   &                  \\
35     & 16:36:44.83 & +40:56:51.0 & $ 5.7 \pm 1.9\phantom{0}$ & 3.10 & central & S02 (N2.35) & $\phantom{0}3.02$ & $\phantom{0}0.3$ \\
36     & 16:36:45.44 & +41:04:52.0 & $ 5.6 \pm 1.9\phantom{0}$ & 3.08 & central &                 &                   &                  \\
37     & 16:36:35.90 & +41:01:37.9 & $ 9.3 \pm 3.2\phantom{0}$ & 3.06 & central & S02 (N2.19) & $\phantom{0}3.49$ & $\phantom{0}0.1$ \\
38     & 16:36:32.02 & +41:00:04.9 & $ 9.2 \pm 3.1\phantom{0}$ & 3.06 & central & S02 (N2.26) & $\phantom{0}3.26$ & $\phantom{0}0.2$ \\
39     & 16:36:53.14 & +41:03:46.0 & $ 5.7 \pm 2.0\phantom{0}$ & 3.05 & central &                 &                   &                  \\
40     & 16:37:08.62 & +41:04:48.9 & $ 4.3 \pm 1.5\phantom{0}$ & 3.05 & central &                 &                   &                  \\
41     & 16:36:25.57 & +41:00:35.8 & $ 7.8 \pm 2.7\phantom{0}$ & 3.04 & central &                 &                   &                  \\
42     & 16:36:52.87 & +41:02:52.0 & $ 3.9 \pm 1.3\phantom{0}$ & 3.03 & central & S02 (N2.36) & $\phantom{0}3.00$ & $\phantom{0}0.3$ \\
43     & 16:36:27.00 & +40:58:14.8 & $ 6.8 \pm 2.4\phantom{0}$ & 3.02 & central & S02 (N2.34) & $\phantom{0}3.05$ & $\phantom{0}0.2$ \\
44     & 16:37:01.81 & +41:06:22.9 & $ 6.1 \pm 2.1\phantom{0}$ & 3.01
       & edge    &                 &                   &
       \\ 
45     & 16:37:09.13 & +41:01:59.9 & $ 4.3 \pm 1.5\phantom{0}$ & 3.01 & central &                 &                   &                  \\ \hline \hline
       & \it{16:36:50.48} & \it{+40:58:54.0} & $\mathit{4.6 \pm 1.6\phantom{0}}$ & $\mathit{2.90}$ & \it{central} & S02 (N2.33) & $\phantom{0}3.06$ & $\phantom{0}0.2$ \\
       & \it{16:36:28.21} & \it{+41:01:41.9} & $\mathit{6.8 \pm 2.5\phantom{0}}$ & $\mathit{2.88}$ & \it{central} & S02 (N2.31) & $\phantom{0}3.07$ & $\phantom{0}1.5$ \\
       & \it{16:36:47.21} & \it{+41:08:48.0} & $\mathit{4.0 \pm 1.9\phantom{0}}$ & $\mathit{2.14}$ & \it{central} & S02 (N2.28) & $\phantom{0}3.24$ & $\phantom{0}0.1$ \\
       & \it{16:36:18.34} & \it{+40:59:11.7} & $\mathit{10.5 \pm 5.2\phantom{0}}$ & $\mathit{2.07}$ & \it{edge}  & S02 (N2.25) & $\phantom{0}3.37$ & $\phantom{0}0.5$ \\ 
       & \it{16:36:51.99} & \it{+41:05:54.0} & $\mathit{6.8 \pm 3.6\phantom{0}}$ & $\mathit{1.93}$ & \it{edge}   & S02 (N2.16) &  $\phantom{0}3.51$ & $\phantom{0}4.0$ \\
       & \it{16:36:11.36} & \it{+40:59:25.6} & $\mathit{13.3 \pm 11.8}$ & $\mathit{1.13}$ & \it{edge} &  S02 (N2.18) &  $\phantom{0}3.49$ & $\phantom{0}0.6$ \\ \hline
\end{tabular}
\label{table:n2}\caption{\small $\rm 850 \, \mu m$ source list for the
       ELAIS N2 field of the ``SCUBA 8\,mJy Survey''. Sources are marked in Fig
8.}
\end{table*}

\begin{table*}
\begin{tabular}{|l|c|c|r|c|c|l|c|c|} \hline
 & RA    & DEC   & $\rm S_{850}\phantom{000}$ & S/N & Noise & Previous & Prev. & Sep. \\
       &(J2000)&(J2000)&     /mJy\phantom{00}  &     & Region &
       Reference& S/N      & /arcsec \\ \hline        
01     & 10:52:01.33 & +57:24:43.3 & $9.6 \pm 1.6\phantom{0}$ & 7.68 & deep & S02 (LH.01) & $\phantom{0}8.10$ & $\phantom{0}0.6$ \\
02     & 10:52:38.21 & +57:24:35.1 & $11.0\pm 2.3\phantom{0}$ & 5.33 & central & S02 (LH.02) & $\phantom{0}5.22$ & $\phantom{0}0.9$ \\  
03     & 10:51:58.39 & +57:18:00.3 & $ 7.9\pm 1.7\phantom{0}$ & 5.31 & deep    & S02 (LH.03) & $\phantom{0}5.06$ & $\phantom{0}1.0$ \\
04     & 10:52:04.05 & +57:25:29.3 & $7.8 \pm 1.8\phantom{0}$ & 4.88 & deep    & S02 (LH.04) & $\phantom{0}5.03$ & $\phantom{0}1.4$ \\
05     & 10:52:30.39 & +57:22:13.2 & $10.8 \pm 2.6\phantom{0}$ & 4.58 & central & S02 (LH.06)& $\phantom{0}4.50$ & $\phantom{0}2.1$ \\
06     & 10:52:22.71 & +57:19:32.3 & $14.0 \pm 3.4\phantom{0}$ & 4.52 & central & S02 (LH.09)& $\phantom{0}4.20$ & $\phantom{0}0.3$ \\
07     & 10:51:51.54 & +57:26:35.2 & $8.0 \pm 2.0\phantom{0}$ & 4.45  & deep    & S02 (LH.07)& $\phantom{0}4.50$ & $\phantom{0}0.4$ \\
08     & 10:51:59.60 & +57:24:21.3 & $4.8 \pm 1.2\phantom{0}$ & 4.26  & deep    & S02 (LH.08)& $\phantom{0}4.38$ & $\phantom{0}3.3$ \\
09     & 10:51:59.26 & +57:17:18.3 & $8.0 \pm 2.1\phantom{0}$ & 4.12  & deep    & S02 (LH.05)& $\phantom{0}4.57$ & $\phantom{0}0.4$ \\
10     & 10:51:42.39 & +57:24:45.1 & $12.0 \pm 3.2\phantom{0}$ & 4.04 & central & S02 (LH.10)& $\phantom{0}4.18$ & $\phantom{0}0.1$ \\
11     & 10:51:53.82 & +57:18:47.3 & $6.7 \pm 1.8\phantom{0}$ & 4.01  & deep    & S02 (LH.27)& $\phantom{0}3.38$ & $\phantom{0}0.3$ \\ \hline
12     & 10:52:16.78 & +57:19:23.3 & $10.1 \pm 2.8\phantom{0}$ & 3.80 & central & S02 (LH.17)& $\phantom{0}3.55$ & $\phantom{0}0.3$ \\
13     & 10:51:33.57 & +57:26:41.0 & $9.5 \pm 2.7\phantom{0}$ & 3.73 &  central & S02 (LH.13)& $\phantom{0}3.69$ & $\phantom{0}0.2$ \\
14     & 10:52:07.77 & +57:19:07.3 & $5.6 \pm 1.6\phantom{0}$ & 3.71 &  deep    & S02 (LH.12)& $\phantom{0}4.01$ & $\phantom{0}0.6$ \\
15     & 10:51:30.46 & +57:20:37.9 & $12.0 \pm 3.5\phantom{0}$ & 3.70 & edge    & S02 (LH.11)& $\phantom{0}4.01$ & $\phantom{0}1.1$ \\ 
16     & 10:52:36.37 & +57:25:15.1 & $5.7 \pm 1.7\phantom{0}$ & 3.67 & central  &                &                   &               \\
17     & 10:52:04.30 & +57:27:01.3 & $9.2 \pm 2.7\phantom{0}$ & 3.64 & central  & S02 (LH.14)& $\phantom{0}3.61$ & $\phantom{0}2.3$ \\ 
18     & 10:52:05.67 & +57:20:53.3 & $4.6 \pm 1.4\phantom{0}$ & 3.52 & deep     & S02 (LH.22)& $\phantom{0}3.61$ & $\phantom{0}0.4$ \\
19     & 10:52:24.58 & +57:21:19.3 & $11.7 \pm 3.5\phantom{0}$ & 3.50 & central & S02 (LH.15)& $\phantom{0}3.60$ & $\phantom{0}0.3$ \\  \hline
20     & 10:52:27.18 & +57:22:21.2 & $12.1 \pm 3.7\phantom{0}$ & 3.40 & central &                &                   &               \\
21     & 10:52:37.67 & +57:20:30.1 & $9.8 \pm 3.1\phantom{0}$ & 3.38 & edge     & S02 (LH.20)& $\phantom{0}3.51$ & $\phantom{0}0.3$ \\
22     & 10:51:46.97 & +57:24:51.2 & $7.1 \pm 2.2\phantom{0}$ & 3.38 & central  & S02 (LH.23)& $\phantom{0}3.48$ & $\phantom{0}0.3$ \\ 
23     & 10:52:28.07 & +57:25:09.2 & $5.8 \pm 1.8\phantom{0}$ & 3.37 & central  & S02 (LH.16)& $\phantom{0}3.56$ & $10.4$ \\
24     & 10:52:03.94 & +57:20:07.3 & $4.0 \pm 1.3\phantom{0}$ & 3.36 & deep     & S02 (LH.31)& $\phantom{0}3.24$ & $\phantom{0}0.4$ \\
25     & 10:51:48.36 & +57:21:48.2 & $12.1 \pm 3.8\phantom{0}$ & 3.34 & central &                &                   &               \\
26     & 10:52:09.87 & +57:20:40.3 & $9.0 \pm 2.8\phantom{0}$ & 3.33 & central & S02 (LH.34)& $\phantom{0}3.16$ & $\phantom{0}0.4$ \\
27     & 10:52:34.57 & +57:20:02.1 & $10.0 \pm 3.2\phantom{0}$ & 3.29 & central & S02 (LH.28)& $\phantom{0}3.31$ & $\phantom{0}0.3$ \\
28     & 10:51:42.89 & +57:24:12.1 & $11.0 \pm 3.6\phantom{0}$ & 3.26 & central & S02 (LH.24)& $\phantom{0}3.47$ & $\phantom{0}0.1$ \\   
29     & 10:52:01.71 & +57:19:16.3 & $4.0 \pm 1.3\phantom{0}$ & 3.23 & deep & S02 (LH.21)& $\phantom{0}3.50$ & $\phantom{0}0.3$ \\ 
30     & 10:51:59.99 & +57:20:39.3 & $4.2 \pm 1.4\phantom{0}$ & 3.21 & deep & S02 (LH.32)& $\phantom{0}3.22$ & $\phantom{0}0.3$ \\
31     & 10:52:27.28 & +57:19:06.2 & $7.7 \pm 2.5\phantom{0}$ & 3.20 & central & S02 (LH.26)& $\phantom{0}3.31$ & $\phantom{0}0.3$ \\
32     & 10:52:34.51 & +57:25:34.1 & $4.1 \pm 1.4\phantom{0}$ & 3.18 & central &                &                &               \\
33     & 10:52:10.86 & +57:24:13.3 & $8.4 \pm 2.8\phantom{0}$ & 3.17 & central &                &                &               \\
34     & 10:51:33.81 & +57:19:29.0 & $7.8 \pm 2.6\phantom{0}$ & 3.12 & central & S02 (LH.33)& $\phantom{0}3.20$ & $\phantom{0}0.1$ \\
35     & 10:51:55.77 & +57:23:12.3 & $4.0 \pm 1.3\phantom{0}$ & 3.12 & deep    & S02 (LH.18)& $\phantom{0}3.55$ & $\phantom{0}0.6$ \\
36     & 10:51:23.29 & +57:20:33.8 & $13.9 \pm 4.7\phantom{0}$& 3.09 & edge    &                &                   &               \\
37     & 10:52:29.57 & +57:26:20.2 & $4.7 \pm 1.6\phantom{0}$& 3.09 & central & S02 (LH.19)& $\phantom{0}3.54$ & $\phantom{0}1.6$ \\
38     & 10:52:41.14 & +57:21:47.1 & $14.3 \pm 4.9\phantom{0}$& 3.06 & edge   &                 &                  &               \\
39     & 10:52:03.45 & +57:16:54.3 & $7.9 \pm 2.7\phantom{0}$ & 3.06 & central & S02 (LH.36)& $\phantom{0}3.00$ & $\phantom{0}0.5$ \\ 
40     & 10:51:52.56 & +57:22:29.2 & $4.4 \pm 1.5\phantom{0}$ & 3.06 & deep    &                &                   &               \\ \hline \hline
 & \it{10:52:42.21} & \it{+57:18:28.0} & $\mathit{10.0 \pm 3.5\phantom{0}}$ & $\mathit{2.99}$  & \it{edge} & S02 (LH.30)& $\phantom{0}3.25$ & $\phantom{0}0.1$ \\
       & \it{10:52:16.43} & \it{+57:25:04.3} & $\mathit{5.0 \pm 1.8 \phantom{0}}$ & $\mathit{2.95}$ & \it{central} & S02 (LH.29)& $\phantom{0}3.30$ & $\phantom{0}2.7$ \\
       & \it{10:52:36.03} & \it{+57:18:20.1} & $\mathit{10.5 \pm 3.8 \phantom{0}}$ & $\mathit{2.91}$ & \it{edge}  & S02 (LH.25)& $\phantom{0}3.46$ & $\phantom{0}0.3$ \\
       & \it{10:51:57.61} & \it{+57:26:03.3} & $\mathit{6.5 \pm 2.3 \phantom{0}}$ & $\mathit{2.90}$ & \it{central} & S02 (LH.35)& $\phantom{0}3.02$ & $\phantom{0}0.3$ \\ \hline
\end{tabular}
\label{table:lh}\caption{\small $\rm 850 \, \mu m$ source list for the
       Lockman Hole East field of the ``SCUBA 8\,mJy Survey''. Sources are marked in Fig 9.}
\end{table*}

More up-to-date $\rm 850\, \mu m$ source lists for the ``SCUBA 8\,mJy
Survey'' than those given in Scott et al. (2002) are presented here. 
There are two reasons for the differences between this list and
that given in Scott et al (2002). Firstly, $\simeq 20$ sq. arcmin of
additional data were taken after the publication of the original
paper, most of it in the ELAIS N2 field, and this has now been added
into the survey maps. Secondly, rather than doing a fit to only those
peaks identified at $>3$\,mJy in the smoothed images, the simultaneous
maximum likelihood source extraction algorithm was applied to every
peak in the maps. 

Column 1 gives the source number in order of
decreasing signal-to-noise ratio as derived from the new IDL-reduction and
simultaneous maximum-likelihood source extraction method, and corresponds to
the labelling of the circled sources in Figs. 8 and 9. Columns 2 and 3 give
the right ascension and declination of the source in J2000
coordinates. Column 4 gives the simultaneously fitted $\rm 850 \, \mu
m$ flux densities of the sources. The error includes a 10\%
calibration error combined in quadrature. Column 5 gives the measured
signal-to-noise ratio of the source from the simultaneously-fitted
model. Column 6 defines the noise region in which the source was
found; `deep' corresponds to the deep linear strip at the centre of
the Lockman Hole East field, `central' corresponds to the parts of the
map which have seen the full integration time (outside of the deep area), and `edge' corresponds to the rather
noisier regions near the perimeter which have not seen the full integration
time. Column 7 gives any previous reference to the $\rm 850 \, \mu m$
source. Reference S02 is an abbreviation for Scott et al. (2002). 
Column 8 gives the previously recorded signal-to-noise
ratio where applicable, and Column 9 gives the distance between the
listed and previously referenced positions. The table listings given
in italics correspond to previously referenced sources with $\rm S/N >
3.00$, which did not meet this criterion in this analysis.

Table 5 shows that the top 7 sources originally detected at $\rm >4.00
\sigma$ in the ELAIS N2 field (Scott et al. 2002) remain robustly
detected at this signal to noise threshold. One can also see that the
top 15 sources presented in Scott et al. (2002) are confirmed with
$\rm S/N > 3.50$. Similarly, Table 6 shows that the top 10 sources
presented in Scott et al. for the Lockman Hole East field are
confirmed with $\rm S/N > 4.00$ and the original top 15 sources have
all been detected at $\rm > 3.50 \sigma$.
 
\subsection{The Canada-UK Deep Submillimetre Survey}

\begin{figure}
 \centering
   \vspace*{10cm}
   \leavevmode
   \includegraphics{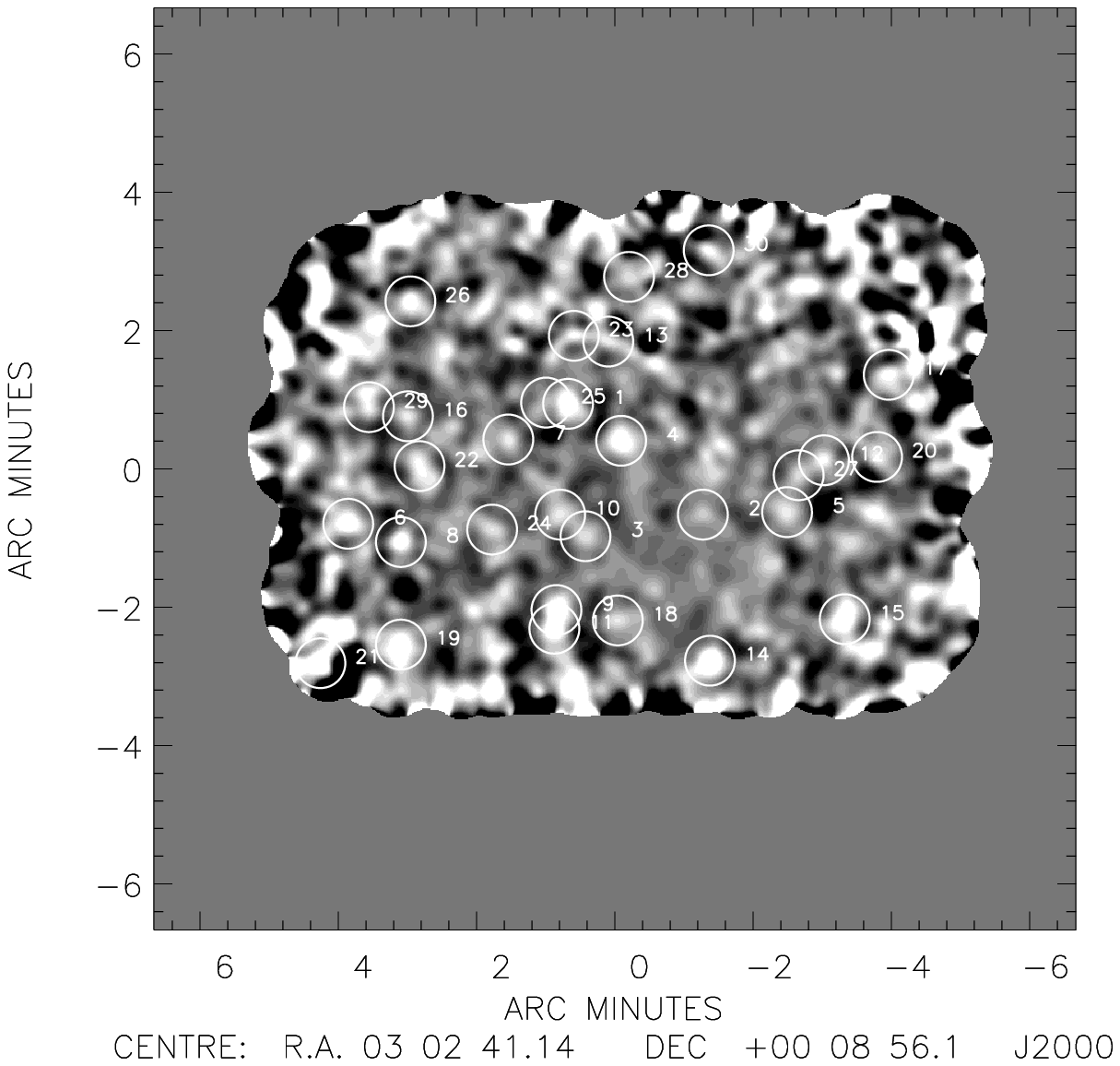}
\label{fig:cudss03_850} 
\caption{\small{The $\rm 850 \, \mu m$ image of the CUDSS 03-Hour field,
smoothed with a beam-size Gaussian (14.5 arcsec FWHM). The
numbered circles highlight those sources found at a significance of
$>3.00$. The labelling
corresponds to the numbers in Table 7.}}

 \centering
   \vspace*{10cm}
   \leavevmode
   \includegraphics{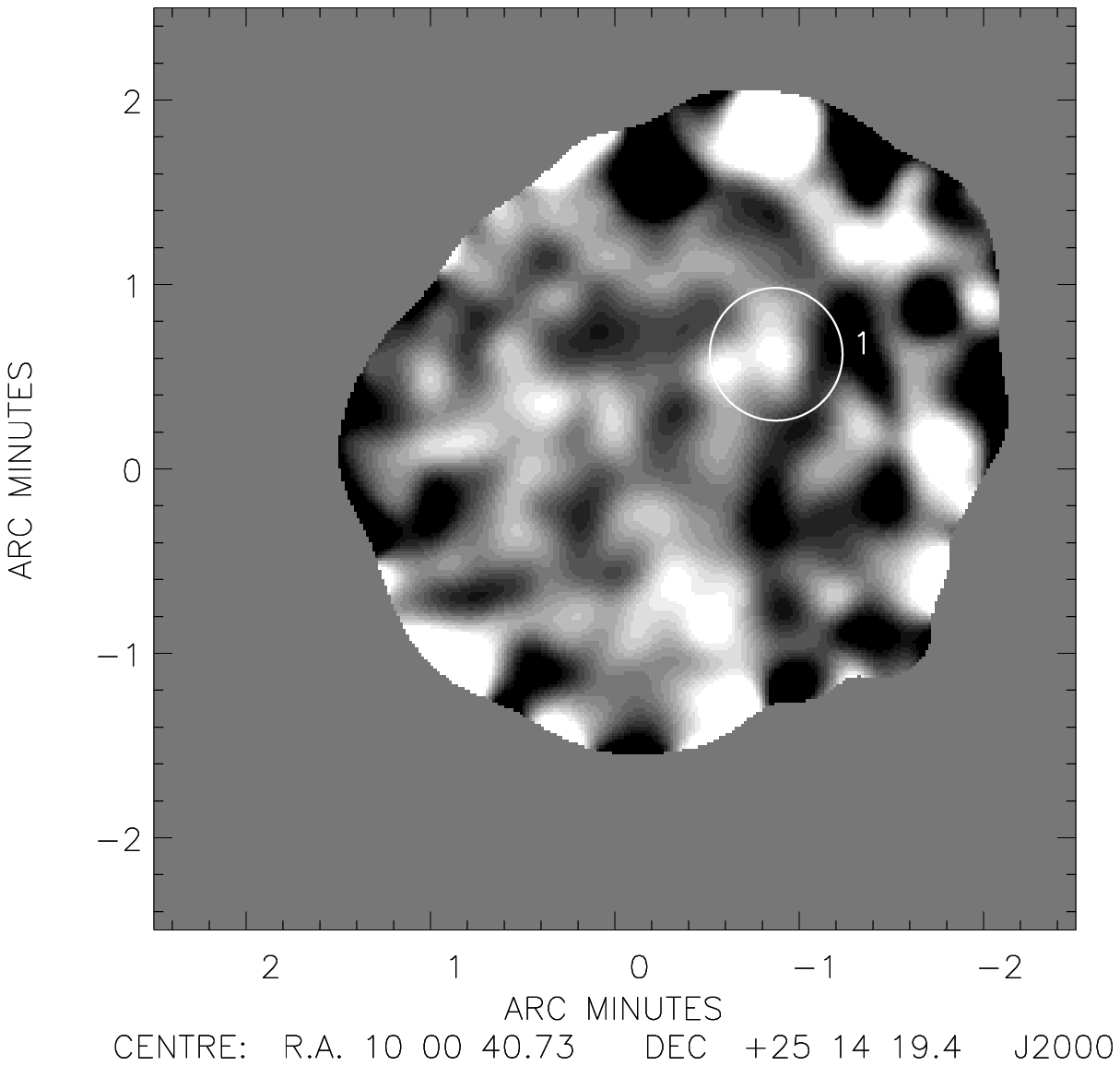}
\label{fig:cudss10_850} 
\caption{\small{The $\rm 850 \, \mu m$ image of the CUDSS 10-Hour field,
smoothed with a beam-size Gaussian (14.5 arcsec FWHM). The
numbered circles highlight those sources found at a significance of
$>3.00$. The labelling
corresponds to the numbers in Table 8.}}
 \end{figure}

\begin{figure}
 \centering
   \vspace*{10cm}
   \leavevmode
   \includegraphics{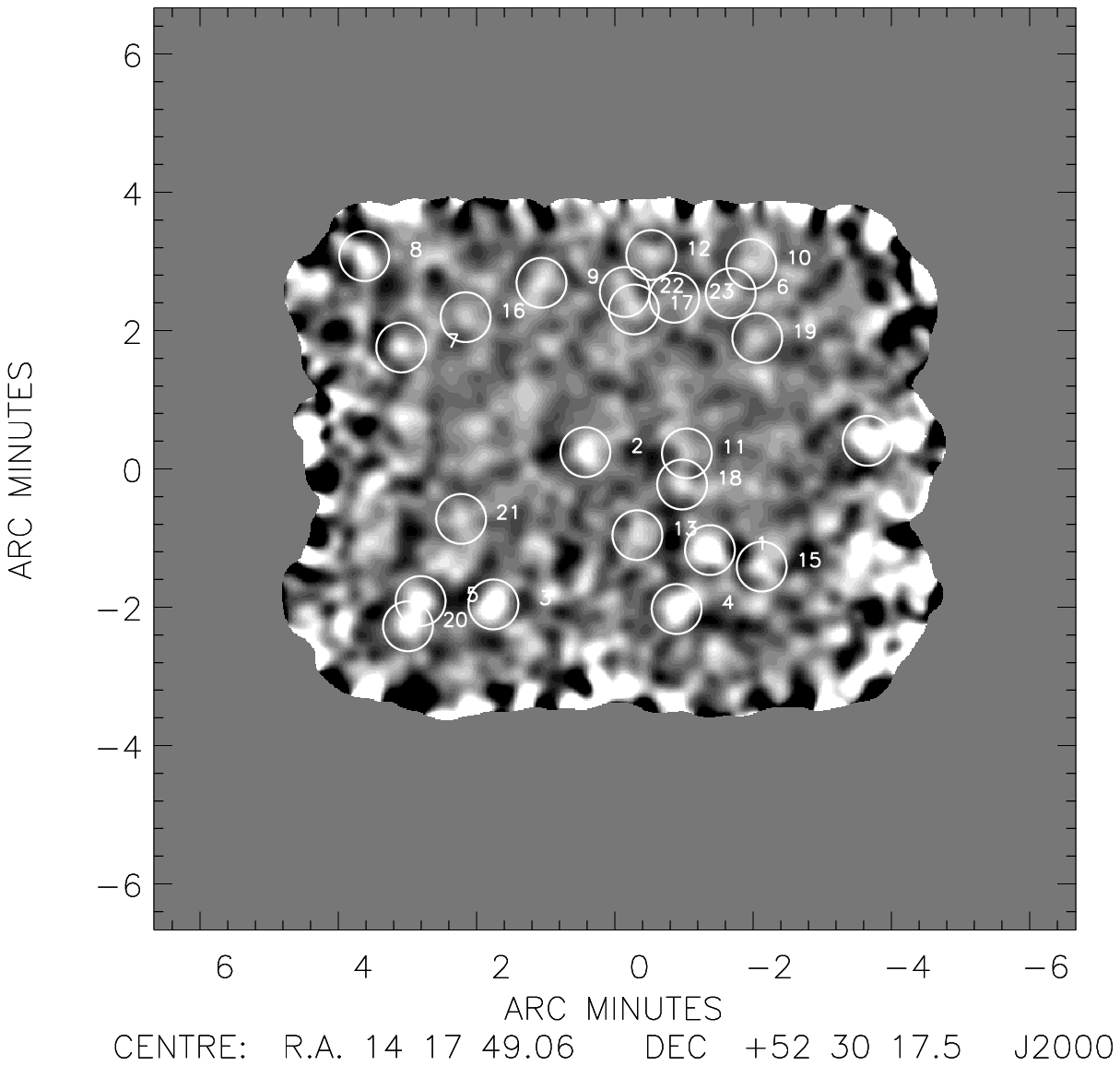}
\label{fig:cudss14_850} 
\caption{\small{The $\rm 850 \, \mu m$ image of the CUDSS 14-Hour field,
smoothed with a beam-size Gaussian (14.5 arcsec FWHM). The
numbered circles highlight those sources found at a significance of
$>3.00$. The labelling
corresponds to the numbers in Table 9.}}

 \centering
   \vspace*{10cm}
   \leavevmode
   \includegraphics{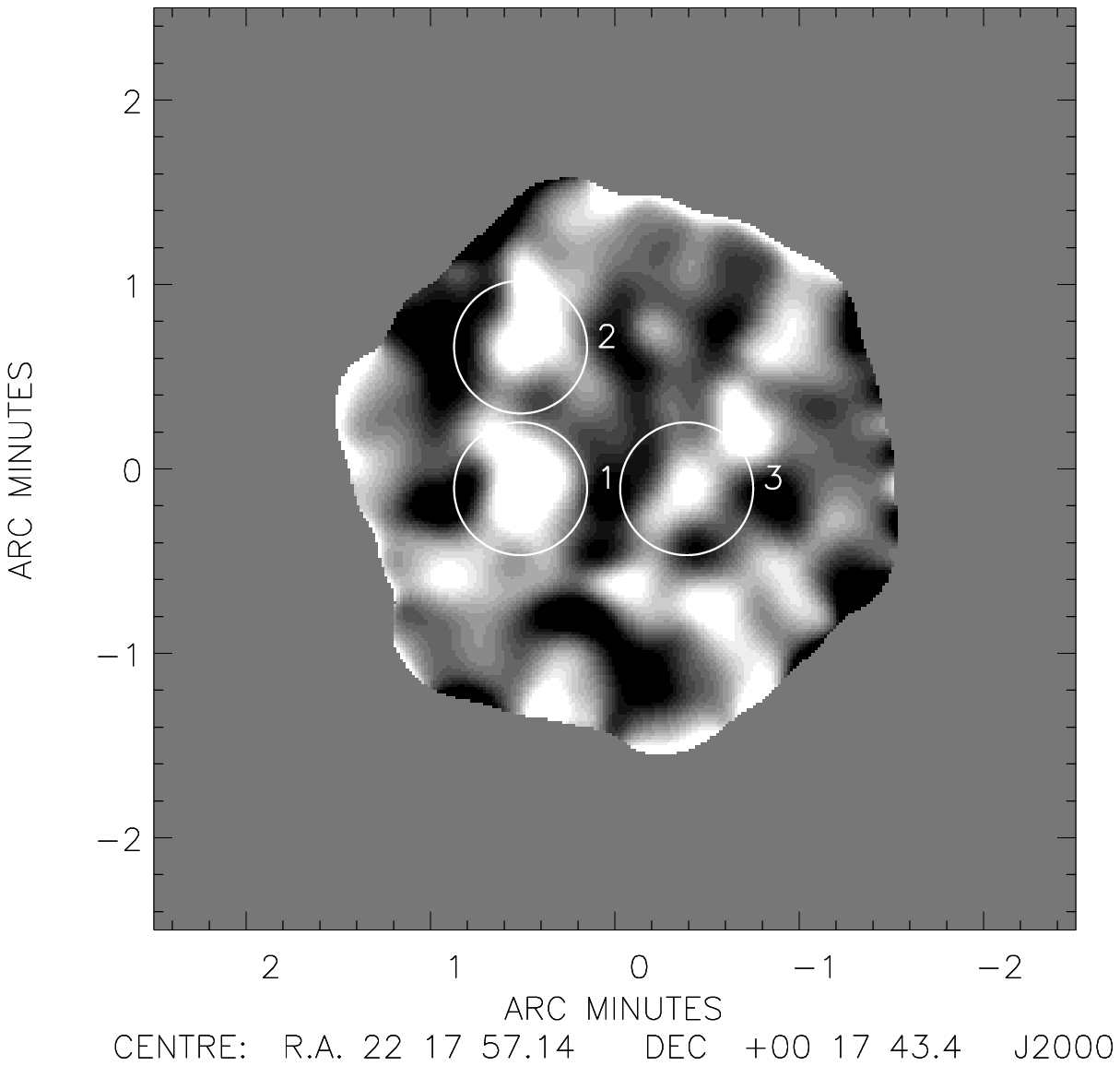}
\label{fig:cudss22_850} 
\caption{\small{The $\rm 850 \, \mu m$ image of the CUDSS 22-Hour field,
smoothed with a beam-size Gaussian (14.5 arcsec FWHM). The
numbered circles highlight those sources found at a significance of
$>3.00$. The labelling
corresponds to the numbers in Table 10.}}
 \end{figure}

The ``Canada UK Deep Submillimetre Survey (CUDSS)'' (Eales et
al. 1999, Lilly et al. 1999, Gear et al. 2000, Eales et al. 2000, Webb
et al. 2003a), covers a total
of $\simeq 130$\, sq. arcmin over 4 regions of sky, selected to
coincide with areas observed in the ``Canada-France Redshift Survey (CFRS)''
(Lilly et al. 1995). The 03-Hour field is composed of a deep pencil
beam area ($\simeq 8$\, sq. arcmin in size, $\rm
\sigma_{850}=1.1$\,mJy/beam; Eales et al. 1999, Lilly et al. 1999),
embedded in a wider-area, shallower map covering an additional
55\,sq. arcmin with a typical rms noise level of $\rm
\sigma_{850}=1.8$\,mJy/beam (Webb et al. 2003a). The 14-Hour field
(Eales et al. 2000) is similar in size, the uniform noise region
covering approximately 57\,sq. arcmin, but to a slightly deeper
uniform noise level of $\rm \sigma_{850}=1.5$\,mJy/beam.
The 10-Hour and 22-Hour fields
(Eales et al. 1999, Lilly et al. 1999) are small in area,
each having a uniform noise region of $\simeq 7$\, sq. arcmin, with
rms noise levels of $\rm \sigma_{850}=1.3$ and 1.5\,mJy/beam respectively.
\begin{table*}
\begin{tabular}{|l|c|c|r|c|c|l|c|c|} \hline
 & RA    & DEC   & $\rm S_{850}\phantom{000}$ & S/N & Noise & Previous & Prev. & Sep. \\
       &(J2000)&(J2000)&     /mJy\phantom{00}  &     & Region&
       Reference& S/N      &/arcsec \\ \hline        
01     & 03:02:43.84 & +00:09:52.6 & $7.0 \pm 1.4\phantom{0}$ & 5.90 & central & W03 (03h.19) & $\phantom{0}3.2$ & $\phantom{0}1.8$     \\
02     & 03:02:36.04 & +00:08:16.6 & $3.4 \pm 0.8\phantom{0}$ & 4.89 & deep    & W03 (03h.06) & $\phantom{0}5.4$ & $\phantom{0}1.3$     \\
03     & 03:02:42.84 & +00:07:57.6 & $3.6 \pm 0.9\phantom{0}$ & 4.61 & deep    & W03 (03h.02) & $\phantom{0}6.7$ & $\phantom{0}3.9$     \\
04     & 03:02:40.77 & +00:09:20.6 & $6.9 \pm 1.7\phantom{0}$ & 4.51 & central &                           &     &         \\
05     & 03:02:31.17 & +00:08:18.6 & $4.8 \pm 1.2\phantom{0}$ & 4.31 & central & W03 (03h.03) & $\phantom{0}6.4$ & $\phantom{0}5.1$     \\
06     & 03:02:56.57 & +00:08:08.6 & $7.0 \pm 1.8\phantom{0}$ & 4.21 & central & W03 (03h.24) & $\phantom{0}3.0$ & $\phantom{0}3.5$     \\
07     & 03:02:47.31 & +00:09:21.6 & $4.7 \pm 1.3\phantom{0}$ & 4.18 & central &                           &     &         \\
08     & 03:02:53.51 & +00:07:52.6 & $9.0 \pm 2.3\phantom{0}$ & 4.16 & central &                           &     &         \\
09     & 03:02:44.51 & +00:06:53.6 & $5.6 \pm 1.5\phantom{0}$ & 4.14 & central & W03 (03h.04) & $\phantom{0}6.2$ & $\phantom{0}2.2$     \\
10     & 03:02:44.31 & +00:08:16.6 & $4.0 \pm 1.0\phantom{0}$ & 4.14 & central & W03 (03h.05) & $\phantom{0}5.8$ & $\phantom{0}5.3$     \\
11     & 03:02:44.64 & +00:06:37.6 & $6.0 \pm 1.6\phantom{0}$ & 4.03 & central & W03 (03h.01) & $\phantom{0}7.4$ & $\phantom{0}3.4$     \\ \hline
12     & 03:02:29.04 & +00:09:03.6 & $5.3 \pm 1.5\phantom{0}$ & 3.87 & central &                           &     &         \\
13     & 03:02:41.51 & +00:10:46.6 & $9.8 \pm 2.8\phantom{0}$ & 3.81 & central &                           &     &         \\
14     & 03:02:35.64 & +00:06:09.6 & $7.9 \pm 2.2\phantom{0}$ & 3.79 & central & W03 (03h.07) & $\phantom{0}5.3$ & $\phantom{0}2.2$     \\
15     & 03:02:27.84 & +00:06:45.6 & $6.0 \pm 1.8\phantom{0}$ & 3.63 & central & W03 (03h.15)* & $\phantom{0}3.5$ & $\phantom{0}7.8$     \\
       & & & & &         & W03 (03h.27)* & $\phantom{0}3.0$ &   13.5      \\           
16     & 03:02:53.11 & +00:09:41.6 & $4.8 \pm 1.4\phantom{0}$ & 3.50 & central & W03 (03h.20) & $\phantom{0}3.2$ & $\phantom{0}3.3$     \\ \hline
17     & 03:02:25.31 & +00:10:17.6 & $7.2 \pm 2.2\phantom{0}$ & 3.49 & central &                           &     &         \\
18     & 03:02:40.97 & +00:06:44.6 & $3.0 \pm 0.9\phantom{0}$ & 3.48 & deep    &                           &     &         \\
19     & 03:02:53.51 & +00:06:23.6 & $7.3 \pm 2.2\phantom{0}$ & 3.47 & central & W03 (03h.23) & $\phantom{0}3.0$ & $10.9$    \\
20     & 03:02:25.97 & +00:09:06.6 & $6.3 \pm 1.9\phantom{0}$ & 3.47 & central & W03 (03h.14) & $\phantom{0}3.5$ & $\phantom{0}3.0$     \\
21     & 03:02:58.17 & +00:06:07.6 & $13.9 \pm 4.3\phantom{0}$& 3.44 & edge    &                           &     &         \\
22     & 03:02:52.44 & +00:08:58.6 & $4.3 \pm 1.4\phantom{0}$ & 3.32 & central & W03 (03h.10) & $\phantom{0}4.5$ & $\phantom{0}1.4$     \\
23     & 03:02:43.51 & +00:10:51.6 & $9.0 \pm 2.9\phantom{0}$ & 3.23 & central &                           &     &         \\ 
24     & 03:02:48.24 & +00:08:03.6 & $3.9 \pm 1.3\phantom{0}$ & 3.15 & central &                           &     &         \\
25     & 03:02:45.11 & +00:09:53.6 & $4.0 \pm 1.3\phantom{0}$ & 3.14 & central &                           &     &         \\
26     & 03:02:52.97 & +00:11:21.6 & $6.2 \pm 2.1\phantom{0}$ & 3.08 & central & W03 (03h.11) & $\phantom{0}4.0$ & $\phantom{0}1.1$     \\
27     & 03:02:30.51 & +00:08:50.6 & $4.8 \pm 1.6\phantom{0}$ & 3.07 & central &                           &     &         \\
28     & 03:02:40.31 & +00:11:42.6 & $6.5 \pm 2.2\phantom{0}$ & 3.04 & central &                           &     &         \\
29     & 03:02:55.37 & +00:09:49.6 & $5.3 \pm 1.8\phantom{0}$ & 3.03 & central &                           &     &         \\
30     & 03:02:35.71 & +00:12:05.6 & $7.2 \pm 2.5\phantom{0}$ & 3.00 &
       central &                           &     &         \\ \hline
       \hline 
  &\it{03:02:35.91} & \it{+00:09:57.6} & $\mathit{3.7 \pm 1.3\phantom{0}}$ & $\mathit{2.97}$ &\it{central} & W03 (03h.13) & $\phantom{0}3.8$ & $\phantom{0}4.4$     \\
       &\it{03:02:38.77} & \it{+00:10:27.6} & $\mathit{5.2 \pm 1.9\phantom{0}}$ & $\mathit{2.80}$ &\it{central} & W03 (03h.12) & $\phantom{0}4.0$ & $\phantom{0}1.9$     \\
       &\it{03:02:26.24} & \it{+00:06:18.6} & $\mathit{4.8 \pm 1.9\phantom{0}}$ & $\mathit{2.67}$ &\it{central} & W03 (03h.08) & $\phantom{0}5.0$ & $\phantom{0}2.3$     \\
       &\it{03:02:26.11} & \it{+00:08:17.6} & $\mathit{3.8 \pm 1.5\phantom{0}}$ & $\mathit{2.66}$ &\it{central} & W03 (03h.21) & $\phantom{0}3.1$ & $\phantom{0}3.4$     \\
       &\it{03:02:32.77} & \it{+00:10:20.6} & $\mathit{4.0 \pm 1.7\phantom{0}}$ & $\mathit{2.51}$ &\it{central} & W03 (03h.18) & $\phantom{0}3.3$ & $\phantom{0}5.8$     \\
       &\it{03:02:31.37} & \it{+00:10:33.6} & $\mathit{4.5 \pm 2.0\phantom{0}}$ & $\mathit{2.33}$ &\it{central} & W03 (03h.17) & $\phantom{0}3.4$ & $\phantom{0}5.2$     \\
       &\it{03:02:34.97} & \it{+00:09:16.6} & $\mathit{3.4 \pm 1.5\phantom{0}}$ & $\mathit{2.33}$ &\it{central} & W03 (03h.26) & $\phantom{0}3.0$ & $\phantom{0}4.5$     \\
       &\it{03:02:28.31} & \it{+00:10:19.6} & $\mathit{3.5 \pm 1.7\phantom{0}}$ & $\mathit{2.09}$ &\it{central} & W03 (03h.09) & $\phantom{0}4.6$ & $\phantom{0}8.9$     \\
       &\it{03:02:35.44} & \it{+00:08:51.6} & $\mathit{2.3 \pm 1.1\phantom{0}}$ & $\mathit{2.08}$ &\it{central} & W03 (03h.16) & $\phantom{0}3.4$ & $\phantom{0}9.5$     \\
       &\it{03:02:39.17} & \it{+00:06:14.6} & $\mathit{2.2 \pm 1.3\phantom{0}}$ & $\mathit{1.77}$ &\it{central} & W03 (03h.22) & $\phantom{0}3.1$ & $12.5$    \\
       &\it{03:02:38.11} & \it{+00:11:10.6} & $\mathit{2.9 \pm
       2.4\phantom{0}}$ & $\mathit{1.22}$ &\it{central} & W03 (03h.25)
       & $\phantom{0}3.0$ & $\phantom{0}8.2$     \\ \hline
\end{tabular}
\label{table:cudss03}\caption{\small $\rm 850 \, \mu m$ source list for the
       03h field of the ``Canada-UK Deep Submillimetre Survey''. Sources are marked in Fig 10.}
\end{table*}

\begin{table*}
\begin{tabular}{|l|c|c|r|c|c|l|c|c|} \hline
 & RA    & DEC   & $\rm S_{850}\phantom{000}$ & S/N & Noise & Previous & Prev. & Sep. \\
       &(J2000)&(J2000)&     /mJy\phantom{00}  &     & Region&
       Reference& S/N      & /arcsec \\ \hline        
01     & 10:00:36.86 & +25:14:56.9 & $3.3 \pm 1.1\phantom{0}$ & 3.12 & central & E99 (10h.B)* & $\phantom{0}5.0$ & $\phantom{0}3.4$     \\
 & & & & & & E99 (10h.C)* & $\phantom{0}3.8$ & $15.9$     \\ 
 & & & & & & E99 (10h.D)* & $\phantom{0}3.6$ & $15.0$     \\ \hline \hline 
       & \it{10:00:38.12} & \it{+25:14:51.9} & $\mathit{2.8 \pm
       1.0\phantom{0}}$ & $\mathit{2.79}$ & \it{central} & E99 (10h.A)
       & $\phantom{0}6.1$ & $\phantom{0}2.1$     \\ \hline
\end{tabular} 
\label{table:cudss10}\caption{\small $\rm 850 \, \mu m$ source list for the
       10h field of the ``Canada-UK Deep Submillimetre Survey''. Sources
are marked in Fig 11.}
\end{table*}

\begin{table*}
\hspace{-0.4cm}
\small
\begin{tabular}{|l|c|c|r|c|c|l|c|c|} \hline
 & RA    & DEC   & $\rm S_{850}\phantom{000}$ & S/N & Noise & Previous & Prev. & Sep. \\
       &(J2000)&(J2000)&     /mJy\phantom{00}  &     & Region&
       Reference& S/N      &/arcsec  \\ \hline        
01     & 14:17:40.03 & +52:29:07.0 & $8.5 \pm 1.4\phantom{0}$ & 7.62 & central & E00 (14h.01) & $10.1$ & $\phantom{0}2.1$     \\
02     & 14:17:51.86 & +52:30:32.0 & $6.0 \pm 1.1\phantom{0}$ & 6.01 & central & E00 (14h.02) & $\phantom{0}6.3$ & $\phantom{0}2.1$ \\
03     & 14:18:00.61 & +52:28:20.0 & $7.2 \pm 1.5\phantom{0}$ & 5.45 & central & E00 (14h.03) & $\phantom{0}5.4$ & $\phantom{0}3.6$ \\
04     & 14:17:43.21 & +52:28:16.0 & $5.7 \pm 1.4\phantom{0}$ & 4.57 & central & E00 (14h.04) & $\phantom{0}5.3$ & $\phantom{0}2.0$ \\
05     & 14:18:07.51 & +52:28:22.9 & $5.8 \pm 1.4\phantom{0}$ & 4.42 & central & E00 (14h.05) & $\phantom{0}4.5$ & $\phantom{0}2.3$ \\ \hline
06     & 14:17:38.05 & +52:32:50.0 & $4.9 \pm 1.4\phantom{0}$ & 3.90 & central &              &                  &                  \\
07     & 14:18:09.39 & +52:32:02.9 & $5.4 \pm 1.5\phantom{0}$ & 3.76 & central &              &                  &                  \\
08     & 14:18:12.91 & +52:33:21.9 & $9.3 \pm 2.7\phantom{0}$ & 3.71 & edge    &              &                  &                  \\
09     & 14:17:56.03 & +52:32:59.0 & $3.7 \pm 1.1\phantom{0}$ & 3.59 & central &              &                  &                  \\ \hline
10     & 14:17:36.07 & +52:33:15.0 & $4.1 \pm 1.2\phantom{0}$ & 3.47 & central &              &                  &                  \\
11     & 14:17:42.22 & +52:30:31.0 & $3.7 \pm 1.1\phantom{0}$ & 3.46 & central & E00 (14h.18) & $\phantom{0}3.0$ & $\phantom{0}4.5$ \\
12     & 14:17:45.61 & +52:33:23.0 & $3.9 \pm 1.2\phantom{0}$ & 3.37 & central &              &                  &                  \\
13     & 14:17:46.93 & +52:29:20.0 & $4.2 \pm 1.3\phantom{0}$ & 3.31 & central &              &                  &                  \\
14     & 14:17:25.02 & +52:30:41.9 & $9.0 \pm 2.9\phantom{0}$ & 3.25 & edge    & E00 (14h.17) & $\phantom{0}3.3$ & $\phantom{0}4.5$ \\
15     & 14:17:35.11 & +52:28:53.0 & $4.5 \pm 1.4\phantom{0}$ & 3.25 & central &              &                  &                  \\
16     & 14:18:03.26 & +52:32:29.0 & $3.7 \pm 1.2\phantom{0}$ & 3.20 & central &              &                  &                  \\
17     & 14:17:47.25 & +52:32:36.0 & $3.3 \pm 1.1\phantom{0}$ & 3.19 & central & E00 (14h.11) & $\phantom{0}3.5$ & $\phantom{0}2.4$ \\ 
18     & 14:17:42.66 & +52:30:04.0 & $4.5 \pm 1.5\phantom{0}$ & 3.14 & central &              &                  &                  \\
19     & 14:17:35.53 & +52:32:11.0 & $3.2 \pm 1.1\phantom{0}$ & 3.14 & central &              &                  &                  \\
20     & 14:18:08.71 & +52:28:00.9 & $4.4 \pm 1.5\phantom{0}$ & 3.13 & central & E00 (14h.09) & $\phantom{0}4.1$ & $\phantom{0}4.1$ \\ 
21     & 14:18:03.68 & +52:29:33.9 & $3.1 \pm 1.0\phantom{0}$ & 3.05 & central & E00 (14h.10) & $\phantom{0}3.5$ & $\phantom{0}5.0$ \\
22     & 14:17:48.13 & +52:32:51.0 & $3.4 \pm 1.2\phantom{0}$ & 3.04 & central &              &                  &                  \\
23     & 14:17:43.42 & +52:32:46.0 & $3.9 \pm 1.3\phantom{0}$ & 3.01 & central &              &                  &                  \\ \hline \hline
       & \it{14:17:41.57} & \it{+52:28:27.0} & $\mathit{3.7 \pm 1.3\phantom{0}}$ & $\mathit{2.96}$ & \it{central} & E00 (14h.13) & $\phantom{0}3.4$ & $\phantom{0}3.9$ \\ 
       &  \it{14:18:11.68} &  \it{+52:30:05.9} & $\mathit{5.7 \pm 2.0\phantom{0}}$ & $\mathit{2.90}$ &  \it{central} & E00 (14h.19) & $\phantom{0}3.0$ & $\phantom{0}2.5$ \\
       &  \it{14:18:12.44} &  \it{+52:29:13.9} & $\mathit{6.3 \pm 2.4\phantom{0}}$ & $\mathit{2.77}$ &  \it{central} & E00 (14h.16) & $\phantom{0}3.7$ & $\phantom{0}6.3$ \\
       &  \it{14:18:09.28} &  \it{+52:31:01.9} & $\mathit{4.1 \pm 1.6}\phantom{0}$ & $\mathit{2.61}$ &  \it{central} & E00 (14h.14) & $\phantom{0}3.3$ & $\phantom{0}6.0$ \\
       &  \it{14:18:01.61} &  \it{+52:30:28.0} & $\mathit{3.3 \pm 1.3\phantom{0}}$ & $\mathit{2.56}$ &  \it{central} & E00 (14h.08) & $\phantom{0}4.0$ & $13.8$ \\
       &  \it{14:17:56.45} &  \it{+52:29:12.0} & $\mathit{3.0 \pm 1.2\phantom{0}}$ & $\mathit{2.50}$ &  \it{central} & E00 (14h.06) & $\phantom{0}4.2$ & $\phantom{0}5.2$ \\
       &  \it{14:18:05.21} &  \it{+52:28:55.9} & $\mathit{3.3 \pm 1.4\phantom{0}}$ & $\mathit{2.41}$ &  \it{central} & E00 (14h.12) & $\phantom{0}3.4$ & $\phantom{0}0.9$ \\
       &  \it{14:17:29.53} &  \it{+52:28:17.9} & $\mathit{4.2 \pm 1.9\phantom{0}}$ & $\mathit{2.26}$ &  \it{central} & E00 (14h.15) & $\phantom{0}3.1$ & $\phantom{0}2.4$ \\
       &  \it{14:18:01.60} &  \it{+52:29:44.0} & $\mathit{1.8 \pm 1.0\phantom{0}}$ & $\mathit{1.76}$ &  \it{central} & E00 (14h.07) & $\phantom{0}3.2$ & $\phantom{0}6.8$ \\ \hline       
\end{tabular}
\label{table:cudss14}\caption{\small $\rm 850 \, \mu m$ source list for the
       14h field of the ``Canada-UK Deep Submillimetre Survey''. Sources are 
marked in Fig 12.}
\end{table*}

\begin{table*}
\begin{tabular}{|l|c|c|r|c|c|l|c|c|} \hline
 & RA    & DEC   & $\rm S_{850}\phantom{000}$ & S/N & Noise & Previous & Prev. & Sep. \\
       &(J2000)&(J2000)&     /mJy\phantom{00}  &     & Region&
       Reference& S/N      & /arcsec \\  \hline      
01     & 22:17:59.18 & +00:17:36.9 & $5.6 \pm 1.3\phantom{0}$ & 4.62 & central &             &              &              \\
02     & 22:17:59.18 & +00:18:22.9 & $4.8 \pm 1.3\phantom{0}$ & 4.14 & central &             &              &              \\ \hline
03     & 22:17:55.58 & +00:17:36.9 & $3.6 \pm 1.1\phantom{0}$ & 3.36 & central &             &              &              \\ \hline
\end{tabular}
\label{table:cudss22}\caption{\small $\rm 850 \, \mu m$ source list for the
       22h field of the ``Canada-UK Deep Submillimetre Survey''. Sources
are marked in Fig 13.}
\end{table*}

The data were originally reduced independently by the Cardiff and Toronto
groups, using the standard SURF procedures. They made additional attempts to improve the quality of the final
map, firstly by allowing the residual sky removal to be a linear function of
position (i.e. a planar fit was applied rather than a D.C. offset), and
secondly by examining the Fourier-transform of each bolometer's
measured signal to search for non-white noise profiles (although Eales et
al. (2000) reported this produced negligible improvements to the final
regridded images). The chop throw in all cases was fixed
at 30 arcsec in an east-west direction as observed on the sky. 
Source extraction was carried out by convolving a normalised template of the full
beam-profile, constructed from the many observations of Uranus taken
thoughout the lifetime of this survey, with the raw survey maps. The
``CUDSS'' team used a 
method of noise modelling which is in effect quite similar to the way in
which the noise maps were created in the IDL reduction of the ``SCUBA
8\,mJy Survey''. They began with the basic assumption that the noise
on any bolometer was independent of the noise on every other bolometer,
and then measured the standard deviation of the intensities for each
bolometer in units of one hour (the length of each CUDSS
pointing). Artificial data were then created by replacing the real data
with the output of a
Gaussian random-number generator with the same standard deviation as
the real data, re-running the sky subtraction and clipping routines
from the SURF package to account for any non-Gaussian nature arising
from these processes, and finally rescaling the mock data so
that it had the same standard deviation as the real data. In total,
1000 simulated maps were generated, each of which was convolved with the 
beam template as were the real data. The final noise maps were produced
by measuring the standard deviation of these convolved maps, pixel by pixel.

As previously stated in Section 3, the method of convolving the raw images
with the normalised point spread function (PSF) is formally the best
method of source-extraction, provided that the sources are all well
separated from one another. It does, however, run into difficulties
when dealing with partially confused sources, a problem which is
likely to be fairly common, as inferred from the ELAIS N2 image
(Section 5.1) and the clustering analyses of Section 7. In the
``SCUBA 8\,mJy Survey'' the problem of confusion was tackled by means
of a maximum-likelihood fit of the beam template to all
potential sources simultaneously using the raw data. 
Eales et al. (2000) instead
addressed the problem of confusion by attempting a deconvolution with
the CLEAN algorithm (Hogbom 1974). They created an initial list of
possible sources based on the beam-convolved signal image divided by
the Gaussian generated noise image, then iteratively CLEANed the raw
data in boxes centred on the positions of the potential sources. For
each source the information from CLEAN was then used to remove all
other possible sources from the raw image, before again carrying out a
convolution with the beam template on this new map, and dividing by the
noise to measure the signal-to-noise ratio.

Comparative source lists for detections with $\rm S/N>3.00$ from the
two different reductions and source extraction procedures are given in 
Tables 7 - 10 (see also Figs 10 - 13). 
Columns 1-5, 8 and 9 are analogous to those of Table 5.
Column 6 defines the noise region in which the source was
found; `deep' corresponds to the deep pencil-beam surveys which
constitute part of a wider-area and somewhat shallower image,
`central' corresponds to the parts of the map which have seen 
the full integration time (outside of the deep area), and `edge' corresponds to the rather
noisier regions near the perimeter which have not seen the full integration
time. Column 7 gives any previous reference to the $\rm 850 \, \mu m$
source. Reference E99 is an abbreviation for Eales et
al. (1999), E00 is an abbreviation for Eales et
al. (2000), and W03 is an abbreviation for Webb et al. (2003a). The presence of
a * indicates that a previous reduction found more than one source here, whereas
in this reduction only one was found. 

As can be seen from comparing the two different reduction and source
extraction procedures, a majority of the most highly significant
objects ($\rm > 4.00\sigma$) identified in the initial CUDSS analyses
are also recovered here, but as one considers detections at decreasing
signal-to-noise ratios, the two catalogues increasingly diverge.
The comparative results for the 03-Hour field presented in Table 7
show that the top 6 sources detected by Webb et al. (2003a) are
robustly confirmed at the $\rm > 4.00 \sigma$ level while the top 7
are recovered above a signal-to-noise threshold of 3.50.
We find all but 3 of their 12 $\rm >4.00\sigma$
sources at better than $\rm >3.00\sigma$. In addition we have detected
a further 3 previously unpublished sources at better than $\rm
>4.00\sigma$. Dropping down the list to lower significances, however,
the resulting source catalogues of significant
($\rm >3.00\sigma$) detections are actually markedly different, in particular
only a handful of objects are common to both lists for $\rm
3.00<S/N<4.00$. That is not to say that all detections under a
signal-to-noise ratio of 3.00 are spurious, simply that there is a
rapidly increasing probability of contamination from fake sources on
decreasing the signal-to-noise threshold. A few of those objects
identified at $\rm >3.00\sigma$ by the original analysis of Webb et
al. (2003a) fall only just short of this criterion in this new analysis
adding some credibility to their reality. In the 14-Hour field,
the top 5 sources identified by Eales et al. (2000) are also securely recovered
as the top 5 detections in this analysis with $\rm S/N > 4.00$. 
Again, on dropping to lower
signal-to-noise ratios the two catalogues diverge. Only a further
5 of the remaining 14 objects detected by Eales et al. (2000) at $\rm
>3.00\sigma$ are recovered by this criterion in this
maximum-likelihood analysis, although a few of these objects fall only
just below this threshold. For those potential
sources common to both catalogues, the combination of the
IDL-reduction and simultaneous maximum-likelihood source extraction
algorithm provide much more conservative values of the signal-to-noise
ratio, in most cases by $\rm 1 - 2 \sigma$. At this stage it is not
possible to say which of the two reduction algorithms is the more
accurate. The deep radio
imaging of the ``SCUBA 8\,mJy Survey'' fields (Ivison et al. 2002)
has provided quite a stringent test of the IDL-reduction and the 
simultaneous maximum-likelihood source extraction algorithm technique,
robustly detecting 50\% of the bright SCUBA sources uncovered in the
``8\,mJy Survey'', and less significantly detecting a further 20\% of
the objects, suggesting a $30\%$ upper limit on the contamination from
spurious / confused sources. A similar analysis of the CUDSS fields,
however, is unlikely to be very informative regarding a comparison of the
two independent source lists, since the CUDSS fields are smaller
and deeper than the ``8\,mJy Survey'' fields, designed with the aim of
studying less bright SCUBA sources in the range $\rm 3\,mJy < S_{850}
< 6\,mJy$, and hence only the very brightest (and most significant of
the CUDSS sources) are likely to be detected in a radio image even at
a depth of $\rm \sigma_{1.4GHz}=5 \,\mu Jy/beam$. 
\begin{table*}
\begin{tabular}{|l|c|c|r|c|c|l|c|c|} \hline
 & RA    & DEC   & $\rm S_{850}\phantom{000}$ & S/N & Noise & Previous & Prev. & Sep. \\
       &(J2000)&(J2000)&     /mJy\phantom{00}  &     & Region&
       Reference& S/N      & /arcsec \\ \hline        
01     & 10:34:02.05 & +57:46:27.1 & $4.9 \pm 0.9\phantom{0}$ & 6.45 & central & B99 (LH.1) & $\phantom{0}5.1$ & $\phantom{0}2.1$ \\ \hline
02     & 10:33:55.80 & +57:45:10.1 & $2.6 \pm 0.7\phantom{0}$ & 3.79 & central &     &                  &                  \\ \hline \hline
       & \it{10:33:55.42} & \it{+57:47:38.1} & $\mathit{2.0 \pm 0.8\phantom{0}}$ & 2.59 & \it{central} & B99 (LH.2) & $\phantom{0}3.3$ & $10.6$ \\ \hline 
\end{tabular}
\label{table:lhdeep}\caption{\small $\rm 850 \, \mu m$ source list for the
       Lockman Hole Field of the ``Hawaii Flanking Fields Survey''. Sources
are marked in Fig 14.}

\begin{tabular}{|l|c|c|r|c|c|l|c|c|} \hline
 & RA    & DEC   & $\rm S_{850}\phantom{000}$ & S/N & Noise & Previous
       & Prev. & Sep. \\
       &(J2000)&(J2000)&     /mJy\phantom{00}  &     & Region&
       Reference& S/N      & /arcsec \\ \hline        
01     & 13:12:31.82 & +42:44:28.6 & $3.6 \pm 0.7\phantom{0}$ & 6.52 & deep & B99 (SSA13.1) & $\phantom{0}4.7$ & $\phantom{0}3.4$ \\
02     & 13:12:13.94 & +42:37:00.7 & $10.2 \pm 2.2\phantom{0}$ & 5.19 & central &  & & \\
03     & 13:12:27.56 & +42:45:01.5 & $2.8 \pm 0.6\phantom{0}$ & 5.08 & deep & B99 (SSA13.2)& $\phantom{0}3.8$ & $\phantom{0}6.0$ \\
04     & 13:12:19.93 & +42:39:30.7 & $6.1 \pm 1.6\phantom{0}$ & 4.22 & central &  & & \\
05     & 13:12:08.51 & +42:38:19.7 & $7.0 \pm 1.9\phantom{0}$ & 4.03 & central &  & & \\
06     & 13:12:25.00 & +42:39:56.7 & $7.8 \pm 2.1\phantom{0}$ & 4.01 & central & B99 (SSA13.6)& $\phantom{0}3.4$ & $\phantom{0}1.3$ \\ \hline
07     & 13:12:25.82 & +42:39:38.7 & $8.7 \pm 2.5\phantom{0}$ & 3.75 & central &  & & \\
08     & 13:12:17.66 & +42:42:51.7 & $7.4 \pm 2.1\phantom{0}$ & 3.66 & central &  & & \\
09     & 13:12:22.29 & +42:45:00.7 & $2.7 \pm 0.8\phantom{0}$ & 3.65 & deep &  & & \\ \hline
10     & 13:12:13.94 & +42:39:49.7 & $5.9 \pm 1.8\phantom{0}$ & 3.43 & central &  & & \\
11     & 13:12:31.26 & +42:40:22.7 & $8.1 \pm 2.5\phantom{0}$ & 3.41 & central &  & & \\
12     & 13:12:05.79 & +42:38:52.7 & $10.2 \pm 3.2\phantom{0}$ & 3.36 & central &  & & \\
13     & 13:12:14.40 & +42:43:33.7 & $10.1 \pm 3.2\phantom{0}$ & 3.34 & central &  & & \\
14     & 13:12:27.28 & +42:41:54.7 & $6.6 \pm 2.1\phantom{0}$ & 3.33 & central & B99 (SSA13.7)& $\phantom{0}3.3$ & $13.9$ \\
15     & 13:12:04.88 & +42:37:51.7 & $10.2 \pm 3.4\phantom{0}$ & 3.18 & central  &  & & \\
16     & 13:12:33.71 & +42:40:22.6 & $7.9 \pm 2.6\phantom{0}$ & 3.18 & central  &  & & \\
17     & 13:12:28.99 & +42:40:14.7 & $7.1 \pm 2.4\phantom{0}$ & 3.14 & central  &  & & \\
18     & 13:12:11.22 & +42:42:20.7 & $5.9 \pm 2.0\phantom{0}$ & 3.14 & central  &  & & \\
19     & 13:12:27.27 & +42:38:59.7 & $9.1 \pm 3.1\phantom{0}$ & 3.03 & central  &  & & \\ \hline \hline
       & \it{13:12:18.66} & \it{+42:38:25.7} & $\mathit{4.0 \pm 1.4\phantom{0}}$ & $\mathit{2.99}$ & \it{central}  & B99 (SSA13.5)  &$\phantom{0}3.3$ & 12.8 \\
       & \it{13:12:25.65} &\it{+42:43:48.5} & $\mathit{1.6 \pm 0.6\phantom{0}}$ & $\mathit{2.69}$ & \it{deep}  & B99  (SSA13.3) & $\phantom{0}3.2$ & $\phantom{0}1.6$ \\
       & \it{13:12:05.95} & \it{+42:44:37.7} & $\mathit{1.5 \pm 3.2\phantom{0}}$ & $\mathit{0.49}$ & \it{edge}  & B99  (SSA13.9) &$\phantom{0}3.4$ & $10.1$ \\
       & \it{n/a} & \it{n/a} & \it{n/a} & \it{n/a} & \it{bad bol.} & B99 (SSA13.8)& $\phantom{0}3.5$ & \it{n/a} \\
       & \it{n/a} & \it{n/a} & \it{n/a} & \it{n/a} & \it{edge} & B99 (SSA13.4)& $\phantom{0}3.3$ & \it{n/a} \\ \hline
\end{tabular}
\label{table:SSA13}\caption{\small $\rm 850 \, \mu m$ source list for the
       SSA13 Field of the ``Hawaii Flanking Fields Survey''. Sources are 
marked in Fig 15.}

\begin{tabular}{|l|c|c|r|c|c|l|c|c|} \hline
 & RA    & DEC   & $\rm S_{850}\phantom{000}$ & S/N & Noise & Previous & Prev. & Sep. \\
       &(J2000)&(J2000)&     /mJy\phantom{00}  &     & Region&
       Reference& S/N      & /arcsec \\ \hline        
01     & 17:06:37.03 & +43:55:31.8 & $3.2 \pm 1.0\phantom{0}$ & 3.51 & central & B99 (SSA17.3)& $\phantom{0}3.7$ & $\phantom{0}2.0$ \\ \hline
02     & 17:06:29.53 & +43:55:08.8 & $4.0 \pm 1.2\phantom{0}$ & 3.36 & central &              &                  &                  \\
03     & 17:06:25.08 & +43:57:40.8 & $5.6 \pm 1.8\phantom{0}$ & 3.33 & central & B99 (SSA17.1)& $\phantom{0}4.2$ & $\phantom{0}2.0$ \\
04     & 17:06:32.86 & +43:54:05.8 & $5.1 \pm 1.7\phantom{0}$ & 3.16 & central & B99 (SSA17.4)& $\phantom{0}3.6$ & $\phantom{0}3.4$ \\ \hline \hline
       & \it{17:06:25.55} & \it{+43:54:39.8} & $\mathit{3.1 \pm 1.4}$ & $\mathit{2.35}$ & \it{central} & B99 (SSA17.2)& $\phantom{0}3.9$ & $\phantom{0}0.6$ \\
       & \it{17:06:20.37} & \it{+43:54:09.8} &$\mathit{2.6 \pm 2.6}$ & $\mathit{1.02}$ & \it{edge} & B99 (SSA17.5)& $\phantom{0}3.1$ & $\phantom{0}5.8$ \\ \hline 
\end{tabular}
\label{table:SSA17}\caption{\small $\rm 850 \, \mu m$ source list for the
       SSA17 Field of the ``Hawaii Flanking Fields Survey''. Sources
are marked in Fig 16.}
\end{table*}
\begin{table*}
\begin{tabular}{|l|c|c|r|c|c|l|c|c|} \hline
 & RA    & DEC   & $\rm S_{850}\phantom{000}$ & S/N & Noise & Previous & Prev. & Sep. \\
       &(J2000)&(J2000)&     /mJy\phantom{00}  &     & Region&
       Reference& S/N      & /arcsec \\ \hline        
01     & 22:17:33.96 & +00:13:53.4 & $4.7 \pm 0.8\phantom{0}$ & 6.97 & central & B99 (SSA22.1)& $\phantom{0}6.9$ & $\phantom{0}2.2$ \\
02     & 22:17:35.03 & +00:15:36.4 & $2.7 \pm 0.7\phantom{0}$ & 4.29 & central & B99 (SSA22.2)& $\phantom{0}5.3$ & $\phantom{0}5.1$ \\
03     & 22:17:31.23 & +00:16:07.4 & $3.1 \pm 0.8\phantom{0}$ & 4.04 & central &  & &  \\ \hline
04     & 22:17:41.56 & +00:16:04.4 & $3.7 \pm 1.0\phantom{0}$ & 3.96 & central & B99 (SSA22.5)& $\phantom{0}3.1$ & $\phantom{0}7.0$ \\ 
05     & 22:17:40.90 & +00:14:56.4 & $3.0 \pm 0.8\phantom{0}$ & 3.92 & central &  & &  \\
06     & 22:17:35.96 & +00:15:56.4 & $2.5 \pm 0.7\phantom{0}$ & 3.77 & central & B99 (SSA22.3)& $\phantom{0}4.0$ & $13.1$ \\
07     & 22:17:37.36 & +00:16:21.4 & $3.6 \pm 1.0\phantom{0}$ & 3.75 & central &  & &  \\ 
08     & 22:17:29.30 & +00:13:57.4 & $2.6 \pm 0.7\phantom{0}$ & 3.66 & central &  & &  \\ \hline
09     & 22:17:19.96 & +00:15:25.4 & $8.1 \pm 2.7\phantom{0}$ & 3.22 & edge &  & &  \\ 
10     & 22:17:33.43 & +00:16:13.4 & $2.4 \pm 0.8\phantom{0}$ & 3.11 & central &  & &  \\
11     & 22:17:33.76 & +00:15:42.4 & $1.8 \pm 0.6\phantom{0}$ & 3.10 & central & B99 (SSA22.4)& $\phantom{0}3.6$ & $\phantom{0}3.6$ \\
12     & 22:17:41.03 & +00:13:32.4 & $3.2 \pm 1.1\phantom{0}$ & 3.08 & central &  & &  \\ \hline    
\end{tabular}
\label{table:SSA22}\caption{\small $\rm 850 \, \mu m$ source list for the
       SSA22 Field of the ``Hawaii Flanking Fields Survey''. Sources are marked in Fig 17.}

\begin{tabular}{|l|c|c|r|c|c|l|c|c|} \hline
 & RA    & DEC   & $\rm S_{850}\phantom{000}$ & S/N & Noise & Previous & Prev. & Sep. \\
       &(J2000)&(J2000)&     /mJy\phantom{00}  &     & Region&
       Reference& S/N      & /arcsec \\ \hline        
01     & 12:36:52.01 & +62:12:27.0 & $5.5 \pm 0.7\phantom{0}$ & 12.13
       & central & H98 (HDF.1)& $14.0$ & $\phantom{0}2.2$ \\
       &             &             &                           & 
       & & Serj03 (HDF.1)& $15.3$ & $\phantom{0}1.6$ \\
02     & 12:36:56.59 & +62:12:06.0 & $3.6 \pm 0.6\phantom{0}$ &
       $\phantom{0}7.06$ & central & H98 (HDF.2)&
       $\phantom{0}5.4$ & $\phantom{0}2.3$ \\
       & & & & &  & Serj03 (HDF.2)&
       $\phantom{0}7.6$ & $\phantom{0}2.6$ \\
03     & 12:36:53.16 & +62:13:55.0 & $2.7 \pm 0.6\phantom{0}$ & $\phantom{0}4.79$ & central & Serj03 (HDF.8)& $\phantom{0}3.5$ & $\phantom{0}0.8$ \\

04     & 12:36:50.58 & +62:13:18.0 & $2.1 \pm 0.5\phantom{0}$ &
$\phantom{0}4.35$ & central & H98 (HDF.4)*& $\phantom{0}4.6$ & $2.6$ \\
  & & & & & & H98 (HDF.5)*& $\phantom{0}4.2$ & $9.9$ \\
  & & & & & & Serj03 (HDF.4\&5)& $\phantom{0}5.1$& n/a\\ \hline

05     & 12:36:44.87 & +62:11:40.0 & $2.9 \pm 0.8\phantom{0}$ & $\phantom{0}3.87$ & central &         &        &       \\
06     & 12:37:03.17 & +62:13:04.0 & $2.8 \pm 0.8\phantom{0}$ & $\phantom{0}3.66$ & central &         &        &       \\ \hline \hline
       & \it{12:36:44.00} &\it{+62:13:09.0} & $\mathit{1.2 \pm 0.6\phantom{0}}$ &$\mathit{\phantom{0}2.13}$ & \it{central} & H98 (HDF.3) & $\phantom{0}5.0$ & $\phantom{0}7.5$ \\ 
       &  & & & &  & Serj03 (HDF.3) & $\phantom{0}2.1$ & $\phantom{0}2.9$ \\ 
       &\it{n/a}         &\it{n/a}        & $\mathit{n/a\phantom{000}}$              &\it{n/a}      & \it{edge}        & Serj03 (HDF.6) & $\phantom{0}3.8$ &   \it{n/a}      \\ 
       &\it{n/a}         &\it{n/a}        & $\mathit{n/a\phantom{000}}$              &\it{n/a}      & \it{edge}        & Serj03 (HDF.7) & $\phantom{0}3.7$ &   \it{n/a}      \\ \hline
\end{tabular}
\label{table:hdf}\caption{\small $\rm 850 \, \mu m$ source list for the
       ``Hubble Deep Field Survey''. Sources are marked in Fig 18.}
\end{table*}

\subsection{The Hawaii Submillimetre Survey}
\begin{figure}
 \centering
   \vspace*{10cm}
   \leavevmode
   \includegraphics{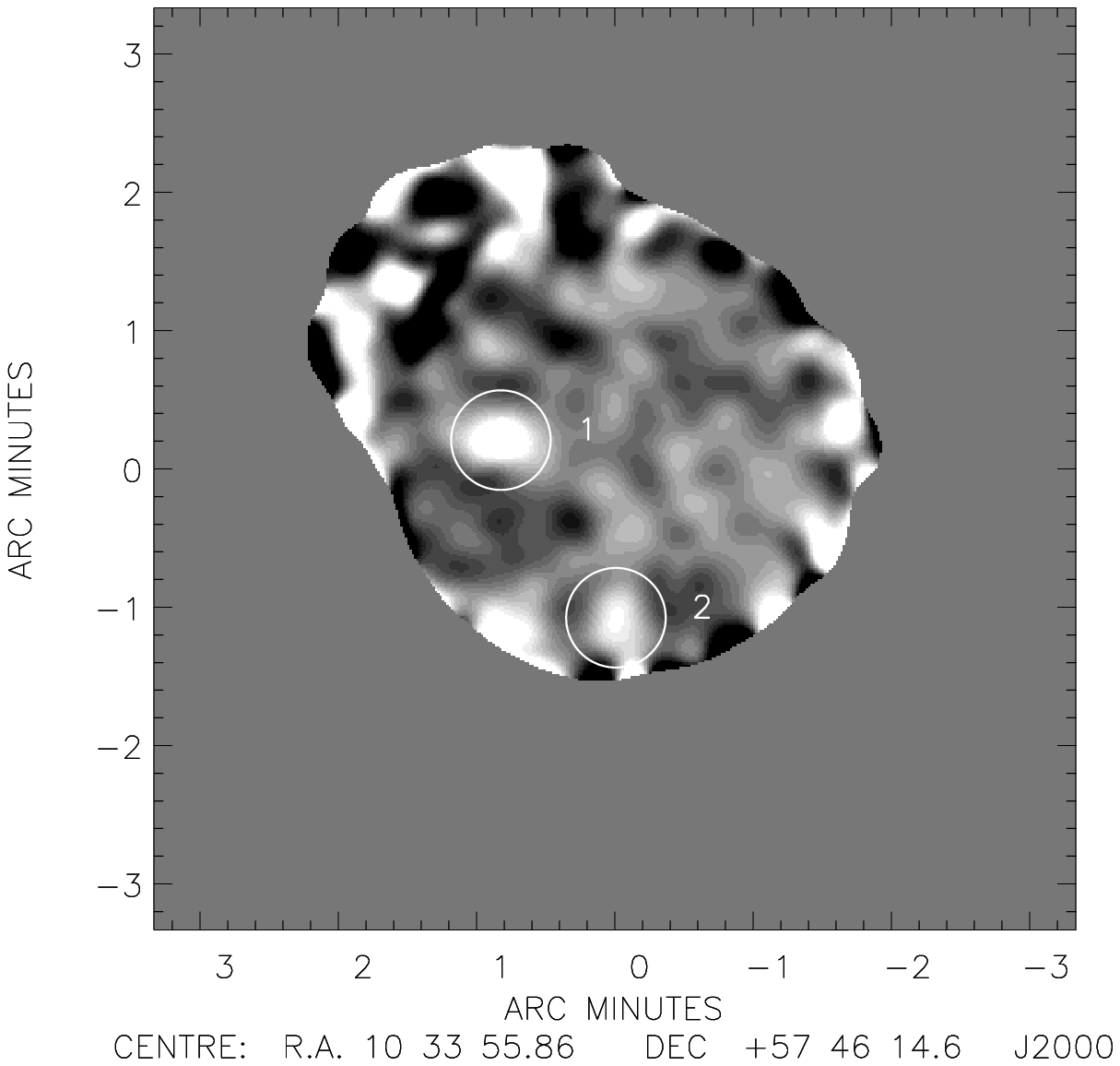}
\label{fig:lhdeep} 
\caption{\small{The $\rm 850 \, \mu m$ image of the Lockman Hole field
   from the Hawaii Flanking Fields Survey,
smoothed with a beam-size Gaussian (14.5 arcsec FWHM). The
numbered circles highlight those sources found at a significance of
$>3.00$. The labelling
corresponds to the numbers in Table 11.}}

 \centering
   \vspace*{10cm}
   \leavevmode
   \includegraphics{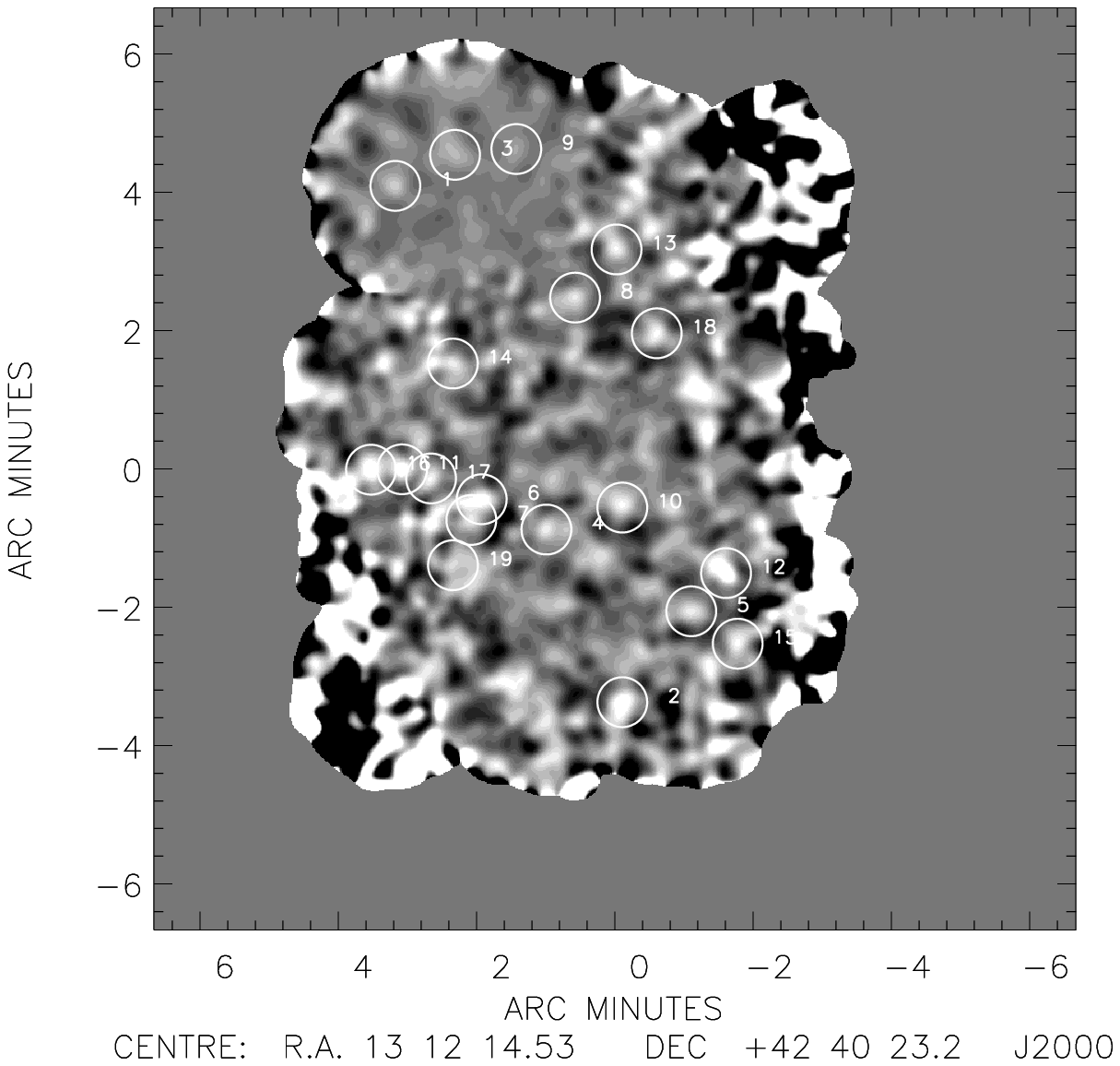}
\label{fig:lhdeep} 
\caption{\small{The $\rm 850 \, \mu m$ image of the SSA13 field
   from the Hawaii Flanking Fields Survey,
smoothed with a beam-size Gaussian (14.5 arcsec FWHM). The
numbered circles highlight those sources found at a significance of
$>3.00$. The labelling
corresponds to the numbers in Table 12.}}
\end{figure}

\begin{figure}
 \centering
   \vspace*{10cm}
   \leavevmode
   \includegraphics{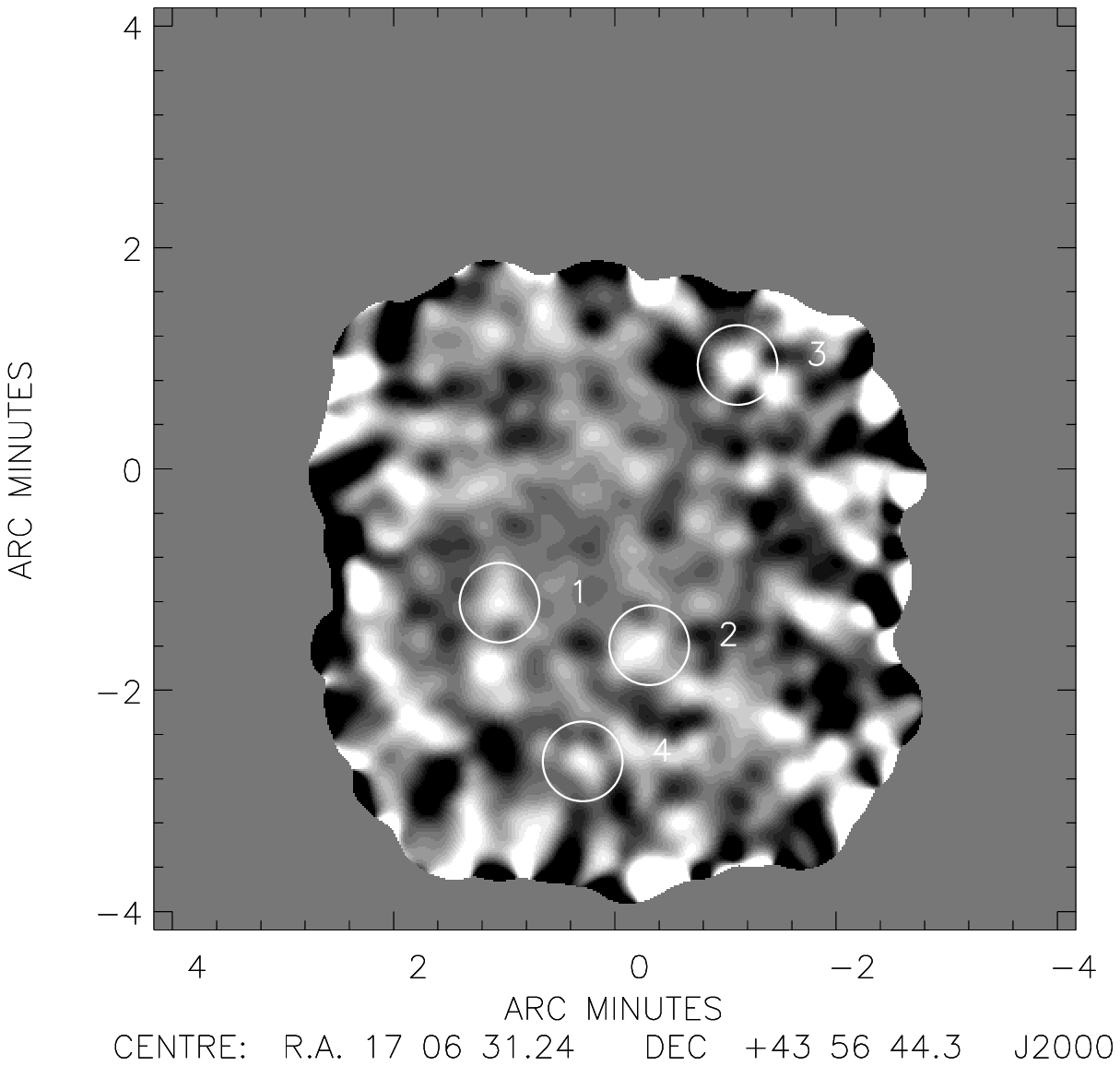}
\label{fig:lhdeep} 
\caption{\small{The $\rm 850 \, \mu m$ image of the SSA17 field
   from the Hawaii Flanking Fields Survey,
smoothed with a beam-size Gaussian (14.5 arcsec FWHM). The
numbered circles highlight those sources found at a significance of
$>3.00$. The labelling
corresponds to the numbers in Table 13.}}

 \centering
   \vspace*{10cm}
   \leavevmode
   \includegraphics{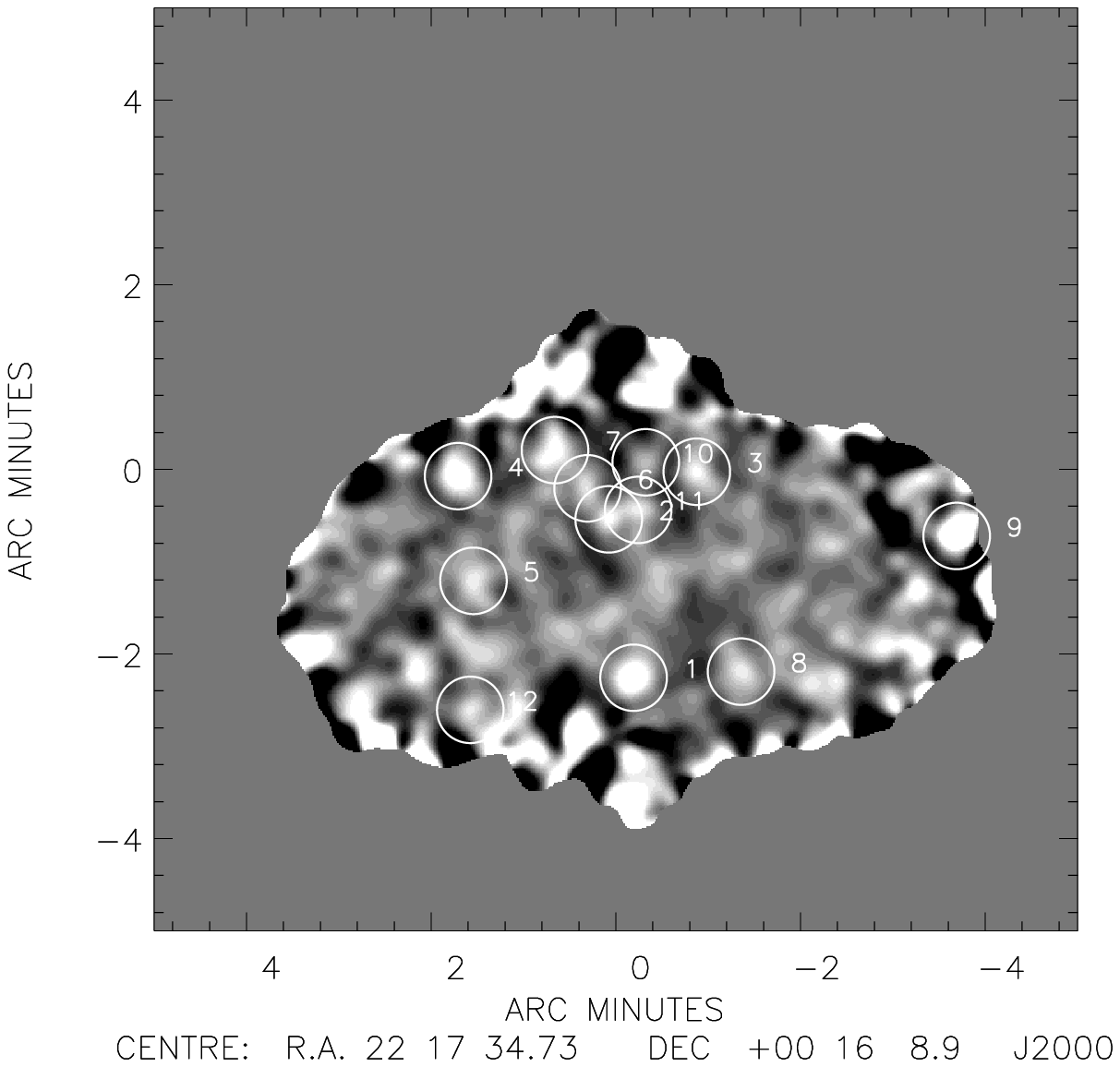}
\label{fig:lhdeep} 
\caption{\small{The $\rm 850 \, \mu m$ image of the SSA22 field
   from the Hawaii Flanking Fields Survey,
smoothed with a beam-size Gaussian (14.5 arcsec FWHM). The
numbered circles highlight those sources found at a significance of
$>3.00$. The labelling
corresponds to the numbers in Table 14.}}
\end{figure}

The ``Hawaii Flanking Fields Survey (HFFS)'' (Barger et al. 1998,
Barger, Cowie \& Sanders 1999, Barger, Cowie \& Richards 2000), covers a total
of $\simeq 110$\, sq. arcmin over 4 regions of sky. 
The Lockman Hole deep field is a small pencil beam map
(Barger et al. 1998), covering $\simeq 8$\,
sq. arcmin of sky to an rms noise level of $\rm \sigma_{850}=0.8\,mJy/beam$.
The SSA13 field is composed of a deep pencil
beam area ($\simeq 8$\, sq. arcmin in size, $\rm
\sigma_{850}=0.7$\,mJy/beam; Barger et al. 1998),
embedded in a wider-area, shallower map covering an additional
45\,sq. arcmin with a typical rms noise level of $\rm
\sigma_{850}=2.5$\,mJy/beam (Barger, Cowie \& Sanders 1999). 
The SSA17 and SSA22 fields (Barger, Cowie \& Sanders 1999) both have
regions of uniform noise
covering approximately 20\,sq. arcmin of sky, the SSA17 field to
$\rm \sigma_{850}=1.6$\,mJy/beam, and the SSA22 field to $\rm
\sigma_{850}=0.9$\,mJy/beam. 

The data were originally reduced using the standard SURF pipeline, 
and source extraction was carried out by
convolving the signal maps with the beam. In order to determine the
absolute noise levels, Barger, Cowie \& Sanders (1999) first
eliminated any significant sources (estimated to be at the $\rm \ge 2.8
\sigma$ level) by subtracting appropriately normalised versions of the
beam profile. They then placed beam-sized apertures at random
positions on the residual signal map, using the standard deviation
between the enclosed pixels to estimate the noise. These values were
then used to iteratively adjust the normalization of the variance
array values, until the dispersion of the signal-to-noise ratio was
approximately one. 

Comparative source lists for detections with $\rm S/N>3.00$ from the
two different reductions and source extraction procedures are given in 
Tables 11 - 14 (see also Figs 14 - 17). 
Columns 1-6, 8 and 9 are analogous to those of Tables 7 to 10. 

In one case, a source previously identified by Barger, Cowie \&
Sanders (1999) has been marked with `bad bol.' as this region
appears to have been observed with a bad bolometer, hence the
``source'' is most likely to be an artefact of SURF's attempt to
interpolate between neighbouring areas of good quality data. Column 7
gives any previous reference to the $\rm 850 \, \mu m$ 
source. Reference B99 is an abbreviation for Barger, Cowie \& Sanders
(1999). 

The two reduction and source-extraction algorithms again produce quite
different results. There is no clear trend in the variations between
the independent measurements of signal-to-noise for those sources
detected significantly in both analyses; in some cases the SURF
reduction yields a higher signal-to-noise estimate whereas for other
sources the situation is reversed. The IDL-based reduction and maximum
likelihood algorithm has identified nearly twice as many peaks at the
$\rm >3.00\sigma$ level as the original catalogue of Barger, Cowie \&
Sanders (1999), including all of their original $\rm >4.00\sigma$
objects. The comparative results for the Lockman Hole deep field given
in Table 11 show that of the two sources published in Barger, Cowie \&
Sanders (1999), their most significant source is confirmed at $> 6\sigma$.
The most discrepant of the fields is the SSA13 field (Table 12), which 
has particulaly uneven noise 
when compared to all of the
other survey fields and it is most likely because of this that only
4/9 of the original detections could be identified at $\rm S/N
>3.00$. In the deep part of the SSA13 field, the top two
sources are confirmed at better than $\rm 5.00 \sigma$ in this
analysis. However, in the wider shallower area only one further source
listed in Barger, Cowie \& Sanders (1999) could be confirmed at $\rm
S/N > 4.00$. Our re-analysis of the SSA17 field (Table 13) found only
one of the original sources with $\rm S/N > 3.50$ and with only 3/5 of
the objects listed in Barger, Cowie \& Sanders at $\rm > 3.00 \sigma$.
In the SSA22 field, all 5 of the $\rm >3.00\sigma$ detected
by Barger, Cowie \& Sanders (1999) are recovered, with firm agreement
on the top 2 sources at better than $\rm 4.00\sigma$ in both catalogues.

\subsection{The Hubble Deep Field Submillimetre Survey}
\begin{figure}
 \centering
   \vspace*{10cm}
   \leavevmode
   \includegraphics{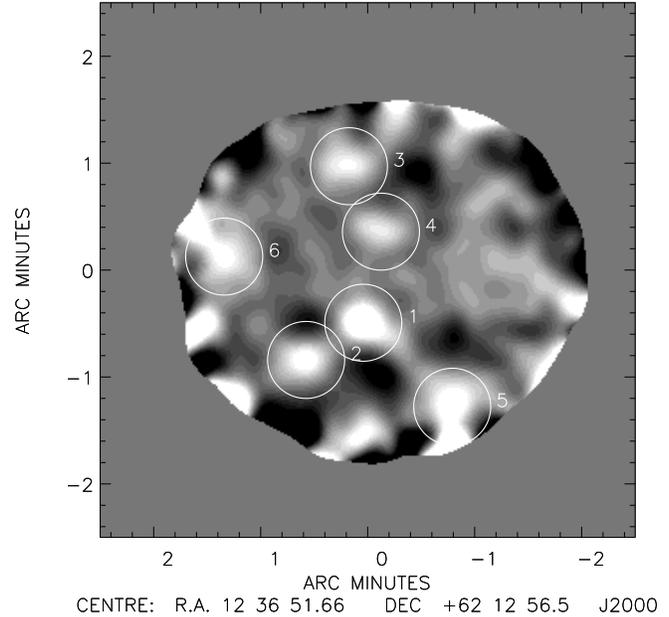}
\label{fig:lhdeep} 
\caption{\small{The $\rm 850 \, \mu m$ image of the Hubble Deep Field North,
smoothed with a beam-size Gaussian (14.5 arcsec FWHM). The
numbered circles highlight those sources found at a significance of
$>3.00$. The labelling
corresponds to the numbers in Table 15.}}
\end{figure}

The Hubble Deep field is the deepest of the submillimetre surveys,
covering approximately 6 sq. arcmin of sky down to
the confusion level of $\rm \sigma_{850}=0.5$\,mJy/beam. It differs
slightly from the other SCUBA surveys in that two different chop
throws of 30'' and 45'', both fixed in celestial coordinates, were
applied to the observations, each for approximately half of the
integration time. Both SURF and IDL reductions have previously been
carried out on this field (Hughes et al. 1998 and Serjeant et al. 2003
respectively), but the maximum-likelihood simultaneous-fitting
algorithm has not been used previously. 

Comparative source lists for detections with $\rm S/N>3.00$ from the
three different reductions and source extraction procedures are given in 
Table 15 (see also Fig 18).
Column 1 gives the source number corresponding to the labelling on
Fig. 18, in order of decreasing signal-to-noise. Columns 2 and 3 give
the right ascension and declination of the source in J2000
coordinates. Column 4 gives the simultaneously fitted $\rm 850 \, \mu
m$ flux densities of the sources. The error includes a 10\%
calibration error combined in quadrature. Column 5 gives the measured
signal-to-noise ratio of the source from the simultaneously fitted
model. Column 6 defines the noise region in which the source was
found; `central' corresponds to the parts of the map which have seen
the full integration time (outside of the deep area), and `edge' corresponds to the rather
noisier regions near the perimeter which have not seen the full integration
time. Column 7 gives any previous reference to the $\rm 850 \, \mu m$
source. Reference H98 is an abbreviation for Hughes et al. (1998), and
Serj03 is an abbreviation for Serjeant et al. (2003). The presence of
a * indicates that Hughes et al. (1998) deconvolved two sources here,
whereas Serjeant et al. (2003) and my own reduction extracted only one.
Column 8 gives the previously recorded signal-to-noise
ratio where applicable, and column 9 gives the distance between the
listed and previously referenced positions. The table listings given
in italics correspond to previously referenced sources with $\rm S/N >
3.00$, which did not meet this criteria in this analysis.

The analysis of Serjeant et al. (2003) combined the deep jigglemap
imaging of the Hubble Deep Field North with additional photometry
which both increased the map size slightly and improved the sampling
at some positions around the edges. Since photometry measurements require
pre-selection of a known object's position this would bias the number
of sources recovered in this region of sky and so this data has not been
included in the analysis presented here. Consequently the sources
labelled as HDF.6 and HDF.7 presented in Serjeant et al. (2003) which
do not peak within the main jigglemap imaging have
not been re-identified in this analysis although they have been
recovered in the ``HDF-N Supermap'' of Borys et al. (2003). However,   
two new sources which have eluded previous analyses, have been
recovered at the $\rm S/N >3.50$ level and all of the original
detections, with the exception of HDF.3 in the Hughes et al. (1998)
analysis, have been confirmed with $\rm S/N > 4.00$.

\section{Source Counts}
\begin{table*}
\begin{tabular}{|c|c|c|} \hline
Flux density & Raw 850~${\rm \mu m}$ source counts & Corrected 850~${\rm \mu
m}$ source counts \\
/mJy & $N (>S) \mathrm{deg^{-2}}$ & $N (>S) \mathrm{deg^{-2}}$ \\
\hline
\phantom{1}2.0 & $2880 \pm 310$ & $3920^{+\phantom{0}550}_{-1120}$ \\
\phantom{1}2.5 & $1680 \pm 180$	& $2140^{+\phantom{0}320}_{- \phantom{0}620}$ \\
\phantom{1}3.0 & $1080 \pm 120$ & $1250^{+\phantom{0}200}_{- \phantom{0}370}$ \\
\phantom{1}3.5 & $\phantom{0}830  \pm 100$ & $ \phantom{0}680^{+\phantom{0}120}_{- \phantom{0}200}$ \\
\phantom{1}4.0 & $\phantom{0}700  \pm \phantom{0}90$ & $ \phantom{0}620^{+\phantom{0}110}_{- \phantom{0}190}$ \\
\phantom{1}4.5 & $\phantom{0}700  \pm \phantom{0}90$ & $ \phantom{0}490^{+\phantom{00}90}_{- \phantom{0}150}$ \\
\phantom{1}5.0 & $\phantom{0}540  \pm \phantom{0}70$ & $ \phantom{0}380^{+\phantom{00}70}_{- \phantom{0}120}$ \\
\phantom{1}5.5 & $\phantom{0}500  \pm \phantom{0}70$ & $ \phantom{0}330^{+\phantom{00}70}_{- \phantom{0}110}$ \\
\phantom{1}6.0 & $\phantom{0}420  \pm \phantom{0}60$ & $ \phantom{0}310^{+\phantom{00}60}_{- \phantom{0}100}$ \\
\phantom{1}6.5 & $\phantom{0}340  \pm \phantom{0}60$ & $ \phantom{0}230^{+\phantom{00}50}_{- \phantom{00}80}$ \\
\phantom{1}7.0 & $\phantom{0}300  \pm \phantom{0}50$ & $ \phantom{0}180^{+\phantom{00}50}_{-\phantom{00}60}$ \\
\phantom{1}7.5 & $\phantom{0}260 \pm \phantom{0}50$ & $ \phantom{0}180^{+\phantom{00}50}_{-\phantom{00}60}$ \\ 
\phantom{1}8.0 & $\phantom{0}210 \pm \phantom{0}40$ &	$ \phantom{0}150^{+\phantom{00}40}_{-\phantom{00}60}$ \\
\phantom{1}8.5 & $\phantom{0}160 \pm \phantom{0}40$ &	$ \phantom{0}100^{+\phantom{00}30}_{-\phantom{00}40}$ \\
\phantom{1}9.0 & $\phantom{0}130 \pm \phantom{0}30$ &	$\phantom{00}70^{+\phantom{00}30}_{-\phantom{00}30}$ \\
\phantom{1}9.5 & $\phantom{00}90  \pm \phantom{0}30$ &	$\phantom{00}60^{+\phantom{00}30}_{-\phantom{00}30}$ \\
10.0 &    $\phantom{00}70\pm  \phantom{0}20$ &	$\phantom{00}40^{+\phantom{00}20}_{-\phantom{00}20}$ \\
10.5 & $\phantom{00}60 \pm \phantom{0}20$ & $\phantom{00}20^{+\phantom{00}10}_{-\phantom{00}10}$ \\
11.0 & $\phantom{00} 30 \pm \phantom{0}20$ & $\phantom{00}10^{+\phantom{00}10}_{-\phantom{00}10}$ \\
11.5 & $\phantom{00}20 \pm \phantom{0}10$ & $\phantom{00}10^{+\phantom{00}10}_{-\phantom{00}10}$ \\ 
12.0 & $\phantom{00}20 \pm \phantom{0}10$ & $\phantom{00}10^{+\phantom{00}10}_{-\phantom{00}10}$ \\
12.5 & $\phantom{00}10 \pm \phantom{0}10$ & $\phantom{00}10^{+\phantom{00}10}_{-\phantom{00}10}$ \\ \hline 
\end{tabular}

\caption{\small The 850~${\rm  \mu m}$ source counts per square
degree based on sources with S/N $>3.50$ in both survey maps, and excluding those detected
in the non-uniform noise regions. Column 1 gives the flux density and column 2 the cumulative raw counts per
square degree with the Poisson error. Column 3 gives the cumulative corrected
counts per square degree, the upper error corresponding to the Poisson
error, and the lower error accounting for both the Poisson error and
the presence of spurious sources based on the simulation data.}
\end{table*}

Previous measurements of the cumulative number counts have differed quite
markedly between the various surveys (eg. Blain et al. 1999, Barger,
Cowie \& Sanders 1999, Eales et al. 2000, Scott et al. 2002),
particularly at bright flux densities ($\rm >5\,mJy$). This
is not surprising given the small area of sky observed by each
individual survey and the low number density of bright
sources. In addition to cosmic variance, if the SCUBA sources have a tendency to cluster on scales of
a few arcminutes (for which evidence is given in the next section)
the problem is further compounded since it is more likely that a
mapping a blank
field of only several tens of square arcminutes in size will fall in
an area of lower source density. Here, the sources detected with
a signal-to-noise ratio $>3.50$ from our reanalysis of the CUDSS,
Hawaii and HDF surveys have been combined with those identified in the
8\,mJy Survey, to produce the most accurate number counts to date,
from $2-12.5$\,mJy at 0.5\,mJy intervals. The regions of non-uniform
noise towards the edge of the maps and any sources they contained 
were excluded. The simulations described in Section 4.1 were used to
correct for the effects of flux-density boosting and incompleteness on
a field by field basis. Estimating
the level of contamination from spurious / confused sources, however,
requires the generation of fully simulated images. The accuracy of
such images is hampered by the lack of knowledge regarding the
clustering properties of the SCUBA population down to the faintest
flux density levels. The simulations described in Section 4.2 imply
that the fraction of spurious / confused sources (defined
as having no input source at least half as bright as the output flux
density) at a significance of $\rm >3.50 \sigma$ is $\simeq30\%$.
This number is in line with the upper limit placed on the confused /
spurious fraction from deep radio follow-up of the ``8\,mJy Survey''
fields (Ivison et al. 2002). The raw and
corrected cumulative number counts are given in Table 16 along with
$\rm 1 \sigma$ Poisson error bars (calculated from the square root of
the number of detections on which the source count was based). 
The estimated 30\% fraction of spurious / confused sources has been
combined in quadrature with the lower Poisson error bar in the
corrected number counts.

\begin{figure*}
 \centering
   \vspace*{13cm}
   \leavevmode
   \includegraphics{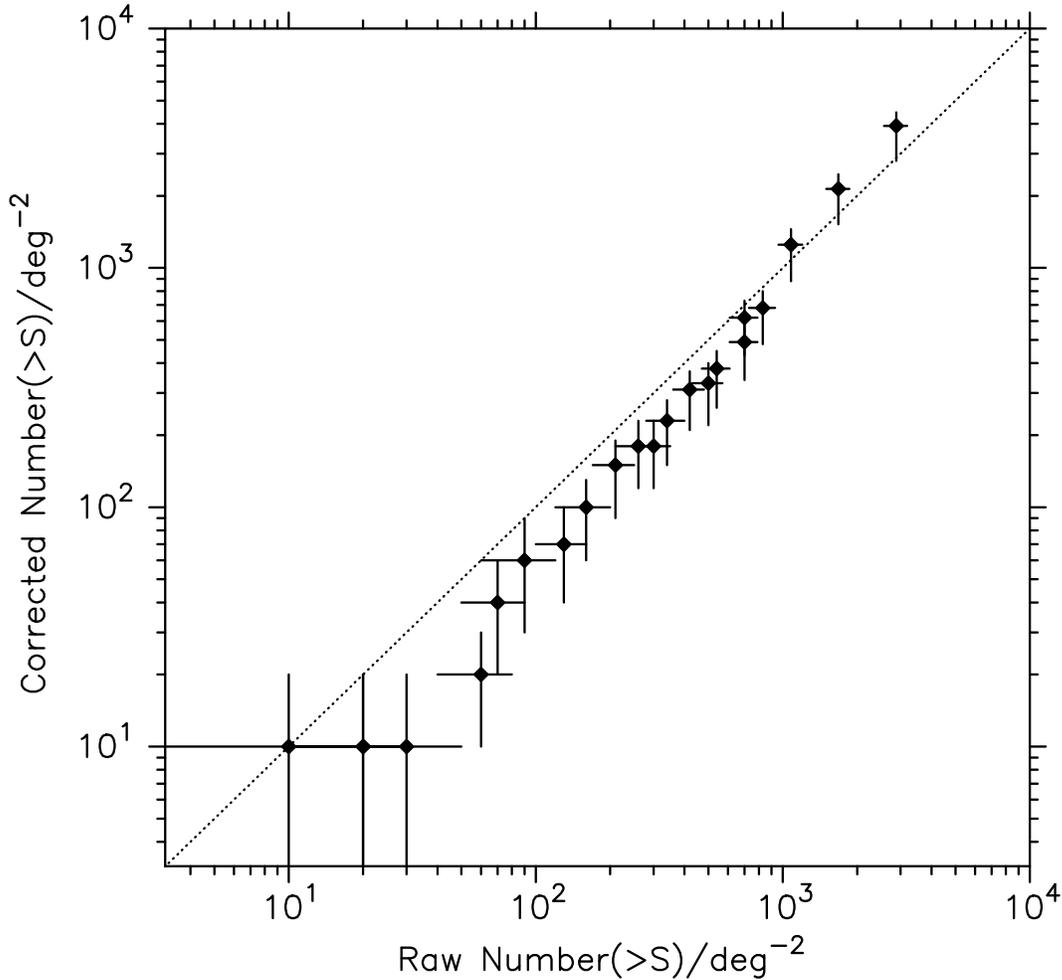}
\label{fig:corr_v_raw_count} 
\caption{\small{Cumulative number counts corrected for the effects of
   boosting and incompleteness versus the raw number counts. The
 dotted line marks the case where the raw and corrected cumulative
 numbers counts are the same. The error bars are as given in Table 16.}}
 \end{figure*}
\begin{figure*}
 \centering
   \vspace*{18.5cm}
   \leavevmode
   \includegraphics{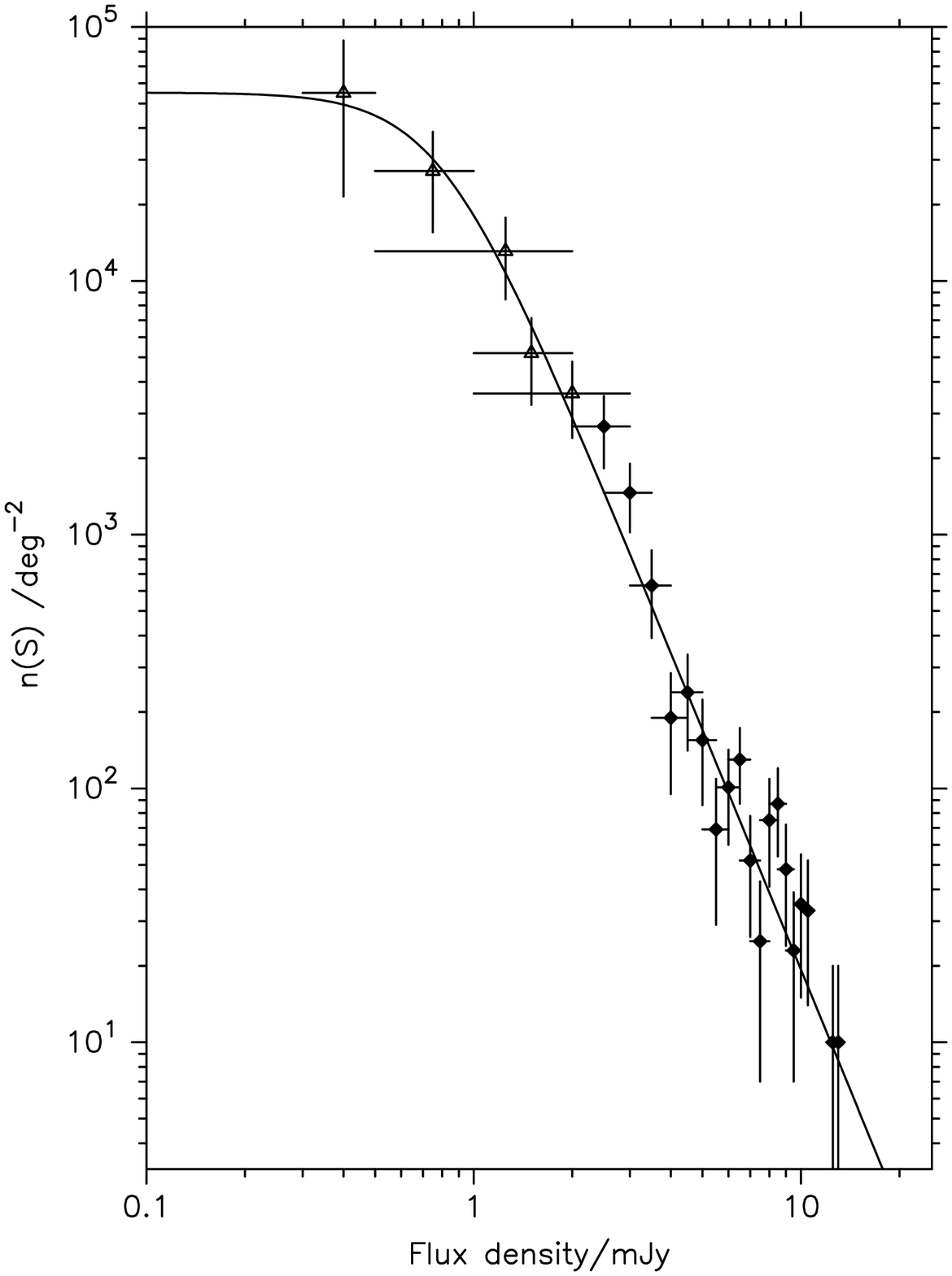}
\label{fig:diff_count} 
\caption{\small{A plot of differential number counts vs. flux
   density. The solid diamonds represent data from the combined blank
 field survey re-analysis only, with $\rm 1\sigma$ Poisson error bars
 on the y-axis, and the flux density range on which the differential
 count is based marked as an error bar on the
 x-axis. The open triangles represent data points from the lensing
 surveys of Blain et al. (1999) and Cowie et al. (2002). The solid curve
 is a best fit parametric model of the form $\rm n(s)=\frac{N_{0}}{(a+ S^{\alpha})}$.}}
 \end{figure*}

Figure 19 shows the corrected versus the raw number counts at each
flux density, together
with a dotted line marking the locus of where the raw and corrected
counts take the same value. An increase in the cumulative number
counts along either axis correponds to a decrease in the flux density
threshold. One can see immediately that at brighter flux densities,
the effect of boosting is stronger than incompleteness and hence the
real number density of sources above a specified flux density threshold is
lower than the directly measured value. This may also be implied from
an examination of the
raw data alone. The area of sky mapped in the ``SCUBA 8\,mJy Survey''
is almost identical to the area of sky mapped by the other deeper
surveys combined. At bright flux densities the deeper surveys are
essentially complete and so one can compare the number of detections
above a specified signal-to-noise ratio and flux density threshold
between the shallower and deeper maps. For a retrieved $\rm S_{850\mu
m}>10\,mJy$ and $\rm S/N >3.50$, the ``SCUBA 8\,mJy Survey'' identified 8
sources, whereas in all the other surveys only 1 source was found
to satisfy this criteria. Similarly, for a retrieved $\rm S_{850\mu
m}>8\,mJy$ and $\rm S/N >3.50$, the ``SCUBA 8\,mJy Survey'' identified
20 objects, whereas only a total of 5 were found in the other
surveys. Given the small area of sky considered, part of the
discrepancy between the number of bright sources recovered in the
``SCUBA 8\,mJy Survey'' and the other deeper blank field surveys can
be explained by small number statistics but it seems highly unlikely
that this is the sole cause. Moving towards fainter flux density thresholds, the increased source
density makes incompleteness more of a problem and around the
confusion limit of $\rm \sim 3\,mJy$ (where the density of sources is
$\simeq 1000$ per square degree) the raw and corrected cumulative
number counts are approximately the same.

One of the simplest ways of describing the number counts is by
carrying out a simple parametric fit of a power-law model to the
differential number counts as a function of flux density. The
differential number counts at a specific flux density were determined
by the difference in cumulative counts between the two values on
either side of the data point, divided by the change in flux density,
which approximates to a measure of the gradient of a tangent to the cumulative
number counts curve at that point. The completeness and boosting
corrected number counts were used for this procedure. Following
Barger, Cowie \& Sanders (1999), a model of the form
\be \rm n(S) = \frac{dN(S)}{dS} = \frac{N_{0}}{(a+ S^{\alpha})} \ee
\noindent was fitted to the 
differential  
data points by means of a minimised $\rm \chi^{2}$
method, where the values of $\rm N_{0}$, a and $\rm \alpha$ were
allowed to vary freely. The best fit values for these parameters were $\rm N_{0} = 2.67 \times 10^{4}$, $\rm a=0.49$ and
$\rm \alpha=3.14$, predicting a total $\rm 850 \, \mu m$
background of $\rm 3.8 \times 10^{4}\,mJy\,deg^{-2}$, mid-way between
the $\rm 850 \, \mu m$ extragalactic values of $\rm 3.1 \times
10^{4}\,mJy\,deg^{-2}$ and $\rm 4.4 \times 10^{4}\,mJy\,deg^{-2}$ as
measured from COBE-FIRAS by Puget et al. (1996) and Fixsen et
al. (1998) for the lower and upper values respectively. The values of
the fitted parameters are in fact very close to the values
originally determined by Barger, Cowie \& Sanders (1999) who found $\rm N_{0} = 3.0 \times 10^{4}$, $\rm a=0.5$ and
$\rm \alpha=3.2$. Figure 20 shows a plot of the differential number
counts against flux density. The solid diamonds represent data from
the combined blank field survey re-analysis only, with $\rm 1\sigma$
Poisson error bars on the y-axis and a flux density range of 1\,mJy on the
x-axis, corresponding to the change in flux density between the two
data points on either side of the point at which the differential
counts has been determined. The open triangles represent data points
from the lensing surveys of Blain et al. 1999 and Cowie et al. 2002,
and the solid curve is the best fit parametric model.

\begin{figure*}
 \centering
   \vspace*{18.5cm}
   \leavevmode
   \includegraphics{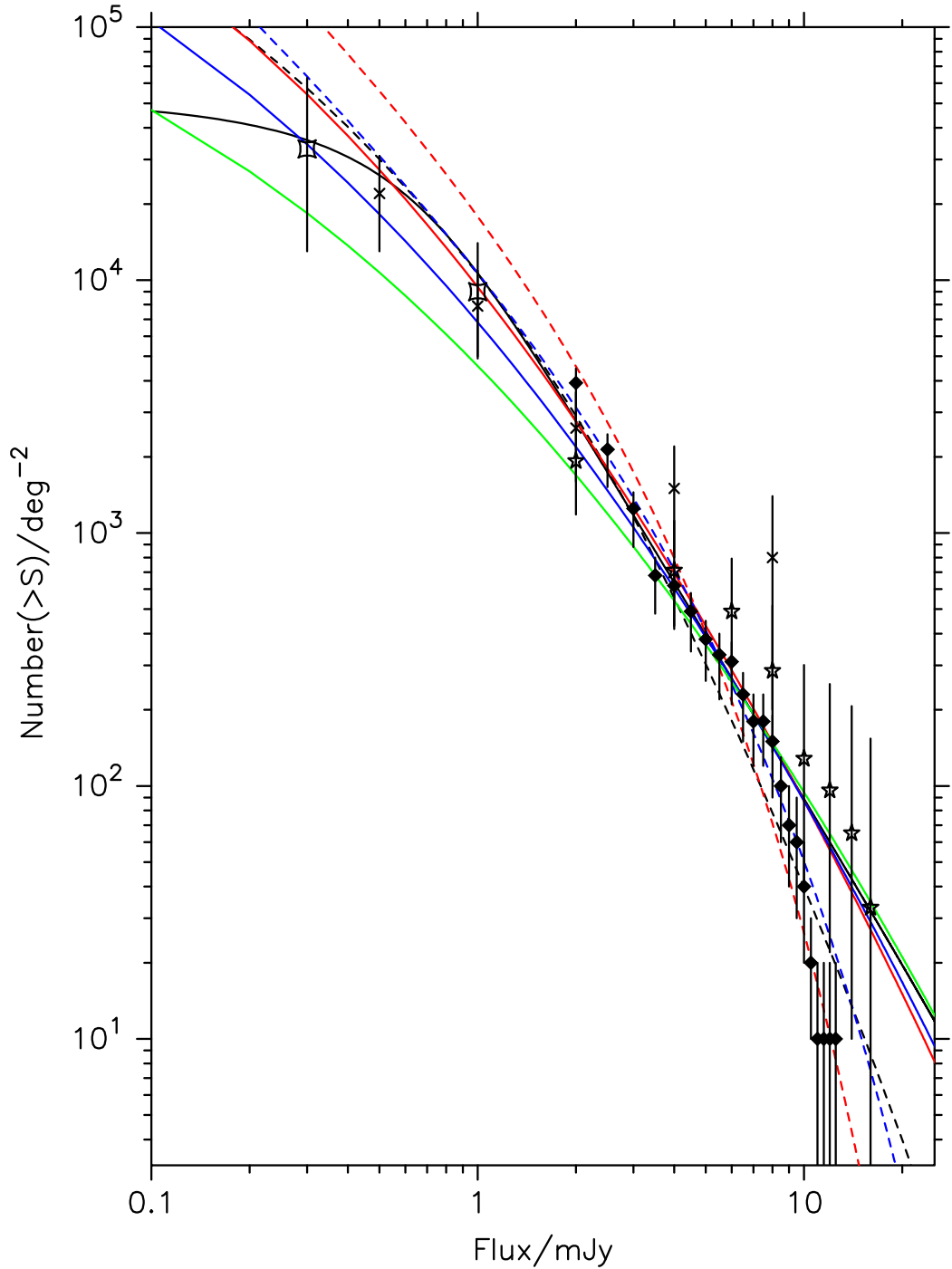}
\label{fig:cum_count_0307} 
\caption{\small{A plot of cumulative number counts vs. flux
   density, along with a series of models assuming an $\rm
   \Omega_{M}=0.3$, $\rm \Omega_{\Lambda}=0.7$ cosmology.
The solid diamonds represent the cumulative source counts from the
   completeness and boosting corrected counts derived from the combined blank
 field survey re-analysis only, with error bars as given in Table 16.
  The crosses and round-edged squares represent data points from the lensing
 surveys of Blain et al. (1999) and Cowie et al. (2002) respectively,
while the stars indicate the results of and Borys et al. (2003). 
The various models are
   described in the main text.}}
 \end{figure*}

Fig. 21 shows the corrected cumulative number counts, as
well as a series of models predicting the $\rm 850 \, \mu m$ source
counts for an adopted $\rm \Omega_{M}=0.3$, $\rm \Omega_{\Lambda}=0.7$ 
cosmology, in line with the most
recent cosmological parameter results from the supernova cosmology
project (Perlmutter et al. 1999), X-ray clusters (Allen, Schmidt \&
Fabian 2002a, 2002b), and WMAP observations of the cosmic microwave
background (Spergel et al. 2003). A value of $\rm H_{0}=67\,
km\,s^{-1}\,Mpc^{-1}$ was assumed throughout. The black lines are
parameterised models, the solid lines following a simple power-law
description of the data points and are the same for both figures. The
dashed lines represent work by Rowan-Robinson (2000,2001) who generated
best-fit
models of the measured far-infrared through to submillimetre number counts, as
constrained by the extragalactic background, for
a number of assumed cosmologies. In the remaining models (red, blue
and green curves), values of dust temperature $\rm T_{d}$
and emissivity index $\rm \beta$  based on optically thin greybody
emission, were allowed to vary where necessary so that as good a fit
as possible could be obtained for the  $\rm \Omega_{M}=0.3$, $\rm
\Omega_{\Lambda}=0.7$ cosmology. 

The solid black line is the
integrated $\rm dN/dS$ best-fit parametric model given in equation
13, providing a good description of the cumulative number counts for
$\rm S_{850}<10\,mJy$.
Rowan-Robinson (2001) has also taken a parameterised approach to
modelling the infrared through to submillimetre counts and
backgrounds, in this case by using multiwavelength observational data
in these wavelength regimes to place constraints on models made up of
four spectral components: infrared cirrus, an M82-like starburst, an
Arp220-like starburst and an AGN torus. The model assumes that the
evolution  of the star formation rate manifests itself as pure
luminosity evolution. The models, consistent with infrared and
submillimetre counts and backgrounds, showed a flat star formation rate
from $z=1-3$, in agreement with other studies of the star formation
history such as the HDF (Hughes et al. 1998). The most striking
difference between other modelling work and Rowan-Robinson (2001) is
the dominant role of the cirrus component at submillimetre
wavelengths. The black dashed line in Fig. 21 shows the
model of Rowan-Robinson (2001) which appears to be the best-fit of
those presented here to the cumulative $\rm 850 \, \mu m$ source count 
data points over the whole flux density range.

The red lines on Fig. 21 represent a pure luminosity evolution of
the form $(1+z)^{3}$ out to a threshold redshift, and constant
thereafter. The solid line began with the $\rm 60\,\mu m$ luminosity
function of Saunders et al. (1990), interpolated to $\rm 850\,\mu
m$ assuming an optically thin greybody with a single dust temperature
and emissivity index, to describe the far-infrared / submm dust
SED. The same values for $\rm T_{d}$ and $\rm \beta$ were assumed in
calculating the K-correction for consistency. The dashed line began
with the $\rm 850\,\mu m$ luminosity function measured by Dunne et
al. (2000). A dust temperature of $\rm 40\,K$ and emissivity index of
$\rm \beta=1.3$ appeared to best satisfy the data points, using
threshold redshifts of $z=2.0$ and $z=1.5$ for the Saunders et
al. (1990) and Dunne et al. (2000) based luminosity functions
respectively.
This simple description of luminosity evolution appeared
to work well for the interpolated Saunders et al. (1990) $\rm 60 \,\mu
m$ luminosity function for $\rm S_{850}<8\,mJy $, but predicted too
many sources brighter than this. Conversely, using the Dunne et
al. (2000) $\rm 850\,\mu m$ luminosity function, a good fit to the
data points was found for $\rm S_{850}>2\,mJy $ but overpredicted the
number of sources fainter than this by a factor of 2-3.

The solid and dashed blue lines on the cumulative number density plot also
assume pure luminosity evolution of the interpolated $\rm 60\, \mu m$
Saunders et al. (1990) and directly measured $\rm 850\, \mu m$ Dunne
et al. (2000) luminosity functions respectively. In this case,
however, the luminosity evolution $\rm g(\mathit{z})$ takes a more
realistic form, which is fully compatible with models of cosmic
chemical evolution and naturally includes a peak in the evolution
function (Jameson et al. 1999, Smail et al. 2002):
\be \rm g(\mathit{z})=(1+\mathit{z})\rm^{3/2} sech^{2}[bln(1+\mathit{z})\rm
- c]cosh^{2}c \ee
\noindent where best fits values for the parameters b and c based on
multiwavelength far-infrared to submillimetre counts and constraints
on the probable redshift distribution of the $\rm 850 \,\mu m$
population (Smail et al. 2002) are $\rm b=2.2 \pm 0.1$ and $\rm c=1.84
\pm 0.1$. The interpolated $\rm 60\, \mu m$
luminosity function combined with this description of pure luminosity
evolution fits the data well for $\rm S_{850}<8\,mJy$ assuming an
optically thin greybody description of the thermal dust emission with
parameters $\rm T_{d}=39\,K$ and $\rm \beta=1.3$. The data points
across the whole flux density range are fitted well using the Dunne et
al. (2000) $\rm 850 \, \mu m$ luminosity function assuming dust
emission parameters of  $\rm T_{d}=30\,K$ and $\rm \beta=1.0$,
although it should be noted that this combination of dust temperature
and emissivity index does not describe the spectral energy
distribution well for any known local galaxies based on the ``SCUBA
Local Universe Galaxy Survey (SLUGS)'' (Dunne et al. 2000).

The solid green line on Fig. 21 uses
the $\rm 60 \,\mu m$ luminosity function of Saunders et al. (1990)
combined with luminosity evolution taking the form $(1+z)^{4}$ up to a
threshold redshift $z_{\mathrm{thresh}}$, and $(1+z)^{-4}$ thereafter
(Chapman et al. 2002a). They find a best fit of
$z_{\mathrm{thresh}}=2.6$ for the transitional redshift, based on
current constraints on the redshift distribution as well as the source
counts. This particular $\rm 850 \, \mu m$ counts
model has a rather shallower gradient than the others and could only
be made to match number counts of this combined re-analysis over the
flux density range $\rm \sim 3-7\, mJy$ for $\rm T_{d}=36\,K$ and a very
extreme $\rm \beta=2.0$. Possibly a higher value of the dust
emissivity index could produce a steeper counts model, however, such
high values are not observed in in any known class of objects. No
satisfactory fit could be obtained beginning with the Dunne et
al. (2000) $\rm 850 \,\mu m$ luminosity function. The proposed Chapman et
al. (2002a) luminosity evolution does not seem consistent
with the blank field survey source counts, although the number counts
derived from the lensing survey of Blain et al. 1999 are much
shallower, predicting rather higher number densities of bright sources, and could plausibly be fit by this scenario for less
extreme values of $\rm beta$.

One particularly interesting feature of this source counts re-analysis
is the apparent steepening of the cumulative number counts beyond $\rm
S_{850}> 8\,mJy$. This could at least in part be due to small number
statistics of the brightest sources, however, if real has some
important implications. Firstly, it could indicate an intrinsic
turn-over in the underlying luminosity function. In turn, this would
suggest a very interesting upper limit on the luminosity of a high
redshift galaxy, perhaps reflecting an upper limit on the overall mass of the
system. This could place useful constraints on galaxy formation
theories. Secondly, a steepening of the 
bright 
source counts could make the
SCUBA population much more prone to the effects of gravitational
lensing. For weak lensing scenarios, the ratio of the number of
observed sources brighter than a flux density threshold to the true
number of sources brighter than that flux density threshold is
given by
\be \rm \frac{N_{obs}(>S)}{N_{true}(>S)}=\mu^{\gamma-1} \ee
\noindent where $\rm \mu$ is the magnification amplitude and $\gamma$
is the slope of the cumulative source counts. For $\rm
1<S_{850}<8\,mJy$ the counts slope $\rm \gamma \simeq 2.5$, whereas for
$\rm S_{850}>8\,mJy$ this increases to $\rm \gamma \simeq 5.5$. No attempt
has been made to correct the $\rm 850 \, \mu m$ source counts for the
effects of lensing, but there is some evidence to suggest that some
blank field sources have been gravitationally lensed. Almaini et
al. (2003) found a strong cross-correlation in the ELAIS N2 field
between the distribution of SCUBA and Chandra sources even though the
coincidence of detections in the submillimetre and X-ray wavelengths
was only $\sim 5\%$. One proposed explanation for this effect is that
the SCUBA and Chandra sources trace the same large scale structure at
high redshift ($z>1$),
however, there also appears to be a similarly strong cross-correlation
between the low redshift $I-$band sources and the higher redshift SCUBA
and Chandra sources in this field (Almaini et al. 2005) suggesting
an alternative explanation may be valid: the SCUBA and Chandra
detections may have been magnified in certain regions of the field by
the presence of high mass density structure, as traced by the $I-$band
imaging, at $z\simeq 0.5$. Chapman et al. (2002b) have also pointed
out that some submillimetre sources have apparent counterparts which
are optically bright galaxies, $I<21.5$, lying at modest redshifts,
$z<1$. This could of course be explained by these counterparts being
correct, reflecting a population of galaxies which are detected as
very cold, luminous submillimetre sources, however, a second
explanation for such systems is that the optically bright galaxy is a
foreground object acting as a gravitational lens, amplifying the more
distant SCUBA galaxy. The detection of luminous molecular CO emission
at the redshift of the optically bright galaxy would provide a
powerful test to distinguish between these two scenarios. If the
latter explanation  of gravitational lensing is correct, Chapman et
al. (2002b) estimate that up to $\rm 3-5\%$ of the $>10$\,mJy
submillimetre sources detected in blank field surveys could be
gravitationally amplified by foreground galaxies.

Gravitational lensing by clusters of galaxies can of course be used to
study the fainter submillimetre sources, and this technique has been
successfully applied by Blain et al. (1999) and Cowie et al. (2002)
(the data points marked by the crosses and curved-edged squares on
Fig. 21 respectively). The agreement between the source
counts derived from the two cluster-lensing surveys is extremely good at flux
densities below $\rm S_{850} \simeq 2$\,mJy, and there is a smooth transition at
this point between the faint number counts from the cluster-lensing
surveys and the brighter source counts from our combined reanalysis of
the blank field surveys. However, the number counts derived by Blain
et al. (1999) at 4 and 8\,mJy are significantly higher than the
combined blank field survey counts presented in Table 16. The most
likely explanation for this discrepancy is that the numbers quoted by
Blain et al. (1999) suffer badly from small number statistics at
bright flux densities, their lensing survey being composed of a number of
small fields covering a total area of
sky of only a few tens of square arcminutes. In comparison, the total area of
sky observed by the ``8\,mJy Survey'', ``CUDSS'', ``HDF'' and ``Hawaii
Survey'' is 460 sq. arcmin, an order of magnitude larger. If
the bright SCUBA population does strongly cluster on arcminute scales,
as implied by the evidence presented in Section 7, this could also
affect the bright end of the number counts in small area surveys such
as Blain et al. (1999). 

The data points marked on Fig. 21 by open stars are determined
from the 165 sq. arcmin ``Hubble Deep Field North SCUBA Super-map''
presented by Borys et al. (2003). The rms noise level of of this
composite map ranges from $\rm \sigma_{rms}\simeq 0.5-7$\,mJy / beam
(Borys et al. 2003). In general the counts of Borys et al. (2003) are
slightly higher than those determined from this re-analysis, although
they agree to well within the error bars for $\rm
S_{850}<10\,mJy$. At brighter flux densities, the two sets of
cumulative number counts begin to diverge, those of Borys et
al. (2003) continuing to follow an approximately power-law decline
with increasing flux density whereas those presented in this paper
steepen quite markedly. There are several possible reasons for this
discrepancy. Firstly both of these analyses suffer from small number
statistics at the brightest flux densities ($\rm S_{850}>10\,mJy$) and
so cosmic variance may be an issue particularly if these objects tend
to cluster on arcminute scales. Secondly Borys et al. had problems
calibrating their wide area scan-map data which represents the
shallowest regions of the super-map. If they have over-estimated their
flux conversion factors by $\rm \sim 20\%$ the two sets of number
count measurements agree well. Finally 4/5 of the sources
recovered by Borys et al. at $\rm S/N > 3.5$ and $\rm S_{850}>16\,mJy$
have rms noise levels of $\rm \sigma_{rms} >5$\,mJy/beam suggesting
that these may have been recovered in non-uniform edge regions for
which the simulations presented in section 4.2 have shown to have very
poor source reliability. This would obviously increase the cumulative
number count measurements at all flux densities fainter than this.

Overall, the increased accuracy of this new source counts analysis
will allow the evolutionary nature, as well as the dust and
star-forming properties of the submillimetre population to be studied
in much more detail than has been possible before. This will be
particularly useful when combined with knowledge of the redshift
distribution of submillimetre sources, which is slowly being built up
at the present.

\section{Clustering}
If the bright 850~$\rm \mu m$ sources are
indeed the progenitors of massive elliptical galaxies then they should
be strongly clustered, an inevitable result of gravitational collapse
from Gaussian initial density fluctuations since the rare high-mass
peaks are strongly biased with respect to the mass. There is a great
deal of evidence to support the presence of this bias at high
redshift. The correlations of Lyman-break galaxies at $z\simeq 3$
(Steidel et al. 1999) are
almost identical to those of present-day field galaxies, even
though the mass must have been much more uniform at early
times. Furthermore, the correlations increase with UV luminosity
(Giavalisco \& Dickinson 2001) reaching scale lengths of $r_{0} \simeq
7.5 h^{-1}$ Mpc - approximately 1.5 times the present-day value. In
the case of luminous proto-ellipticals an even stronger bias is expected
since one is selecting not just massive galaxies but those that have
collapsed particularly early in order to generate the oldest stellar
populations. This is suggested by studies of the local Universe which
have shown that early-type galaxies are much more clustered than
late-type galaxies (eg. Guzzo et al. 1997, Willmer et al. 1998), and
more recently by the findings of Daddi et al. (2000) who have
investigated the clustering properties of extremely red objects (EROs). They
detect a strong clustering signal of the EROs which is about an order
of magnitude larger than the  clustering of $K-$selected field
galaxies, and also  report a smooth trend of increasing clustering
amplitude with increasing $R-K$ colour, reaching $r_{0} \simeq
11 h^{-1}$ Mpc for $R-K>5$. These results are probably the strongest
evidence to date that the largest fraction of EROs is composed of
ellipticals at $z>1$. There are already some hints of strong clustering 
in the bright sub-mm population from the discovery of a strong excess of 
bright SCUBA sources around high-redshift AGN (Ivison et al. 2000,
Stevens et al. 2003). Blain et al. (2004) have also combined a sample
of 73 submillimetre galaxies over seven fields for which they had
spectroscopic redshifts, finding tentative evidence for strong
clustering of submillimetre galaxies with a correlation length $r_{0}
\sim (6.9 \pm 2.1)h^{-1}$\,Mpc.

\subsection{Angular correlation functions}
The angular correlation function $\rm w(\theta)$ is the
projection of the spatial function on the sky and is defined in terms
of the joint probability $\rm \delta P$ of finding two galaxies
separated by an angular distance $\rm \theta$ with respect to that
expected for a random distribution
\be \rm \delta P = N^{2} [1 + w(\theta)] \delta\Omega_{1} \delta\Omega_{2} \ee
\noindent where $\rm \delta\Omega_{1}$ and $\rm \delta\Omega_{2}$ are
elements of solid angle, and $\rm N$ is the mean surface density of
objects. If $\rm w(\theta)=0$ the distribution is homgeneous. A positive $\rm w(\theta)$, therefore, corresponds to an
over-density of sources separated by distance $\rm \theta$.

There are a variety of possible estimators for $w(\theta)$ as a function of 
pair-count ratios. Following Landy \& Szalay (1993), we have adopted
the estimator:
\be\rm w(\theta) = \frac{(DD -2DR +RR)}{RR} \ee
\noindent in which the variance is minimised to almost Poisson level. DD is the number of distinct data pairs in the real
image within a bin covering a specified range of $\rm \theta$, DR is
the number of cross-pairs between the real and mock catalogues within
the same range of $\rm \theta$, and RR is the number of random-random
pairs. DR and RR are normalised with respect to the total number of
data-data pairs from the real image.

Scott et al. (2002) made the first attempt at measuring an angular
correlation function for the bright SCUBA population. Their
results provided tentative evidence for strong clustering of the
submillimetre sources on scales of 1-2 arcmin but
proved inconclusive due to the small number of sources recovered in
the ``8\,mJy Survey'' alone. More recent attempts to measure the
clustering strength have relied on combining catalogues of
submillimetre sources with Lyman-break galaxies (Webb et al. 2003b) or
Chandra sources (Almaini et al. 2003) and assuming that even though
the objects were rarely coincident, the two populations identified at
different wavelengths were tracing the same large scale
structure. These results further supported
the view that the bright SCUBA galaxies cluster strongly on arcminute
scales, however, in both cases the combined correlation functions were
dominated by the larger number of Lyman-break galaxies or Chandra
sources. 

Further SCUBA galaxy clustering measurements, some 
tentatively detecting clustering, have been made (Blain et al. 2004,
Borys et al. 2004, Webb et al. 2003a). 

Here, the 2-point angular correlation function analysis 
combines the sources identified in the ``8\,mJy Survey'' fields with
those objects detected in this re-reduction of the other blank field
surveys, effectively creating a master catalogue with sources from
double the area of the ``8\,mJy Survey'' alone.

A catalogue of randomly placed fake sources was
created for each of the survey fields, with the
number of fake objects contained in each field's random catalogue
chosen to be directly proportional to the area of the image (number in
mock catalogue $\rm = 0.01 \times Area(/arcsec^{2})$), so as to
reflect a uniform number density across the sky. The wider area fields
which dominate this analysis are largely signal-to-noise
limited even at bright flux densities, and hence there is an increasing
probability of finding significant detections in deeper regions of the
maps. Although the positions were allocated randomly, the Gaussian
convolved noise maps were used to weight the number density of sources
across the image, since a larger density of sources above a
specified signal-to-noise threshold would be expected in regions of
lower noise. In practice, this meant dividing the full image into a
series of sub-images, 20 arcsec by 20 arcsec in size, and
calculating the mean noise level in each of these grid sections.
For sources brighter than $\rm \simeq 5 \,mJy$, the number of sources expected above a constant signal-to-noise
threshold increases approximately as $\rm N(>S_{850\mu m}) \propto
S_{850\mu m}^{-1.5}$, and consequently the relative number of sources in each
sub-image scales as $\langle \mathrm{noise} \rangle^{-1.5}$. The
fake source positions were then allocated according to a Poisson
distribution, masking any positions which were covered by a negative
sidelobe accompanying a significant source in the real survey
data. If the number of fake sources included in the catalogue is very
much larger than the number recovered from the real data (as is the
case here), this is in
essence equivalent to choosing positions by combining the results from
source extraction on a series of fully simulated survey
fields. Detections 
lying in non-uniform regions of the
maps in both the real and fake catalogues were omitted from the
correlation function calculations since many of the objects recovered
in the real data in these high noise regions were likely to be
spurious (see Section 4.2).

Although combining all existing blank field survey data together does increase the number of sources on which a
correlation function can be based, the numbers are still fairly
small. Above a signal-to-noise ratio of 3.50, there are a total of 53
sources detected brighter than 5\,mJy, of which 51 may be used in a
correlation function analysis (the other 2 are lone sources in two of
the deep single SCUBA pointing fields). This number increases to 104
sources brighter than 5\,mJy at $\rm >3.00 \sigma$, of which 100 may
be used in a correlation function analysis. However, this will also
increase the contribution from spurious sources in the catalogue and
may therefore dilute the measured signal.
Figs. 22 and 23 show the directly measured angular correlation
data points for detections brighter than 5\,mJy, with $\rm 1 \sigma$
Poisson error bars, and for signal-to-noise thresholds of $>3.50$ and
$>3.00$ respectively. The bin-size used in both plots is 29 arcsec
(twice the beam-size). Rather surprisingly, there is little change in
\clearpage
\begin{figure}
 \centering
   \vspace*{7.3cm}
   \leavevmode
   \includegraphics{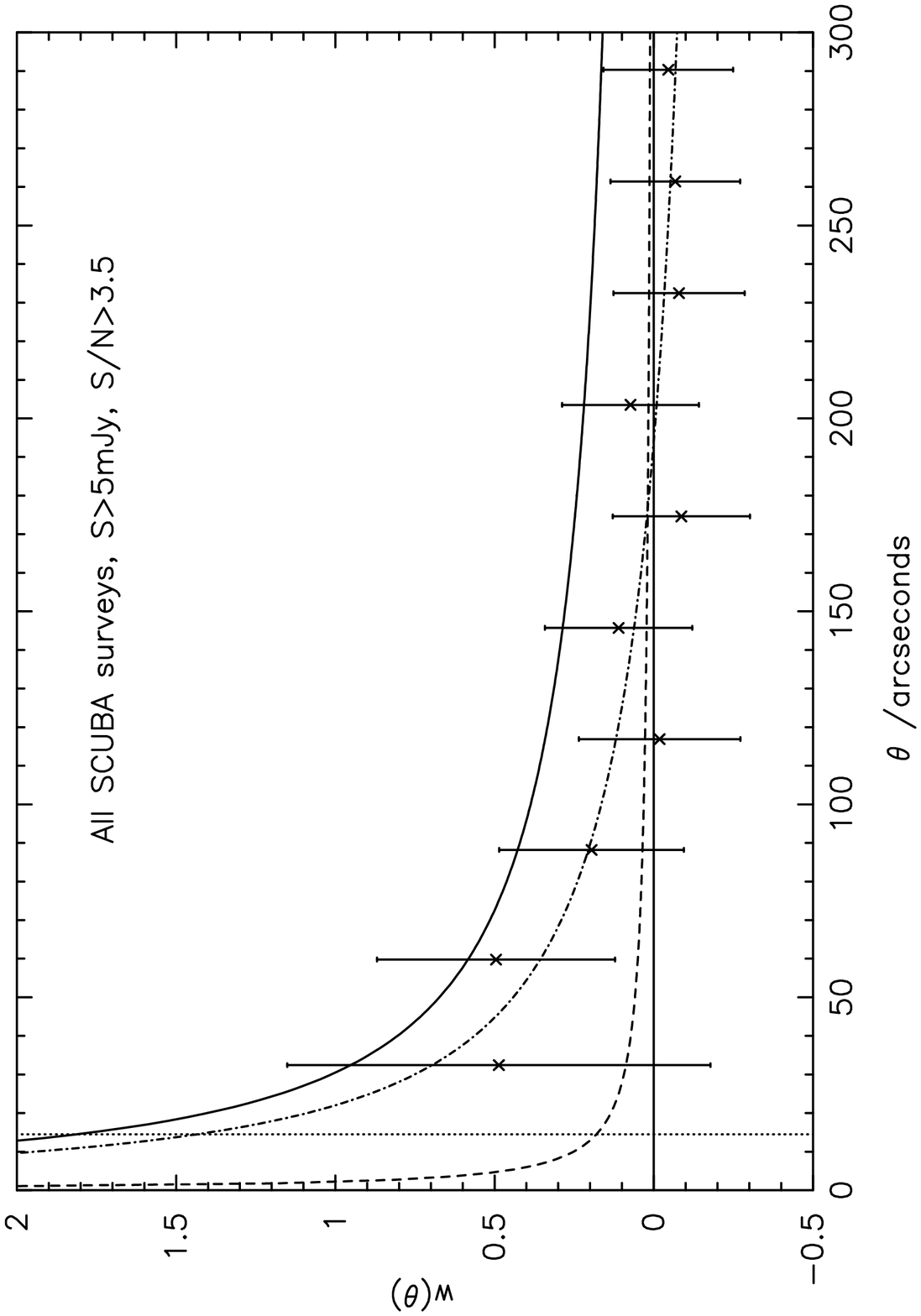}
\label{fig:w_3_5sig_5mjy} 
\caption{\small 2-point angular correlation function for
sources brighter than 5\,mJy, detected at a significance of $\rm >
3.50\sigma$, over all of the survey fields. The error bars are $\rm
1\sigma$ Poisson errors. The solid
power-law line indicates the correlation function found by Daddi et
al. (2000) for EROs with $R-K > 5$ and $K < 18.5$, and the
dashed power-law line indicates the correlation function found by
Giavalisco et al. (1998) for Lyman break galaxies at $z\sim3$. The
dot-dash line shows the best-fit power-law to the data points. The
vertical dotted line indicates the size of the JCMT beam at $\rm 850\,\mu m$.}
 \centering
   \vspace*{7.3cm}
   \leavevmode
   \includegraphics{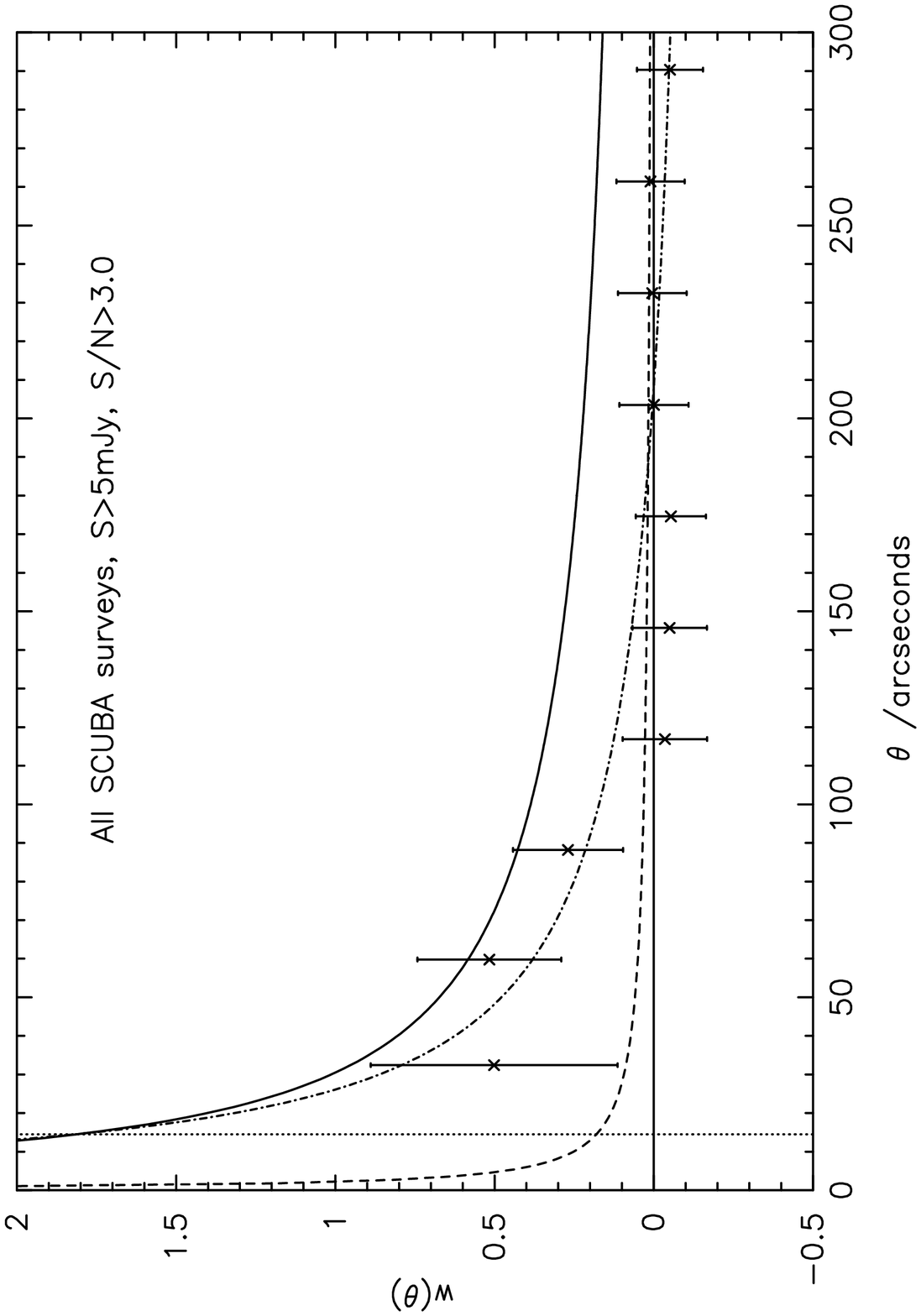}
\label{fig:w_3sig_5mjy} 
\caption{\small 2-point angular correlation function for
sources brighter than 5\,mJy, detected at a significance of $\rm >
3.00\sigma$, over all of the survey fields. The error bars are $\rm
1\sigma$ Poisson errors. The solid
power-law line indicates the correlation function found by Daddi et
al. (2000) for EROs with $R-K > 5$ and $K < 18.5$, and the
dashed power-law line indicates the correlation function found by
Giavalisco et al. (1998) for Lyman break galaxies at $z\sim3$. The
dot-dash line shows the best-fit power-law to the data points. The
vertical dotted line indicates the size of the JCMT beam at $\rm 850\,\mu m$.}
\end{figure}

\begin{figure}
 \centering
   \vspace*{7.3cm}
   \leavevmode
   \includegraphics{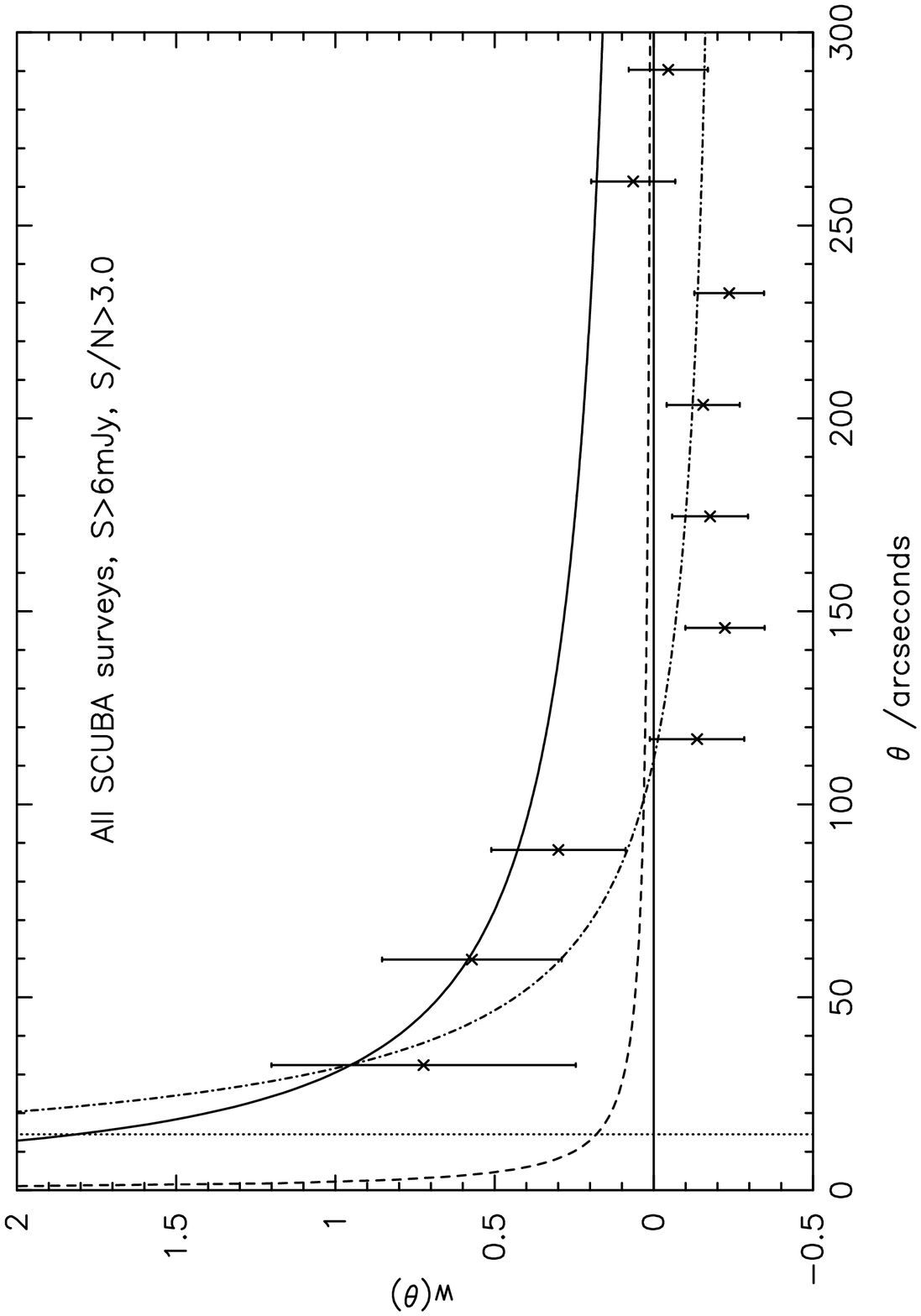}
\label{fig:w_3sig_6mjy} 
\caption{\small 2-point angular correlation function for
sources brighter than 6\,mJy, detected at a significance of $\rm > 3.00\sigma$, over all of the survey fields. The error bars are $\rm
1\sigma$ Poisson errors. The solid
power-law line indicates the correlation function found by Daddi et
al. (2000) for EROs with $R-K > 5$ and $K < 18.5$, and the
dashed power-law line indicates the correlation function found by
Giavalisco et al. (1998) for Lyman break galaxies at $z\sim3$. The
dot-dash line shows the best-fit power-law to the data points. The
vertical dotted line indicates the size of the JCMT beam at $\rm 850\,\mu m$.}
 \centering
   \vspace*{7.3cm}
   \leavevmode
   \includegraphics{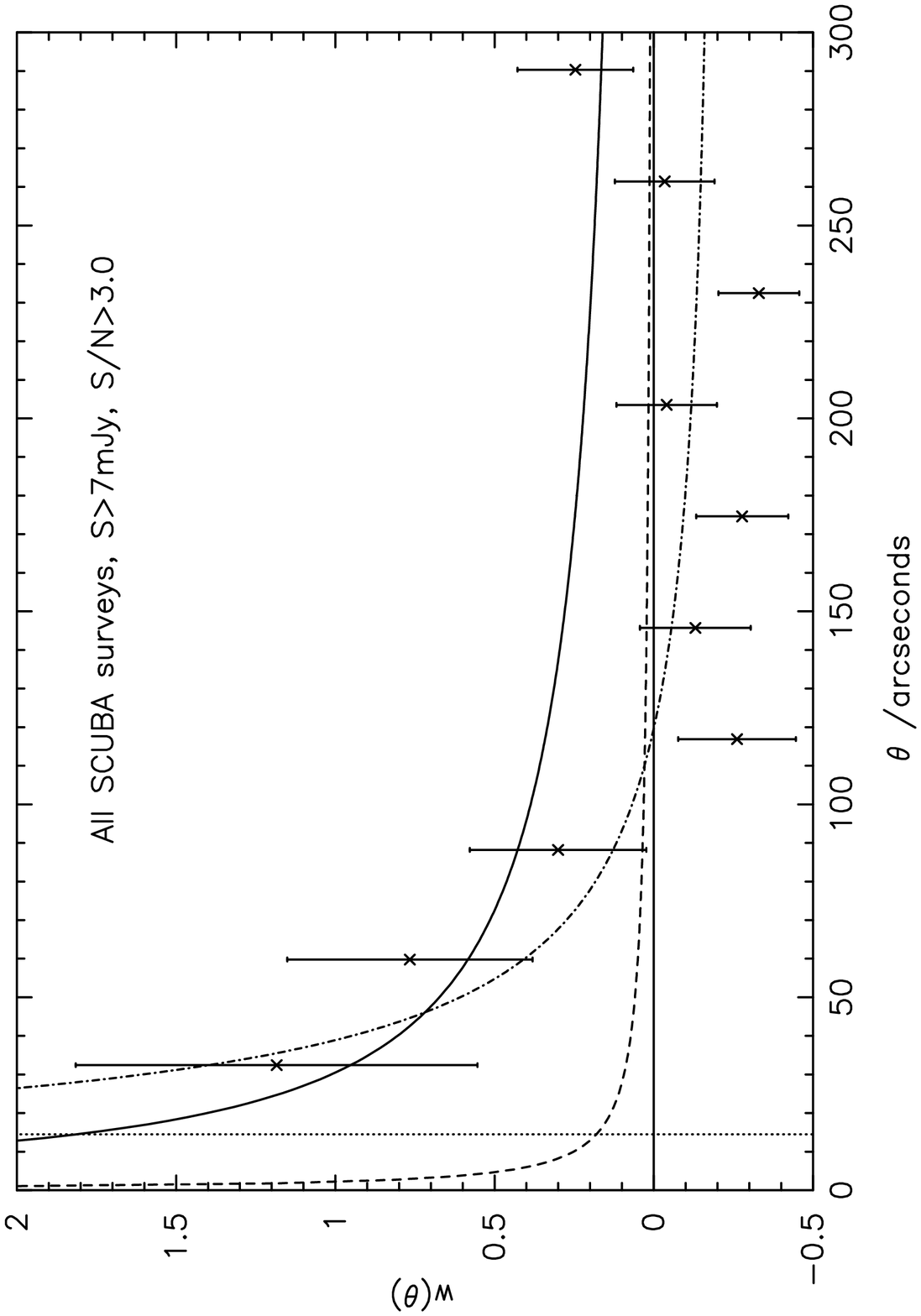}
\label{fig:w_3sig_7mjy} 
\caption{\small 2-point angular correlation function for
sources brighter than 7\,mJy, detected at a significance of $\rm > 3.00\sigma$, over all of the survey fields. The error bars are $\rm
1\sigma$ Poisson errors. The solid
power-law line indicates the correlation function found by Daddi et
al. (2000) for EROs with $R-K > 5$ and $K < 18.5$, and the
dashed power-law line indicates the correlation function found by
Giavalisco et al. (1998) for Lyman break galaxies at $z\sim3$. The
dot-dash line shows the best-fit power-law to the data points. The
vertical dotted line indicates the size of the JCMT beam at $\rm 850\,\mu m$.}
\end{figure}
\clearpage
\begin{table*}
\centering
\begin{tabular}{|c|c|c|c|c|c|}\hline
Flux density & S/N & A($\rm \theta /deg$) & $\rm \delta$ & $\rm \sigma^{2}$ & $\rm \theta_{0}$ \\
threshold & threshold & & & & /arcsec\\ \hline
$\rm >5\,mJy$ & $>3.50$ & $5.00 \times 10^{-2}$ & 0.70 & 0.386 & 49.9 \\
$\rm >5\,mJy$ & $>3.00$ & $1.82 \times 10^{-2}$ & 0.89 & 0.230 & 39.9 \\
$\rm >6\,mJy$ & $>3.00$ & $2.49 \times 10^{-3}$ & 1.36 & 0.280 & 43.8 \\
$\rm >7\,mJy$ &	$>3.00$ & $1.31 \times 10^{-3}$ & 1.56 & 0.265 & 46.7 \\ \hline
\end{tabular}
\label{table:ang_corr}\caption{\small Best-fit parameters to the
angular correlation functions shown in Figs. 22 to 25. Column 1 gives the
flux density cutoff as measured from the raw images and Column 2 gives
the signal-to-noise cutoff for source detection. Colum 3 gives the
amplitude A, Column 4 gives the slope of the power-law $\rm \delta$,
Column 5 gives the integral constraint $\rm \sigma^{2}$ and Column 6
gives the characteristic angular separation $\rm \theta_{0}$.}
\end{table*}

\noindent the proportion of excess pairs between corresponding data points in the
first 100 arcsec of the two plots. The Gaussian-convolved noise
images did not imply that there was anything unusual about the survey
data in regions of higher than average  $\rm 3.00 \sigma$ source density, indicating
that elevated noise where there is a concentration of significant
detections is unlikely to be the cause. There are two likely
explanations as to why the signal from the $\rm > 3.50 \sigma$
catalogue is not diluted in the $\rm > 3.00 \sigma$ plot. Firstly,
the simulation results in Section 4.2, which made estimates of the fraction
of spurious / confused sources at various signal-to-noise thresholds,
have been pessimistic. It is difficult to generate accurate realisations of
the SCUBA maps on which to make estimates of contamination in a source
catalogue due to confusion and noise without knowing the clustering
properties of the SCUBA population in the first place, and the fact that
there is very little difference in the proportion of excess pairs on
scales of $\sim1$ arcmin between the two signal-to-noise datasets
may be suggesting that most of the $\rm 3.00-3.49 \sigma$ sources are
real. Perhaps a more likely scenario, however, is that the
clustering properties of the SCUBA population are strong over a
significant range of flux
densities, and thus detections arising from the confusion of faint
objects are more likely to be found on the same scale and in the
vicinity of the real bright SCUBA sources. The major difference
between the $\rm > 3.00$ and $\rm > 3.50 \sigma$ plots is the size of
the Poisson error bars. As one might expect, the error bars are a factor of $\sim \sqrt 2$
smaller in Fig. 23, reflecting the increase by a factor of 2 of the
number of sources considered, increasing the significance of the
number of excess pairs contained in the first 3 bins (14.5 - 101.5
arcsec) from $\rm 2.5$ to $\rm 3.7\sigma$.

Cutting at higher flux densities decreases the number of sources
available for producing a 2-point angular correlation function even
further, thus using the smaller ``safer'' catalogue of sources above a
signal-to-noise threshold of 3.50 leads to tentative but non-significant
measures of the clustering strength above the noise level. Instead,
the analysis has been repeated using the full $\rm >3.00\sigma$ source
lists for 
flux density cut-offs of $>6$ and $>7$\,mJy shown in Figs. 24 and
25 respectively, pushing the available data to its limits. It should
be noted that these flux density thresholds are the raw values as
measured directly from the maps and should therefore not be taken as
``absolute'' because of the boosting effects described in Section
4.1. However, it is fair to say that cutting at $\rm S_{850}>7$\,mJy
defines a set of objects with generally higher star formation rates
than cutting at $\rm S_{850}>5$\,mJy and so Figs. 23 to 25 can
still be used to look for any trends in clustering strength with
increasing flux density. For comparison, the number of excess pairs within
$\sim100$ arcsec of each other in the real data when compared
to a random distribution is significant at the $\rm 3.3 \sigma$ level
for $\rm S_{850\mu m}>6$\,mJy, and at the $\rm 3.5 \sigma$ level for $\rm S_{850\mu m}>7$\,mJy.

For all three flux density limits, the measured correlation
functions indicate a clustering strength much larger than that
measured by Giavalisco et al. (1998) for Lyman-break galaxies at
$z\simeq3$ (the dashed lines in the figures), even given the
rather large error bars on the SCUBA data points. Could the apparent
difference in clustering strength simply be due to projection effects
over redshift space?
The Lyman-break technique uses colour selection to identify
high-redshift galaxies through multi-band imaging across the $\rm 1216\AA$
line and the $\rm 912\AA$ Lyman break. At $z > 2.5$ the Lyman limit
is redshifted far enough into the optical window to be observable in
broad-band ground-based photometry. By placing filters on either side
of the redshifted Lyman limit one can find high-redshift objects by
their strong spectral breaks. Giavalisco et al. (1998) used a custom
photometric system, $U_{n}GR$ (Steidel \& Hamilton 1993) optimized for
selecting Lyman-break galaxies with $z\simeq 3$. By the nature of this
method, 90\% of the galaxies they used were confined to the redshift
range $2.6 < z < 3.4$, with none at $z<2.2$. The redshift range of the
bright SCUBA population used in these calculations of angular
correlation functions is much more uncertain, but Ivison et al. (2002)
have suggested a median redshift of $z=2.4$ based on the 
radio-to-submillimetre 
spectral indices (Carilli \& Yun, 1999 \& 2000), with inferred
redshifts spanning the range of $z \simeq 1-4$. Therefore, unless the bright SCUBA population
occupies a very much narrower redshift band than implied by the
radio-to-submillimetre spectral indices the dilution of the angular
clustering signal by projection over redshift cannot be the reason for
the large difference in clustering strength. 

Instead, this contrast in clustering properties implies that the
Lyman-break galaxies and bright SCUBA sources are sampling two
different stages or mass domains in galaxy formation (see also Barger,
Cowie \& Richards 2000, Webb et al. 2003b). The stronger clustering
exhibited by the bright $\rm 850\,
\mu m$ sources suggests that these objects are tracing the rarest
high-mass peaks of the Gaussian initial density fluctuations and
are the progenitors of the most massive ellipticals, whereas the
weaker clustering of the Lyman-break galaxies indicates a weaker bias
with respect to mass, detecting the formation of smaller disk or bulge systems.
The strength of clustering, however, is consistent with that
measured by Daddi et al. (2000) for extremely red objects (EROs) with
$K<18.5$ and $R-K>5$ (the solid lines in the figures), perhaps suggesting
an evolutionary sequence from SCUBA source to ERO. The numerical
coincidence between the co-moving number densities of EROs and that
estimated for bright SCUBA sources
is also in line with this idea (Scott et al. 2002).

Attempts have been made to fit a standard power-law describing the
angular correlation function to the data points shown in Figs. 22 to
25. The fit is slightly complicated by the fact that the
global number density is unknown and must be estimated from the sample to hand.
Hence,
\be \rm (1 + w_{true}(\theta)) = (1 + w_{obs}(\theta)) \left( 
\frac{\bar{n}}{<n>} \right)^{2} \ee
\noindent where $\rm w(\theta)_{true}$ is the true correlation
function, $\rm w(\theta)_{obs}$ is the observed correlation function,
$\rm \bar{n}$ is the mean number density of sources and $\rm <n>$ is
the global number density of sources. One can rewrite the observed
mean number density $\rm \bar{n}$ in terms of a pertubation $\rm
\delta n$ on the global number density $\rm <n>$ (ie. $\rm \bar{n}=<n>
\pm \delta n$), such that in
averaging over a set of sky areas the size of the sample 
\be \rm (1 + w_{true}(\theta)) = (1 + w_{obs}(\theta)) (1 +
\sigma^{2}) \ee
\noindent where $\rm \sigma^{2}$ is the rms surface density
variation. If the real correlation functions take the power-law form
$\rm w_{true}(\theta)=A\theta^{-\delta} =
\left(\frac{theta}{theta_0}\right)^{-delta}$, then 
\be \rm w_{obs}(\theta) = \frac{(A\theta^{-\delta}) -
\sigma^2}{1+\sigma^{2}}
=\frac{\left(\frac{\theta}{\theta_{0}}\right)^{-\delta} -
\sigma^{2}}{1 + \sigma^{2}}\ee

\begin{figure}
 \centering
   \vspace*{6.6cm}
   \leavevmode
   \includegraphics{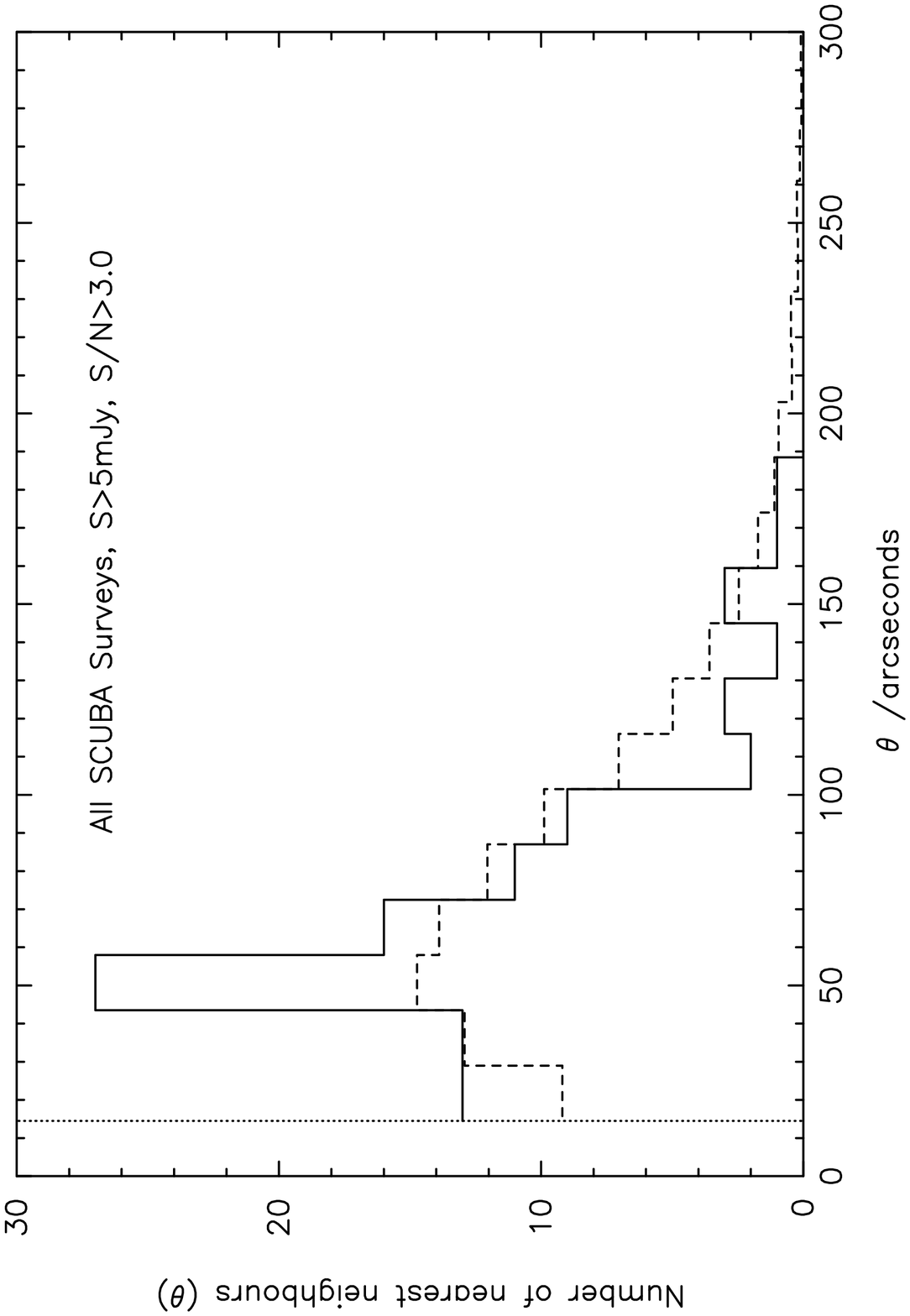}
\label{fig:near_neighb_3sig_5mjy} 
\caption{\small Nearest-neighbour analysis for sources brighter than
   5\,mJy, detected at a significance of $\rm >3.00\sigma$ over all of
   the survey fields.The
vertical dotted line indicates the size of the JCMT beam at $\rm 850\,\mu
   m$. The solid line shows the distribution of
   nearest-neighbour pairs for the actual dataset, whereas the
   dashed line shows the expected nearest-neighbour histogram for
   the same surface density of sources when distributed randomly. The
   probability that the two distributions are the same is $<0.05$.}
\end{figure}
\begin{figure}
 \centering
   \vspace*{6.6cm}
   \leavevmode
   \includegraphics{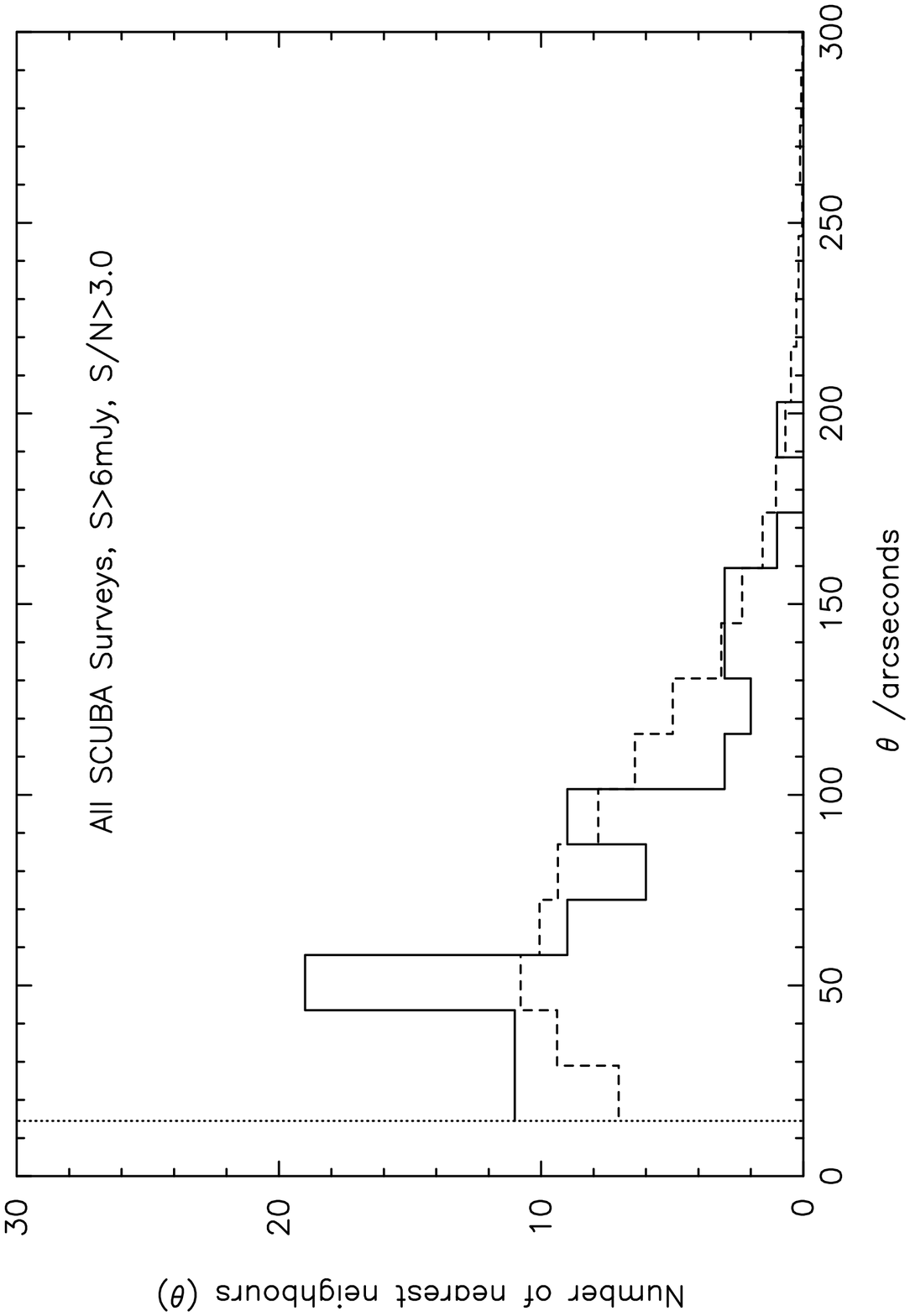}
\label{fig:near_neighb_3sig_6mjy} 
\caption{\small Nearest-neighbour analysis for sources brighter than
   6\,mJy, detected at a significance of $\rm >3.00\sigma$ over all of
   the survey fields.The
vertical dotted line indicates the size of the JCMT beam at $\rm 850\,\mu
   m$. The solid line shows the distribution of
   nearest-neighbour pairs for the actual dataset, whereas the
   dashed line shows the expected nearest-neighbour histogram for
   the same surface density of sources when distributed randomly. The
   probability that the two distributions are the same is $\rm <0.05$.}
\end{figure}

\begin{figure}
 \centering
   \vspace*{6.6cm}
   \leavevmode
   \includegraphics{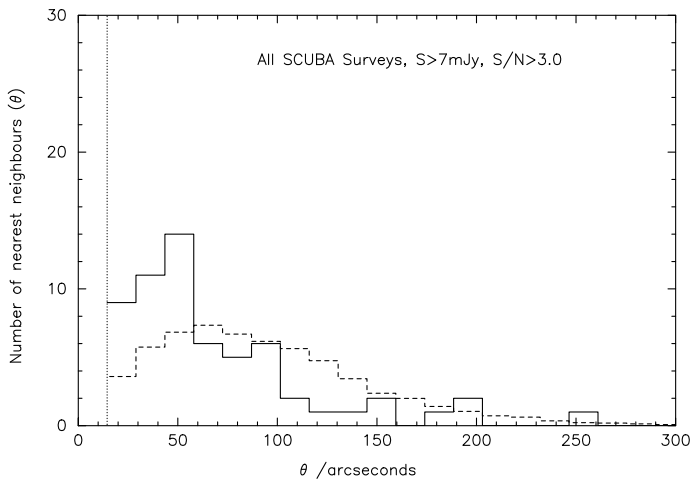}
\label{fig:near_neighb_3sig_7mjy} 
\caption{\small Nearest-neighbour analysis for sources brighter than
   7\,mJy, detected at a significance of $\rm >3.00\sigma$ over all of
   the survey fields.The
vertical dotted line indicates the size of the JCMT beam at $\rm 850\,\mu
   m$. The solid line shows the distribution of
   nearest-neighbour pairs for the actual dataset, whereas the
   dashed line shows the expected nearest-neighbour histogram for
   the same surface density of sources when distributed randomly. The
   probability that the two distributions are the same is $\rm <0.01$.}
\end{figure}

\noindent One can then either determine the values of A, $\rm \delta$
and $\rm \sigma^{2}$ by allowing them all to be free parameters in a
minimised $\rm \chi^{2}$ fit, or alternatively one can rewrite the
integral constraint in terms of the amplitude multiplied by a constant
(ie. $\rm \sigma^{2}= A\times C$) and estimate C by doubly integrating 
an assumed true $\rm w(\theta)$ over the field area $\rm \Omega$,
\be \rm AC= \frac{1}{\Omega^{2}} \int \int
w(\theta)d\Omega_{1}d\Omega_{2} \ee
\noindent This can be done numerically using the random-random correlation,
such that
\be \rm C= \frac{\sum N_{rr}(\theta)\theta^{-\delta}}{\sum N_{rr}(\theta)} \ee

Both of these methods have been attempted for all four sets of data
points, the free parameter fit proving to be the more successful of
the two. The results are given in Table 17.

The large error bars and scatter in the data points
beyond the first 3 fairly robust data point measurements (out to $\sim
100$\,arcsec) means that the best-fit parameters
for a power-law 2-point angular correlation function are not
well constrained. However, it is
interesting to note that in all 4 cases the characteristic angular
scale length $\rm \theta_{0}$ indicates strong clustering of the
bright SCUBA population on scales of $\sim 40-50$ arcsec.

The indication is that a much larger SCUBA survey,
approaching 0.5 square degrees (4 times the area covered here) is
required to obtain a meaningful measure of the correlation power-law
slope and scale length. This is discussed briefly in Section 8 with
more detailed accounts given in van Kampen et al. (2005) and Mortier
et al. (2006).

\subsection{Nearest neighbour statistics}

An alternative method for measuring the strength of clustering is a
nearest-neighbour analysis. This procedure measures the distribution
of the separations between each source and its closest neighbour as
compared to what one would expect from a random distribution, and is
sometimes more informative in deciding whether sources are clustered when
dealing with small datasets like these.

The nearest-neighbour distributions for the bright SCUBA sources were measured from the $\rm S/N
>3.00$ datasets, for flux density thresholds of $\rm S_{850}>5$, $\rm
>6$, and $\rm >7\,mJy$ as measured directly from the raw maps (i.e. no
corrections for flux-boosting effects have been applied). One can readily weight any
uncharacteristically deep areas of an image for calculating a 2-point angular
correlation function by creating a large mock catalogue
of randomly positioned sources, equivalent to a combined catalogue
from conducting many simulations, but which has the local number density of detections
in any given region weighted according to the number counts and the
noise as described in the previous 
Section. A nearest neighbour distribution, however,
requires mock images with the same source number density as the actual
survey fields. In this analysis 100 mock catalogues for each individual field
were generated, the positions allocated randomly according to a
Poisson distribution and constrained to lie within the
boundaries of the original survey data. This was done separately for
the three flux density thresholds, each fake source list
containing the same number of $\rm > 3.00 \sigma$ sources as the real
image. Histograms of the
number of nearest neighbours against nearest-neighbour separation are
shown in Figs 26 - 28 for flux density cut-offs of 5, 6
and 7\,mJy respectively, using bins the size of the
SCUBA beam (14.5 arcsec). The histograms of the real
nearest-neighbour distributions are shown by solid lines, and
the distributions from combining the mock source catalogues are
illustrated by the dashed lines, normalised according to the
total number of real source pairings.

In all three plots there is a clear excess of bright $\rm 850\,\mu m$
sources separated from their nearest neighbour at $\sim50$
arcsec. This is most noticeable in the $\rm S_{850}>7\,mJy$
histogram (Fig. 28) where there are 51 nearest neighbour pairs
compared to 36 expected within the first $\sim 100$\,arcsec (a statistical
excess at the $\rm 2.5 \sigma$ level), and
34 nearest-neighbour pairs compared to 16 expected within the first
$\sim 50$\, arcsec (a statistical excess at the $\rm 4.5\sigma$ level). 

A Kolmogorov-Smirnov test is a simple method for testing the probability
that two distributions are identical. The test statistic, D, is
defined as the maximum absolute difference between an observed ($\rm S_{o}$)
and an expected ($\rm S_{e}$) normalised cumulative distribution, in this
case applied to the cumulative fraction of nearest-neighbour pairs
with angular separation:
\be \rm D= \max |S_{o}(\theta)-S_{e}(\theta)| \ee  
\noindent If the measured D value exceeds a critical value when
compared to the known sampling distribution for D appropriate to the
number of data points, then the two distributions may be rejected as
being the same at that level of significance. 

As shown in Table 18, the distributions of
$\rm >3.00 \sigma$ sources brighter than $\rm >5$ and $\rm >6$\,mJy
were rejected as being consistent with a random distribution at the
95\% confidence level, and the sources brighter than $\rm
>7$\,mJy were inconsistent with a random distribution at better than
the 99\% confidence level. Overall this again suggests that the bright
SCUBA population is strongly clustered on scales of $\sim 1$ arcmin.
\begin{table}
\centering
\begin{tabular}{|c|c|c|c|c|}\hline
Flux density & D & $\rm \theta_{D}$ & Number & P(D) \\
threshold & & /arcsec & sources & \\ \hline
$\rm >5\,mJy$ &	0.1587	& 82.6 & 100 & $<0.05$ \\
$\rm >6\,mJy$ & 0.1659  & 65.5 & $\phantom{0}$78 &  $<0.05$ \\
$\rm >7\,mJy$ &	0.2898  & 65.5 & $\phantom{0}$61 & $<0.01$ \\ \hline
\end{tabular}
\label{table:kstest}\caption{\small Results of applying a Kolmogorov-Smirnov test to the
cumulative fraction of nearest-neighbour pairs. Column 1 gives the
flux density cutoff as measured from the raw images. Column 2 gives
the maximum absolute difference between the observed and expected
normalised cumulative distributions, D, and the nearest-neighbour
separation at which this occurs is given in Column 3. Column 4 gives
the probability that the two distributions are the same.}
\end{table}

\section{Conclusions and Future Work}
Deep blank-field surveys conducted with SCUBA on the JCMT have
successfully resolved $\simeq 30-50\%$ of the far-infrared
extragalactic background into discrete
sources down to the
confusion limit of $\rm S_{850}\simeq 2-3\,mJy$, with deeper surveys
making use of gravitational lensing from intervening massive clusters
probing the very faintest submillimetre sources. This has revealed a
population of heavily dust-enshrouded galaxies at high redshift
($z>1$) undergoing a burst of massive star-forming activity.
The work presented here is an investigation into the nature of the 
most luminous
$\rm 850\,\mu m$ sources ($\rm S_{850}>5\,mJy$), with particular
consideration of their link with the formation and evolution of
the most massive elliptical galaxies visible in the present-day
Universe.

Combining the data from the ``SCUBA 8\,mJy Survey'' with other
existing blank field surveys, re-reduced and analysed in an identical
manner, approximately doubles the area of sky observed with SCUBA 
analysed 
in 
a homogeneous
manner. The combined datasets
allow the determination of the most accurate source counts to date for
$\rm S_{850}>2\,mJy$. The differential
source counts for $\rm S_{850}<8\,mJy$ are well satisfied by a
power-law of the form $\rm
dN(>S)/dS=N_{0}/(a+ S^{\alpha})$ where the best-fit values of the
free parameters were $\rm N_{0} = 2.67 \times 10^{4}$, $\rm a=0.49$ and
$\rm \alpha=3.14$. If one assumes this is true over the full range of
flux densities from the very faint to the very bright, this predicts a
total $\rm 850 \, \mu m$ background of $\rm 3.8 \times 10^{4}\,mJy\,deg^{-2}$, mid-way between
the $\rm 850 \, \mu m$ extragalactic values of $\rm 3.1 \times
10^{4}\,mJy\,deg^{-2}$ and $\rm 4.4 \times 10^{4}\,mJy\,deg^{-2}$.
However, there are some indications that the source counts steepen for
$\rm S_{850}>8\,mJy$. This may simply be a result of small number
statistics, but is possibly indicative of a
high-mass cutoff of the SCUBA population. A steepening of the source
counts would also make the bright SCUBA population more prone to the
effects of gravitational lensing. More physical models of the source
counts, beginning with a local $\rm 850\, \mu m$ luminosity function,
are able to reasonably successfully model the number counts using only strong
pure luminosity evolution with redshift, and assuming a single
temperature optically thin greybody to
represent the thermal spectral energy distribution of the dust with
characteristic dust
temperatures in the range $\rm 30<T_{d}<50\,K$, and emissivity index
in the range $\rm 1.0 < \beta < 2.0$.

Tentative indications that the bright SCUBA population
clusters on scales of $\rm 1-2$ arcmin were first identified in
the ``SCUBA 8\,mJy Survey''
(Scott et al. 2002), but these results proved inconclusive due to
the small number of significant detections. However, the full source 
catalogue derived from combining
all of the survey fields contains a sufficient number of significant detections
to place more meaningful constraints on the clustering properties of the
bright submillimetre population. Measurements of angular correlation
functions and nearest neighbour statistics for $\rm S_{850}>5$, 6, and
7\, mJy, imply strong clustering on scales of $\sim 40-50$\,arcsec at a
significance level of $3.5 - 4\sigma$. Attempts have been made to fit
the angular correlation function data points with a traditional power
law of the form $\rm w(\theta)=A\theta^{-\delta}$, although the small
number of sources detected even by combining all of the existing
blank field surveys together mean that the values of A and $\rm
\delta$ are not well constrained.
However, the data points are in fairly good agreement with
the clustering strength measured for extremely red objects by Daddi et
al. (2000), suggesting a strong link between the SCUBA and ERO
populations.

The results from this work have provided a major part of the
motivation for a new wider-field extragalactic submillimetre survey,
which has now been underway for three years at the JCMT. The aim of this
survey is to map 0.5 square degrees of sky, $\sim7$ times the total area
of the ``SCUBA 8\,mJy Survey'' to a comparable depth of $\rm
\sigma_{rms}\simeq 2.5\,mJy/beam$ at $\rm 850\, \mu m$. The ``SCUBA Half Degree
Extragalactic Survey (SHADES)'' is split over two regions
of sky; the Lockman Hole East (a continuation of the 8\,mJy Survey),
and the Subaru-XMM Deep Field. Based on the $\rm 850
\, \mu m$ source counts, a survey of this nature is expected to yield $200-300$
sources detected at the $\rm >3.5\sigma$ level. This survey also
has associated 
deep radio, mid-infrared, near-infrared, optical and X-ray imaging. 
Early example results from SHADES are discussed by Dunlop (2005), while
full description of the SHADES science goals, design and data
analysis techniques is given in Mortier et al. (2005). In van Kampen
et al. (2005) a variety of theoretical models describing galaxy
formation and evolution are presented with particular emphasis on 
their predictions regarding the redshift distribution and clustering
properties of the SCUBA population. The first results on the sub-millimetre 
number counts from SHADES will be presented by Coppin et al. (2006),
while the results of a search for radio and mid-infrared identifications
in deep VLA and Spitzer data 
will be presented by Ivison et al. (in preparation).

\section*{Acknowledgements}

SES acknowledges the support of a PPARC PDRA. The JCMT is operated by the Joint
Astronomy Center on behalf of the UK 
Particle Physics and Astronomy Research
Council, the Canadian National Research Council 
and the Netherlands Organization for Scientific Research.

\renewcommand{\topfraction}{0.1}
\clearpage
\appendix
\section{Simulations building on the real survey data}
\subsection{Differential completeness plots}

The following plots show the percentage of sources recovered with
signal-to-noise ratio $>3.50$ against input flux density level for
each of the survey fields. Those fields composed of a deep pencil
beam survey within a wider-area shallower survey have had these two
components treated separately. The plotted curves are the best-fit
solutions to the empirical functional form 
\be \rm differential \phantom{0} completeness = 100(1-e^{-a(x-b)}) \ee
where x is the input flux density, and the values of a and b were
determined by a minimised $\rm \chi^{2}$ fit to the simulation results 
for each $\rm 850\, \mu m$ survey field. The values of a and b
determined from these fits are given in Table 2.

\begin{figure}
 \centering
   \vspace*{3.7cm}
   \leavevmode
   \includegraphics{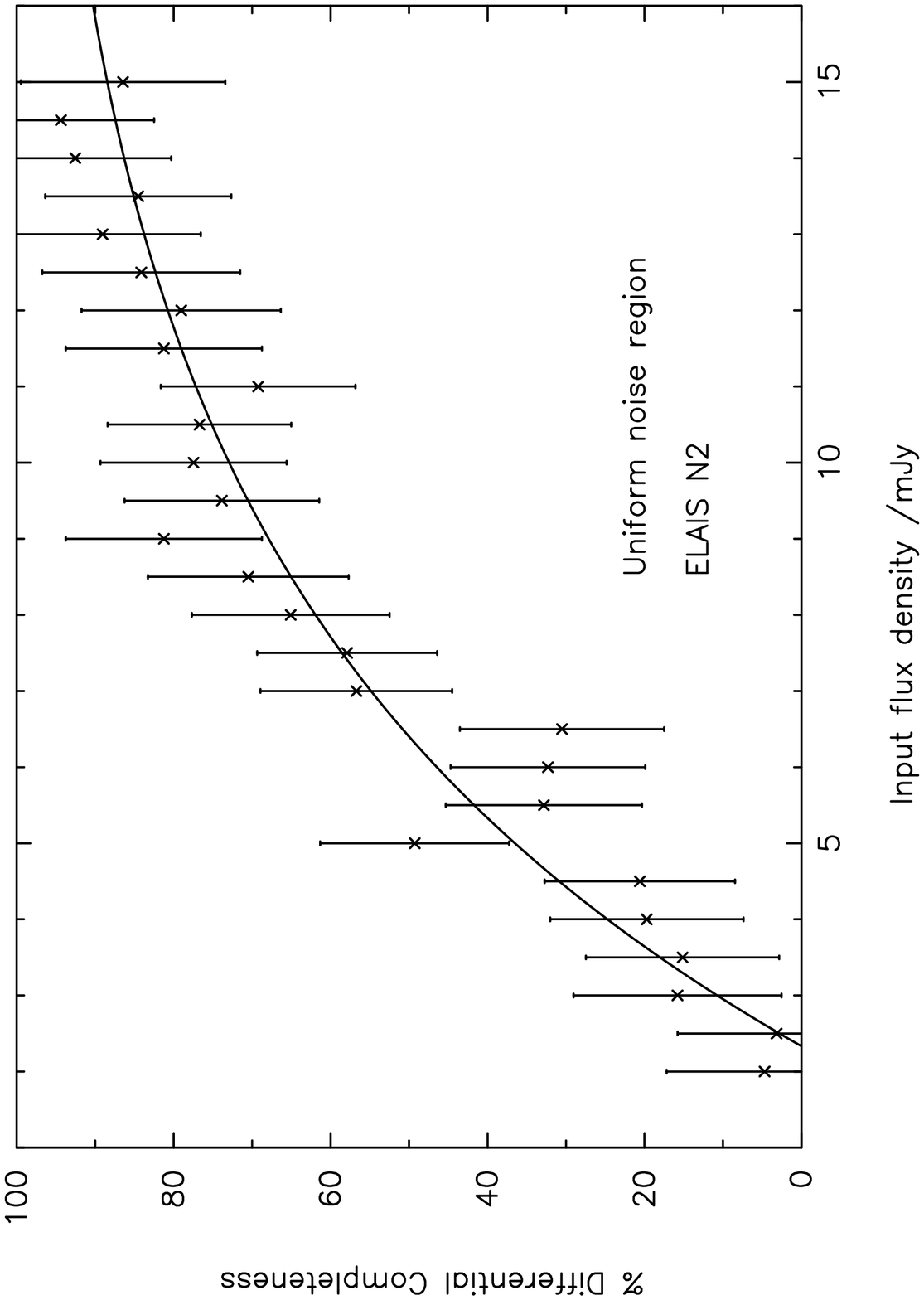}
\label{fig:n2_uni_comp} 
\caption{\small{Percentage of sources recovered against input flux
   density, for the uniform noise regions of the ELAIS N2 field
 from the ``SCUBA 8\,mJy Survey''.}}
 \centering
   \vspace*{3.7cm}
   \leavevmode
   \includegraphics{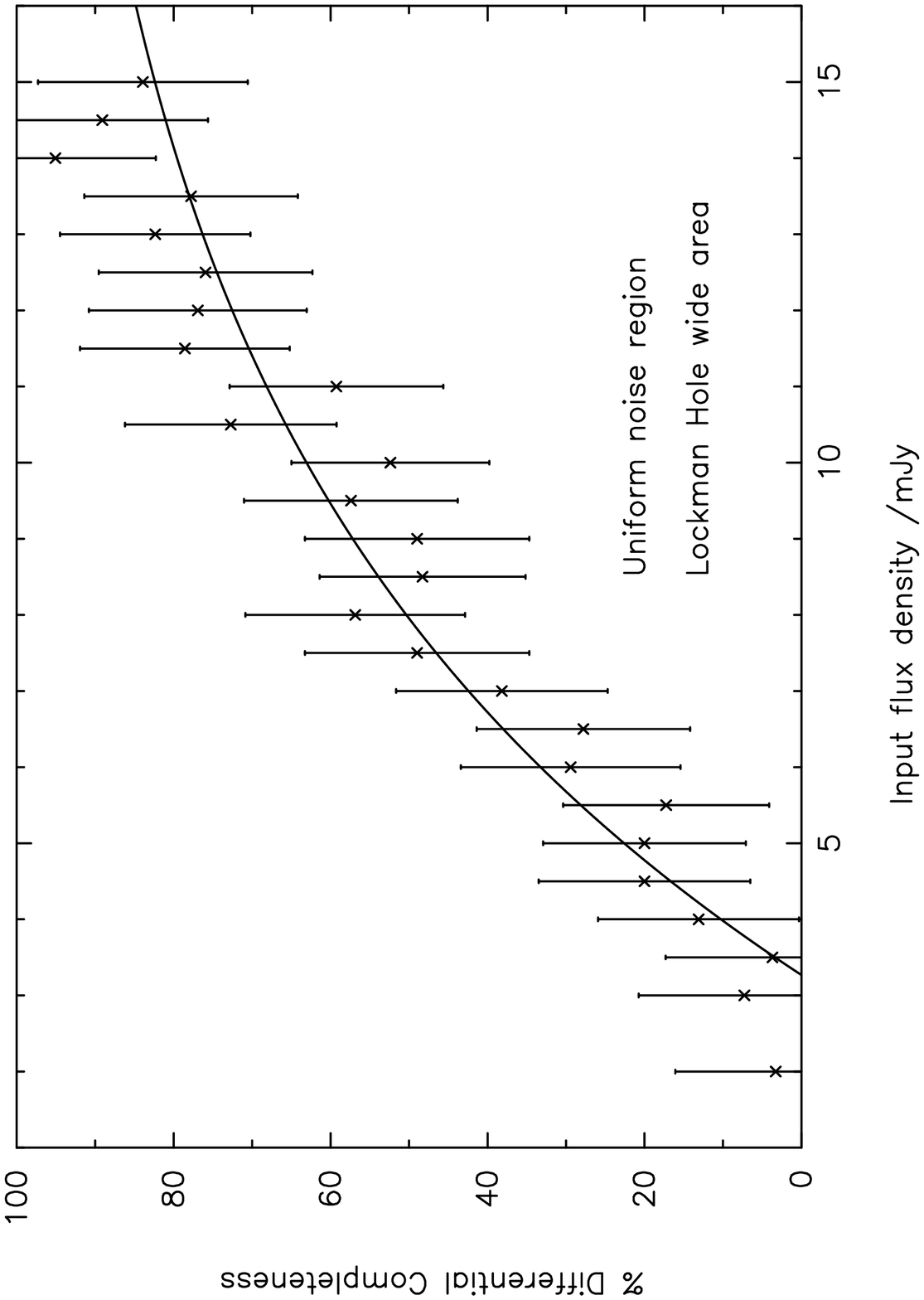}
\label{fig:lhwide_uni_comp} 
\caption{\small{Percentage of sources recovered against input flux
   density, for the uniform noise regions of the Lockman Hole East
 wide area field from the ``SCUBA 8\,mJy Survey''.}}
\centering
   \vspace*{3.7cm}
   \leavevmode
   \includegraphics{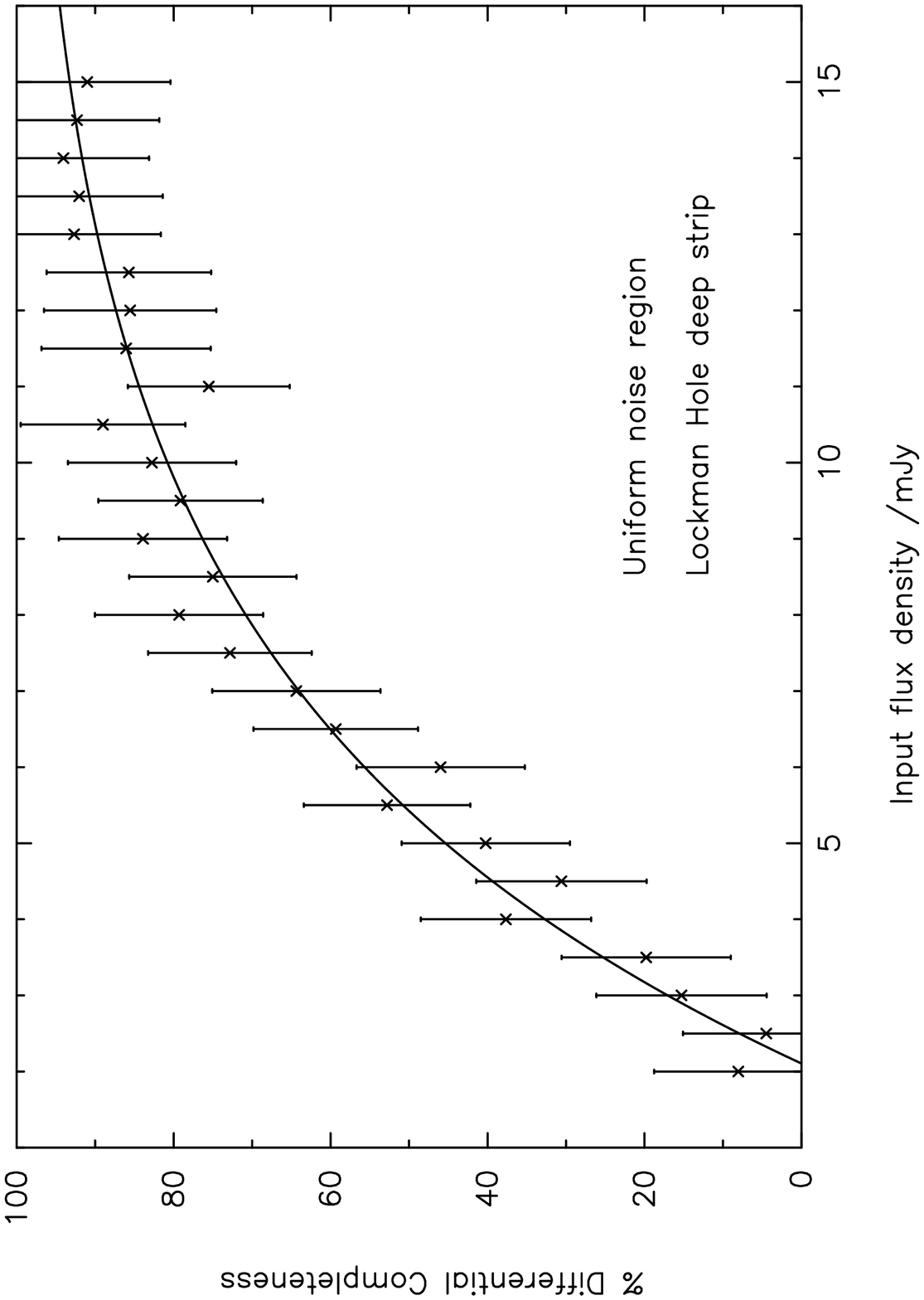}
\label{fig:lhstrip_uni_comp} 
\caption{\small{Percentage of sources recovered against input flux
   density, for the uniform noise regions of the Lockman Hole East
 deep strip from the ``SCUBA 8\,mJy Survey''.}}
\end{figure}
\renewcommand{\topfraction}{0.95}
\newpage
\begin{figure}
 \centering
   \vspace*{5.8cm}
   \leavevmode
   \includegraphics{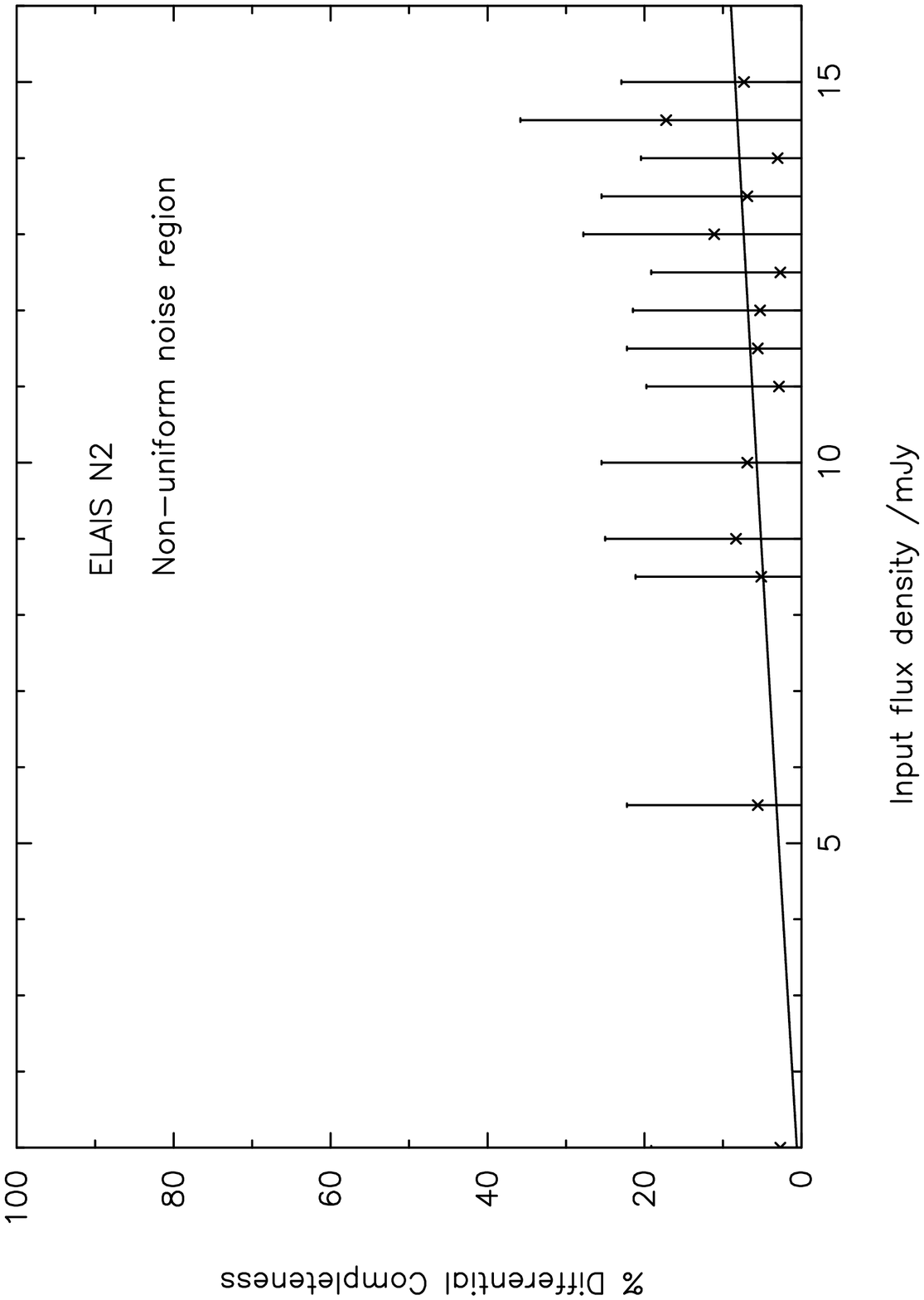}
\label{fig:n2_nonuni_comp} 
\caption{\small{Percentage of sources recovered against input flux
   density, for the non-uniform noise regions of the ELAIS N2 field
 from the ``SCUBA 8\,mJy Survey''.}}
 \centering
   \vspace*{3.7cm}
   \leavevmode
   \includegraphics{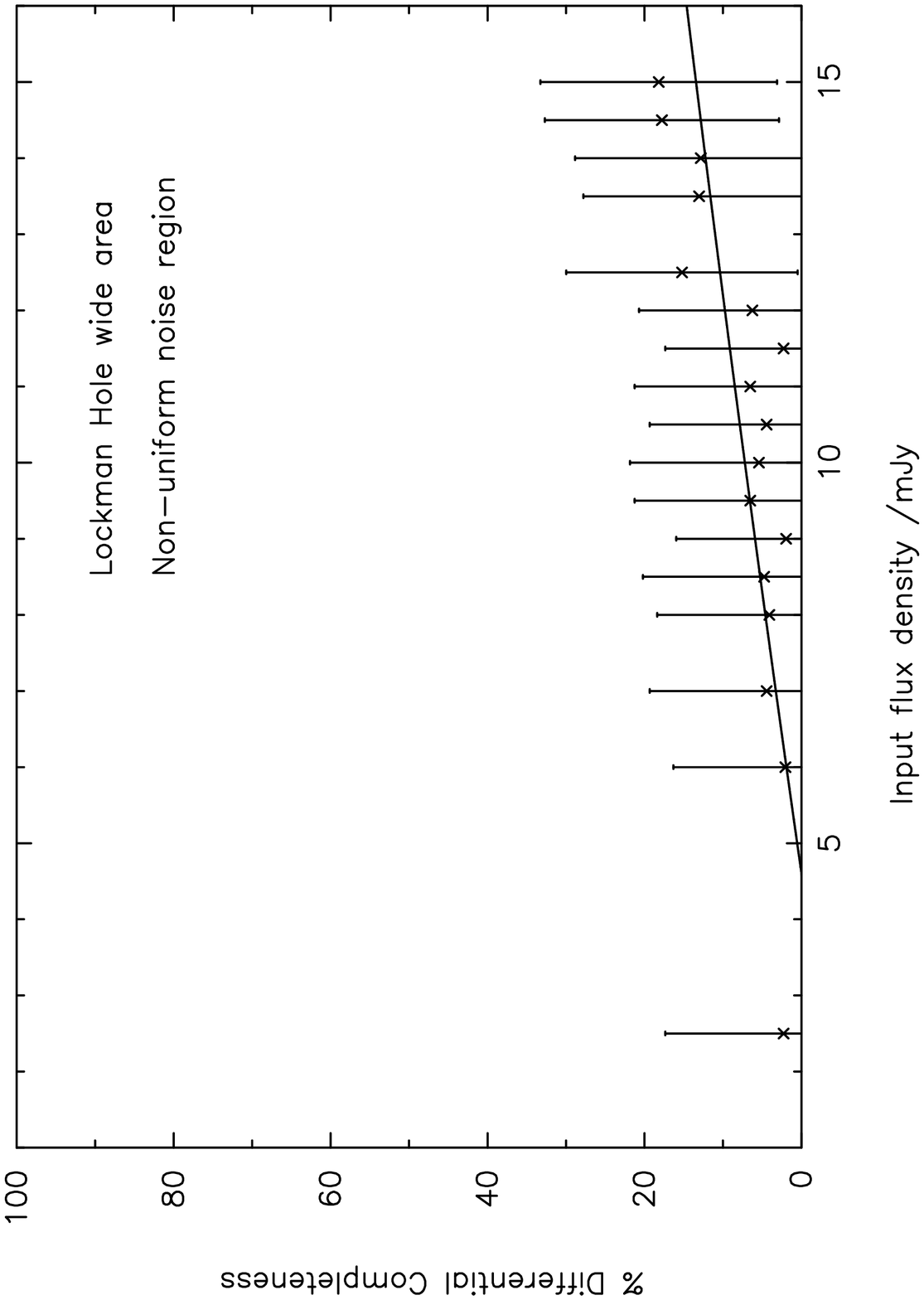}
\label{fig:lhwide_nonuni_comp} 
\caption{\small{Percentage of sources recovered against input flux
   density, for the non-uniform noise regions of the Lockman Hole East
 wide area field from the ``SCUBA 8\,mJy Survey''.}}
 \centering
   \vspace*{3.7cm}
   \leavevmode
   \includegraphics{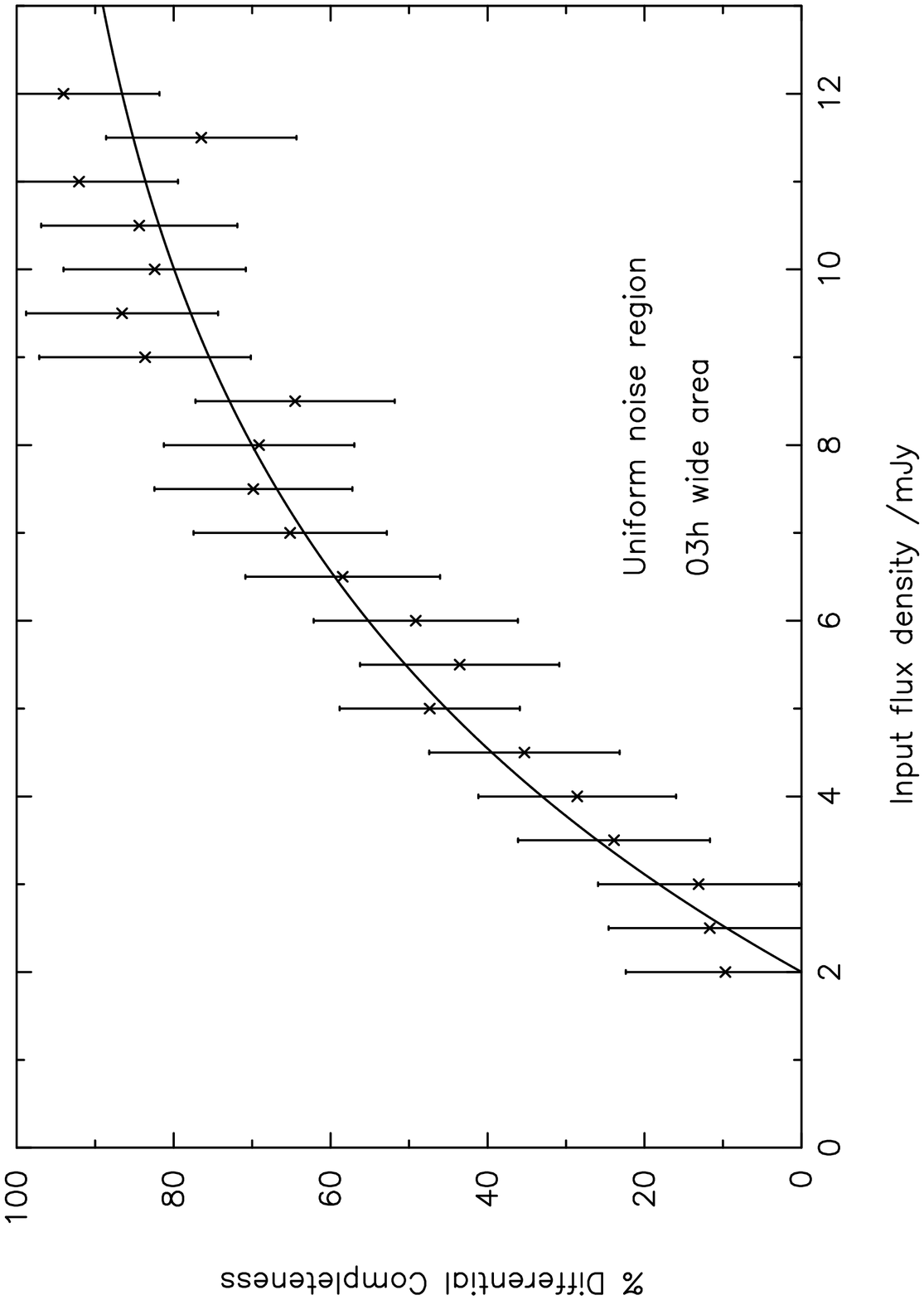}
\label{fig:03wide_uni_comp} 
\caption{\small{Percentage of sources recovered against input flux
   density, for the uniform noise regions of the 03 hour
 wide area field from the ``CUDSS''.}}
 \centering
   \vspace*{3.7cm}
   \leavevmode
   \includegraphics{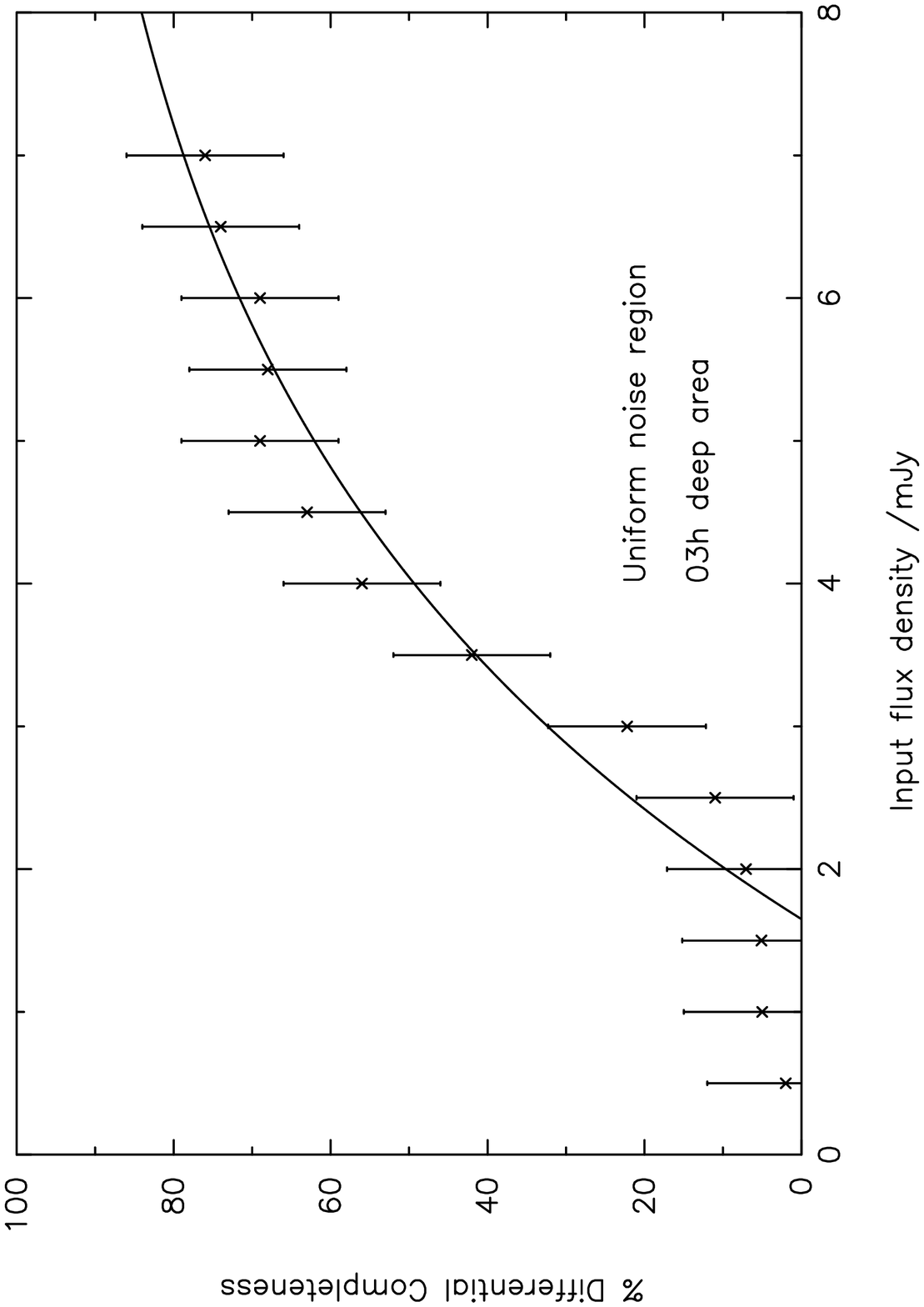}
\label{fig:03deep_uni_comp} 
\caption{\small{Percentage of sources recovered against input flux
   density, for the uniform noise regions of the 03
 hour deep area from the ``CUDSS''.}}
 \end{figure}
\newpage
\begin{figure}
 \centering
   \vspace*{4.8cm}
   \leavevmode
   \includegraphics{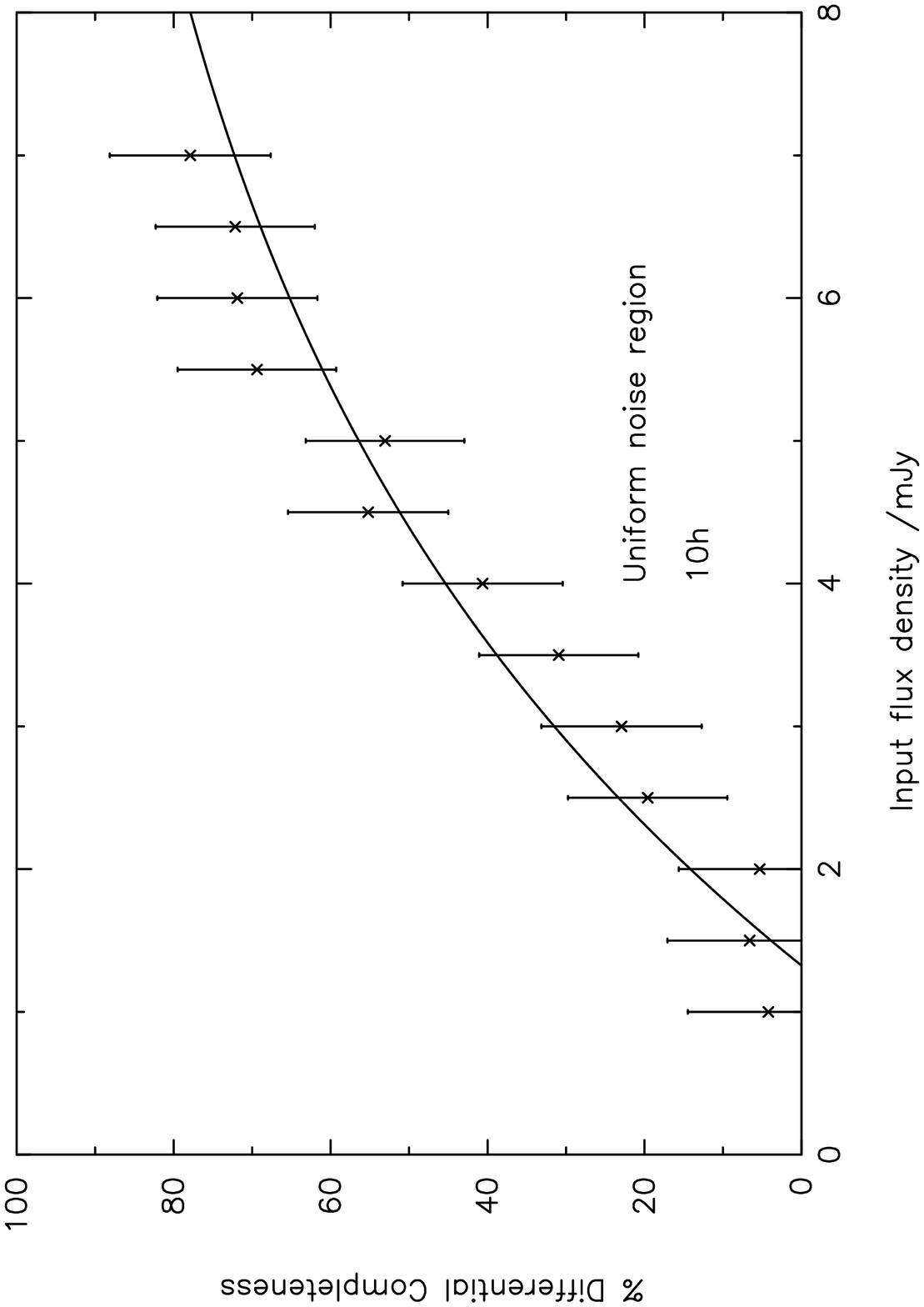}
\label{fig:10_uni_comp} 
\caption{\small{Percentage of sources recovered against input flux
   density, for the uniform noise regions of the 10 hour field from
 the ``CUDSS''.}}
 \centering
   \vspace*{3.7cm}
   \leavevmode
   \includegraphics{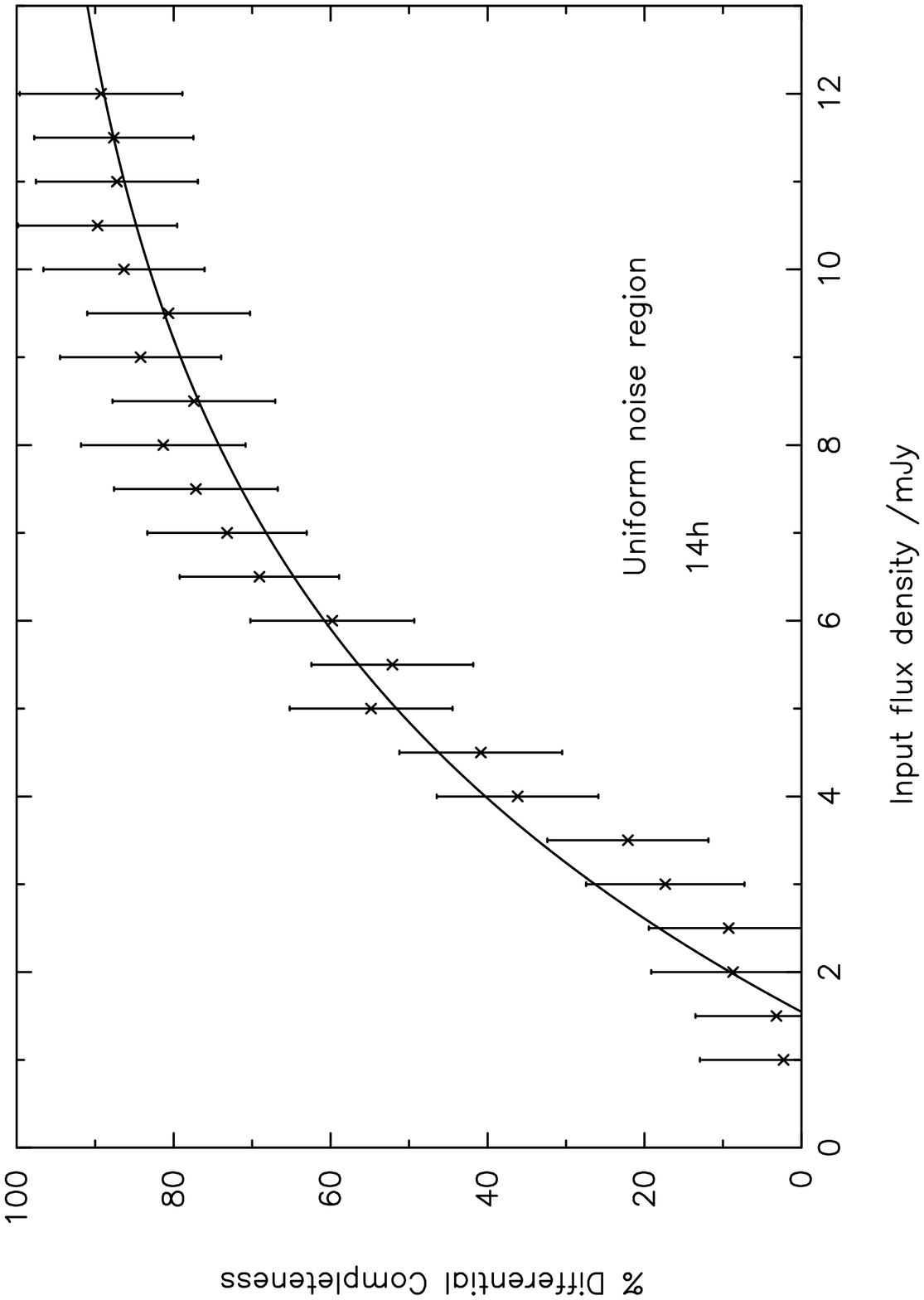}
\label{fig:14_uni_comp} 
\caption{\small{Percentage of sources recovered against input flux
   density, for the uniform noise regions of the 14 hour field
 from the ``CUDSS''.}}
 \centering
   \vspace*{3.7cm}
   \leavevmode
   \includegraphics{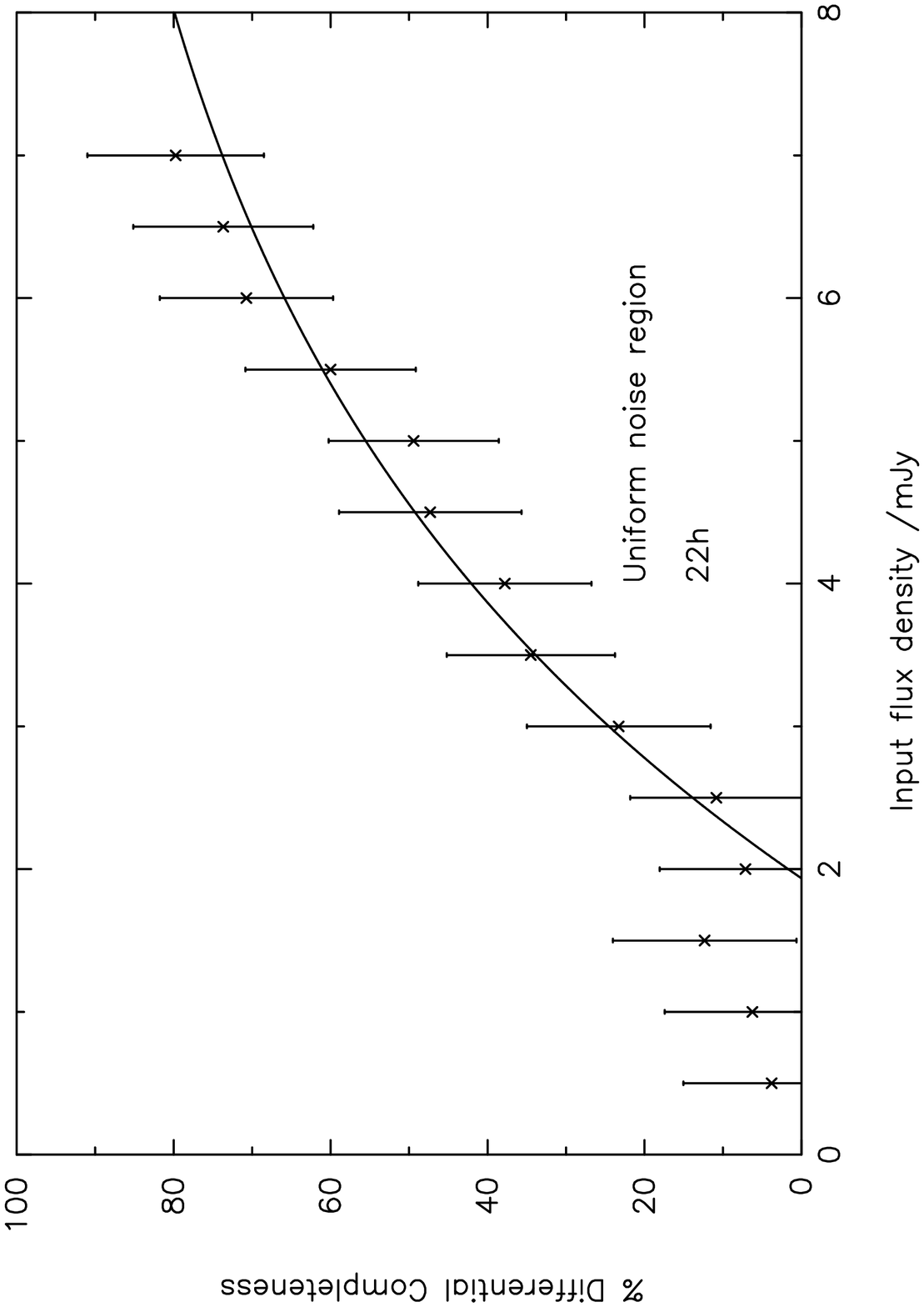}
\label{fig:22_uni_comp} 
\caption{\small{Percentage of sources recovered against input flux
   density, for the uniform noise regions of the 22 hour field
 from the `CUDSS''.}}
 \centering
   \vspace*{3.7cm}
   \leavevmode
   \includegraphics{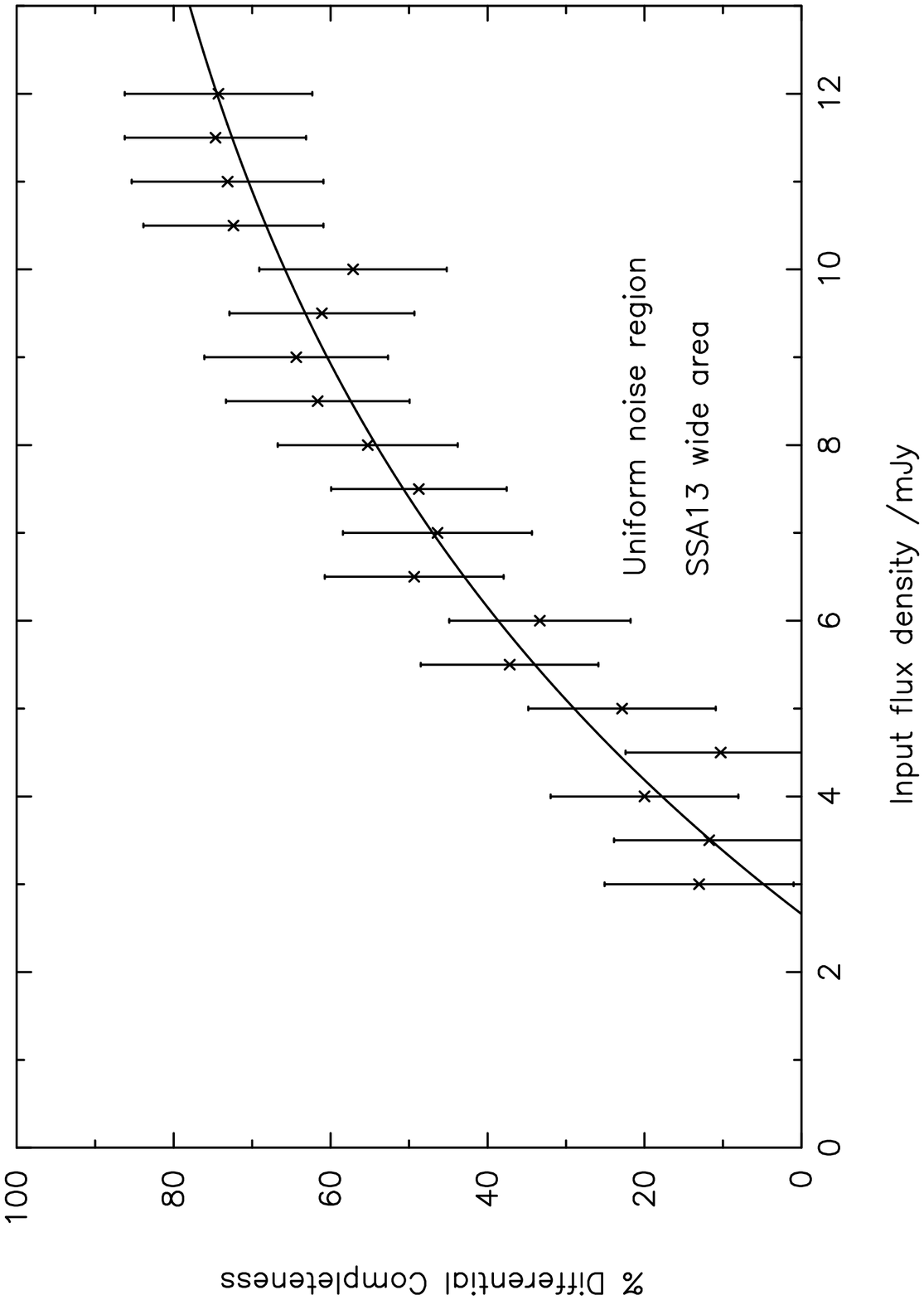}
\label{fig:ssa13wide_uni_comp} 
\caption{\small{Percentage of sources recovered against input flux
   density, for the uniform noise regions of the SSA13
 wide area field from the ``Hawaii Submm Survey''.}}
 \end{figure}

\newpage
\begin{figure}
 \centering
   \vspace*{4.8cm}
   \leavevmode
   \includegraphics{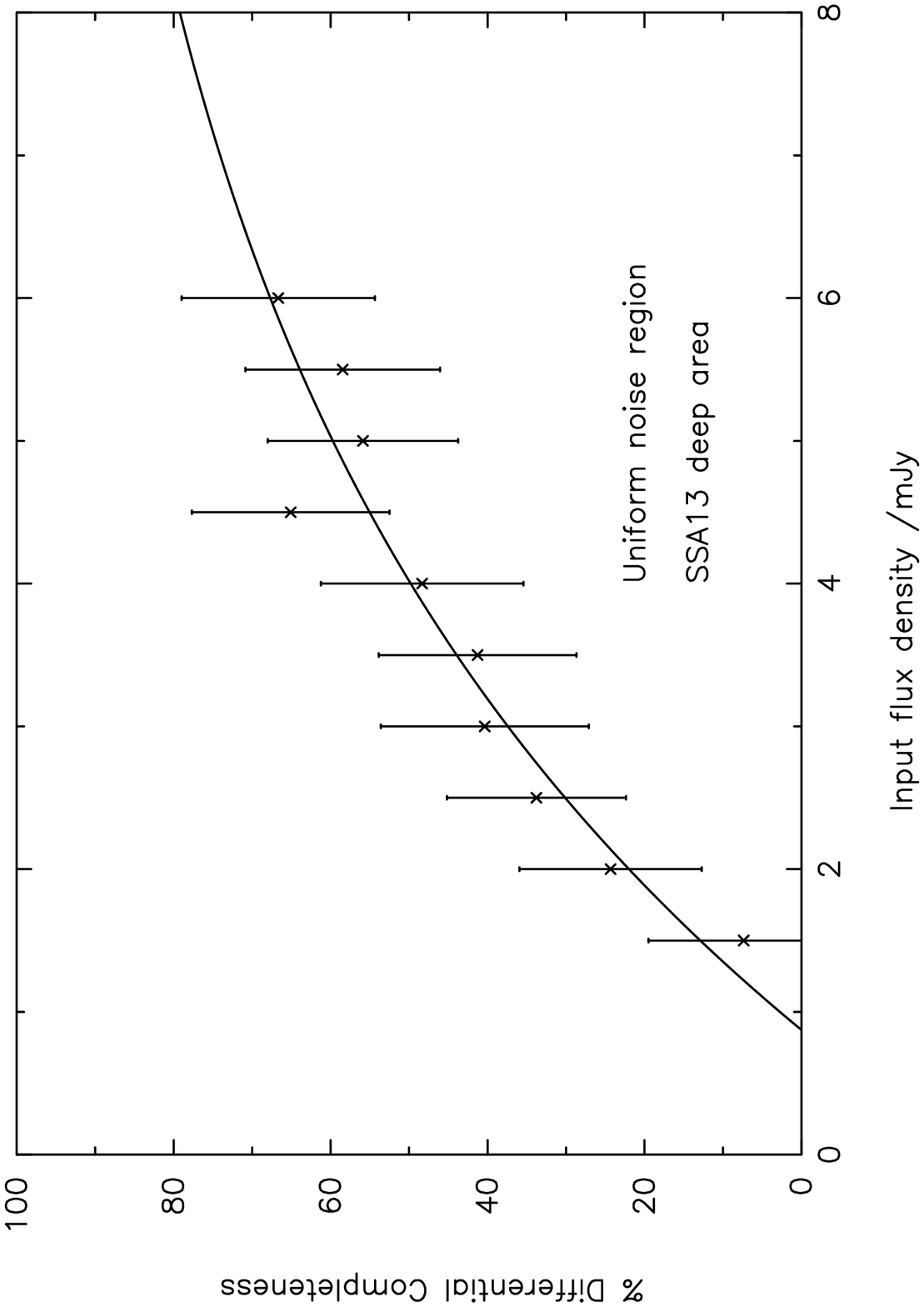}
\label{fig:ssa13deep_uni_comp} 
\caption{\small{Percentage of sources recovered against input flux
   density, for the uniform noise regions of the SSA13
 hour deep area from the ``Hawaii Submm Survey''.}}
 \centering
   \vspace*{3.7cm}
   \leavevmode
   \includegraphics{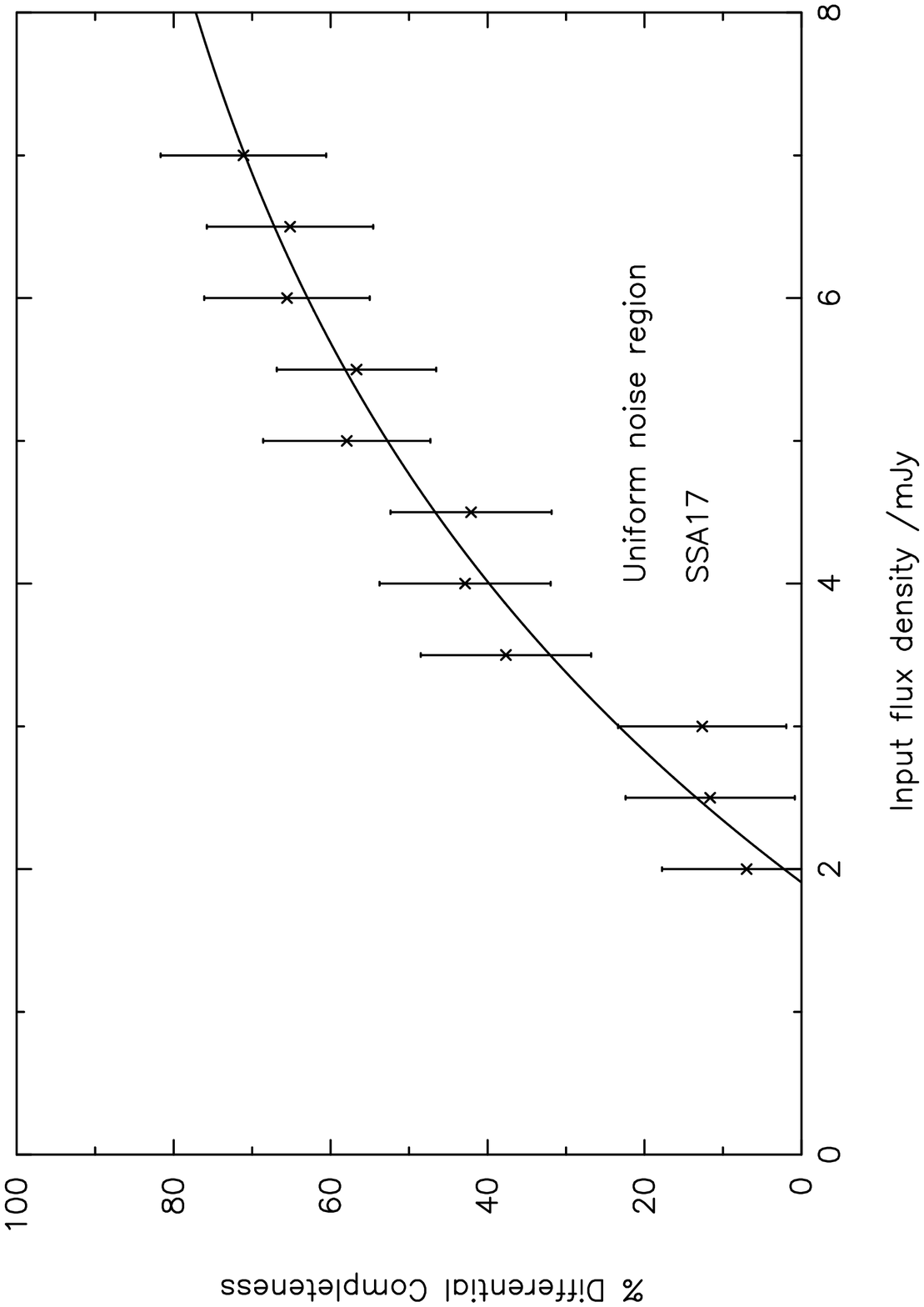}
\label{fig:ssa17_uni_comp} 
\caption{\small{Percentage of sources recovered against input flux
   density, for the uniform noise regions of the SSA17 field from
 the ``Hawaii Submm Survey''.}}
 \centering
   \vspace*{3.7cm}
   \leavevmode
   \includegraphics{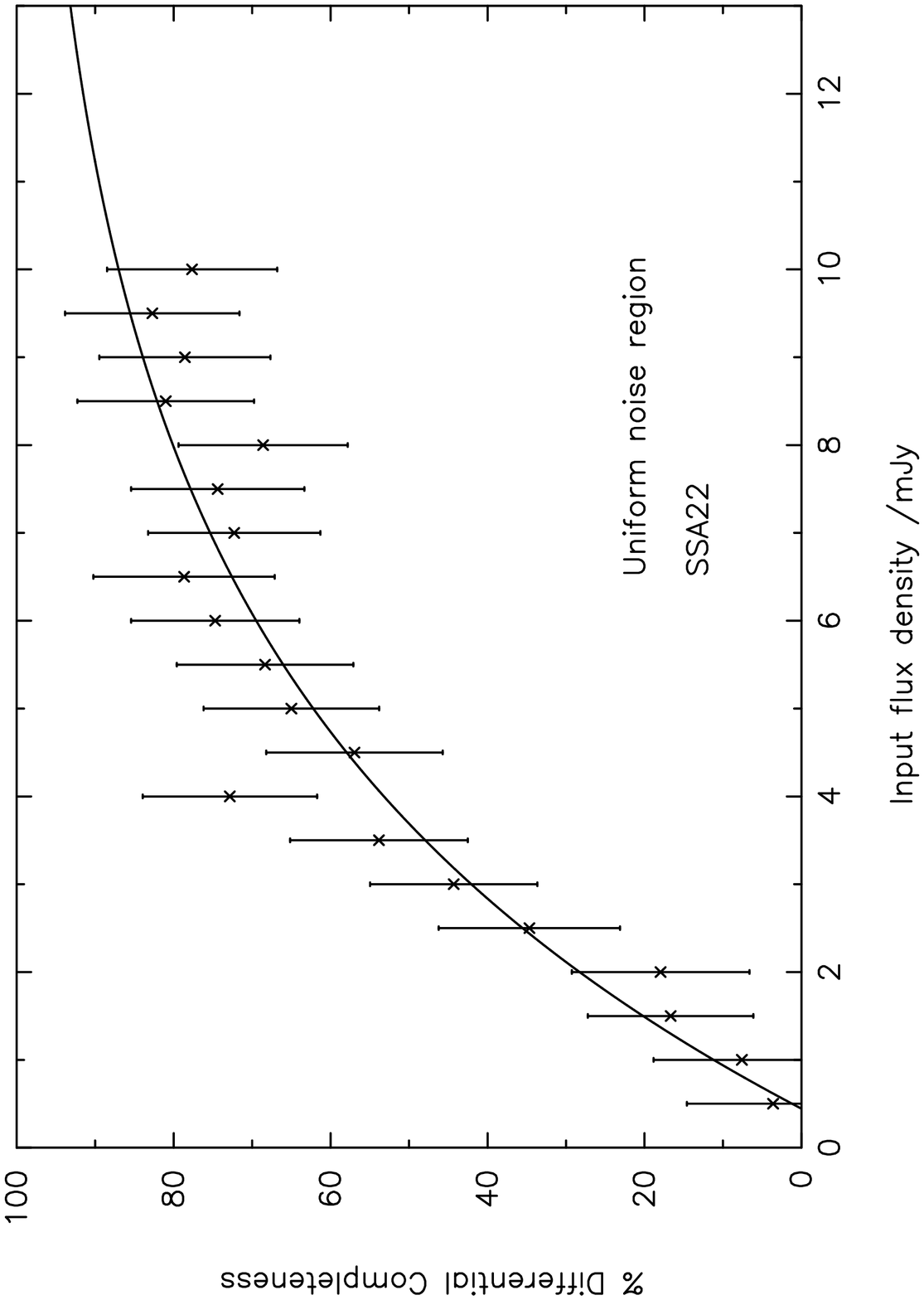}
\label{fig:ssa22_uni_comp} 
\caption{\small{Percentage of sources recovered against input flux
   density, for the uniform noise regions of the SSA22 field
 from the ``Hawaii Submm Survey''.}}
 \centering
   \vspace*{3.7cm}
   \leavevmode
   \includegraphics{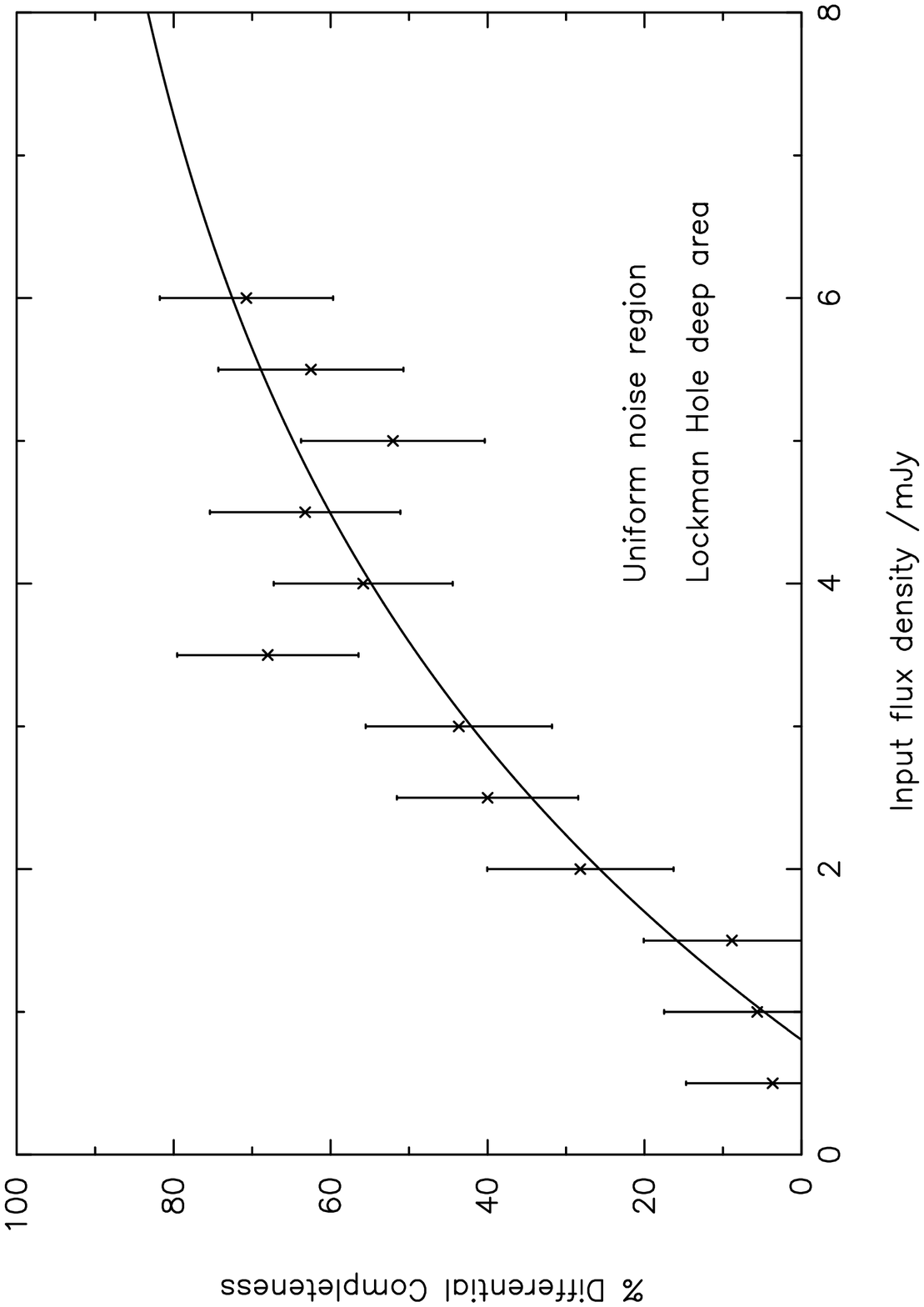}
\label{fig:lhdeep_uni_comp} 
\caption{\small{Percentage of sources recovered against input flux
   density, for the uniform noise regions of the Lockman Hole deep area
 from the ``Hawaii Submm Survey''.}}
 \end{figure}
\renewcommand{\topfraction}{0.3}
\clearpage
\begin{figure}
 \centering
   \vspace*{4.8cm}
   \leavevmode
   \includegraphics{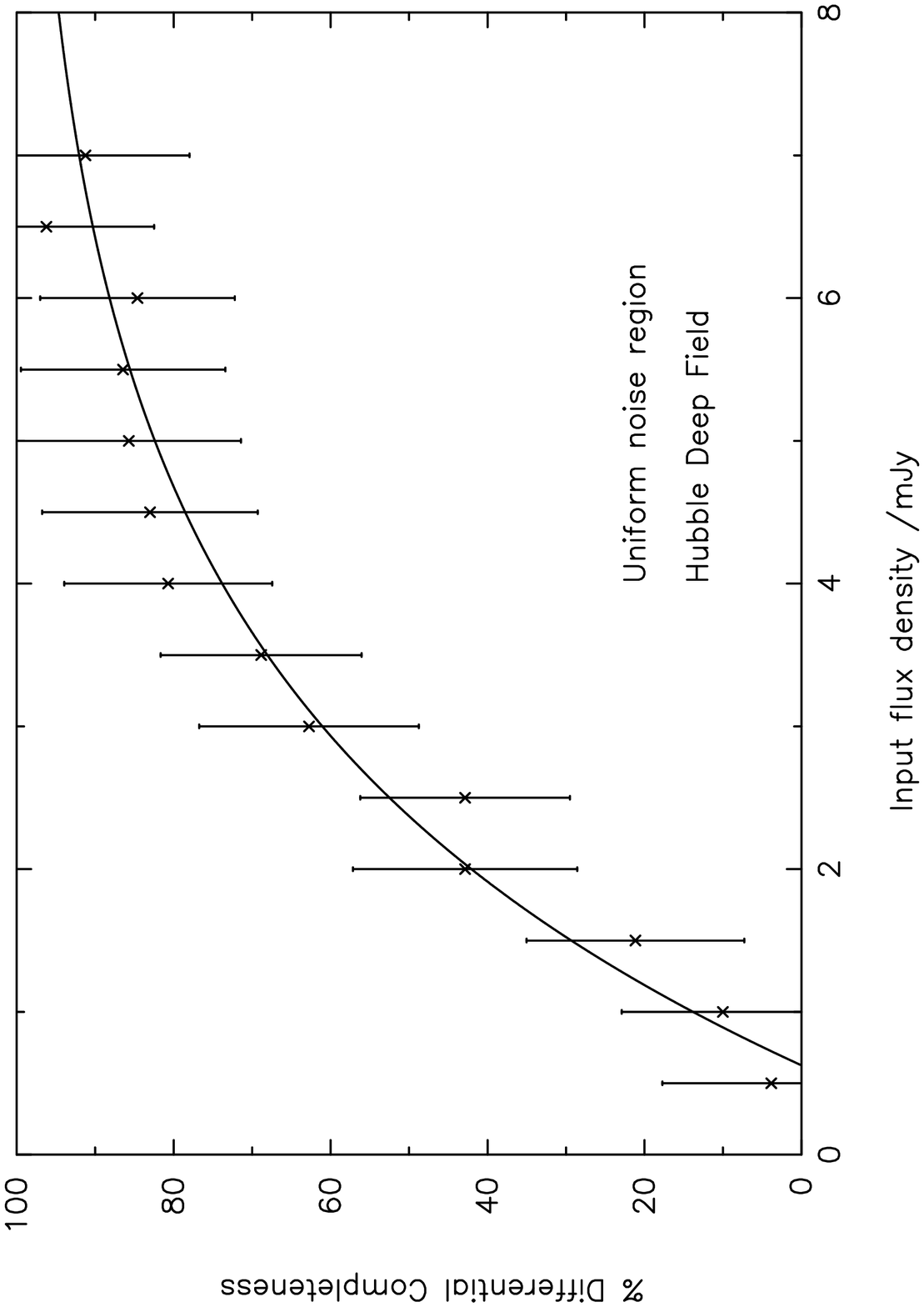}
\label{fig:hdf_uni_comp} 
\caption{\small{Percentage of sources recovered against input flux
   density, for the uniform noise regions of the Hubble Deep Field.}}
 \end{figure}

\subsection{Output versus Input Flux Density}
The following plots show the ratio of output to input flux density 
of sources recovered with
signal-to-noise ratio $>3.50$ against input flux density level for
each of the survey fields. Those fields composed of a deep pencil
beam survey within a wider-area shallower survey have had these two
components treated separately. The plotted curves are the best-fit
solutions to the empirical functional form
\be \rm \frac{output \phantom{0} flux \phantom{0} density}{input
\phantom{0} flux \phantom{0} density} = C e^{-dx} + f \ee
where x is the input flux density, and the values of C, d and f were
determined by a minimised $\rm \chi^{2}$ fit to the simulation results 
for each $\rm 850\, \mu m$ survey field and are given in Table 3.  

\begin{figure}
 \centering
   \vspace*{3.7cm}
   \leavevmode
   \includegraphics{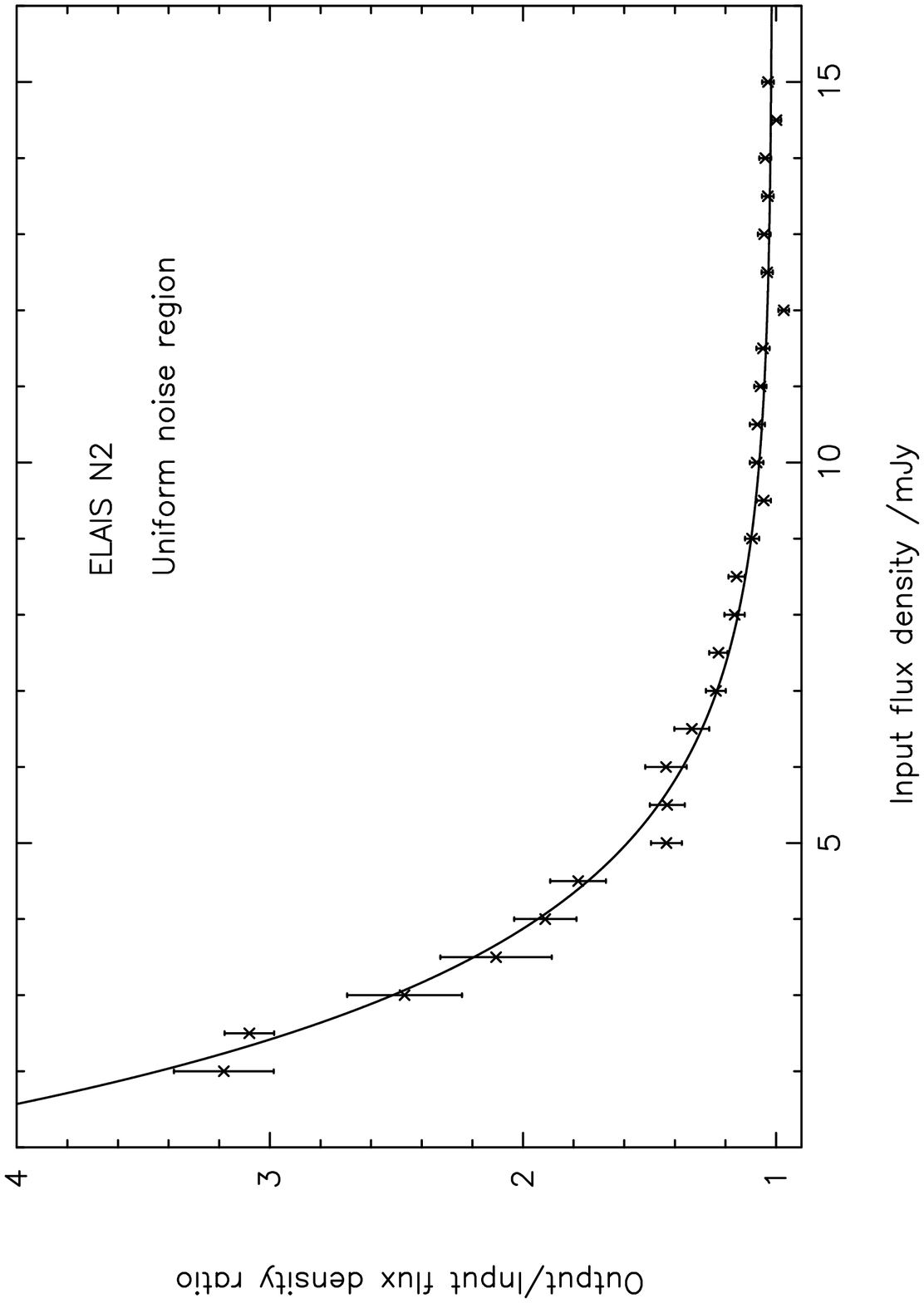}
\label{fig:n2_uni_comp} 
\caption{\small{The ratio of output to input flux density against input flux
   density, for the uniform noise regions of the ELAIS N2 field
 from the ``SCUBA 8\,mJy Survey''.}}
 \centering
   \vspace*{3.7cm}
   \leavevmode
   \includegraphics{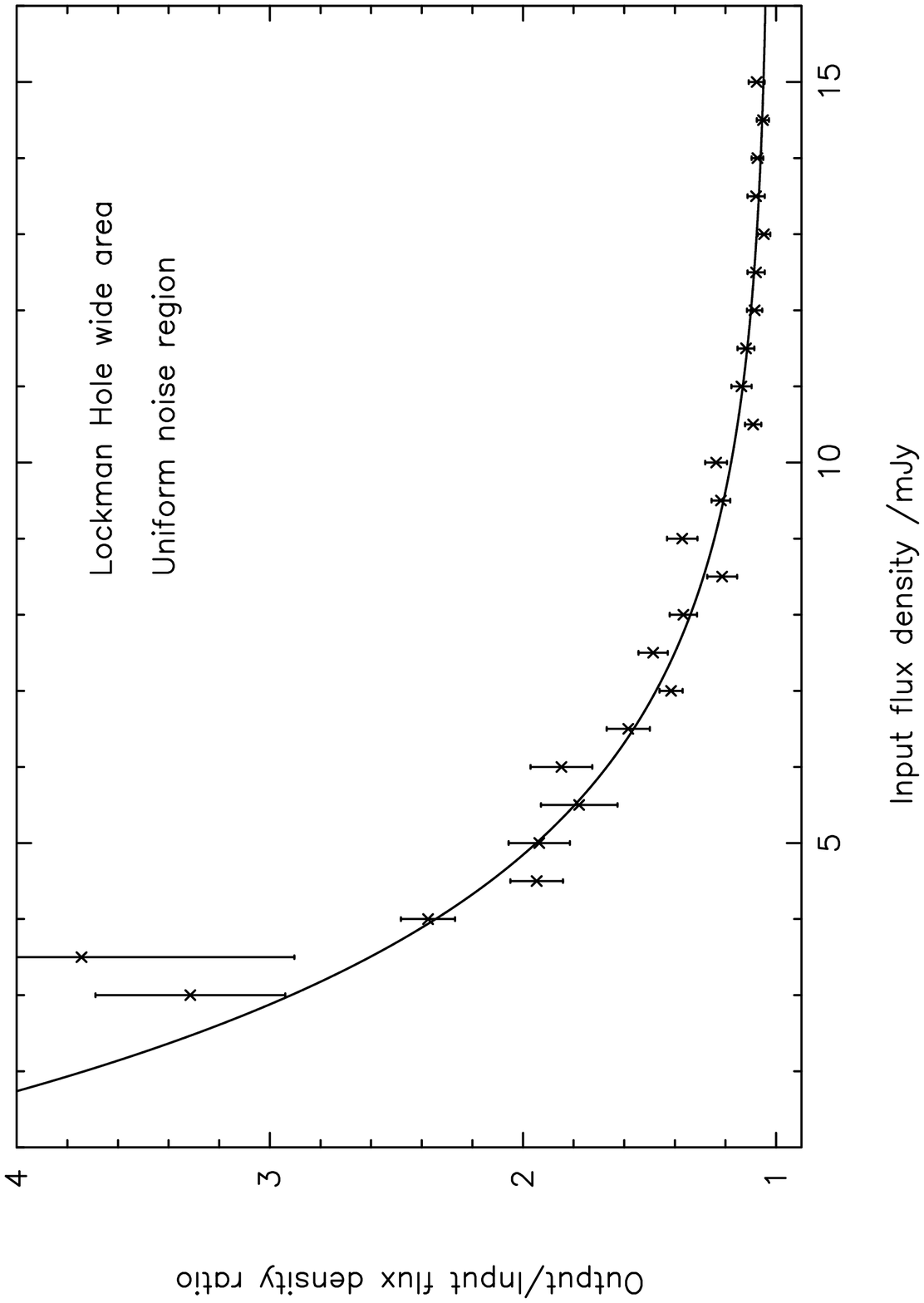}
\label{fig:lhwide_uni_comp} 
\caption{\small{The ratio of output to input flux density against input flux
   density, for the uniform noise regions of the Lockman Hole East
 wide area field from the ``SCUBA 8\,mJy Survey''.}}
 \end{figure}

\renewcommand{\topfraction}{0.3}
\newpage
\begin{figure}
 \centering
   \vspace*{4.8cm}
   \leavevmode
   \includegraphics{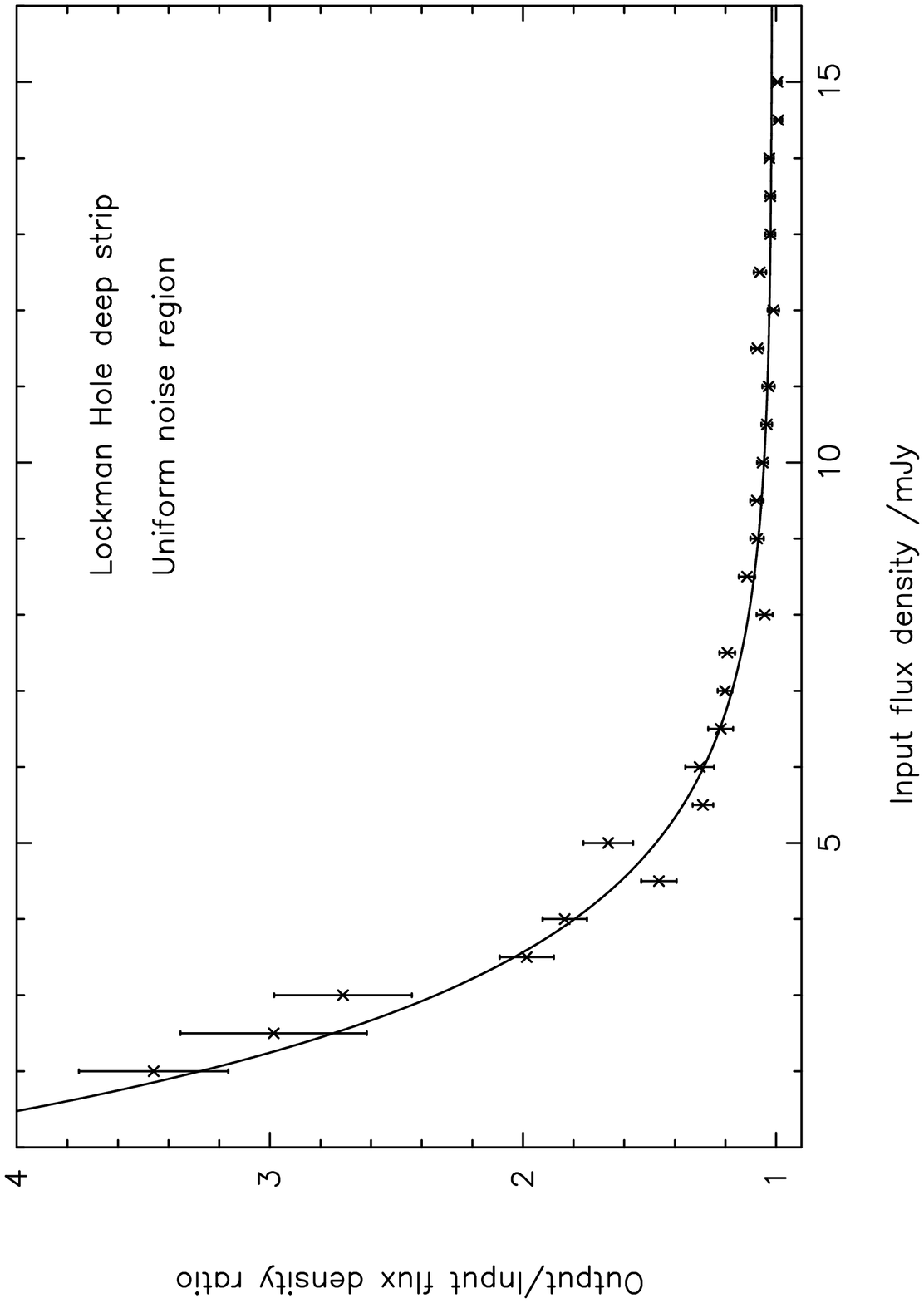}
\label{fig:lhstrip_uni_comp} 
\caption{\small{The ratio of output to input flux density against input flux
   density, for the uniform noise regions of the Lockman Hole East
 deep strip from the ``SCUBA 8\,mJy Survey''.}}
 \centering
   \vspace*{4.21cm}
   \leavevmode
   \includegraphics{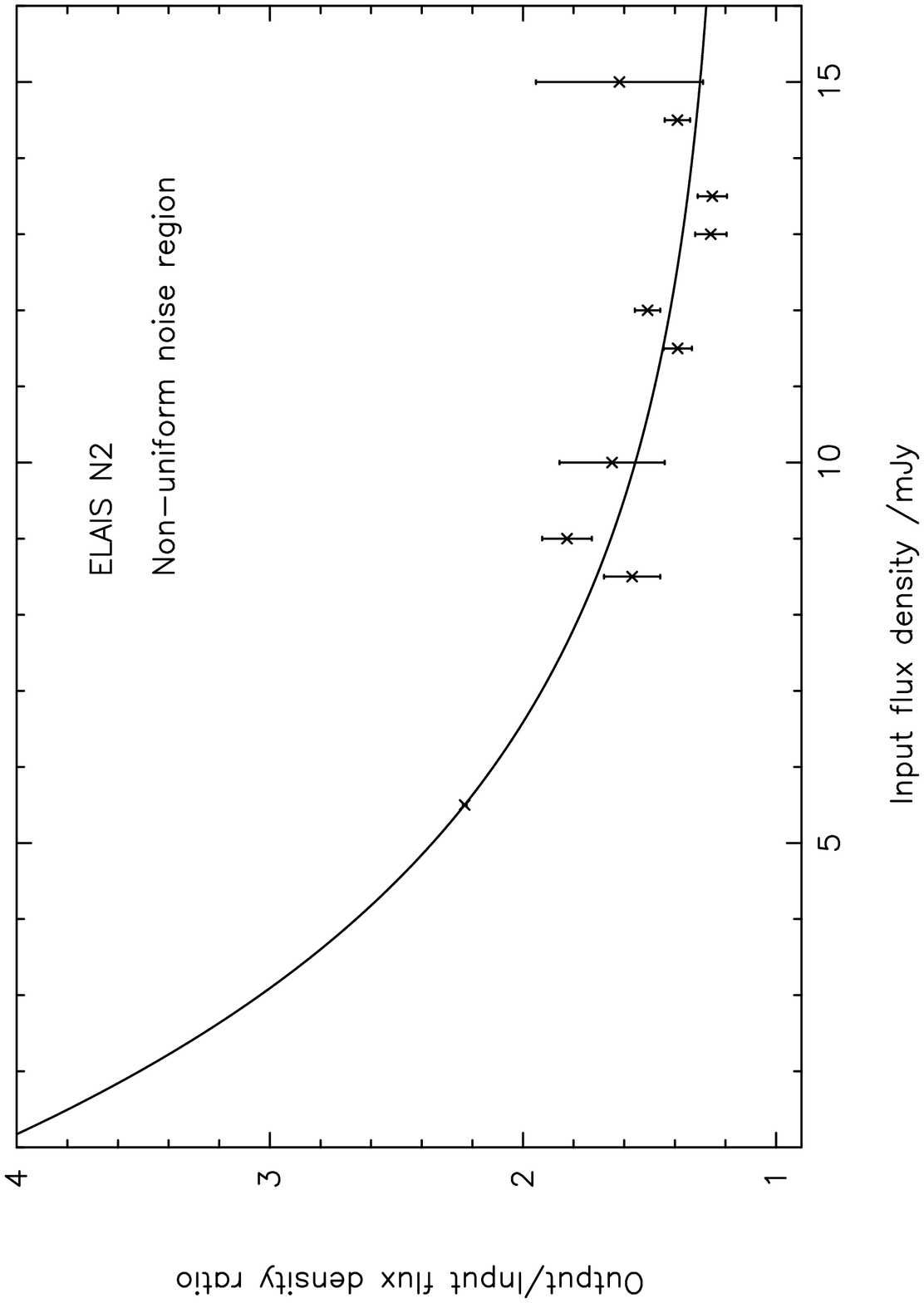}
\label{fig:n2_nonuni_comp} 
\caption{\small{The ratio of output to input flux density against input flux
   density, for the non-uni noise regions of the ELAIS N2 field
 from the ``SCUBA 8\,mJy Survey''.}}
 \centering
   \vspace*{4.21cm}
   \leavevmode
   \includegraphics{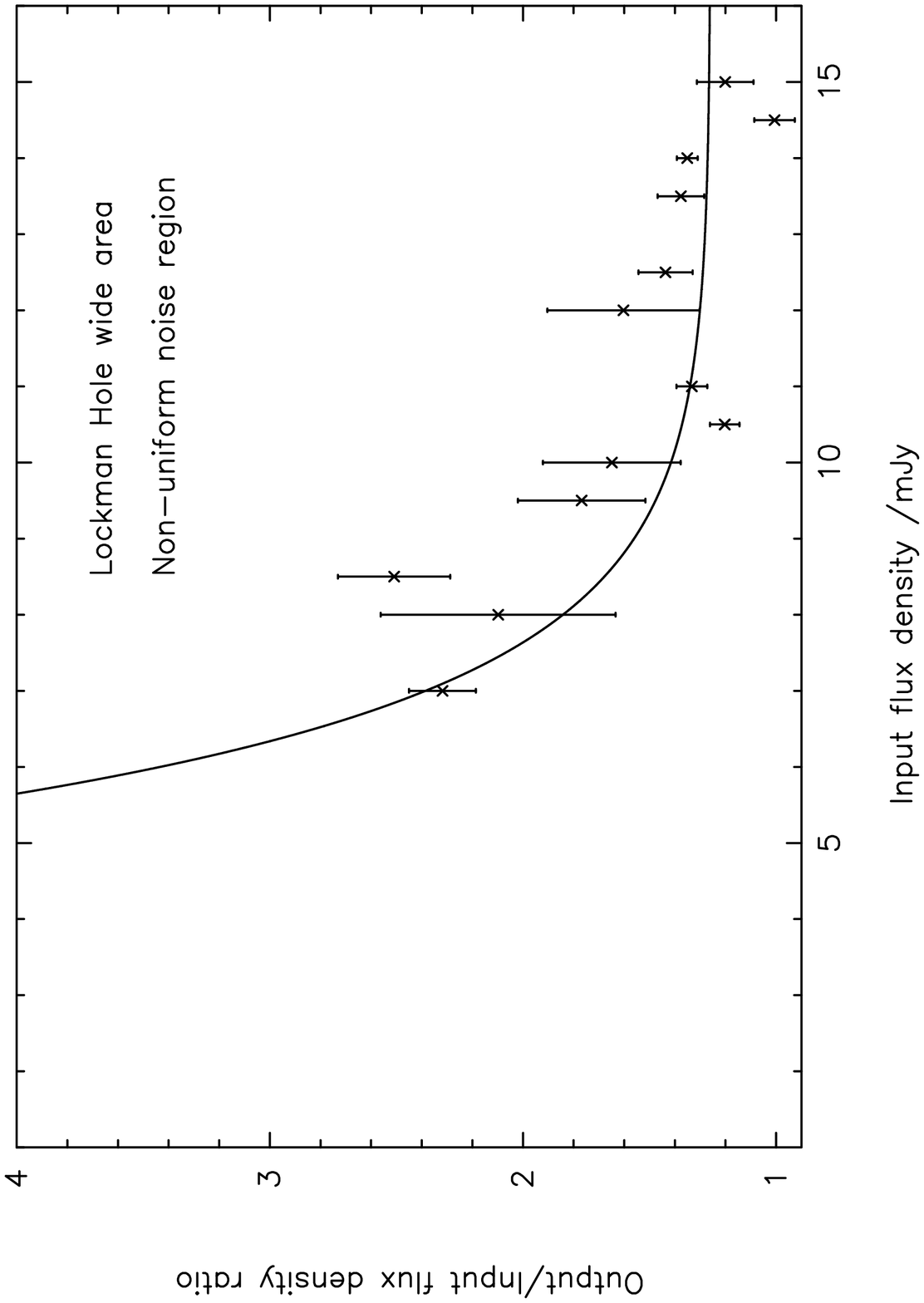}
\label{fig:lhwide_nonuni_comp} 
\caption{\small{The ratio of output to input flux density against input flux
   density, for the non-uni noise regions of the Lockman Hole East
 wide area field from the ``SCUBA 8\,mJy Survey''.}}
 \centering
   \vspace*{3.7cm}
   \leavevmode
   \includegraphics{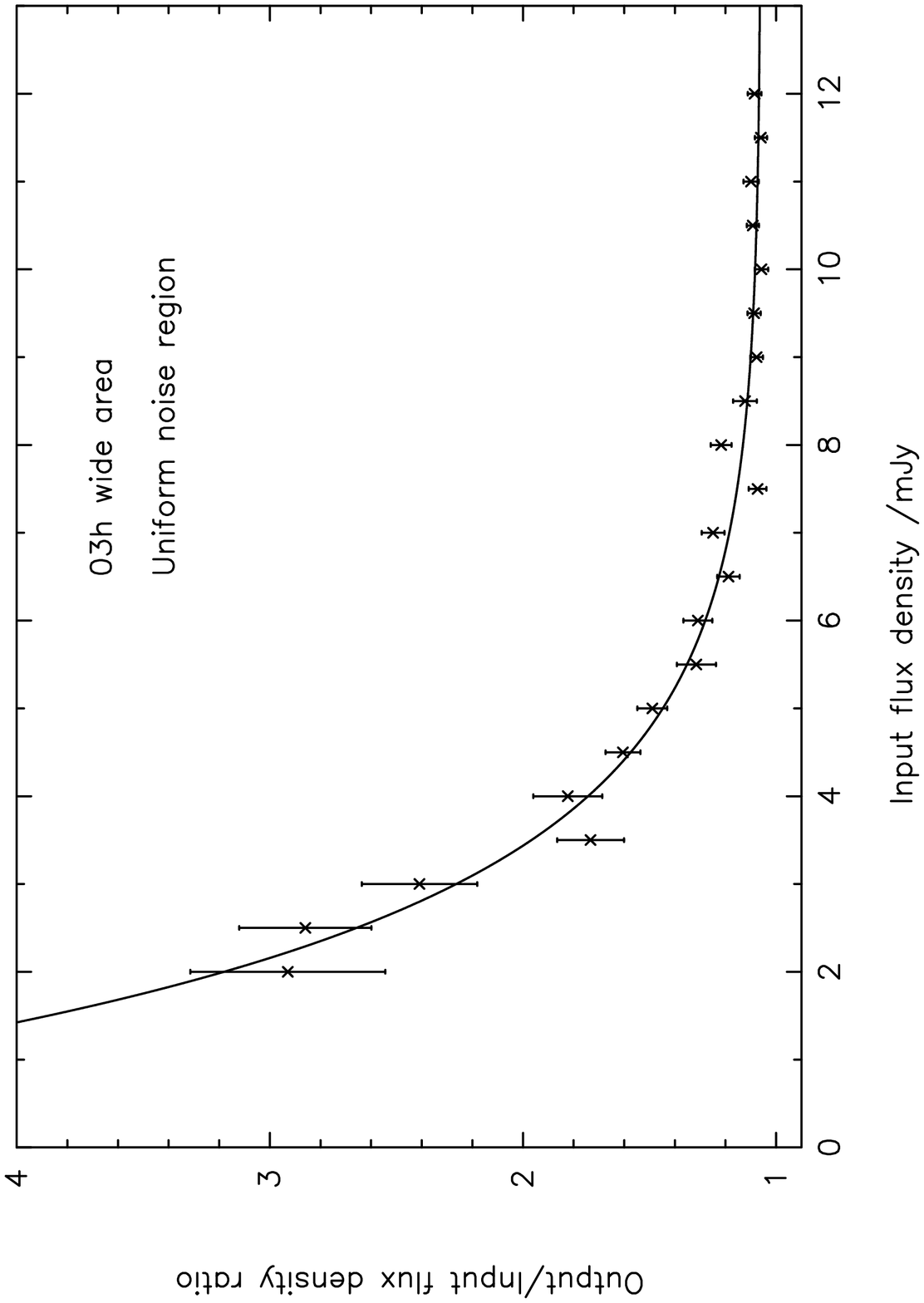}
\label{fig:03wide_uni_comp} 
\caption{\small{The ratio of output to input flux density against input flux
   density, for the uniform noise regions of the 03 hour
 wide area field from the ``CUDSS''.}}
 \end{figure}
\newpage
\begin{figure}
 \centering
   \vspace*{4.8cm}
   \leavevmode
   \includegraphics{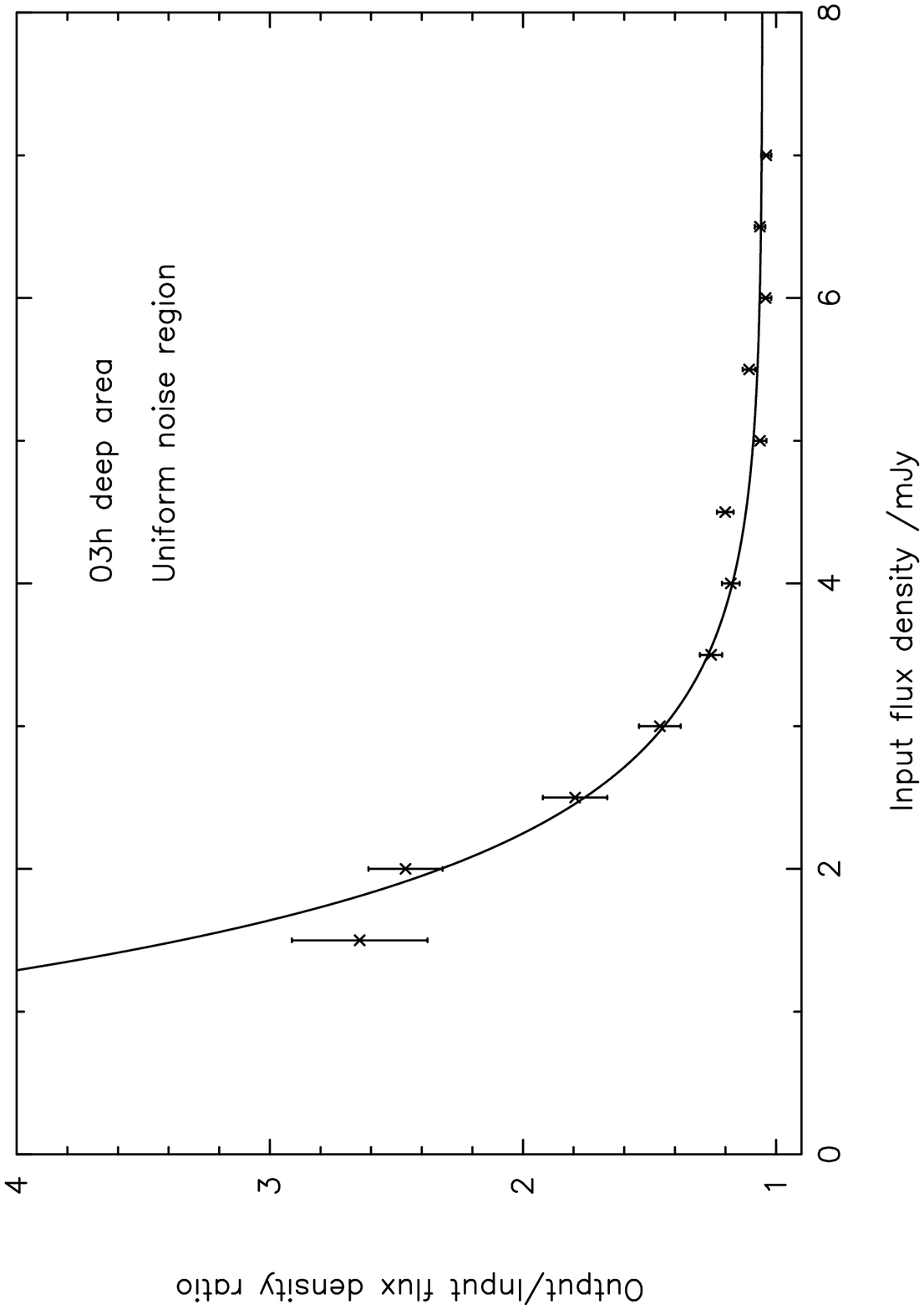}
\label{fig:03deep_uni_comp} 
\caption{\small{The ratio of output to input flux density against input flux
   density, for the uniform noise regions of the 03
 hour deep area from the ``CUDSS''.}}
 \centering
   \vspace*{3.7cm}
   \leavevmode
   \includegraphics{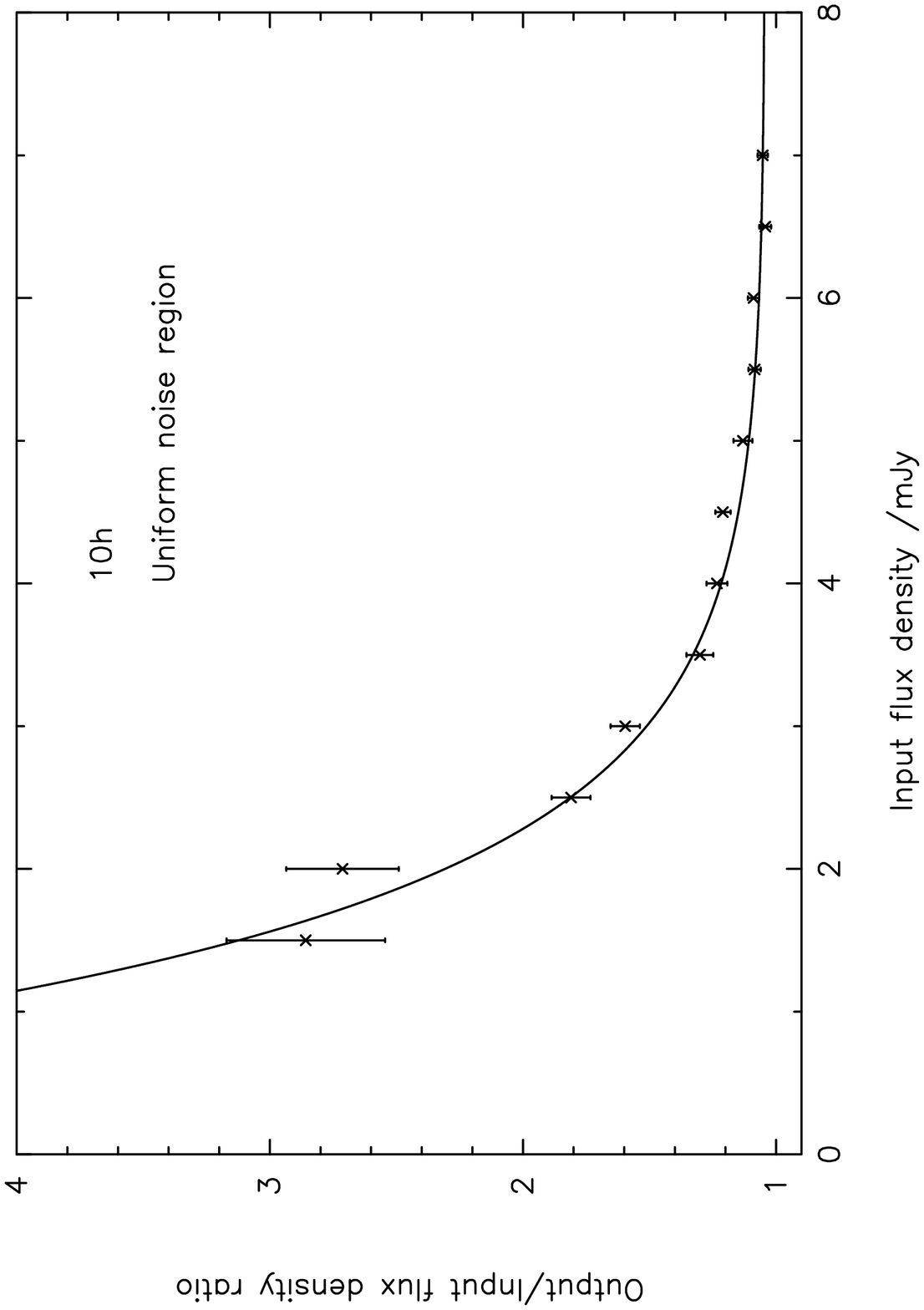}
\label{fig:10_uni_comp} 
\caption{\small{Percentage of sources recovered against input flux
   density, for the uniform noise regions of the 10 hour field from
 the ``CUDSS''.}}
 \centering
   \vspace*{3.7cm}
   \leavevmode
   \includegraphics{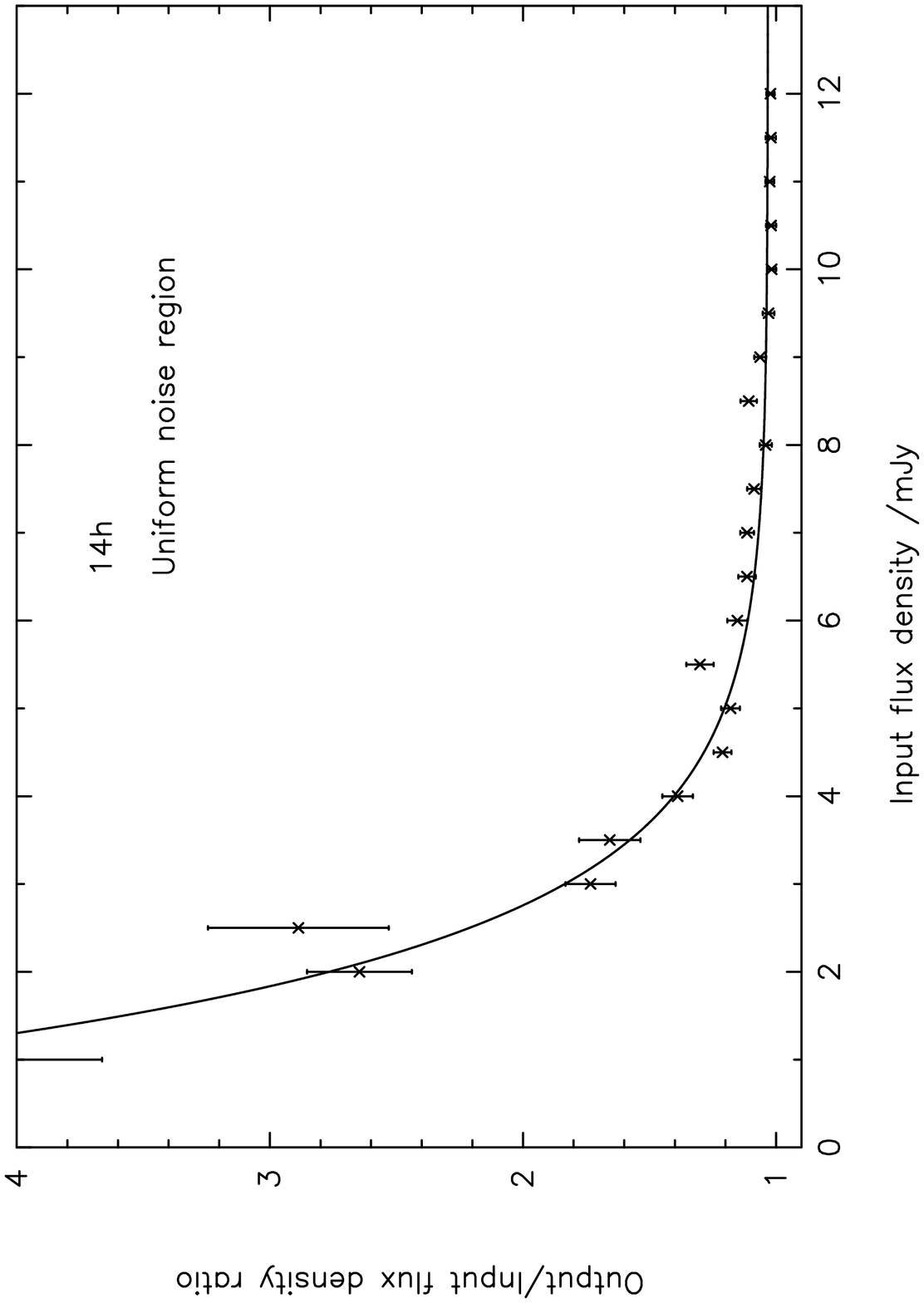}
\label{fig:14_uni_comp} 
\caption{\small{The ratio of output to input flux density against input flux
   density, for the uniform noise regions of the 14 hour field
 from the ``CUDSS''.}}
 \centering
   \vspace*{3.7cm}
   \leavevmode
   \includegraphics{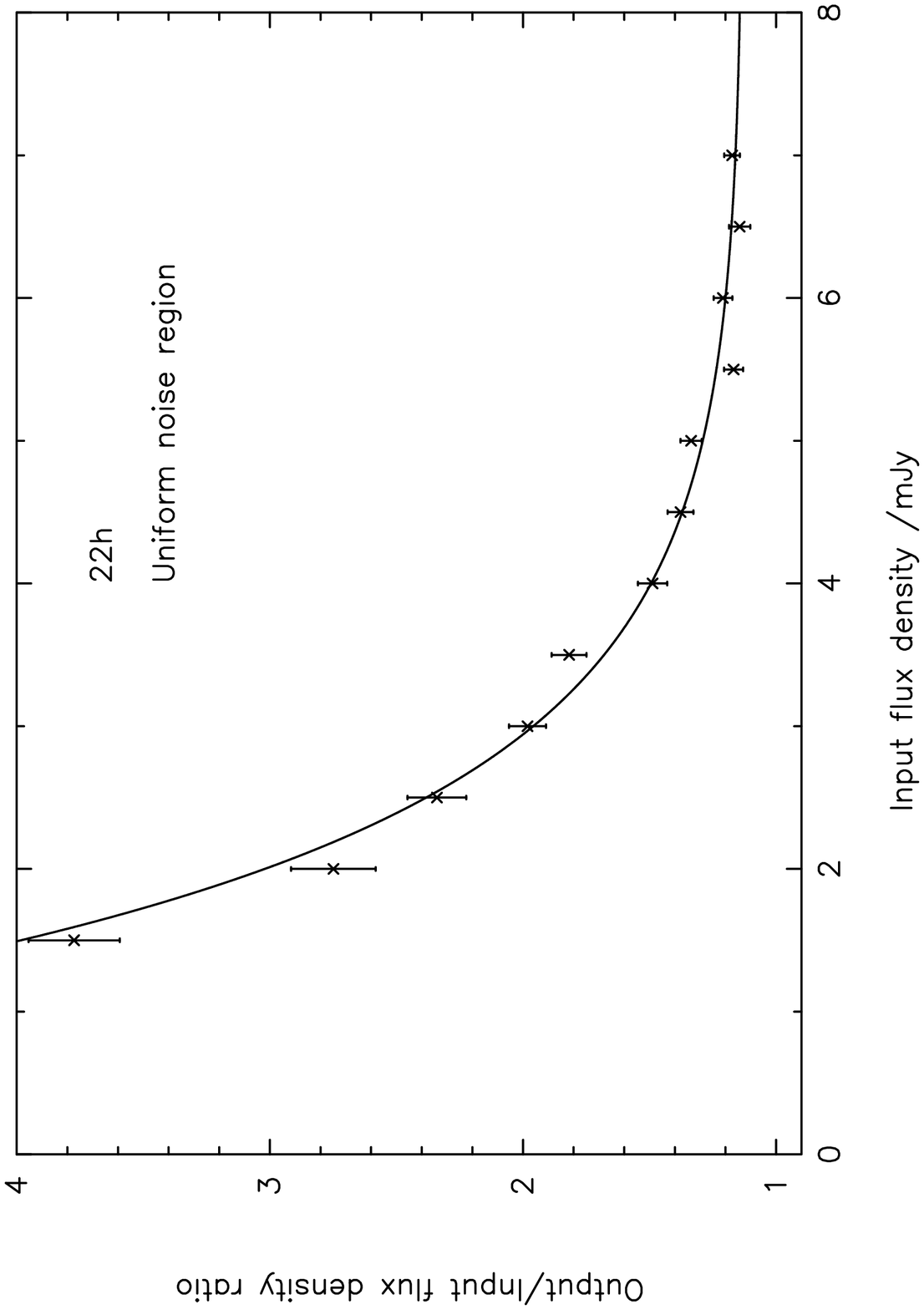}
\label{fig:22_uni_comp} 
\caption{\small{The ratio of output to input flux density against input flux
   density, for the uniform noise regions of the 22 hour field
 from the `CUDSS''.}}
 \end{figure}

\newpage
\begin{figure}
 \centering
   \vspace*{4.8cm}
   \leavevmode
   \includegraphics{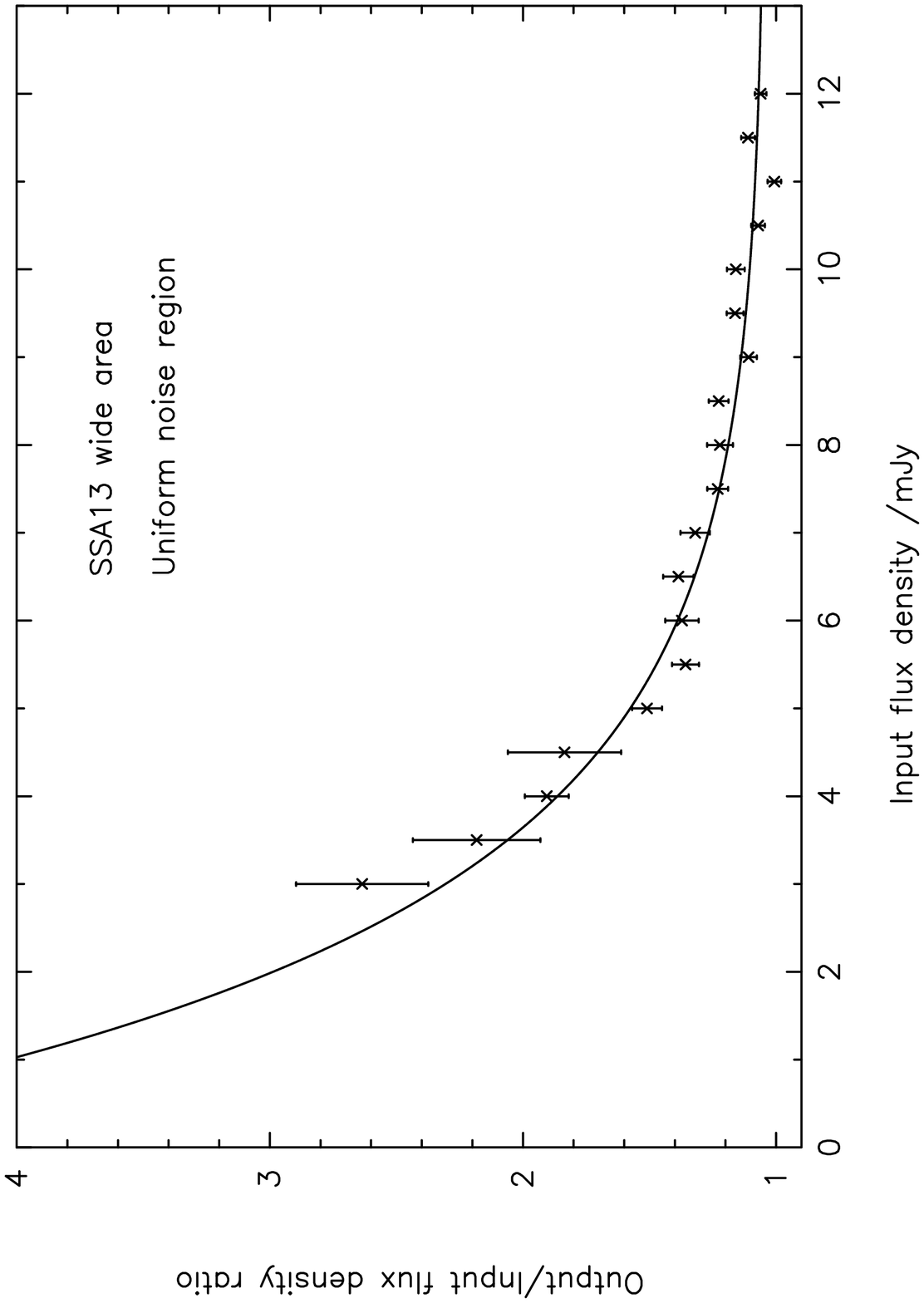}
\label{fig:ssa13wide_uni_comp} 
\caption{\small{The ratio of output to input flux density against input flux
   density, for the uniform noise regions of the SSA13
 wide area field from the ``Hawaii Submm Survey''.}}
 \centering
   \vspace*{3.7cm}
   \leavevmode
   \includegraphics{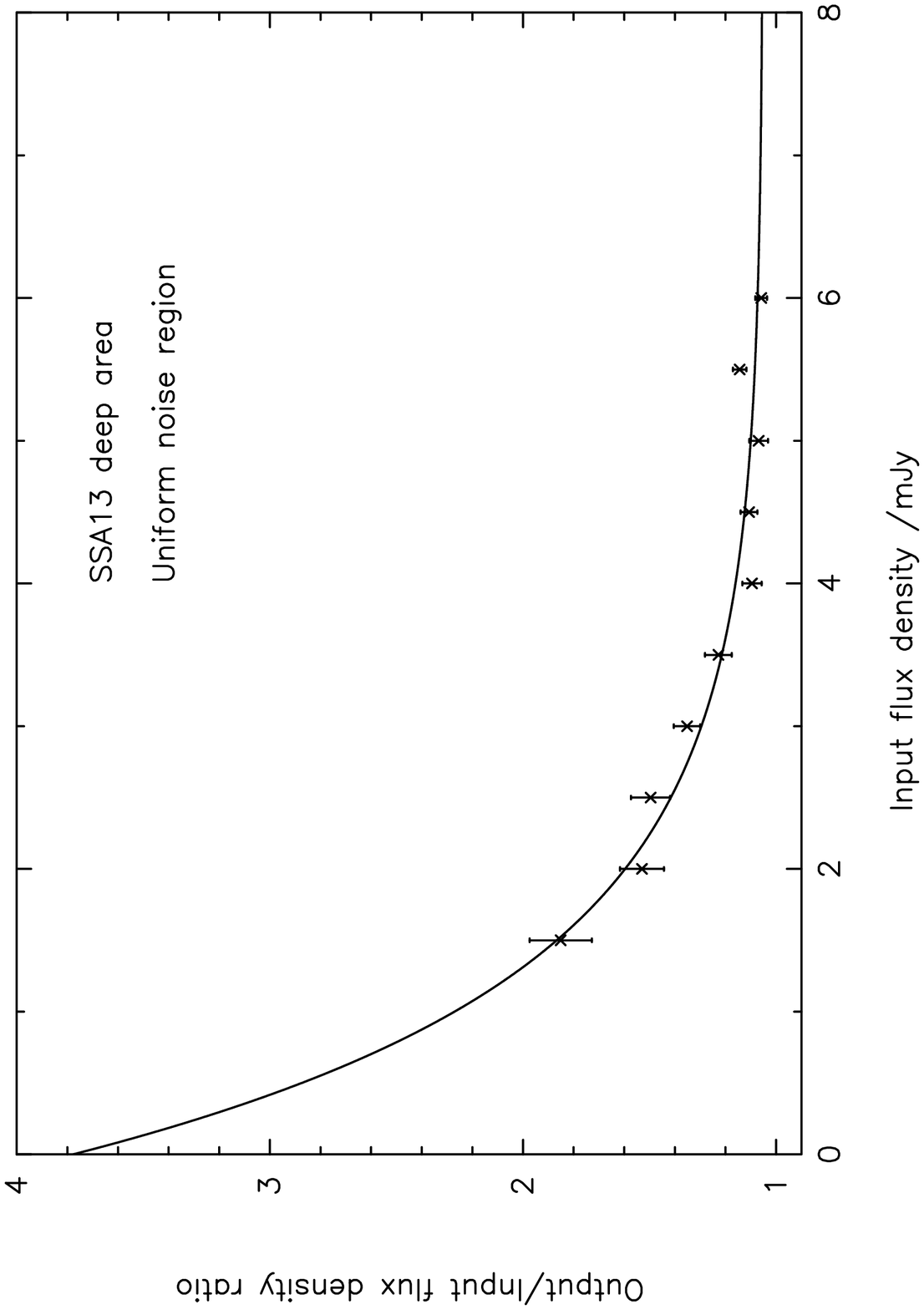}
\label{fig:ssa13deep_uni_comp} 
\caption{\small{The ratio of output to input flux density against input flux
   density, for the uniform noise regions of the SSA13
 hour deep area from the ``Hawaii Submm Survey''.}}
 \centering
   \vspace*{3.7cm}
   \leavevmode
   \includegraphics{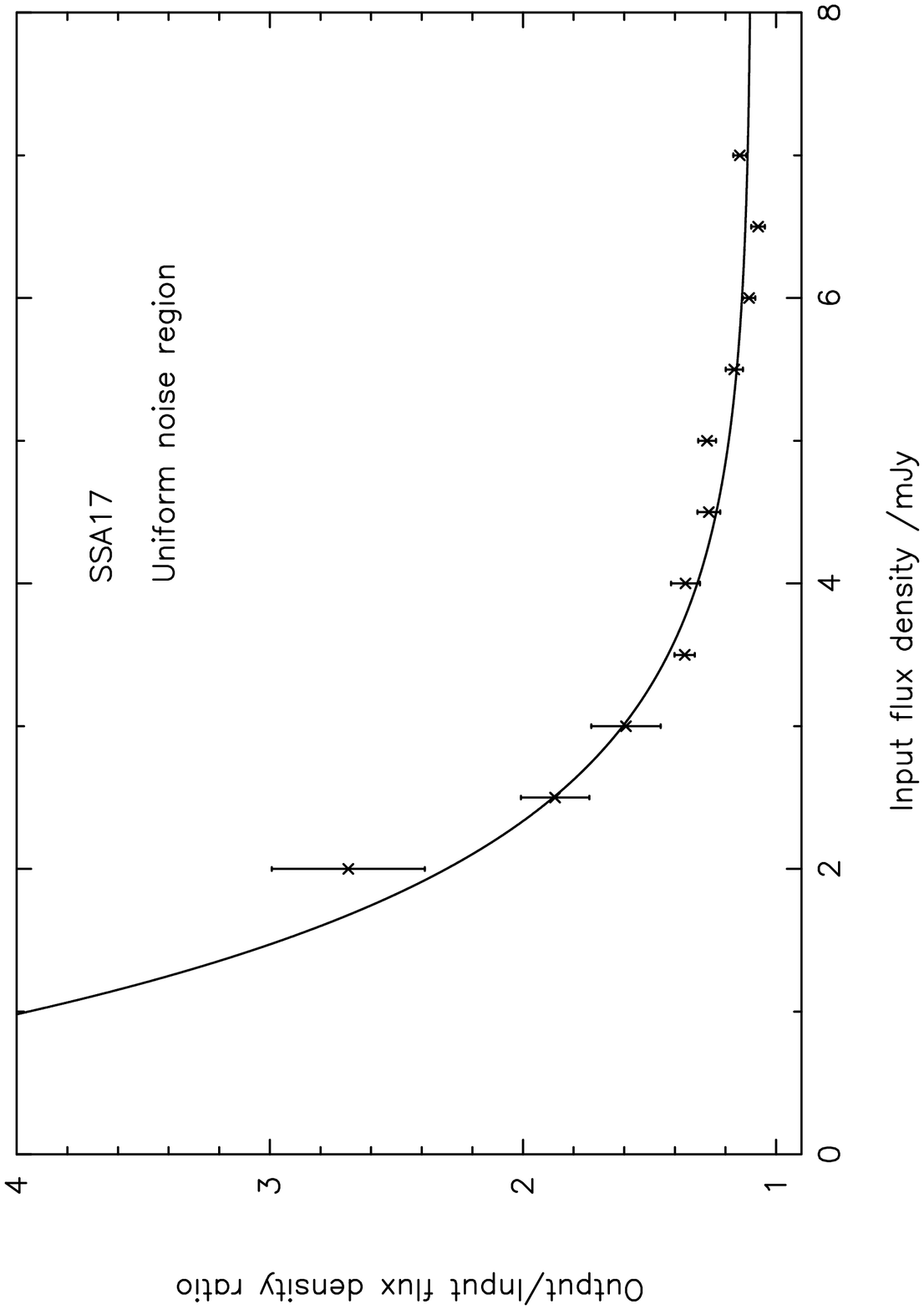}
\label{fig:ssa17_uni_comp} 
\caption{\small{The ratio of output to input flux density against input flux
   density, for the uniform noise regions of the SSA17 field from
 the ``Hawaii Submm Survey''.}}
 \centering
   \vspace*{3.7cm}
   \leavevmode
   \includegraphics{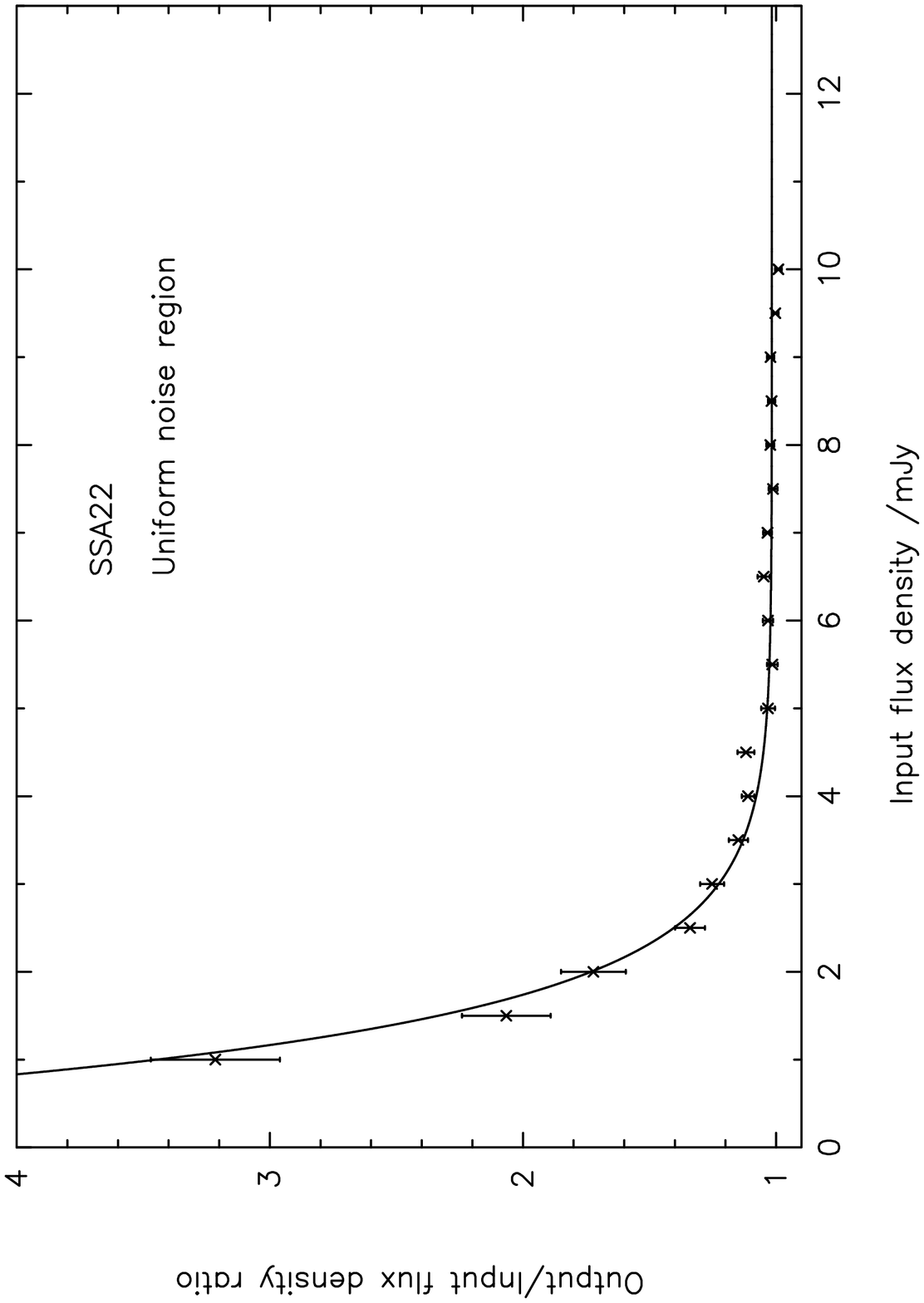}
\label{fig:ssa22_uni_comp} 
\caption{\small{The ratio of output to input flux density against input flux
   density, for the uniform noise regions of the SSA22 field
 from the ``Hawaii Submm Survey''.}}
 \end{figure}

\renewcommand{\topfraction}{0.6}
\clearpage
\begin{figure}
 \centering
   \vspace*{4.8cm}
   \leavevmode
   \includegraphics{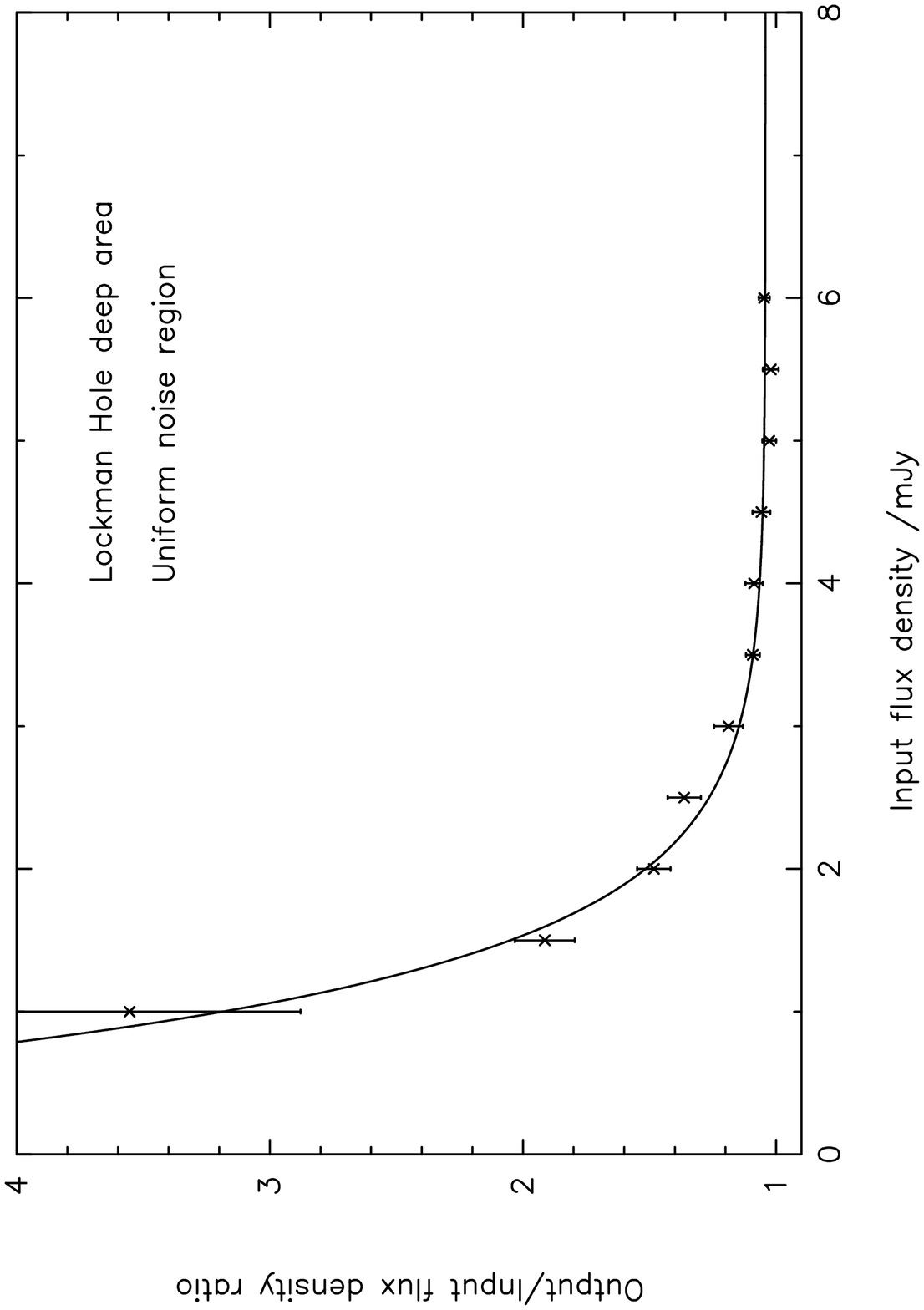}
\label{fig:lhdeep_uni_comp} 
\caption{\small{The ratio of output to input flux density against input flux
   density, for the uniform noise regions of the Lockman Hole deep area
 from the ``Hawaii Submm Survey''.}}
 \centering
   \vspace*{3.7cm}
   \leavevmode
   \includegraphics{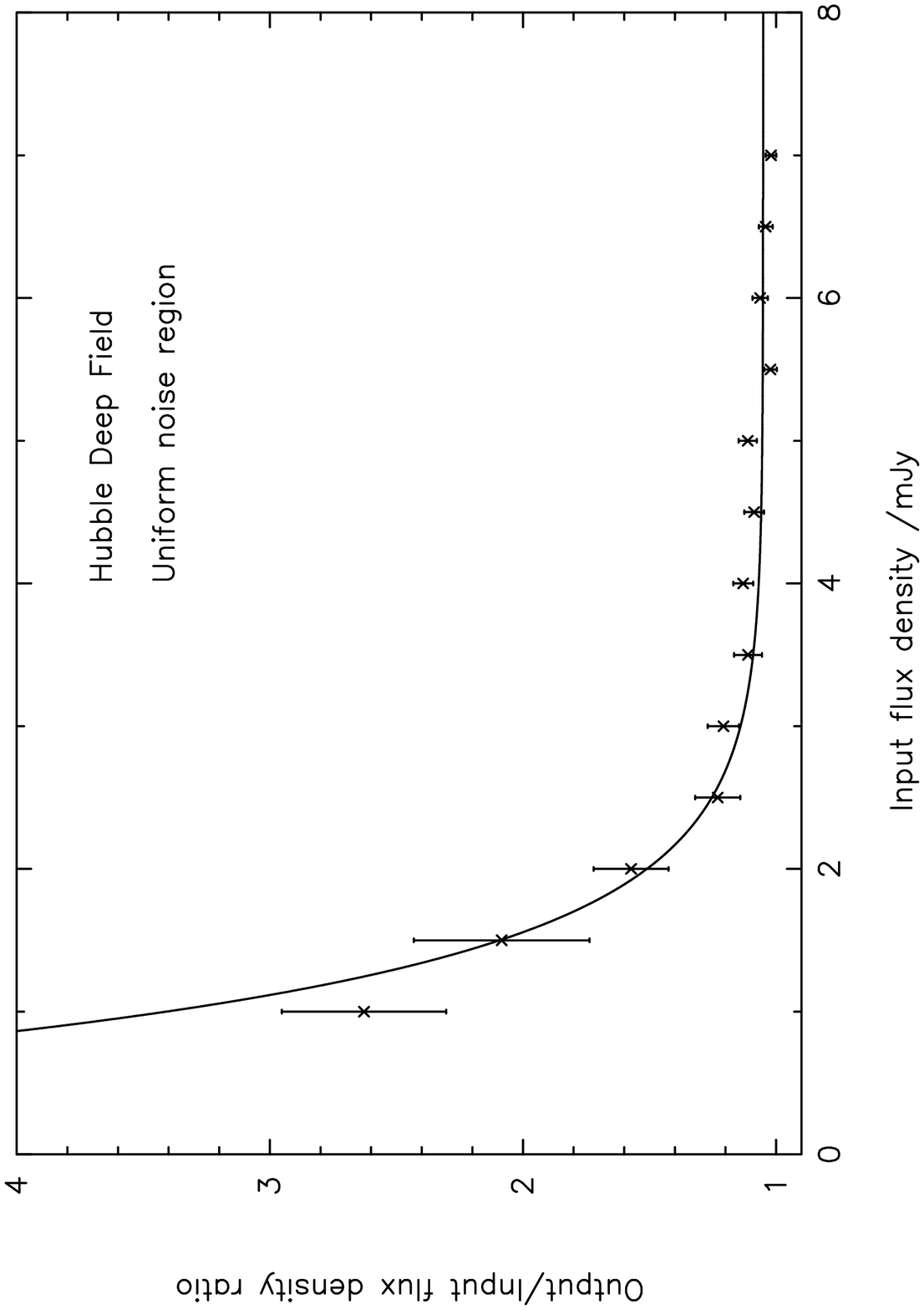}
\label{fig:hdf_uni_comp} 
\caption{\small{The ratio of output to input flux density against input flux
   density, for the uniform noise regions of the Hubble Deep Field.}}
 \end{figure}

\subsection{Positional Uncertainty}
The following plots show the mean positional uncertainty 
of sources recovered with
signal-to-noise ratio $>3.50$ against input flux density level for
each of the survey fields. Those fields composed of a deep pencil
beam survey within a wider-area shallower survey have had these two
components treated separately. The mean positional uncertainty in retrieving the fake sources was
found to be well approximated by a linear dependence on the input flux
density such that
  
\be \rm positional \phantom{0} error = -gx + h \ee
where x is the input flux density, the values of g and h for each 
$\rm 850\, \mu m$ survey field were determined by a minimised $\rm
\chi^{2}$ fit to the simulation results (given in Table 4), and the
positional uncertainty is given in arcseconds. Note these simulations
do not account for pointing drifts whilst observing.

\renewcommand{\topfraction}{0.95}
\begin{figure}
 \centering
   \vspace*{4.8cm}
   \leavevmode
   \includegraphics{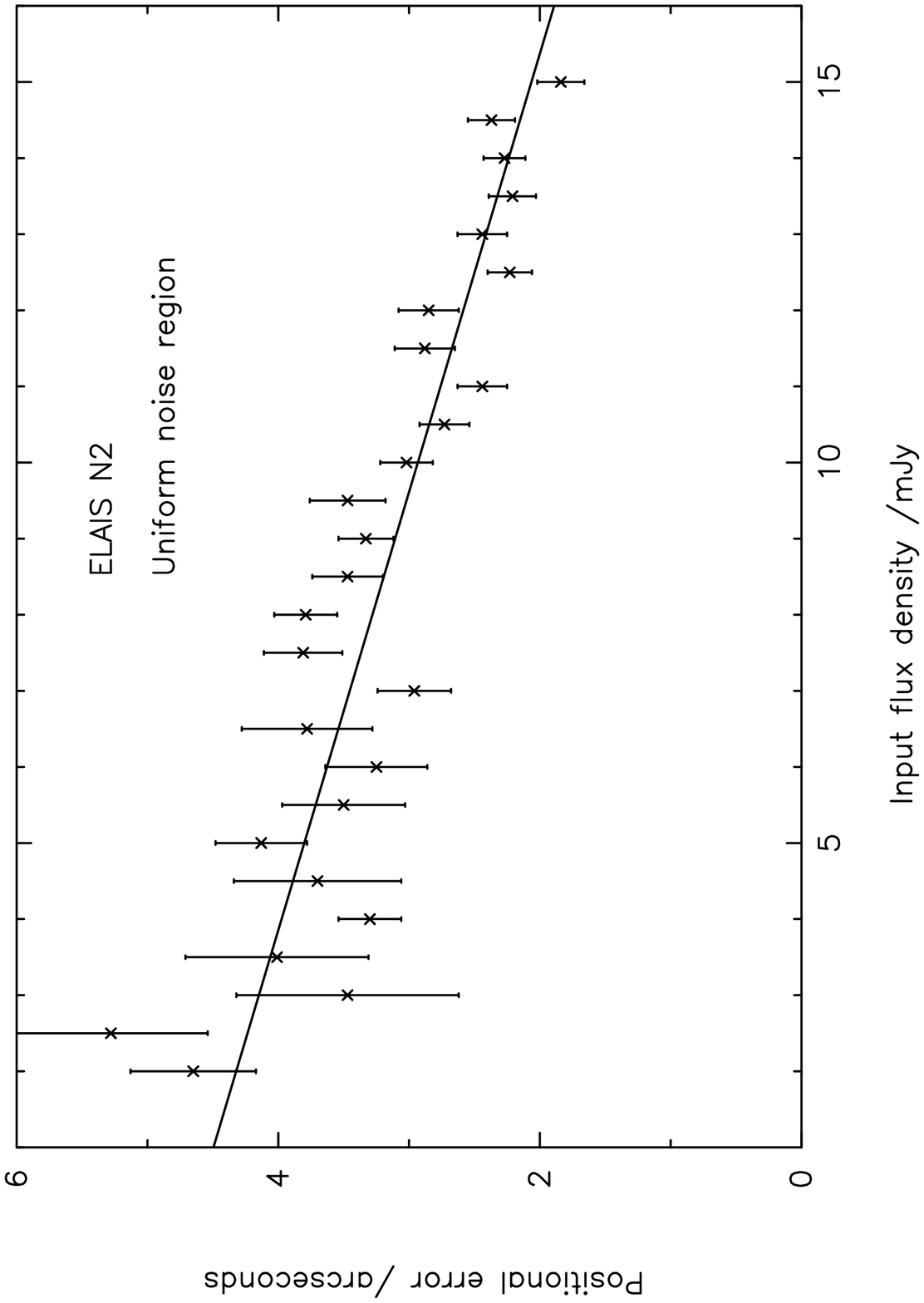}
\label{fig:n2_uni_poserr} 
\caption{\small{Mean positional uncertainty against input flux
   density, for the uniform noise regions of the ELAIS N2 field
 from the ``SCUBA 8\,mJy Survey''.}}
 \centering
   \vspace*{3.7cm}
   \leavevmode
   \includegraphics{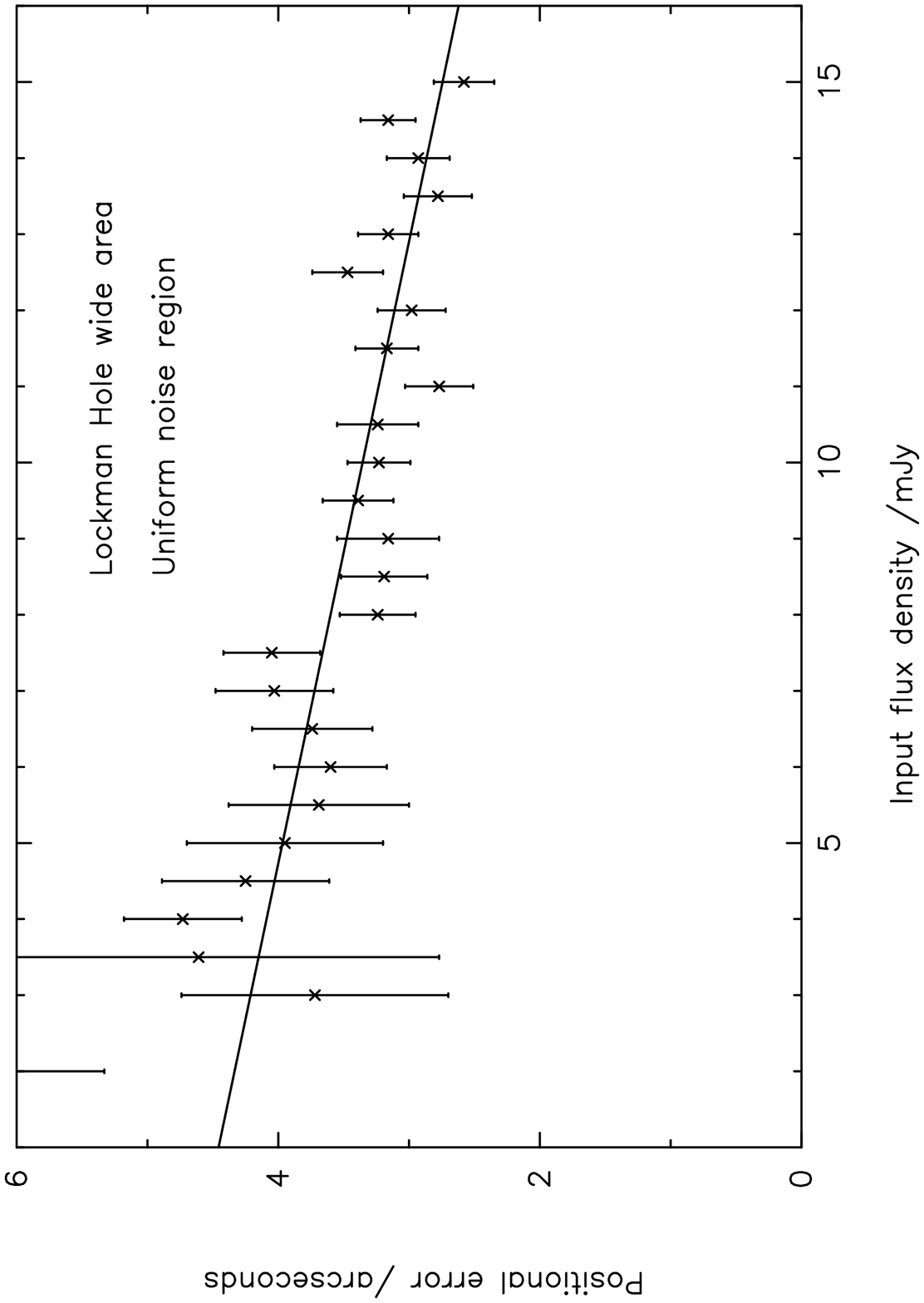}
\label{fig:lhwide_uni_poserr} 
\caption{\small{Mean positional uncertainty against input flux
   density, for the uniform noise regions of the Lockman Hole East
 wide area field from the ``SCUBA 8\,mJy Survey''.}}
 \centering
   \vspace*{3.7cm}
   \leavevmode
   \includegraphics{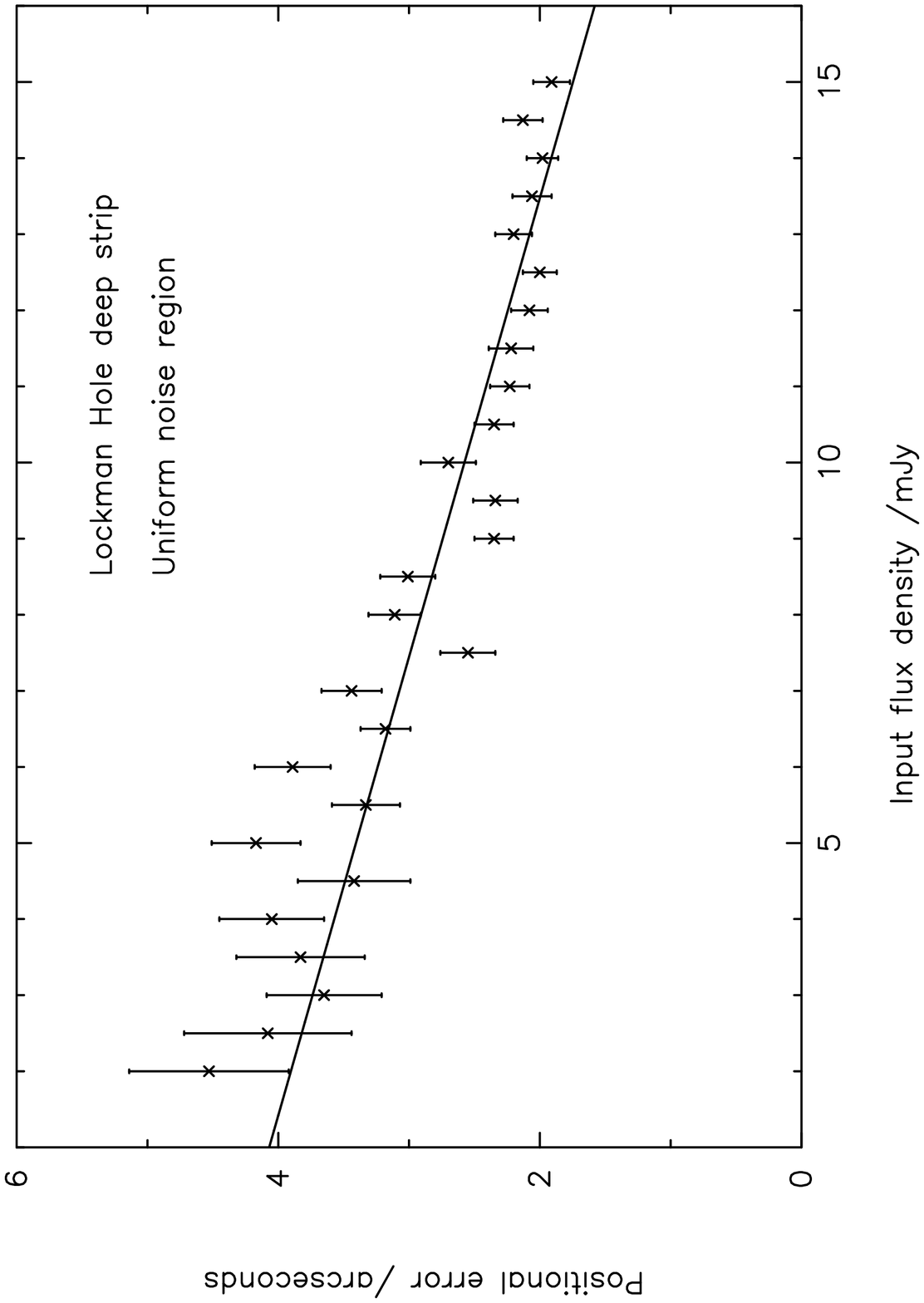}
\label{fig:lhstrip_uni_poserr} 
\caption{\small{Mean positional uncertainty against input flux
   density, for the uniform noise regions of the Lockman Hole East
 deep strip from the ``SCUBA 8\,mJy Survey''.}}
 \centering
   \vspace*{3.7cm}
   \leavevmode
   \includegraphics{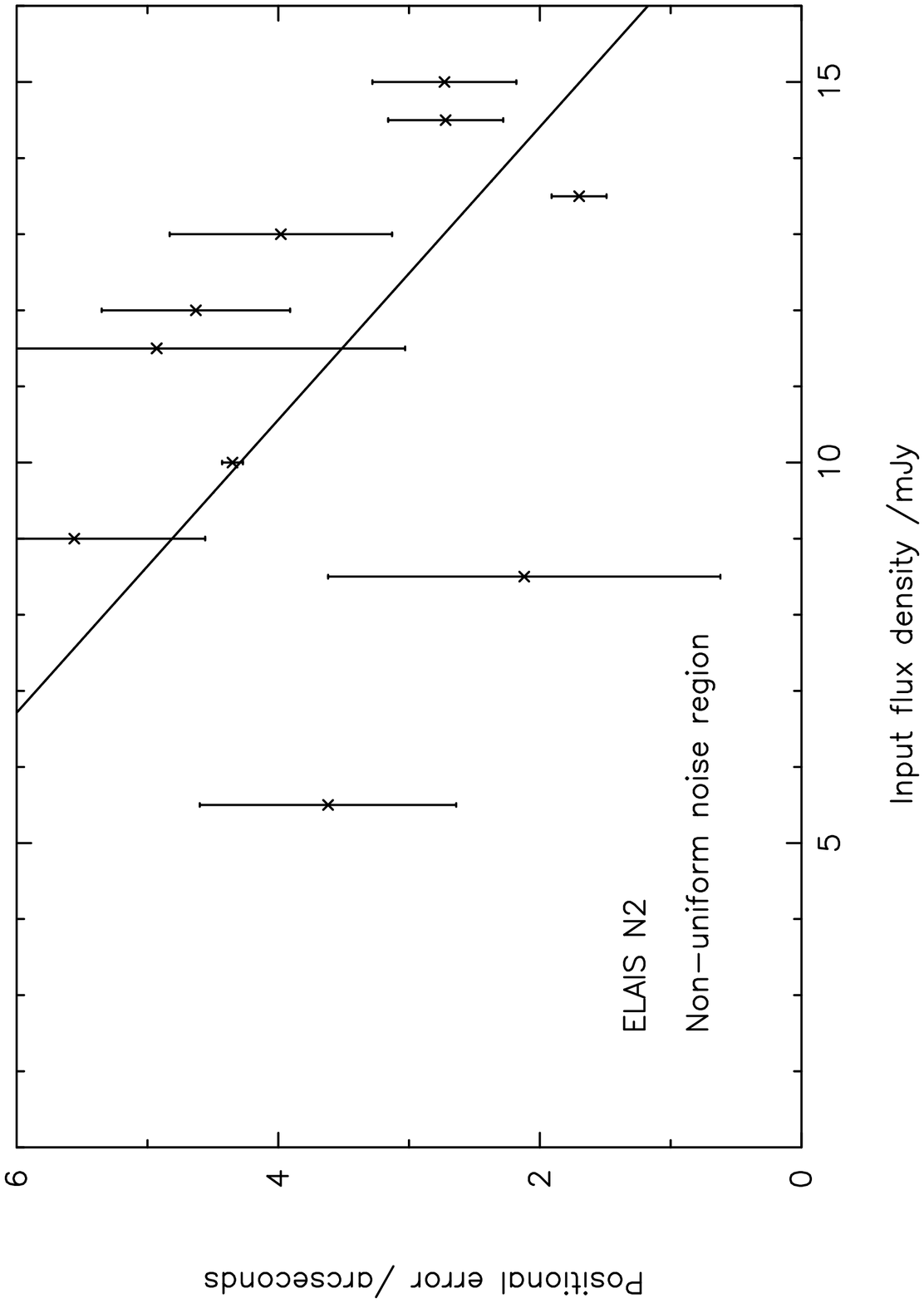}
\label{fig:n2_nonuni_poserr} 
\caption{\small{Mean positional uncertainty against input flux
   density, for the non-uniform noise regions of the ELAIS N2 field
 from the ``SCUBA 8\,mJy Survey''.}}
 \end{figure}

\newpage
\begin{figure}
 \centering
   \vspace*{4.8cm}
   \leavevmode
   \includegraphics{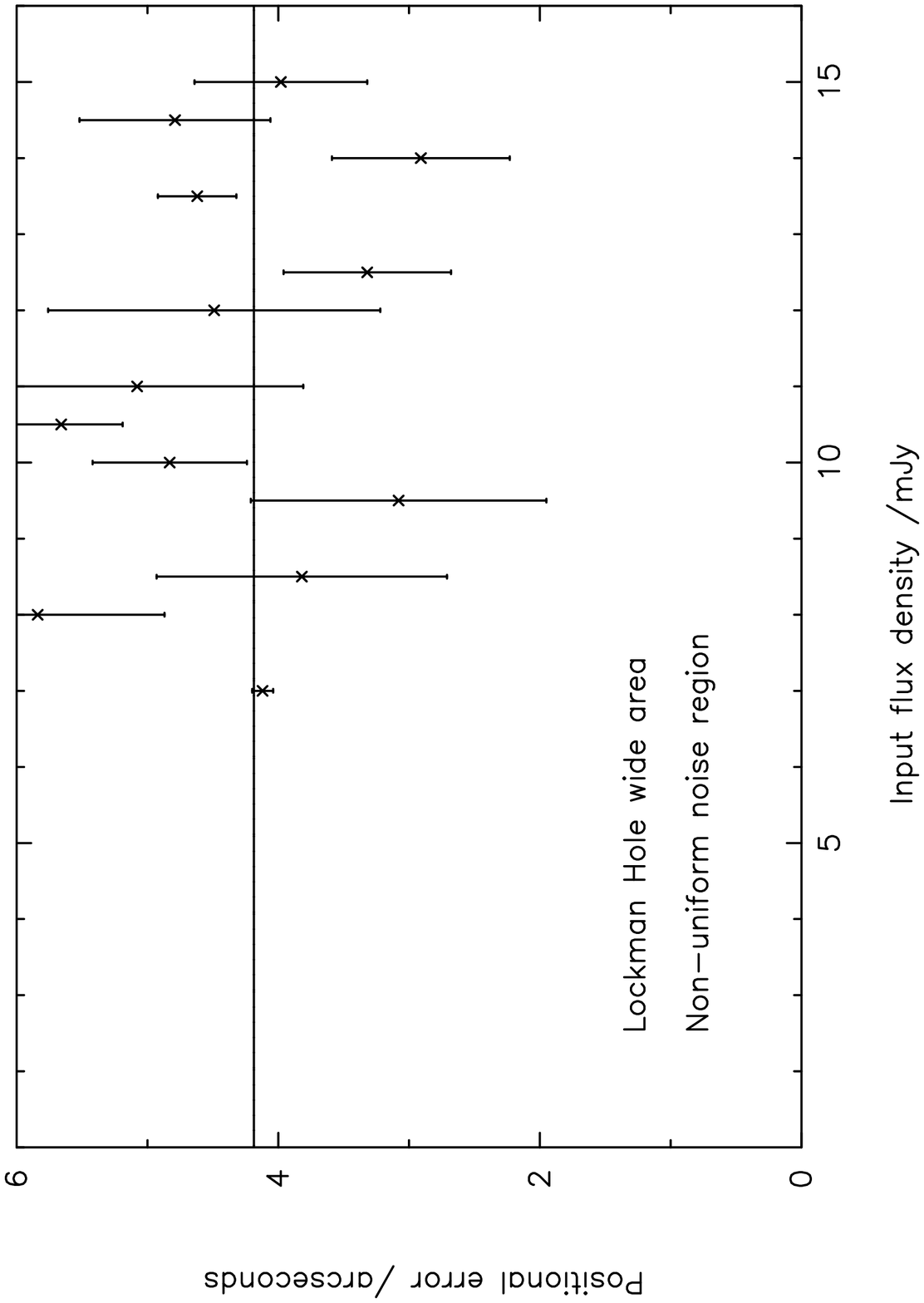}
\label{fig:lhwide_nonuni_poserr} 
\caption{\small{Mean positional uncertainty against input flux
   density, for the non-uniform noise regions of the Lockman Hole East
 wide area field from the ``SCUBA 8\,mJy Survey''.}}
 \centering
   \vspace*{3.7cm}
   \leavevmode
   \includegraphics{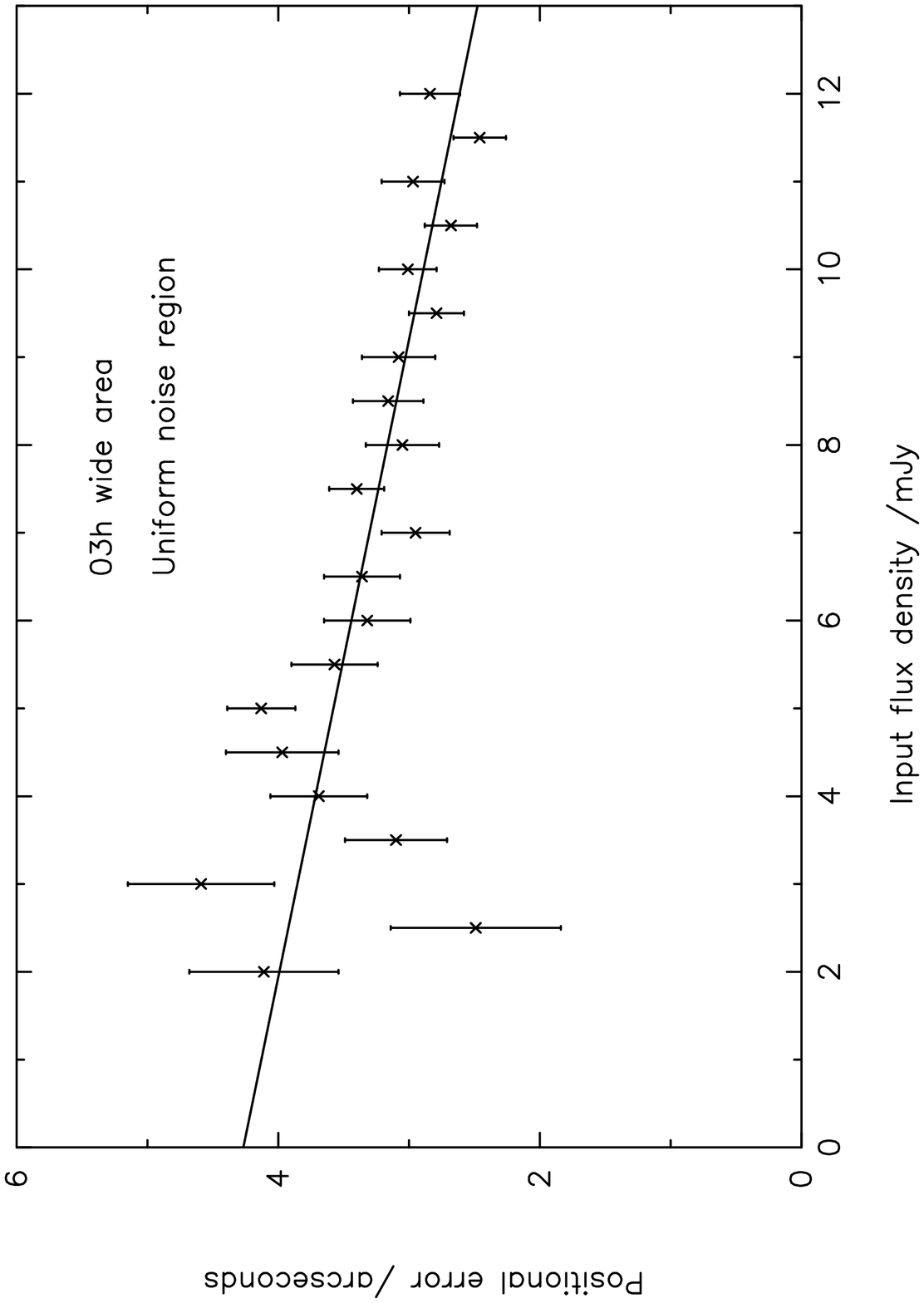}
\label{fig:03wide_uni_poserr} 
\caption{\small{Mean positional uncertainty against input flux
   density, for the uniform noise regions of the 03 hour
 wide area field from the ``CUDSS''.}}
 \centering
   \vspace*{3.7cm}
   \leavevmode
   \includegraphics{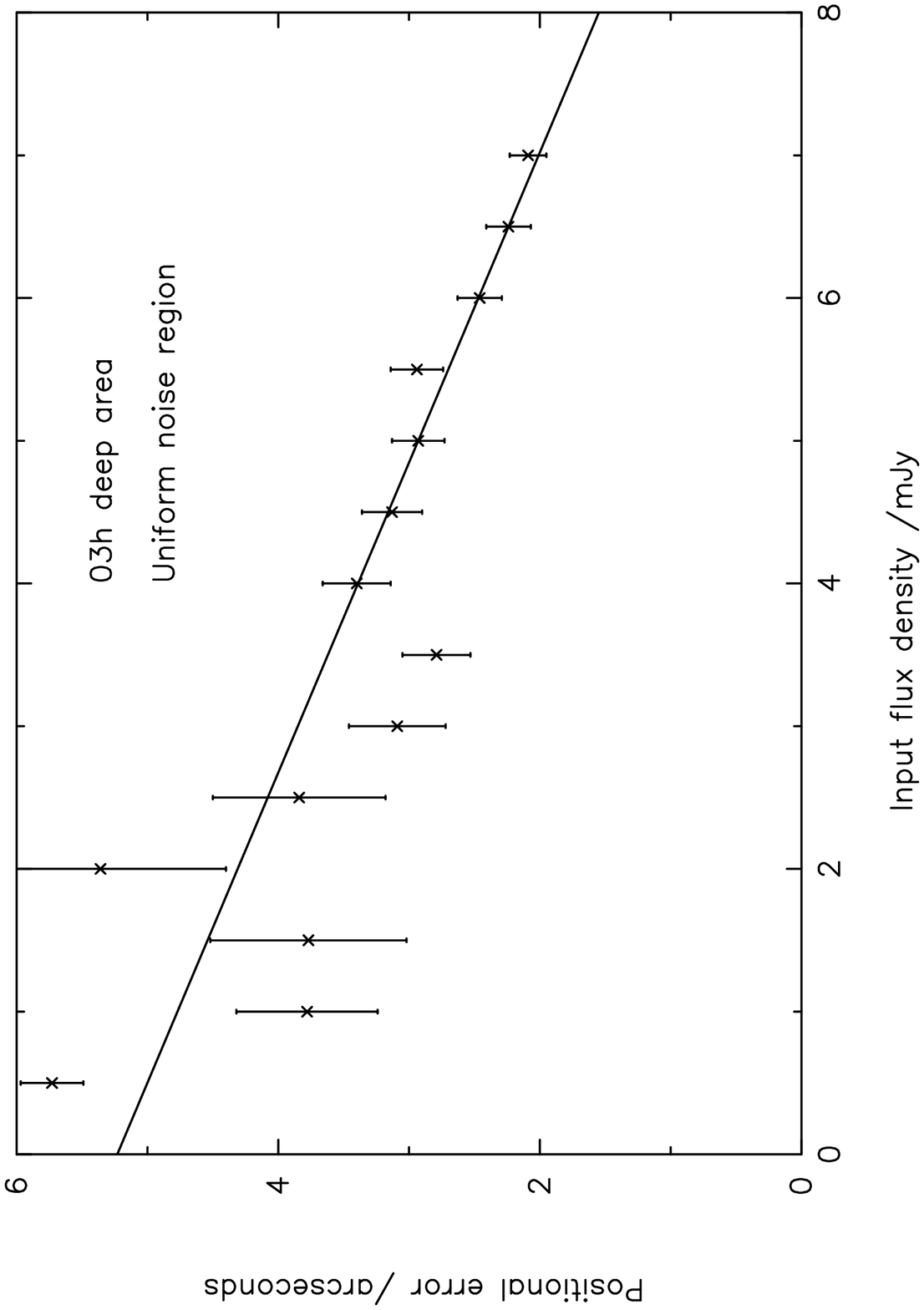}
\label{fig:03deep_uni_poserr} 
\caption{\small{Mean positional uncertainty against input flux
   density, for the uniform noise regions of the 03
 hour deep area from the ``CUDSS''.}}
 \centering
   \vspace*{3.7cm}
   \leavevmode
   \includegraphics{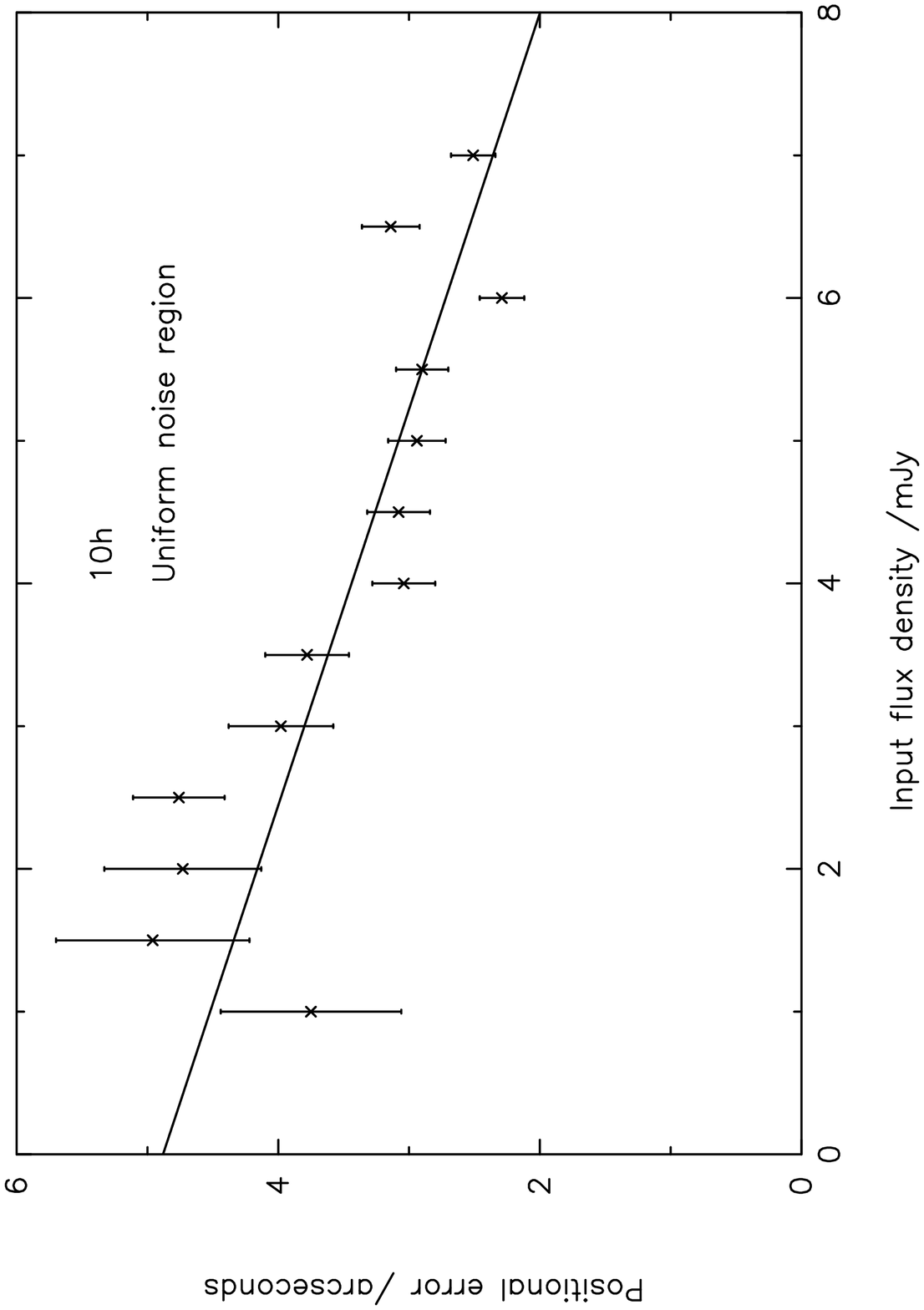}
\label{fig:10_uni_poserr} 
\caption{\small{Mean positional uncertainty against input flux
   density, for the uniform noise regions of the 10 hour field from
 the ``CUDSS''.}}
 \end{figure}

\newpage

\begin{figure}
 \centering
   \vspace*{4.8cm}
   \leavevmode
   \includegraphics{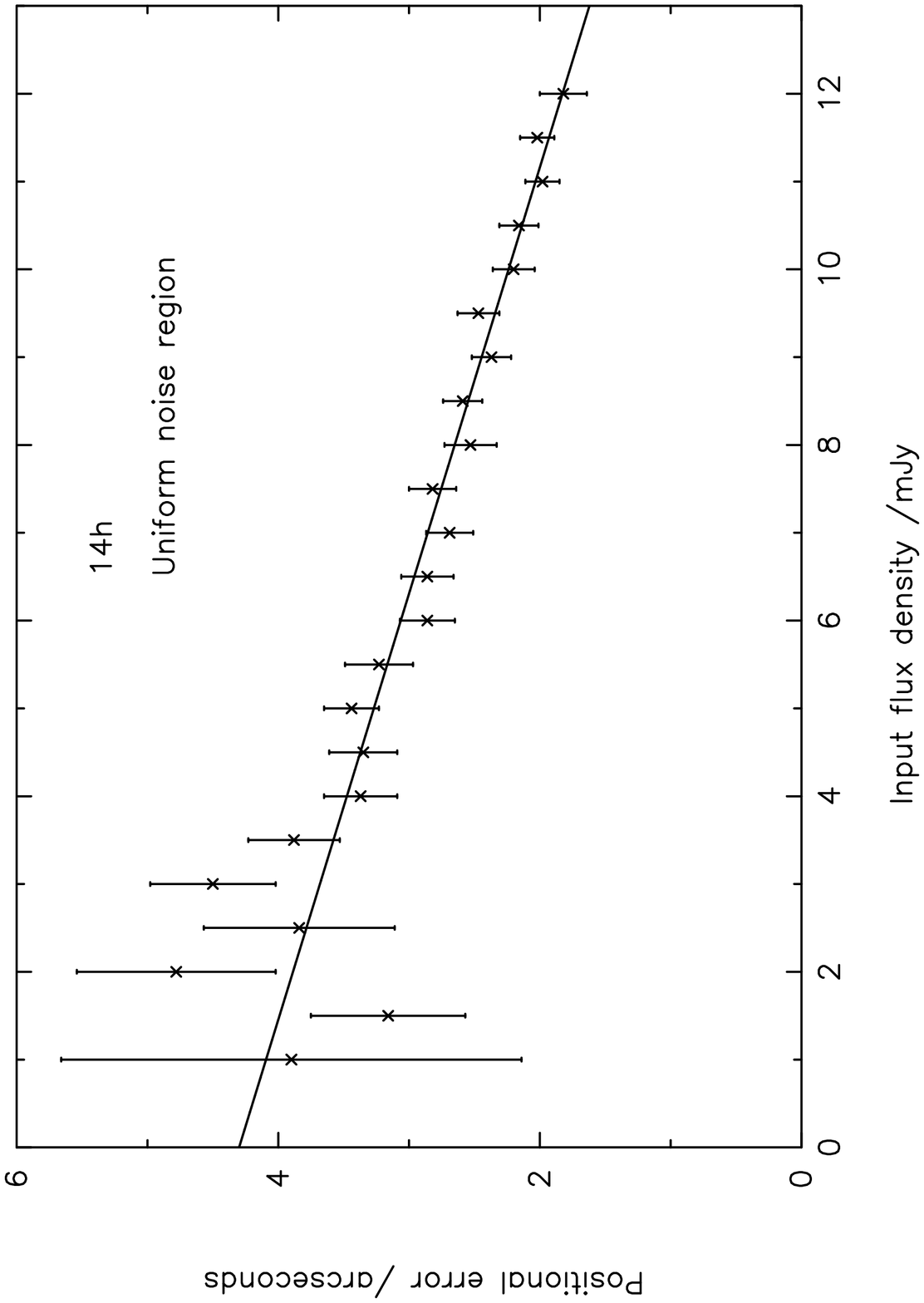}
\label{fig:14_uni_poserr} 
\caption{\small{Mean positional uncertainty against input flux
   density, for the uniform noise regions of the 14 hour field
 from the ``CUDSS''.}}
 \centering
   \vspace*{3.7cm}
   \leavevmode
   \includegraphics{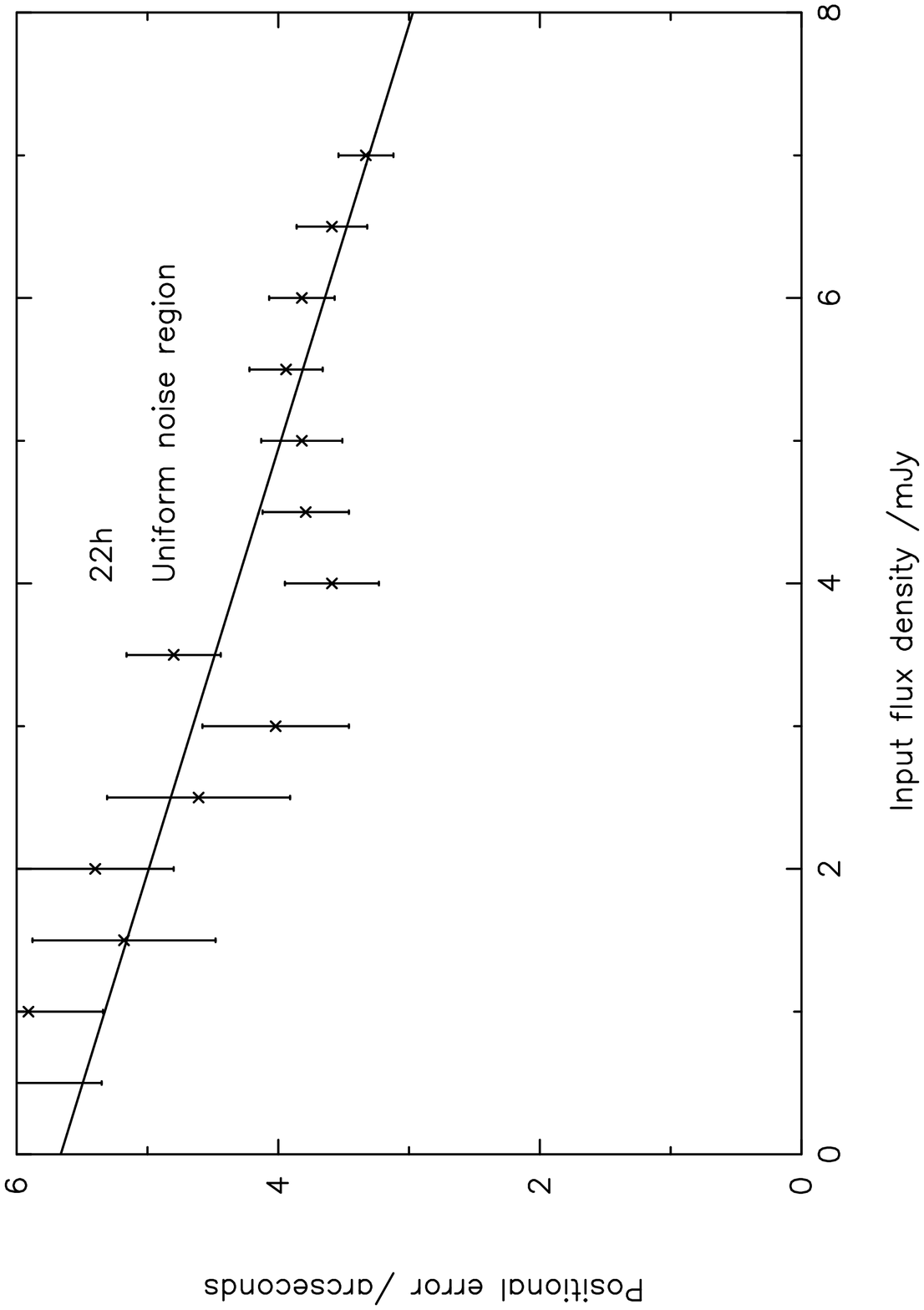}
\label{fig:22_uni_poserr} 
\caption{\small{Mean positional uncertainty against input flux
   density, for the uniform noise regions of the 22 hour field
 from the `CUDSS''.}}
 \centering
   \vspace*{3.7cm}
   \leavevmode
   \includegraphics{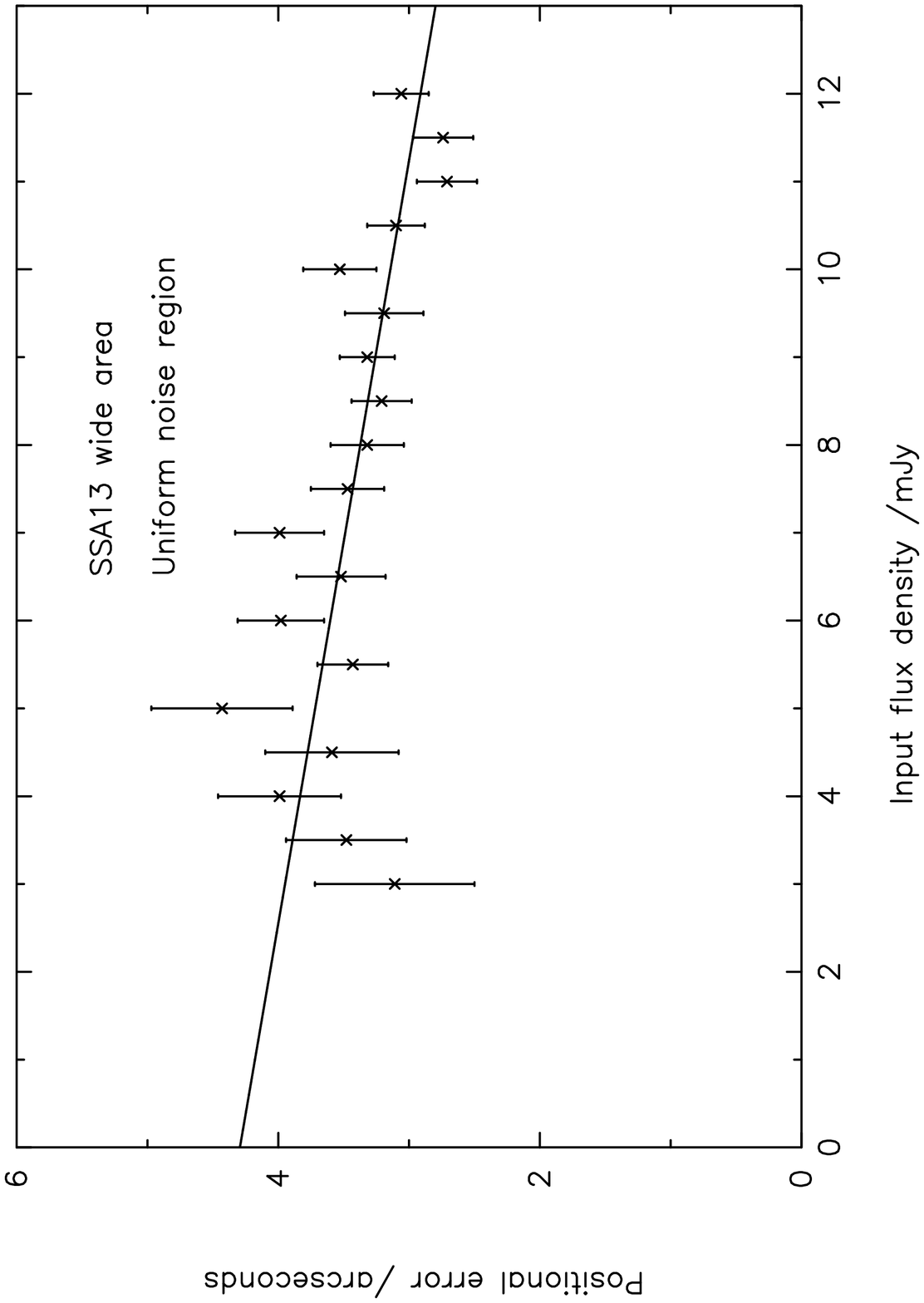}
\label{fig:ssa13wide_uni_poserr} 
\caption{\small{Mean positional uncertainty against input flux
   density, for the uniform noise regions of the SSA13
 wide area field from the ``Hawaii Submm Survey''.}}
 \centering
   \vspace*{3.7cm}
   \leavevmode
   \includegraphics{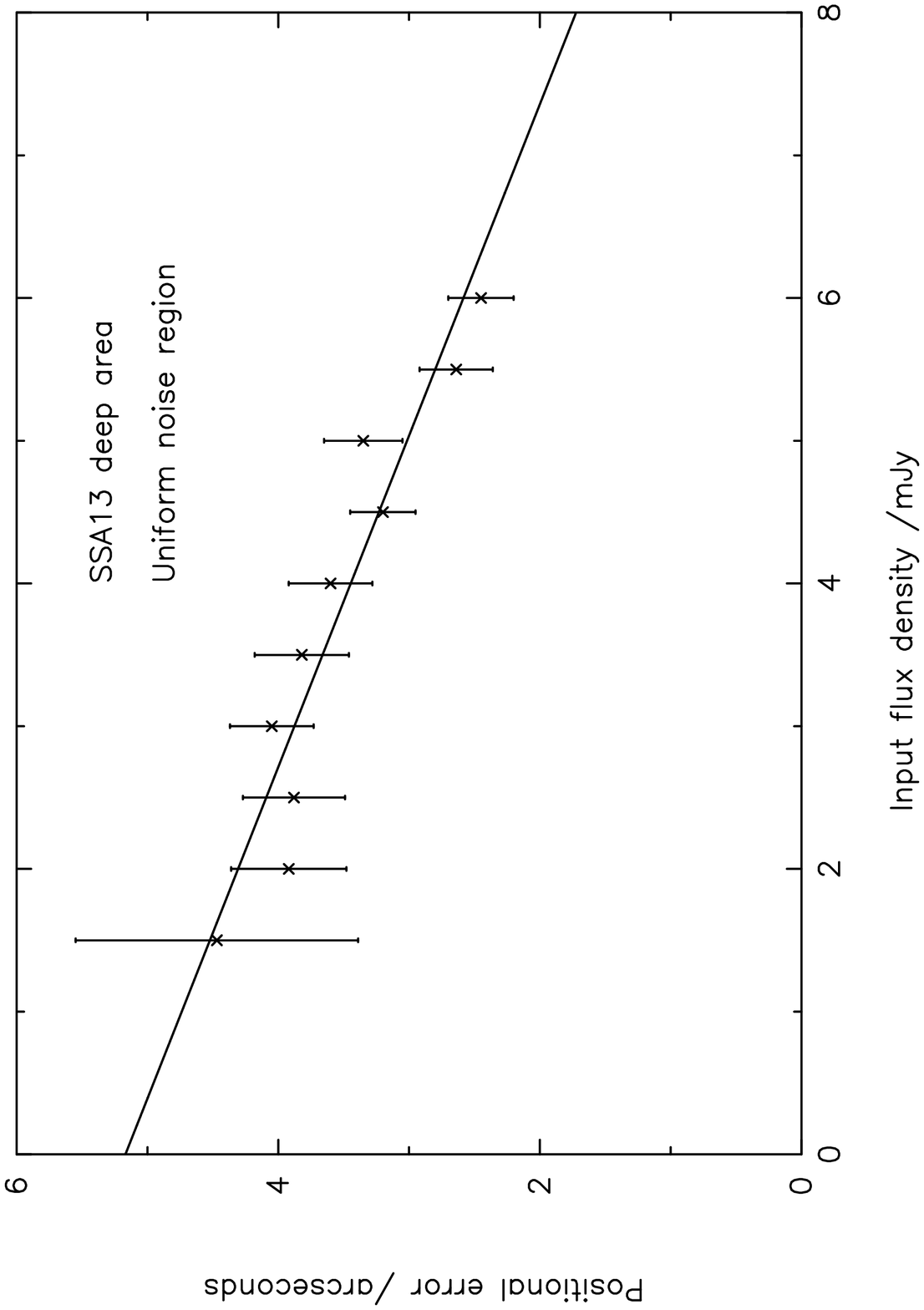}
\label{fig:ssa13deep_uni_poserr} 
\caption{\small{Mean positional uncertainty against input flux
   density, for the uniform noise regions of the SSA13
 hour deep area from the ``Hawaii Submm Survey''.}}
 \end{figure}
\newpage
\begin{figure}
 \centering
   \vspace*{4.8cm}
   \leavevmode
   \includegraphics{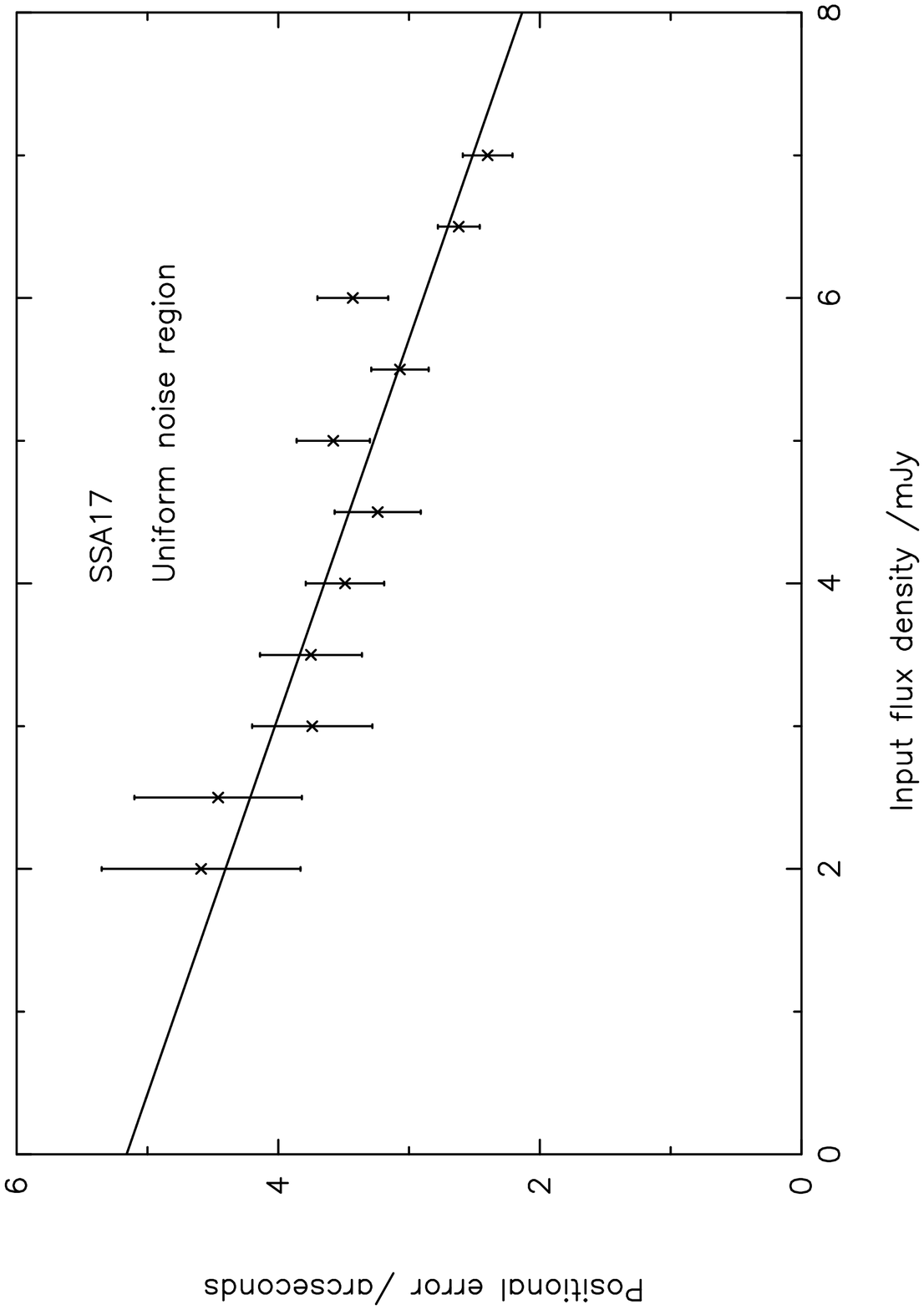}
\label{fig:ssa17_uni_poserr} 
\caption{\small{Mean positional uncertainty against input flux
   density, for the uniform noise regions of the SSA17 field from
 the ``Hawaii Submm Survey''.}}
 \centering
   \vspace*{3.7cm}
   \leavevmode
   \includegraphics{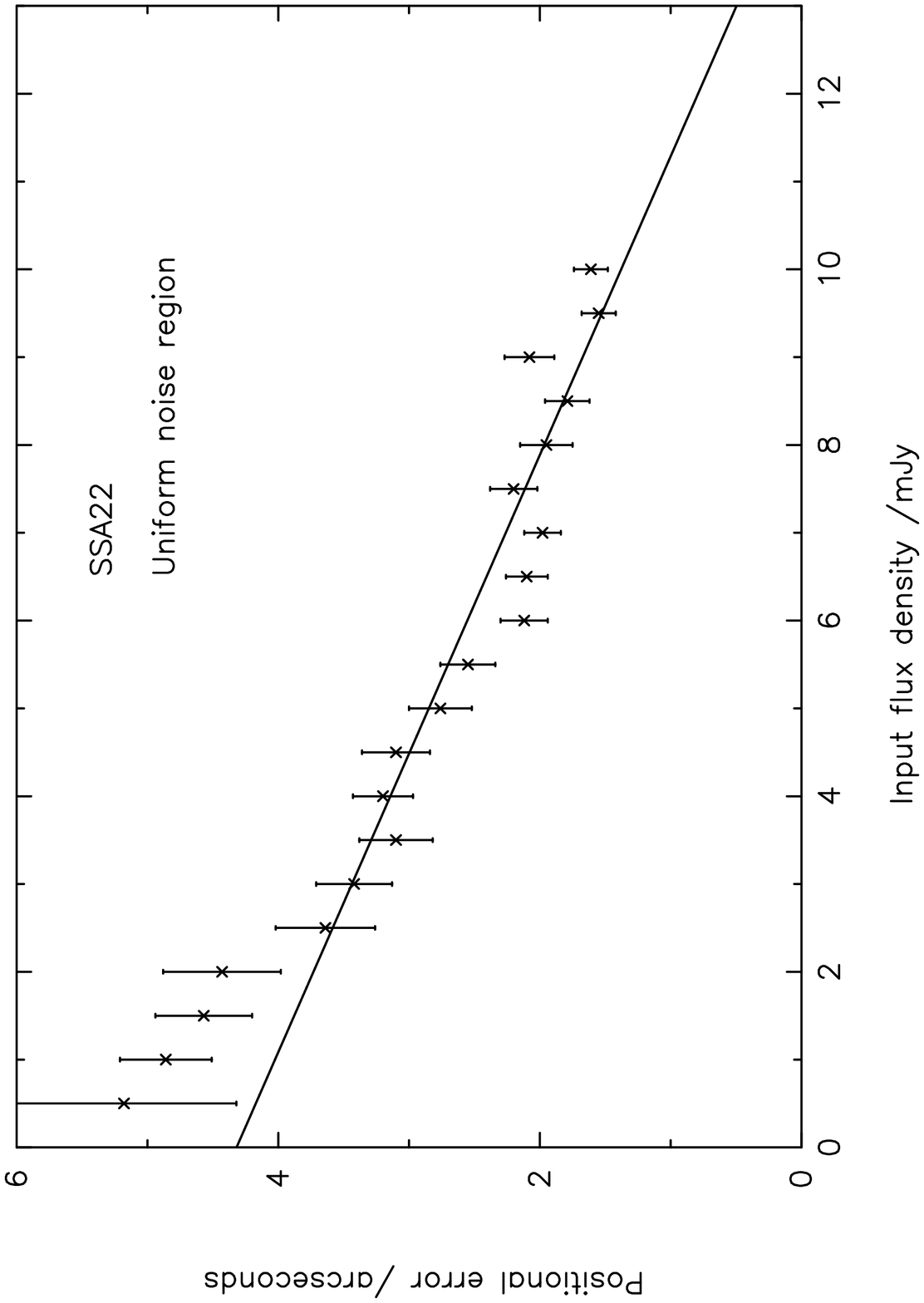}
\label{fig:ssa22_uni_poserr} 
\caption{\small{Mean positional uncertainty against input flux
   density, for the uniform noise regions of the SSA22 field
 from the ``Hawaii Submm Survey''.}}
 \centering
   \vspace*{3.7cm}
   \leavevmode
   \includegraphics{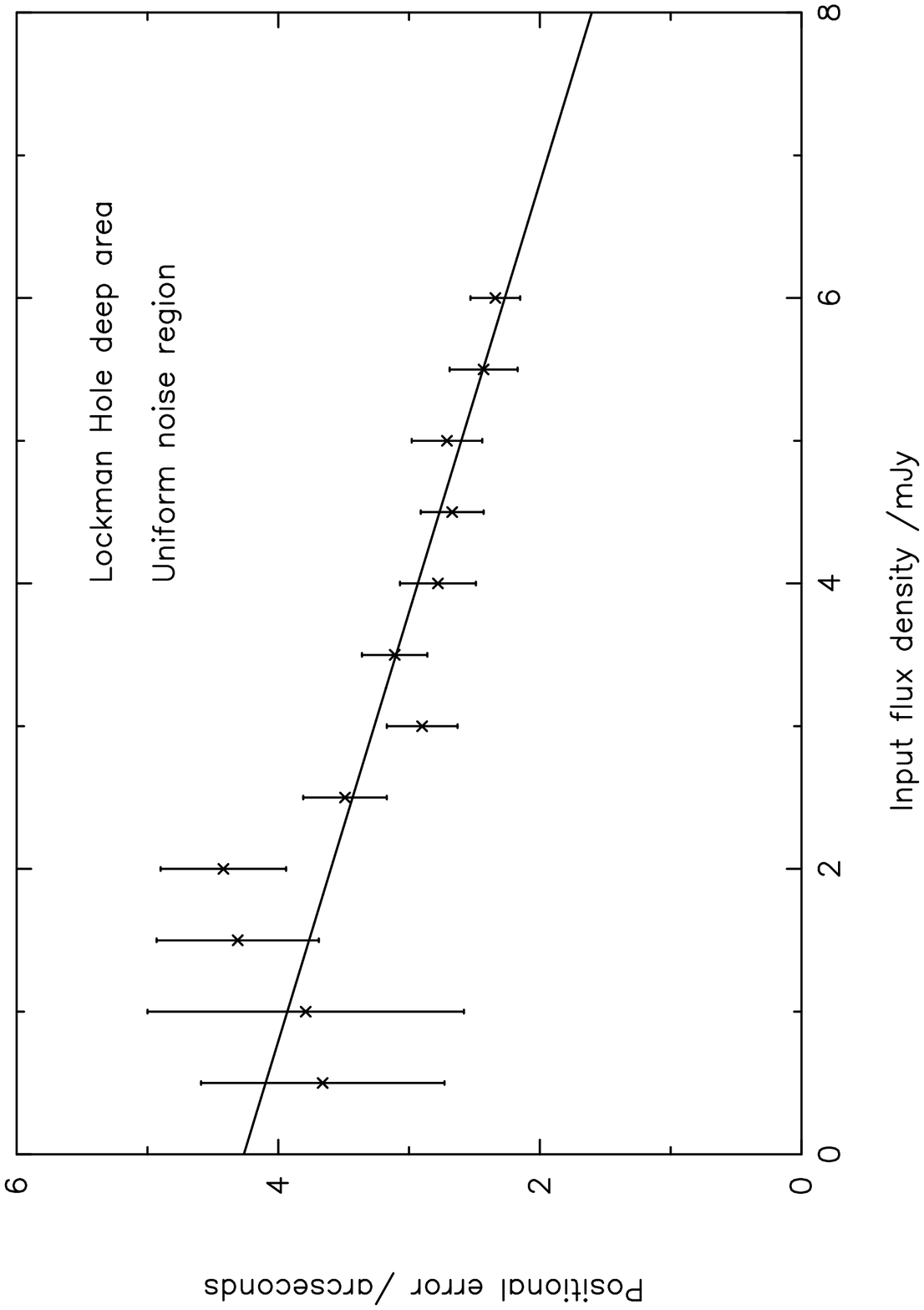}
\label{fig:lhdeep_uni_poserr} 
\caption{\small{Mean positional uncertainty against input flux
   density, for the uniform noise regions of the Lockman Hole deep area
 from the ``Hawaii Submm Survey''.}}
 \centering
   \vspace*{3.7cm}
   \leavevmode
   \includegraphics{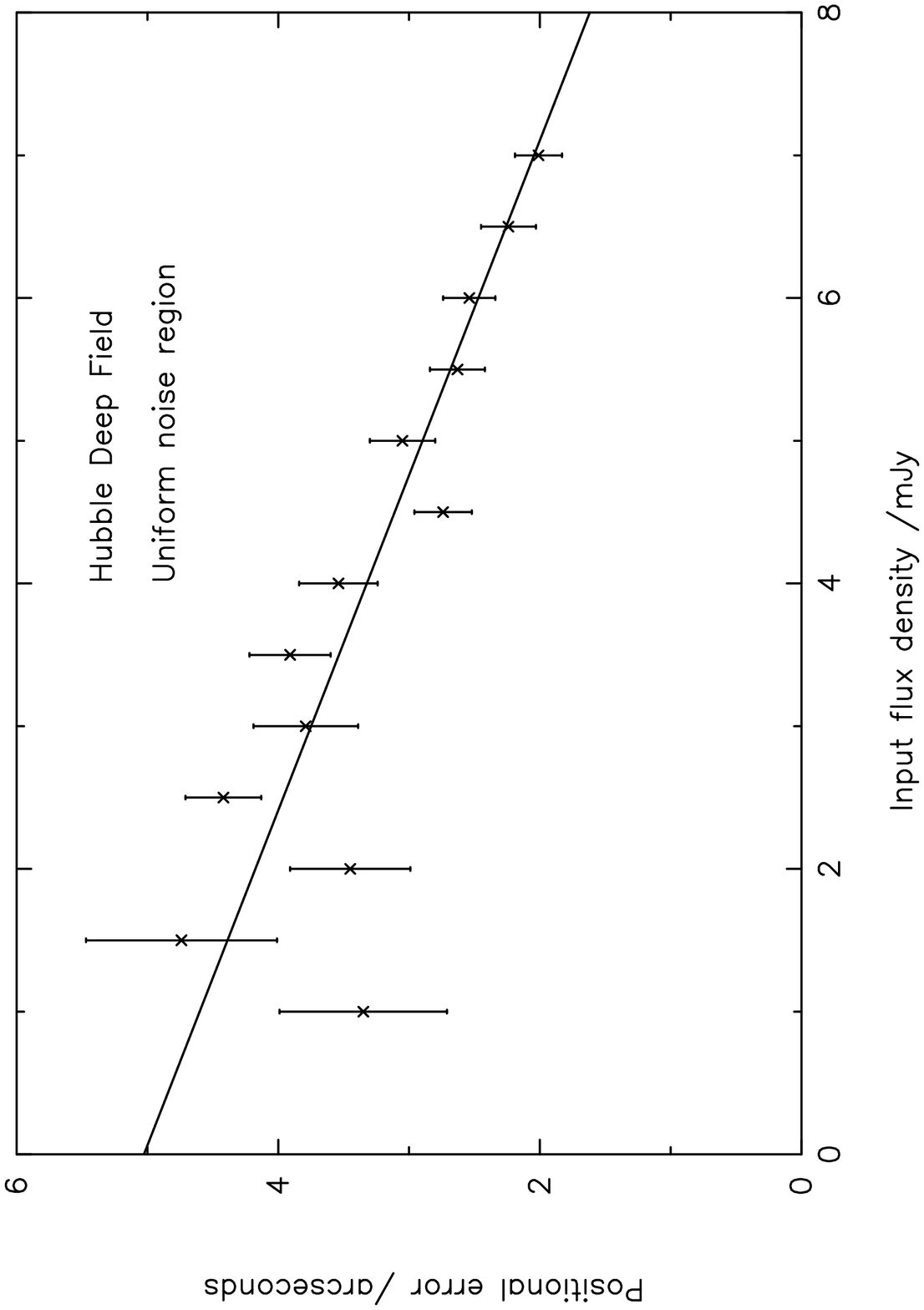}
\label{fig:hdf_uni_poserr} 
\caption{\small{Mean positional uncertainty against input flux
   density, for the uniform noise regions of the Hubble Deep Field.}}
 \end{figure}

\renewcommand{\topfraction}{0.1}
\onecolumn
\clearpage

\section{Completely Simulated Maps}
\subsection{Integral completeness tables}
The following tables give the percentage integral completeness for
each of the survey fields over 
the flux density range $2-10$\,mJy, and for signal-to-noise
thresholds of $\rm >1.50\sigma$ to $\rm >4.00\sigma$. Those fields 
composed of a deep pencil
beam survey within a wider-area shallower survey have had these two
components treated separately.

\begin{table*}
\footnotesize
\begin{tabular}{|c|c|c|c|c|c|c|} \hline
\multicolumn{7}{|c|}{ELAIS N2 field, uniform noise region} \\ \hline
\multicolumn{1}{|c|}{Flux} & 
\multicolumn{6}{c|}{ \% Integral Completeness} \\ \cline{2-7} 
Density & $\rm >1.50\sigma$  & $\rm >2.00\sigma$ & $\rm >2.50\sigma$ & $\rm
 >3.00\sigma$ &  $\rm >3.50\sigma$ & $\rm >4.00\sigma$ \\ \hline
 $\rm>\phantom{0}2\,mJy$ &   $45.7 \pm \phantom{0}0.5$ & $36.3 \pm \phantom{0}0.4$ & $25.6 \pm \phantom{0}0.3$ &
$17.6 \pm \phantom{0}0.2$ & $11.5 \pm \phantom{0}0.1$  & $\phantom{0}7.6 \pm \phantom{0}0.1$ \\ 
 $\rm >\phantom{0}4\,mJy$ & $71.2 \pm \phantom{0}1.7$ & $68.9 \pm \phantom{0}1.6$ & $60.8 \pm \phantom{0}1.4$ &
$50.9 \pm \phantom{0}1.2$ & $38.7 \pm \phantom{0}0.9$ & $28.8 \pm \phantom{0}0.7$ \\
$\rm >\phantom{0}6\,mJy$ & $79.9 \pm \phantom{0}3.1$ & $79.6 \pm \phantom{0}3.1$ & $77.3 \pm \phantom{0}3.0$ &
$72.8 \pm \phantom{0}2.8$ & $64.7 \pm \phantom{0}2.5$ & $55.4 \pm \phantom{0}2.2$ \\
$\rm >\phantom{0}8\,mJy$ & $76.4 \pm \phantom{0}4.4$ & $76.4 \pm \phantom{0}4.4$ & $76.1 \pm
\phantom{0}4.4$ & $74.4 \pm \phantom{0}4.3$ & $71.1 \pm \phantom{0}4.1$ & $ 65.2 \pm \phantom{0}3.7$ \\
$\rm >10\,mJy$ & $76.8 \pm \phantom{0}6.4$ & $76.8 \pm \phantom{0}6.4$ & $76.8 \pm \phantom{0}6.4$ &
$76.8 \pm \phantom{0}6.4$ & $74.6 \pm \phantom{0}6.3$ & $71.1 \pm \phantom{0}6.0$ \\ \hline
\end{tabular}
\label{table:intcomp_n2uni}\caption{\small Percentage integral
completeness results for the uniform noise region of the ELAIS N2
field, over the flux density range $2-10$\,mJy, and for signal-to-noise
thresholds of $\rm >1.50\sigma$ to $\rm >4.00\sigma$.}

\begin{tabular}{|c|c|c|c|c|c|c|} \hline
\multicolumn{7}{|c|}{ELAIS N2 field, non-uniform noise region} \\ \hline
\multicolumn{1}{|c|}{Flux} & 
\multicolumn{6}{c|}{ \% Integral Completeness} \\ \cline{2-7} 
Density & $\rm >1.50\sigma$  & $\rm >2.00\sigma$ & $\rm >2.50\sigma$ & $\rm
 >3.00\sigma$ &  $\rm >3.50\sigma$ & $\rm >4.00\sigma$ \\ \hline
 $\rm>\phantom{0}2\,mJy$ & $12.6  \pm \phantom{0}0.3$ & $\phantom{0}8.2 \pm \phantom{0}0.2$
& $\phantom{0}4.7 \pm \phantom{0}0.1$ & $\phantom{0}2.4 \pm \phantom{0}0.1$ &
$\phantom{0}1.0 \pm \phantom{0}0.1$ & $\phantom{0}0.4 \pm \phantom{0}0.1$ \\ 
 $\rm >\phantom{0}4\,mJy$ & $24.2 \pm \phantom{0}1.1$ & $18.3 \pm \phantom{0}0.8$&
$12.7\pm \phantom{0}0.6$ & $\phantom{0}7.7 \pm \phantom{0}0.3$ & $\phantom{0}4.0 \pm \phantom{0}0.2$ & $\phantom{0}1.9 \pm
\phantom{0}0.1$ \\
$\rm >\phantom{0}6\,mJy$ & $35.6 \pm \phantom{0}2.7$ & $31.6 \pm \phantom{0}2.4$ & $24.9 \pm
\phantom{0}1.9$ & $14.7\pm \phantom{0}1.1$ & $\phantom{0}9.6 \pm \phantom{0}0.7$ & $\phantom{0}5.6
\pm\phantom{0}0.4$ \\
$\rm >\phantom{0}8\,mJy$ & $37.8 \pm \phantom{0}4.4$ & $33.8 \pm \phantom{0}3.9$ & $31.1 \pm
\phantom{0}3.6$ & $23.0 \pm \phantom{0}2.7$ & $14.9 \pm \phantom{0}1.7$ & $10.8 \pm \phantom{0}1.3$ \\
$\rm >10\,mJy$ & $38.1 \pm \phantom{0}5.9$ & $35.7 \pm \phantom{0}5.5$ & $33.3 \pm \phantom{0}5.1$ &
$28.6 \pm \phantom{0}4.4$ & $19.0 \pm \phantom{0}2.9$ & $11.9 \pm \phantom{0}1.8$ \\ \hline
\end{tabular}
\label{table:intcomp_n2non}\caption{\small Percentage integral
completeness results for the non-uniform noise region of the ELAIS N2
field, over the flux density range $2-10$\,mJy, and for signal-to-noise
thresholds of $\rm >1.50\sigma$ to $\rm >4.00\sigma$.}

\begin{tabular}{|c|c|c|c|c|c|c|} \hline
\multicolumn{7}{|c|}{Lockman Hole wide area field, uniform noise region} \\ \hline
\multicolumn{1}{|c|}{Flux} & 
\multicolumn{6}{c|}{ \% Integral Completeness} \\ \cline{2-7} 
Density & $\rm >1.50\sigma$  & $\rm >2.00\sigma$ & $\rm >2.50\sigma$ & $\rm
 >3.00\sigma$ &  $\rm >3.50\sigma$ & $\rm >4.00\sigma$ \\ \hline
 $\rm>\phantom{0}2\,mJy$ & $41.8 \pm \phantom{0}0.5$ & $30.0 \pm \phantom{0}0.4$ & $19.2 \pm
\phantom{0}0.2$ & $11.9 \pm \phantom{0}0.1$ & $\phantom{0}7.1 \pm \phantom{0}0.1$ &  $\phantom{0}4.3
\pm \phantom{0}0.1$ \\
 $\rm >\phantom{0}4\,mJy$ & $72.0 \pm \phantom{0}2.0$ & $65.3  \pm \phantom{0}1.8$ & $52.7
\pm \phantom{0}1.4$ & $38.8 \pm \phantom{0}1.1$ & $26.9 \pm \phantom{0}0.7$ & $18.2 \pm \phantom{0}0.5$ \\
$\rm >\phantom{0}6\,mJy$ & $78.1  \pm \phantom{0}3.5$ & $77.6 \pm \phantom{0}3.4$ & $74.4
\pm \phantom{0}3.3$ & $64.0 \pm \phantom{0}2.8$ & $51.4 \pm \phantom{0}2.3$ & $38.2 \pm \phantom{0}1.7$ \\
$\rm >\phantom{0}8\,mJy$ & $81.3  \pm \phantom{0}5.4$ & $81.3 \pm \phantom{0}5.4$ & $80.9
\pm \phantom{0}5.4$ & $79.6  \pm \phantom{0}5.3$ & $73.8 \pm \phantom{0}4.9$ & $60.9 \pm \phantom{0}4.1$ \\
$\rm >10\,mJy$ & $73.0 \pm \phantom{0}6.8$ & $73.0 \pm \phantom{0}6.8$ & $73.0 \pm \phantom{0}6.8$ &
$73.0 \pm \phantom{0}6.8$ & $72.2 \pm \phantom{0}6.7$ & $67.0 \pm \phantom{0}6.2$ \\ \hline
\end{tabular}
\label{table:intcomp_lhuni}\caption{\small Percentage integral
completeness results for the uniform noise region of the Lockman Hole
wide area field, over the flux density range $2-10$\,mJy, and for
signal-to-noise thresholds of $\rm >1.50\sigma$ to $\rm >4.00\sigma$.}

\begin{tabular}{|c|c|c|c|c|c|c|} \hline
\multicolumn{7}{|c|}{Lockman Hole wide area field, non-uniform noise region} \\ \hline
\multicolumn{1}{|c|}{Flux} & 
\multicolumn{6}{c|}{ \% Integral Completeness} \\ \cline{2-7} 
Density & $\rm >1.50\sigma$  & $\rm >2.00\sigma$ & $\rm >2.50\sigma$ & $\rm
 >3.00\sigma$ &  $\rm >3.50\sigma$ & $\rm >4.00\sigma$ \\ \hline
 $\rm>\phantom{0}2\,mJy$ & $19.2 \pm \phantom{0}0.3$ & $13.1 \pm
\phantom{0}0.2$ & $\phantom{0}7.9 \pm \phantom{0}0.1$ &
$\phantom{0}3.8 \pm  \phantom{0}0.1$ & $\phantom{0}1.8 \pm
\phantom{0}0.1$ & $\phantom{0}0.7 \pm \phantom{0}0.1$ \\
 $\rm >\phantom{0}4\,mJy$ & $34.6 \pm \phantom{0}1.2$ &$25.7 \pm
\phantom{0}0.9$ & $18.0 \pm \phantom{0}0.6$ & $\phantom{0}9.8 \pm
\phantom{0}0.4$& $\phantom{0}5.4 \pm \phantom{0}0.2$ & $\phantom{0}2.5
\pm \phantom{0}0.1$ \\
$\rm >\phantom{0}6\,mJy$ & $39.6 \pm \phantom{0}2.3$ & $34.0 \pm
\phantom{0}2.0$ & $25.4 \pm \phantom{0}1.5$ & $16.2 \pm
\phantom{0}0.9$ & $\phantom{0}8.6 \pm \phantom{0}0.5$ &
$\phantom{0}4.0 \pm \phantom{0}0.2$ \\
$\rm >\phantom{0}8\,mJy$ & $36.8 \pm \phantom{0}3.1$ & $34.7 \pm
\phantom{0}2.9$ & $29.9 \pm \phantom{0}2.5$ & $19.4 \pm
\phantom{0}1.6$ & $10.4 \pm \phantom{0}0.9$ & $\phantom{0}4.9 \pm
\phantom{0}0.4$ \\
$\rm >10\,mJy$ & $30.3 \pm \phantom{0}3.5$ & $30.3 \pm \phantom{0}3.5$
& $28.9 \pm \phantom{0}3.3$ & $21.1 \pm \phantom{0}2.4$ & $14.5 \pm
\phantom{0}1.7$ & $\phantom{0}7.9 \pm \phantom{0}0.9$ \\ \hline
\end{tabular}
\label{table:intcomp_lhwuni}\caption{\small Percentage integral
completeness results for the non-uniform noise region of the Lockman
Hole wide area
field, over the flux density range $2-10$\,mJy, and for signal-to-noise
thresholds of $\rm >1.50\sigma$ to $\rm >4.00\sigma$.}
\vspace{1.2cm}
\end{table*}

\renewcommand{\topfraction}{0.95}
\clearpage
\begin{table*}
\footnotesize
\begin{tabular}{|c|c|c|c|c|c|c|} \hline
\multicolumn{7}{|c|}{Lockman Hole deep strip field, uniform noise region} \\ \hline
\multicolumn{1}{|c|}{Flux} & 
\multicolumn{6}{c|}{ \% Integral Completeness} \\ \cline{2-7} 
Density & $\rm >1.50\sigma$  & $\rm >2.00\sigma$ & $\rm >2.50\sigma$ & $\rm
 >3.00\sigma$ &  $\rm >3.50\sigma$ & $\rm >4.00\sigma$ \\ \hline
 $\rm>\phantom{0}2\,mJy$ & $57.2 \pm \phantom{0}1.5$ & $48.6 \pm \phantom{0}1.2$ & $38.0 \pm
\phantom{0}1.0$ & $29.6 \pm \phantom{0}0.8$ & $21.7 \pm \phantom{0}0.6$ & $15.0 \pm \phantom{0}0.4$ \\
 $\rm >\phantom{0}4\,mJy$ & $82.3 \pm \phantom{0}4.7$ & $82.3 \pm
\phantom{0}4.7$ & $80.0 \pm \phantom{0}4.6$ & $76.1 \pm
\phantom{0}4.4$ & $63.3 \pm \phantom{0}3.6$ & $50.2 \pm
\phantom{0}2.9$ \\
$\rm >\phantom{0}6\,mJy$ & $85.2 \pm \phantom{0}7.9$ & $85.2 \pm
\phantom{0}7.9$ & $85.2 \pm \phantom{0}7.9$ & $85.2 \pm
\phantom{0}7.9$ & $83.5 \pm \phantom{0}7.8$ & $78.3 \pm
\phantom{0}7.3$ \\
$\rm >\phantom{0}8\,mJy$ & $84.6 \pm 11.7$ & $84.6 \pm 11.7$ & $84.6
\pm 11.7$ & $84.6 \pm 11.7$ & $84.6 \pm 11.7$ & $84.6  \pm 11.7$ \\
$\rm >10\,mJy$ & $65.4 \pm 12.8$ & $65.4 \pm 12.8$ & $65.4 \pm  12.8$
& $65.4  \pm 12.8$ & $65.4 \pm 12.8$ & $65.4 \pm 12.8$ \\ \hline
\end{tabular}
\label{table:intcomp_lhstripuni}\caption{\small Percentage integral
completeness results for the uniform noise region of the Lockman Hole
deep strip, over the flux density range $2-10$\,mJy, and for
signal-to-noise thresholds of $\rm >1.50\sigma$ to $\rm >4.00\sigma$.}

\begin{tabular}{|c|c|c|c|c|c|c|} \hline
\multicolumn{7}{|c|}{03h deep area field, uniform noise region} \\ \hline
\multicolumn{1}{|c|}{Flux} & 
\multicolumn{6}{c|}{ \% Integral Completeness} \\ \cline{2-7} 
Density & $\rm >1.50\sigma$  & $\rm >2.00\sigma$ & $\rm >2.50\sigma$ & $\rm
 >3.00\sigma$ &  $\rm >3.50\sigma$ & $\rm >4.00\sigma$ \\ \hline
 $\rm>\phantom{0}2\,mJy$ & $58.7 \pm \phantom{0}2.6$ & $58.5 \pm
\phantom{0}2.5$ & $57.2 \pm \phantom{0}2.5$ & $53.8 \pm
\phantom{0}2.3$ & $46.6 \pm \phantom{0}2.0$ & $38.4 \pm
\phantom{0}1.7$ \\
 $\rm >\phantom{0}4\,mJy$ & $82.2 \pm \phantom{0}7.6$ & $82.2 \pm
\phantom{0}7.6$ & $82.2 \pm \phantom{0}7.6$ & $82.2 \pm
\phantom{0}7.6$ & $82.2 \pm \phantom{0}7.6$ & $82.2 \pm
\phantom{0}7.6$ \\
$\rm >\phantom{0}6\,mJy$ & $88.6 \pm 13.4$ & $88.6 \pm 13.4$ & $88.6
\pm 13.4$ & $88.6 \pm 13.4$ & $88.6 \pm 13.4$ & $88.6 \pm 13.4$ \\
$\rm >\phantom{0}8\,mJy$ & $82.6 \pm 17.2$ & $82.6 \pm 17.2$ & $82.6
\pm 17.2$ & $82.6 \pm 17.2$ & $82.6 \pm 17.2$ & $82.6 \pm 17.2$ \\
$\rm >10\,mJy$ & $90.0 \pm 28.5$ & $90.0 \pm 28.5$ & $90.0 \pm 28.5$ &
$90.0 \pm 28.5$ & $90.0 \pm 28.5$ & $90.0 \pm 28.5$ \\ \hline
\end{tabular}
\label{table:intcomp_03deepuni}\caption{\small Percentage integral
completeness results for the uniform noise region of the 03h deep area
field, over the flux density range $2-10$\,mJy, and for signal-to-noise
thresholds of $\rm >1.50\sigma$ to $\rm >4.00\sigma$.}

\begin{tabular}{|c|c|c|c|c|c|c|} \hline
\multicolumn{7}{|c|}{03h wide area field, uniform noise region} \\ \hline
\multicolumn{1}{|c|}{Flux} & 
\multicolumn{6}{c|}{ \% Integral Completeness} \\ \cline{2-7} 
Density & $\rm >1.50\sigma$  & $\rm >2.00\sigma$ & $\rm >2.50\sigma$ & $\rm
 >3.00\sigma$ &  $\rm >3.50\sigma$ & $\rm >4.00\sigma$ \\ \hline
 $\rm>\phantom{0}2\,mJy$ & $43.8 \pm \phantom{0}0.6$ & $35.5 \pm
\phantom{0}0.5$ & $26.5 \pm \phantom{0}0.4$ & $19.2 \pm
\phantom{0}0.3$ & $13.1 \pm \phantom{0}0.2$ & $\phantom{0}8.9 \pm
\phantom{0}0.1$ \\
 $\rm >\phantom{0}4\,mJy$ & $68.4 \pm \phantom{0}2.2$ & $65.8 \pm
\phantom{0}2.1$ & $60.4 \pm \phantom{0}2.0$ & $52.9 \pm
\phantom{0}1.7$ & $42.3 \pm \phantom{0}1.4$ & $32.6 \pm
\phantom{0}1.1$ \\
$\rm >\phantom{0}6\,mJy$ & $73.8 \pm \phantom{0}4.0$ & $73.2 \pm
\phantom{0}4.0$ & $70.9 \pm \phantom{0}3.8$ & $66.8 \pm
\phantom{0}3.6$ & $60.3 \pm \phantom{0}3.3$ & $52.4 \pm
\phantom{0}2.8$ \\
$\rm >\phantom{0}8\,mJy$ & $71.5 \pm \phantom{0}6.0$ & $71.5 \pm
\phantom{0}6.0$ & $71.5 \pm \phantom{0}6.0$ & $71.5 \pm
\phantom{0}6.0$ & $68.8 \pm \phantom{0}5.7$ & $63.9 \pm
\phantom{0}5.3$ \\
$\rm >10\,mJy$ & $69.1 \pm \phantom{0}8.4$ & $69.1 \pm \phantom{0}8.4$
& $69.1 \pm \phantom{0}8.4$ & $69.1 \pm \phantom{0}8.4$ & $69.1 \pm
\phantom{0}8.4$ & $67.6 \pm \phantom{0}8.2$ \\ \hline
\end{tabular}
\label{table:intcomp_03wuni}\caption{\small Percentage integral
completeness results for the uniform noise region of the 03h wide area
field, over the flux density range $2-10$\,mJy, and for signal-to-noise
thresholds of $\rm >1.50\sigma$ to $\rm >4.00\sigma$.}

\begin{tabular}{|c|c|c|c|c|c|c|} \hline
\multicolumn{7}{|c|}{10h field, uniform noise region} \\ \hline
\multicolumn{1}{|c|}{Flux} & 
\multicolumn{6}{c|}{ \% Integral Completeness} \\ \cline{2-7} 
Density & $\rm >1.50\sigma$  & $\rm >2.00\sigma$ & $\rm >2.50\sigma$ & $\rm
 >3.00\sigma$ &  $\rm >3.50\sigma$ & $\rm >4.00\sigma$ \\ \hline
 $\rm>\phantom{0}2\,mJy$ & $44.4 \pm \phantom{0}1.7$ & $39.0 \pm
\phantom{0}1.5$ & $33.6 \pm \phantom{0}1.3$ & $26.5 \pm
\phantom{0}1.0$ & $19.4 \pm \phantom{0}0.8$ & $14.8 \pm
\phantom{0}0.6$ \\ 
 $\rm >\phantom{0}4\,mJy$ & $57.1 \pm \phantom{0}5.0$ & $56.4 \pm
\phantom{0}4.9$ & $55.6 \pm \phantom{0}4.8$ & $52.6 \pm
\phantom{0}4.6$ & $50.4 \pm \phantom{0}4.4$ & $47.4 \pm
\phantom{0}4.1$ \\
$\rm >\phantom{0}6\,mJy$ & $68.1 \pm \phantom{0}9.9$ & $68.1 \pm
\phantom{0}9.9$ & $68.1 \pm \phantom{0}9.9$ & $66.0 \pm
\phantom{0}9.6$ & $66.0 \pm \phantom{0}9.6$ & $63.8 \pm
\phantom{0}9.3$ \\
$\rm >\phantom{0}8\,mJy$ & $70.0 \pm 15.7$ & $70.0 \pm 15.7$ & $70.0
\pm 15.7$ & $70.0 \pm 15.7$ & $70.0 \pm 15.7$ & $70.0 \pm 15.7$ \\
$\rm >10\,mJy$ & $72.7 \pm 21.9$ & $72.7 \pm 21.9$ & $72.7 \pm 21.9$ &
$72.7 \pm 21.9$ & $72.7 \pm 21.9$ & $72.7 \pm 21.9$ \\ \hline
\end{tabular}
\label{table:intcomp_10uni}\caption{\small Percentage integral
completeness results for the uniform noise region of the 10h
field, over the flux density range $2-10$\,mJy, and for signal-to-noise
thresholds of $\rm >1.50\sigma$ to $\rm >4.00\sigma$.}

\begin{tabular}{|c|c|c|c|c|c|c|} \hline
\multicolumn{7}{|c|}{14h field, uniform noise region} \\ \hline
\multicolumn{1}{|c|}{Flux} & 
\multicolumn{6}{c|}{ \% Integral Completeness} \\ \cline{2-7} 
Density & $\rm >1.50\sigma$  & $\rm >2.00\sigma$ & $\rm >2.50\sigma$ & $\rm
 >3.00\sigma$ &  $\rm >3.50\sigma$ & $\rm >4.00\sigma$ \\ \hline
 $\rm>\phantom{0}2\,mJy$ & $51.0 \pm \phantom{0}0.8$ & $45.1 \pm
\phantom{0}0.7$ & $36.2 \pm \phantom{0}0.5$ & $27.5 \pm
\phantom{0}0.4$ & $20.6 \pm \phantom{0}0.3$ & $15.3 \pm
\phantom{0}0.2$ \\
 $\rm >\phantom{0}4\,mJy$ & $70.3 \pm \phantom{0}2.3$ & $69.1 \pm
\phantom{0}2.2$ & $65.9 \pm \phantom{0}2.1$ & $61.5 \pm
\phantom{0}2.0$ & $55.4 \pm \phantom{0}1.8$ & $48.8 \pm
\phantom{0}1.6$ \\
$\rm >\phantom{0}6\,mJy$ & $74.5 \pm \phantom{0}3.9$ & $74.2 \pm
\phantom{0}3.9$ & $73.4 \pm \phantom{0}3.8$ & $70.9 \pm
\phantom{0}3.7$ & $69.0 \pm \phantom{0}3.6$ & $67.4 \pm
\phantom{0}3.5$ \\
$\rm >\phantom{0}8\,mJy$ & $77.8 \pm \phantom{0}6.0$ & $77.8 \pm
\phantom{0}6.0$ & $77.8 \pm \phantom{0}6.0$ & $76.0 \pm
\phantom{0}5.9$ & $75.4 \pm \phantom{0}5.8$ & $74.9 \pm
\phantom{0}5.8$ \\
$\rm >10\,mJy$ & $78.3 \pm \phantom{0}9.4$ & $78.3 \pm \phantom{0}9.4$
& $78.3 \pm \phantom{0}9.4$ & $76.8 \pm \phantom{0}9.2$ & $76.8 \pm
\phantom{0}9.2$ & $75.4 \pm \phantom{0}9.1$ \\ \hline
\end{tabular}
\label{table:intcomp_14uni}\caption{\small Percentage integral
completeness results for the uniform noise region of the 14h
field, over the flux density range $2-10$\,mJy, and for signal-to-noise
thresholds of $\rm >1.50\sigma$ to $\rm >4.00\sigma$.}
\end{table*}

\newpage
\begin{table*}
\footnotesize
\begin{tabular}{|c|c|c|c|c|c|c|} \hline
\multicolumn{7}{|c|}{22h field, uniform noise region} \\ \hline
\multicolumn{1}{|c|}{Flux} & 
\multicolumn{6}{c|}{ \% Integral Completeness} \\ \cline{2-7} 
Density & $\rm >1.50\sigma$  & $\rm >2.00\sigma$ & $\rm >2.50\sigma$ & $\rm
 >3.00\sigma$ &  $\rm >3.50\sigma$ & $\rm >4.00\sigma$ \\ \hline
 $\rm>\phantom{0}2\,mJy$ & $49.9 \pm \phantom{0}2.4$ & $45.1 \pm
\phantom{0}2.2$ & $36.8 \pm \phantom{0}1.8$ & $27.7 \pm
\phantom{0}1.4$ & $20.5 \pm \phantom{0}1.0$ & $14.3 \pm
\phantom{0}0.7$ \\ 
 $\rm >\phantom{0}4\,mJy$ & $72.2 \pm \phantom{0}8.1$ & $72.2 \pm
\phantom{0}8.1$ & $72.2 \pm \phantom{0}8.1$ & $72.2 \pm
\phantom{0}8.1$ & $65.8 \pm \phantom{0}7.4$ & $55.7 \pm
\phantom{0}6.3$ \\
$\rm >\phantom{0}6\,mJy$ & $80.0 \pm 16.0$ & $80.0 \pm 16.0$ & $80.0
\pm 16.0$ & $80.0 \pm 16.0$ & $80.0 \pm 16.0$ & $80.0 \pm 16.0$ \\
$\rm >\phantom{0}8\,mJy$ & $60.0 \pm 15.5$ & $60.0 \pm 15.5$ & $60.0
\pm 15.5$ & $60.0 \pm 15.5$ & $60.0 \pm 15.5$ & $60.0 \pm 15.5$ \\
$\rm >10\,mJy$ & $75.0 \pm 37.5$ & $75.0 \pm 37.5$ & $75.0 \pm 37.5$ &
$75.0 \pm 37.5$ & $75.0 \pm 37.5$ & $75.0 \pm 37.5$ \\ \hline
\end{tabular}
\label{table:intcomp_22uni}\caption{\small Percentage integral
completeness results for the uniform noise region of the 22h
field, over the flux density range $2-10$\,mJy, and for signal-to-noise
thresholds of $\rm >1.50\sigma$ to $\rm >4.00\sigma$.}

\begin{tabular}{|c|c|c|c|c|c|c|} \hline
\multicolumn{7}{|c|}{SSA13 deep area field, uniform noise region} \\ \hline
\multicolumn{1}{|c|}{Flux} & 
\multicolumn{6}{c|}{ \% Integral Completeness} \\ \cline{2-7} 
Density & $\rm >1.50\sigma$  & $\rm >2.00\sigma$ & $\rm >2.50\sigma$ & $\rm
 >3.00\sigma$ &  $\rm >3.50\sigma$ & $\rm >4.00\sigma$ \\ \hline
 $\rm>\phantom{0}2\,mJy$ & $51.3 \pm \phantom{0}2.0$ & $50.6 \pm
\phantom{0}1.9$ & $48.4 \pm \phantom{0}1.9$ & $47.1 \pm \phantom{0}1.8$ & $42.9
\pm \phantom{0}1.6$ & $38.8 \pm \phantom{0}1.5$ \\
 $\rm >\phantom{0}4\,mJy$ & $67.9 \pm \phantom{0}5.9$ & $67.9 \pm
\phantom{0}5.9$ & $67.2 \pm \phantom{0}5.8$ & $67.2 \pm
\phantom{0}5.8$ & $67.2 \pm \phantom{0}5.8$ & $67.2 \pm
\phantom{0}5.8$ \\
$\rm >\phantom{0}6\,mJy$ & $84.1 \pm 12.7$ & $84.1 \pm 12.7$ & $84.1
\pm 12.7$ & $84.1 \pm 12.7$ & $84.1 \pm 12.7$ & $84.1 \pm 12.7$ \\
$\rm >\phantom{0}8\,mJy$ & $68.0 \pm 13.6$ & $68.0 \pm 13.6$ & $68.0
\pm 13.6$ & $68.0 \pm 13.6$ & $68.0 \pm 13.6$ & $68.0 \pm 13.6$ \\
$\rm >10\,mJy$ & $75.0 \pm 21.7$ & $75.0 \pm 21.7$ & $75.0 \pm 21.7$ &
$75.0 \pm 21.7$ & $75.0 \pm 21.7$ & $75.0 \pm 21.7$ \\ \hline
\end{tabular}
\label{table:intcomp_ssa13deepuni}\caption{\small Percentage integral
completeness results for the uniform noise region of the SSA13 deep area
field, over the flux density range $2-10$\,mJy, and for signal-to-noise
thresholds of $\rm >1.50\sigma$ to $\rm >4.00\sigma$.}

\begin{tabular}{|c|c|c|c|c|c|c|} \hline
\multicolumn{7}{|c|}{SSA13 wide area field, uniform noise region} \\ \hline
\multicolumn{1}{|c|}{Flux} & 
\multicolumn{6}{c|}{ \% Integral Completeness} \\ \cline{2-7} 
Density & $\rm >1.50\sigma$  & $\rm >2.00\sigma$ & $\rm >2.50\sigma$ & $\rm
 >3.00\sigma$ &  $\rm >3.50\sigma$ & $\rm >4.00\sigma$ \\ \hline
 $\rm>\phantom{0}2\,mJy$ & $38.6 \pm \phantom{0}0.6$ & $29.2 \pm
\phantom{0}0.5$ & $21.0 \pm \phantom{0}0.3$ & $14.2 \pm
\phantom{0}0.2$ & $\phantom{0}9.4 \pm \phantom{0}0.2$ &
$\phantom{0}6.2 \pm \phantom{0}0.1$ \\
 $\rm >\phantom{0}4\,mJy$ & $61.2 \pm \phantom{0}2.1$ & $56.6 \pm
 \phantom{0}1.9$ & $49.5 \pm \phantom{0}1.7$ & $39.3 \pm
 \phantom{0}1.4$ & $29.2 \pm \phantom{0}1.0$ & $21.8 \pm
 \phantom{0}0.7$ \\
$\rm >\phantom{0}6\,mJy$ & $70.7 \pm \phantom{0}3.9$ & $67.0 \pm
 \phantom{0}3.7$ & $63.6 \pm \phantom{0}3.5$ & $59.6 \pm
 \phantom{0}3.3$ & $50.6 \pm \phantom{0}2.8$ & $42.0 \pm
 \phantom{0}2.3$ \\
$\rm >\phantom{0}8\,mJy$ & $77.0 \pm \phantom{0}6.5$ & $76.3 \pm
 \phantom{0}6.5$ & $73.4 \pm \phantom{0}6.2$ & $72.7 \pm
 \phantom{0}6.2$ & $66.2 \pm \phantom{0}5.6$ & $60.4 \pm
 \phantom{0}5.1$ \\ 
$\rm >10\,mJy$ & $73.0 \pm \phantom{0}9.2$ & $73.0 \pm \phantom{0}9.2$
 & $73.0 \pm \phantom{0}9.2$ & $73.0 \pm \phantom{0}9.2$ & $69.8 \pm
 \phantom{0}8.8$ & $65.1 \pm \phantom{0}8.2$ \\ \hline
\end{tabular}
\label{table:intcomp_ssa13wuni}\caption{\small Percentage integral
completeness results for the uniform noise region of the SSA13 wide area
field, over the flux density range $2-10$\,mJy, and for signal-to-noise
thresholds of $\rm >1.50\sigma$ to $\rm >4.00\sigma$.}

\begin{tabular}{|c|c|c|c|c|c|c|} \hline
\multicolumn{7}{|c|}{SSA17 field, uniform noise region} \\ \hline
\multicolumn{1}{|c|}{Flux} & 
\multicolumn{6}{c|}{ \% Integral Completeness} \\ \cline{2-7} 
Density & $\rm >1.50\sigma$  & $\rm >2.00\sigma$ & $\rm >2.50\sigma$ & $\rm
 >3.00\sigma$ &  $\rm >3.50\sigma$ & $\rm >4.00\sigma$ \\ \hline
 $\rm>\phantom{0}2\,mJy$ & $49.6 \pm \phantom{0}1.2$ & $42.7 \pm
\phantom{0}1.1$ & $35.3 \pm \phantom{0}0.9$ & $29.2 \pm
\phantom{0}0.7$ & $22.9 \pm \phantom{0}0.6$ & $18.3 \pm
\phantom{0}0.5$ \\
 $\rm >\phantom{0}4\,mJy$ & $70.9 \pm \phantom{0}3.6$ & $69.0 \pm
\phantom{0}3.6$ & $68.3 \pm \phantom{0}3.5$ & $65.1 \pm
\phantom{0}3.3$ & $57.1 \pm \phantom{0}2.9$ & $50.3 \pm
\phantom{0}2.6$ \\
$\rm >\phantom{0}6\,mJy$ & $78.9 \pm \phantom{0}6.5$ & $78.9 \pm
\phantom{0}6.5$ & $78.9 \pm \phantom{0}6.5$ & $78.2 \pm
\phantom{0}6.5$ & $74.1 \pm \phantom{0}6.1$ & $67.3 \pm
\phantom{0}5.6$ \\
$\rm >\phantom{0}8\,mJy$ & $71.8 \pm \phantom{0}8.5$ & $71.8 \pm
\phantom{0}8.5$ & $71.8 \pm \phantom{0}8.5$ & $71.8 \pm
\phantom{0}8.5$ & $71.8 \pm \phantom{0}8.5$ & $71.8 \pm
\phantom{0}8.5$ \\
$\rm >10\,mJy$ & $69.7 \pm 12.1$ & $69.7 \pm 12.1$ & $69.7 \pm 12.1$ &
$69.7 \pm 12.1$ & $69.7 \pm 12.1$ & $69.7 \pm 12.1$ \\ \hline
\end{tabular}
\label{table:intcomp_ssa17uni}\caption{\small Percentage integral
completeness results for the uniform noise region of the SSA17
field, over the flux density range $2-10$\,mJy, and for signal-to-noise
thresholds of $\rm >1.50\sigma$ to $\rm >4.00\sigma$.}

\begin{tabular}{|c|c|c|c|c|c|c|} \hline
\multicolumn{7}{|c|}{SSA22 field, uniform noise region} \\ \hline
\multicolumn{1}{|c|}{Flux} & 
\multicolumn{6}{c|}{ \% Integral Completeness} \\ \cline{2-7} 
Density & $\rm >1.50\sigma$  & $\rm >2.00\sigma$ & $\rm >2.50\sigma$ & $\rm
 >3.00\sigma$ &  $\rm >3.50\sigma$ & $\rm >4.00\sigma$ \\ \hline
 $\rm>\phantom{0}2\,mJy$ & $50.3 \pm \phantom{0}1.2$ & $49.0 \pm
\phantom{0}1.2$ & $46.3 \pm \phantom{0}1.2$ & $43.1 \pm
\phantom{0}1.1$ & $39.0 \pm \phantom{0}1.0$ & $34.6 \pm
\phantom{0}0.9$ \\
 $\rm >\phantom{0}4\,mJy$ & $65.7 \pm \phantom{0}3.6$ & $65.4 \pm
\phantom{0}3.6$ & $64.8 \pm \phantom{0}3.5$ & $64.2 \pm
\phantom{0}3.5$ & $62.7 \pm \phantom{0}3.4$ & $60.9 \pm
\phantom{0}3.3$ \\
$\rm >\phantom{0}6\,mJy$ & $73.9 \pm \phantom{0}6.8$ & $73.9 \pm
\phantom{0}6.8$ & $73.9 \pm \phantom{0}6.8$ & $73.1 \pm
\phantom{0}6.7$ & $73.1 \pm \phantom{0}6.7$ & $73.1 \pm
\phantom{0}6.7$ \\
$\rm >\phantom{0}8\,mJy$ & $69.6 \pm \phantom{0}9.3$ & $69.6 \pm
\phantom{0}9.3$ & $69.6 \pm \phantom{0}9.3$ & $69.6 \pm
\phantom{0}9.3$ & $69.6 \pm \phantom{0}9.3$ & $69.6 \pm \phantom{0}9.3$ \\
$\rm >10\,mJy$ & $78.3 \pm 16.3$ & $78.3 \pm 16.3$ & $78.3 \pm 16.3$ &
$78.3 \pm 16.3$ & $78.3 \pm 16.3$ & $78.3 \pm 16.3$ \\ \hline
\end{tabular}
\label{table:intcomp_ssa22uni}\caption{\small Percentage integral
completeness results for the uniform noise region of the SSA22
field, over the flux density range $2-10$\,mJy, and for signal-to-noise
thresholds of $\rm >1.50\sigma$ to $\rm >4.00\sigma$.}
\end{table*}

\renewcommand{\topfraction}{0.5}
\clearpage

\begin{table*}
\footnotesize
\begin{tabular}{|c|c|c|c|c|c|c|} \hline
\multicolumn{7}{|c|}{Lockman Hole deep field, uniform noise region} \\ \hline
\multicolumn{1}{|c|}{Flux} & 
\multicolumn{6}{c|}{ \% Integral Completeness} \\ \cline{2-7} 
Density & $\rm >1.50\sigma$  & $\rm >2.00\sigma$ & $\rm >2.50\sigma$ & $\rm
 >3.00\sigma$ &  $\rm >3.50\sigma$ & $\rm >4.00\sigma$ \\ \hline
 $\rm>\phantom{0}2\,mJy$ & $45.9 \pm \phantom{0}1.8$ & $45.5 \pm
\phantom{0}1.8$ & $43.6 \pm \phantom{0}1.7$ & $41.8 \pm
\phantom{0}1.6$ & $37.9 \pm \phantom{0}1.5$ & $33.6 \pm
\phantom{0}1.3$ \\
 $\rm >\phantom{0}4\,mJy$ & $60.4 \pm \phantom{0}5.0$ & $60.4 \pm
\phantom{0}5.0$ & $60.4 \pm \phantom{0}5.0$ & $60.4 \pm
\phantom{0}5.0$ & $59.7 \pm \phantom{0}5.0$ & $59.0 \pm
\phantom{0}4.9$ \\
$\rm >\phantom{0}6\,mJy$ & $68.5 \pm \phantom{0}9.3$ & $68.5 \pm
\phantom{0}9.3$ & $68.5 \pm \phantom{0}9.3$ & $68.5 \pm
\phantom{0}9.3$ & $68.5 \pm \phantom{0}9.3$ & $68.5 \pm
\phantom{0}9.3$ \\
$\rm >\phantom{0}8\,mJy$ & $64.3 \pm 12.1$ & $64.3 \pm 12.1$ & $64.3
\pm 12.1$ & $64.3 \pm 12.1$ & $64.3 \pm 12.1$ & $64.3 \pm 12.1$ \\
$\rm >10\,mJy$ & $70.6 \pm 17.1$ & $70.6 \pm 17.1$ & $70.6 \pm 17.1$ &
$70.6 \pm 17.1$ & $70.6 \pm 17.1$ & $70.6 \pm 17.1$ \\ \hline
\end{tabular}
\label{table:intcomp_lhdeepuni}\caption{\small Percentage integral
completeness results for the uniform noise region of the Lockman Hole deep
field, over the flux density range $2-10$\,mJy, and for signal-to-noise
thresholds of $\rm >1.50\sigma$ to $\rm >4.00\sigma$.}

\begin{tabular}{|c|c|c|c|c|c|c|} \hline
\multicolumn{7}{|c|}{Hubble deep field, uniform noise region} \\ \hline
\multicolumn{1}{|c|}{Flux} & 
\multicolumn{6}{c|}{ \% Integral Completeness} \\ \cline{2-7} 
Density & $\rm >1.50\sigma$  & $\rm >2.00\sigma$ & $\rm >2.50\sigma$ & $\rm
 >3.00\sigma$ &  $\rm >3.50\sigma$ & $\rm >4.00\sigma$ \\ \hline
 $\rm>\phantom{0}2\,mJy$ & $61.6 \pm \phantom{0}2.9$ & $61.6 \pm
\phantom{0}2.9$ & $61.6 \pm \phantom{0}2.9$ & $61.3 \pm
\phantom{0}2.9$ & $59.8 \pm \phantom{0}2.8$ & $57.1 \pm
\phantom{0}2.7$ \\
 $\rm >\phantom{0}4\,mJy$ & $77.3 \pm \phantom{0}7.9$ & $77.3 \pm
\phantom{0}7.9$ & $77.3 \pm \phantom{0}7.9$ & $77.3 \pm
\phantom{0}7.9$ & $77.3 \pm \phantom{0}7.9$ & $77.3 \pm
\phantom{0}7.9$ \\
$\rm >\phantom{0}6\,mJy$ & $93.0 \pm 14.2$ & $93.0 \pm 14.2$ & $93.0
\pm 14.2$ & $93.0 \pm 14.2$ & $93.0 \pm 14.2$ & $93.0 \pm 14.2$ \\
$\rm >\phantom{0}8\,mJy$ & $82.4 \pm 20.0$ & $82.4 \pm 20.0$ & $82.4
\pm 20.0$ & $82.4 \pm 20.0$ & $82.4 \pm 20.0$ & $82.4 \pm 20.0$ \\
$\rm >10\,mJy$ & $100\phantom{.}\pm 40.8$ & $100\phantom{.}\pm 40.8$ & $100\phantom{.}\pm
40.8$ & $100\phantom{.}\pm 40.8$ & $100\phantom{.}\pm 40.8$ & $100\phantom{.}\pm 40.8$ \\ \hline
\end{tabular}
\small
\label{table:intcomp_hdfuni}\caption{\small Percentage integral
completeness results for the uniform noise region of the Hubble deep
field, over the flux density range $2-10$\,mJy, and for signal-to-noise
thresholds of $\rm >1.50\sigma$ to $\rm >4.00\sigma$.}
\end{table*}

\subsection{Count correction tables} 
The following tables give the percentage count correction factors for
each of the survey fields over 
the flux density range $2-10$\,mJy, and for signal-to-noise
thresholds of $\rm >1.50\sigma$ to $\rm >4.00\sigma$. Those fields 
composed of a deep pencil
beam survey within a wider-area shallower survey have had these two
components treated separately.

\begin{table*}
\footnotesize
\begin{tabular}{|c|c|c|c|c|c|c|} \hline
\multicolumn{7}{|c|}{ELAIS N2 field, uniform noise region} \\ \hline
\multicolumn{1}{|c|}{Flux} & 
\multicolumn{6}{c|}{ \% Count Correction} \\ \cline{2-7} 
Density & $\rm >1.50\sigma$  & $\rm >2.00\sigma$ & $\rm >2.50\sigma$ & $\rm
 >3.00\sigma$ &  $\rm >3.50\sigma$ & $\rm >4.00\sigma$ \\ \hline
 $\rm>\phantom{0}2\,mJy$ & $\phantom{0}44.0 \pm \phantom{0}0.3$ & $\phantom{0}78.9 \pm
\phantom{0}0.7$ & $160.5 \pm \phantom{0}2.2$ & $334.8 \pm
\phantom{0}6.5$ & $652.6 \pm  17.7$ & $>1000$ \\
 $\rm >\phantom{0}4\,mJy$ & $\phantom{0}14.8 \pm \phantom{0}0.1$ & $\phantom{0}20.8 \pm
\phantom{0}0.2$ & $\phantom{0}36.5 \pm \phantom{0}0.5$ & $\phantom{0}71.1 \pm
\phantom{0}1.4$ & $137.3 \pm \phantom{0}3.7$ & $245.5 \pm
\phantom{00}8.9$ \\
$\rm >\phantom{0}6\,mJy$ & $\phantom{0}13.3 \pm \phantom{0}0.2$ & $\phantom{0}16.6 \pm
\phantom{0}0.3$ & $\phantom{0}24.8 \pm \phantom{0}0.5$ & $\phantom{0}38.5 \pm
\phantom{0}0.9$ & $\phantom{0}61.0 \pm \phantom{0}1.8$ & $\phantom{0}94.3 \pm
\phantom{00}3.6$ \\
$\rm >\phantom{0}8\,mJy$ & $\phantom{0}17.1 \pm \phantom{0}0.4$ & $\phantom{0}18.7 \pm
\phantom{0}0.5$ & $\phantom{0}25.4 \pm \phantom{0}0.7$ & $\phantom{0}37.9 \pm
\phantom{0}1.3$ & $\phantom{0}54.7 \pm \phantom{0}2.3$ & $\phantom{0}75.7 \pm
\phantom{00}3.8$ \\
$\rm >10\,mJy$ & $\phantom{0}26.4 \pm \phantom{0}1.1$ & $\phantom{0}27.1 \pm \phantom{0}1.2$
& $\phantom{0}33.0 \pm \phantom{0}1.6$ & $\phantom{0}45.1 \pm \phantom{0}2.5$ & $\phantom{0}59.7 \pm
\phantom{0}3.9$ & $\phantom{0}75.9 \pm \phantom{00}5.6$ \\ \hline
\end{tabular}
\label{table:countcorr_n2uni}\caption{\small Percentage count correction results for the uniform noise region of the ELAIS N2
field, over the flux density range $2-10$\,mJy, and for signal-to-noise
thresholds of $\rm >1.50\sigma$ to $\rm >4.00\sigma$.}

\begin{tabular}{|c|c|c|c|c|c|c|} \hline
\multicolumn{7}{|c|}{ELAIS N2 field, non-uniform noise region} \\ \hline
\multicolumn{1}{|c|}{Flux} & 
\multicolumn{6}{c|}{ \% Count Correction} \\ \cline{2-7} 
Density & $\rm >1.50\sigma$  & $\rm >2.00\sigma$ & $\rm >2.50\sigma$ & $\rm
 >3.00\sigma$ &  $\rm >3.50\sigma$ & $\rm >4.00\sigma$ \\ \hline
 $\rm>\phantom{0}2\,mJy$ & $\phantom{0}83.0 \pm \phantom{0}1.5$ & $158.4 \pm \phantom{0}4.0$ & $467.8 \pm 20.5$ & $>1000$ & $>1000$ & $>1000$ \\
 $\rm >\phantom{0}4\,mJy$ & $\phantom{0}17.8 \pm \phantom{0}0.3$ & $\phantom{0}33.9 \pm \phantom{0}0.9$ & $100.2 \pm \phantom{0}4.4$ & $366.2 \pm 30.7$ &
 $>1000$ & $>1000$ \\
$\rm >\phantom{0}6\,mJy$ & $\phantom{00}6.8 \pm \phantom{0}0.1$ & $\phantom{0}11.6 \pm \phantom{0}0.3$ & $\phantom{0}34.1 \pm \phantom{0}1.5$ & $124.6 \pm
 10.5$ & $491.7 \pm 81.9$ & $>1000$ \\
$\rm >\phantom{0}8\,mJy$ & $\phantom{00}4.0 \pm \phantom{0}0.1$ & $\phantom{00}5.8 \pm \phantom{0}0.2$ & $\phantom{0}14.7 \pm \phantom{0}0.7$ & $\phantom{0}52.1 \pm \phantom{0}4.4$ & $205.6
 \pm 34.3$ & $616.7 \pm 178.0$ \\
$\rm >10\,mJy$ & $\phantom{00}4.7 \pm \phantom{0}0.2$ & $\phantom{00}6.6 \pm \phantom{0}0.3$ &
 $\phantom{0}14.6 \pm \phantom{0}0.9$ & $\phantom{0}42.0 \pm
\phantom{0}4.2$ & $120.0 \pm 20.3$ & $350.0 \pm 101.0$ \\ \hline
\end{tabular}
\label{table:countcorr_n2non}\caption{\small Percentage count correction results for the non-uniform noise region of the ELAIS N2
field, over the flux density range $2-10$\,mJy, and for signal-to-noise
thresholds of $\rm >1.50\sigma$ to $\rm >4.00\sigma$.}
\vspace{1.0cm}
\end{table*}

\newpage
\begin{table*}
\footnotesize
\begin{tabular}{|c|c|c|c|c|c|c|} \hline
\multicolumn{7}{|c|}{Lockman Hole wide area field, uniform noise region} \\ \hline
\multicolumn{1}{|c|}{Flux} & 
\multicolumn{6}{c|}{ \% Count Correction} \\ \cline{2-7} 
Density & $\rm >1.50\sigma$  & $\rm >2.00\sigma$ & $\rm >2.50\sigma$ & $\rm
 >3.00\sigma$ &  $\rm >3.50\sigma$ & $\rm >4.00\sigma$ \\ \hline
 $\rm>\phantom{0}2\,mJy$ & $\phantom{0}45.7 \pm \phantom{0}0.4$ & $\phantom{0}93.8 \pm
\phantom{0}1.1$ & $207.2 \pm \phantom{0}3.7$ & $459.8 \pm 12.3$ &
$>1000$ & $>1000$ \\
 $\rm >\phantom{0}4\,mJy$ & $\phantom{0}11.4 \pm \phantom{0}0.1$ & $\phantom{0}20.7 \pm
 \phantom{0}0.3$ & $\phantom{0}44.1 \pm \phantom{0}0.8$ & $\phantom{0}96.6 \pm
 \phantom{0}2.6$ & $230.5 \pm \phantom{0}9.5$ & $448.0 \pm \phantom{0}25.8$ \\
$\rm >\phantom{0}6\,mJy$ & $\phantom{00}9.7 \pm \phantom{0}0.1$ & $\phantom{0}12.3
 \pm \phantom{0}0.2$ & $\phantom{0}20.9 \pm \phantom{0}0.4$ & $\phantom{0}41.2 \pm
 \phantom{0}1.2$ & $\phantom{0}91.7 \pm \phantom{0}3.9$ & $171.0 \pm
 \phantom{00}9.9$ \\
$\rm >\phantom{0}8\,mJy$ & $\phantom{0}12.3 \pm \phantom{0}0.3$ & $\phantom{0}13.5 \pm
 \phantom{0}0.3$ & $\phantom{0}18.1 \pm \phantom{0}0.5$ & $\phantom{0}26.6 \pm
 \phantom{0}0.9$ & $\phantom{0}52.2 \pm \phantom{0}2.5$ & $\phantom{0}88.6 \pm
 \phantom{00}5.6$ \\
$\rm >10\,mJy$ & $\phantom{0}22.1 \pm \phantom{0}1.0$ & $\phantom{0}22.7 \pm \phantom{0}1.0$
 & $\phantom{0}28.8 \pm \phantom{0}1.4$ & $\phantom{0}36.4 \pm \phantom{0}2.0$ & $\phantom{0}46.9 \pm
 \phantom{0}3.0$ & $\phantom{0}65.0 \pm \phantom{00}4.9$ \\ \hline
\end{tabular}
\label{table:countcorr_lhuni}\caption{\small Percentage count
correction results for the uniform noise region of the Lockman Hole
wide area
field, over the flux density range $2-10$\,mJy, and for signal-to-noise
thresholds of $\rm >1.50\sigma$ to $\rm >4.00\sigma$.}

\begin{tabular}{|c|c|c|c|c|c|c|} \hline
\multicolumn{7}{|c|}{Lockman Hole wide area field, non-uniform noise region} \\ \hline
\multicolumn{1}{|c|}{Flux} & 
\multicolumn{6}{c|}{ \% Count Correction} \\ \cline{2-7} 
Density & $\rm >1.50\sigma$  & $\rm >2.00\sigma$ & $\rm >2.50\sigma$ & $\rm
 >3.00\sigma$ &  $\rm >3.50\sigma$ & $\rm >4.00\sigma$ \\ \hline
 $\rm>\phantom{0}2\,mJy$ & $\phantom{0}59.3 \pm \phantom{0}0.8$ & $102.7 \pm
\phantom{0}1.7$ & $191.8 \pm \phantom{0}4.4$ & $441.5 \pm 15.2$ &
$>1000$ & $>1000$ \\
 $\rm >\phantom{0}4\,mJy$ & $\phantom{0}12.6 \pm \phantom{0}0.2$ & $\phantom{0}21.8 \pm
 \phantom{0}0.4$ & $\phantom{0}40.7 \pm \phantom{0}0.9$ & $\phantom{0}93.6 \pm
 \phantom{0}3.2$ & $312.7 \pm 19.7$ & $>1000$ \\
$\rm >\phantom{0}6\,mJy$ & $\phantom{00}5.3 \pm \phantom{0}0.1$ & $\phantom{00}8.5 \pm
 \phantom{0}0.1$ & $\phantom{0}15.7 \pm \phantom{0}0.4$ & $\phantom{0}36.1 \pm
 \phantom{0}1.2$ & $120.7 \pm \phantom{0}7.6$ & $673.3 \pm 100.4$ \\
$\rm >\phantom{0}8\,mJy$ & $\phantom{00}3.1 \pm \phantom{0}0.1$ &
 $\phantom{00}4.4 \pm \phantom{0}0.1$ & $\phantom{00}7.6 \pm
 \phantom{0}0.2$ & $\phantom{0}17.2 \pm \phantom{0}0.6$ & $\phantom{0}57.4 \pm
 \phantom{0}3.6$ & $320.0 \pm \phantom{0}47.7$ \\
$\rm >10\,mJy$ & $\phantom{00}2.3 \pm \phantom{0}0.1$ & $\phantom{00}2.9
 \pm \phantom{0}0.1$ & $\phantom{00}4.5 \pm \phantom{0}0.1$ &
 $\phantom{00}9.8 \pm \phantom{0}0.4$ & $\phantom{0}31.4 \pm \phantom{0}2.0$ &
 $172.7 \pm \phantom{0}26.0$ \\ \hline
\end{tabular}
\label{table:countcorr_lhnon}\caption{\small Percentage count
correction results for the nonuniform noise region of the Lockman Hole
wide area
field, over the flux density range $2-10$\,mJy, and for signal-to-noise
thresholds of $\rm >1.50\sigma$ to $\rm >4.00\sigma$.}

\begin{tabular}{|c|c|c|c|c|c|c|} \hline
\multicolumn{7}{|c|}{Lockman Hole deep strip field, uniform noise region} \\ \hline
\multicolumn{1}{|c|}{Flux} & 
\multicolumn{6}{c|}{ \% Count Correction} \\ \cline{2-7} 
Density & $\rm >1.50\sigma$  & $\rm >2.00\sigma$ & $\rm >2.50\sigma$ & $\rm
 >3.00\sigma$ &  $\rm >3.50\sigma$ & $\rm >4.00\sigma$ \\ \hline
 $\rm>\phantom{0}2\,mJy$ & $\phantom{0}41.5 \pm \phantom{0}0.7$ & $\phantom{0}65.2 \pm
\phantom{0}1.3$ & $113.3 \pm \phantom{0}3.1$ & $205.5 \pm
\phantom{0}7.5$ & $352.7 \pm  16.9$ & $606.0 \pm  \phantom{0}38.2$ \\
 $\rm >\phantom{0}4\,mJy$ & $\phantom{0}26.5 \pm \phantom{0}0.8$ & $\phantom{0}27.2 \pm
\phantom{0}0.8$ & $\phantom{0}32.5 \pm \phantom{0}1.1$ & $\phantom{0}45.3 \pm
\phantom{0}1.7$ & $\phantom{0}71.6 \pm \phantom{0}3.5$ & $121.0 \pm
\phantom{00}7.6$ \\
$\rm >\phantom{0}6\,mJy$ & $\phantom{0}50.2 \pm \phantom{0}3.3$ & $\phantom{0}50.2 \pm
\phantom{0}3.3$ & $\phantom{0}50.2 \pm \phantom{0}3.3$ & $\phantom{0}51.8 \pm
\phantom{0}3.5$ & $\phantom{0}56.1 \pm \phantom{0}3.9$ & $\phantom{0}63.9 \pm
\phantom{00}4.8$ \\
$\rm >\phantom{0}8\,mJy$ & $\phantom{0}83.9 \pm 10.7$ & $\phantom{0}83.9 \pm 10.7$ & $\phantom{0}83.9
\pm 10.7$ & $\phantom{0}83.9 \pm 10.7$ & $\phantom{0}83.9 \pm 10.7$ & $\phantom{0}83.9 \pm \phantom{0}10.7$ \\
$\rm >10\,mJy$ & $\phantom{0}81.2 \pm 14.4$ & $\phantom{0}81.2 \pm 14.4$ & $\phantom{0}81.2 \pm 14.4$ &
$\phantom{0}81.2 \pm 14.4$ & $\phantom{0}81.2 \pm 14.4$ & $\phantom{0}81.2 \pm \phantom{0}14.4$ \\ \hline
\end{tabular}
\label{table:countcorr_lhstripuni}\caption{\small Percentage count
correction results for the uniform noise region of the Lockman Hole
deep strip
field, over the flux density range $2-10$\,mJy, and for signal-to-noise
thresholds of $\rm >1.50\sigma$ to $\rm >4.00\sigma$.}

\begin{tabular}{|c|c|c|c|c|c|c|} \hline
\multicolumn{7}{|c|}{03h deep area field, uniform noise region} \\ \hline
\multicolumn{1}{|c|}{Flux} & 
\multicolumn{6}{c|}{ \% Count Correction} \\ \cline{2-7} 
Density & $\rm >1.50\sigma$  & $\rm >2.00\sigma$ & $\rm >2.50\sigma$ & $\rm
 >3.00\sigma$ &  $\rm >3.50\sigma$ & $\rm >4.00\sigma$ \\ \hline
 $\rm>\phantom{0}2\,mJy$ & $\phantom{0}62.3 \pm \phantom{0}2.1$ & $\phantom{0}64.9 \pm
\phantom{0}2.3$ & $\phantom{0}76.5 \pm \phantom{0}2.9$ & $100.4 \pm
\phantom{0}4.4$ & $140.1 \pm \phantom{0}7.2$ & $199.2 \pm \phantom{0}12.2$ \\
 $\rm >\phantom{0}4\,mJy$ & $\phantom{0}58.7 \pm \phantom{0}4.1$ & $\phantom{0}59.9 \pm
\phantom{0}4.3$ & $\phantom{0}62.8 \pm \phantom{0}4.6$ & $\phantom{0}64.8 \pm
\phantom{0}4.8$ & $\phantom{0}67.8 \pm \phantom{0}5.1$ & $\phantom{0}72.8 \pm
\phantom{00}5.7$ \\
$\rm >\phantom{0}6\,mJy$ & $\phantom{0}62.9 \pm \phantom{0}7.5$ & $\phantom{0}63.8 \pm
\phantom{0}7.7$ & $\phantom{0}63.8 \pm \phantom{0}7.7$ & $\phantom{0}64.7 \pm
\phantom{0}7.8$ & $\phantom{0}65.7 \pm \phantom{0}8.0$ & $\phantom{0}65.7 \pm
\phantom{00}8.0$ \\
$\rm >\phantom{0}8\,mJy$ & $\phantom{0}88.5 \pm 17.3$ & $\phantom{0}88.5 \pm 17.3$ & $\phantom{0}88.5
\pm 17.3$ & $\phantom{0}92.0 \pm 18.4$ & $\phantom{0}92.0 \pm 18.4$ & $\phantom{0}92.0 \pm \phantom{0}18.4$ \\
$\rm >10\,mJy$ & $\phantom{0}71.4 \pm 19.1$ & $\phantom{0}71.4 \pm 19.1$ & $\phantom{0}71.4 \pm 19.1$ &
$\phantom{0}76.9 \pm 21.3$ & $\phantom{0}76.9 \pm 21.3$ & $\phantom{0}76.9 \pm \phantom{0}21.3$ \\ \hline
\end{tabular}
\label{table:countcorr_03deepuni}\caption{\small Percentage count
correction results for the uniform noise region of the 03h deep area
field, over the flux density range $2-10$\,mJy, and for signal-to-noise
thresholds of $\rm >1.50\sigma$ to $\rm >4.00\sigma$.}

\begin{tabular}{|c|c|c|c|c|c|c|} \hline
\multicolumn{7}{|c|}{03h wide area field, uniform noise region} \\ \hline
\multicolumn{1}{|c|}{Flux} & 
\multicolumn{6}{c|}{ \% Count Correction} \\ \cline{2-7} 
Density & $\rm >1.50\sigma$  & $\rm >2.00\sigma$ & $\rm >2.50\sigma$ & $\rm
 >3.00\sigma$ &  $\rm >3.50\sigma$ & $\rm >4.00\sigma$ \\ \hline
 $\rm>\phantom{0}2\,mJy$ & $\phantom{0}52.8 \pm \phantom{0}0.6$ & $\phantom{0}89.2 \pm
\phantom{0}1.2$ & $165.6 \pm \phantom{0}3.2$ & $308.9 \pm
\phantom{0}8.0$ & $553.0 \pm  19.3$ &  $919.8 \pm \phantom{0}41.3$ \\
 $\rm >\phantom{0}4\,mJy$ & $\phantom{0}20.4 \pm \phantom{0}0.3$ & $\phantom{0}27.7 \pm
\phantom{0}0.5$ & $\phantom{0}44.0 \pm \phantom{0}0.9$ & $\phantom{0}71.4 \pm
\phantom{0}2.0$ & $116.9  \pm \phantom{0}4.1$ & $190.5 \pm
\phantom{00}8.6$ \\
$\rm >\phantom{0}6\,mJy$ & $\phantom{0}20.7 \pm \phantom{0}0.5$ & $\phantom{0}24.4 \pm
\phantom{0}0.7$ & $\phantom{0}33.1 \pm \phantom{0}1.0$ & $\phantom{0}51.9 \pm
\phantom{0}2.0$ & $\phantom{0}75.4 \pm \phantom{0}3.5$ & $101.8 \pm
\phantom{00}5.6$ \\
$\rm >\phantom{0}8\,mJy$ & $\phantom{0}24.7 \pm \phantom{0}1.0$ & $\phantom{0}25.6 \pm
\phantom{0}1.1$ & $\phantom{0}35.4 \pm \phantom{0}1.8$ & $\phantom{0}53.5 \pm
\phantom{0}3.3$ & $\phantom{0}70.6 \pm \phantom{0}4.9$ & $\phantom{0}87.8 \pm
\phantom{00}6.9$ \\
$\rm >10\,mJy$ & $\phantom{0}32.2 \pm \phantom{0}2.2$ & $\phantom{0}32.5 \pm \phantom{0}2.3$
& $\phantom{0}34.2 \pm \phantom{0}2.4$ & $\phantom{0}49.3 \pm \phantom{0}4.2$ & $\phantom{0}68.0 \pm
\phantom{0}6.8$ & $\phantom{0}85.0 \pm \phantom{00}9.5$ \\ \hline
\end{tabular}
\label{table:countcorr_03wuni}\caption{\small Percentage count
correction results for the uniform noise region of the 03h wide area
field, over the flux density range $2-10$\,mJy, and for signal-to-noise
thresholds of $\rm >1.50\sigma$ to $\rm >4.00\sigma$.}
\end{table*}

\newpage
\begin{table*}
\footnotesize
\begin{tabular}{|c|c|c|c|c|c|c|} \hline
\multicolumn{7}{|c|}{10h field, uniform noise region} \\ \hline
\multicolumn{1}{|c|}{Flux} & 
\multicolumn{6}{c|}{ \% Count Correction} \\ \cline{2-7} 
Density & $\rm >1.50\sigma$  & $\rm >2.00\sigma$ & $\rm >2.50\sigma$ & $\rm
 >3.00\sigma$ &  $\rm >3.50\sigma$ & $\rm >4.00\sigma$ \\ \hline
 $\rm>\phantom{0}2\,mJy$ & $\phantom{0}77.4 \pm \phantom{0}2.6$ & $100.5 \pm
\phantom{0}3.9$ & $147.9 \pm \phantom{0}7.0$ & $231.4 \pm 13.7$ &
$379.4 \pm 28.7$ & $558.0 \pm \phantom{0}51.2$ \\
 $\rm >\phantom{0}4\,mJy$ & $\phantom{0}58.8 \pm \phantom{0}3.9$ & $\phantom{0}63.0 \pm
 \phantom{0}4.3$ & $\phantom{0}73.1 \pm \phantom{0}5.4$ & $\phantom{0}88.7 \pm
 \phantom{0}7.2$ & $104.7 \pm \phantom{0}9.3$ & $129.1 \pm \phantom{0}12.7$ \\
$\rm >\phantom{0}6\,mJy$ & $\phantom{0}64.4 \pm \phantom{0}7.5$ & $\phantom{0}65.3 \pm
 \phantom{0}7.7$ & $\phantom{0}78.3 \pm 10.1$ & $\phantom{0}90.4 \pm 12.5$ & $\phantom{0}92.2 \pm 12.9$
 & $\phantom{0}94.0 \pm \phantom{0}13.3$ \\
$\rm >\phantom{0}8\,mJy$ & $\phantom{0}66.7 \pm 12.2$ & $\phantom{0}66.7 \pm 12.2$ & $\phantom{0}69.0
 \pm 12.8$  & $\phantom{0}76.9 \pm 15.1$ & $\phantom{0}80.0 \pm 16.0$ & $\phantom{0}80.0 \pm \phantom{0}16.0$ \\
$\rm >10\,mJy$ & $122.2 \pm 40.7$ & $122.2 \pm 40.7$ & $122.2 \pm
 40.7$ & $122.2 \pm 40.7$ & $122.2 \pm 40.7$ & $122.2 \pm \phantom{0}40.7$ \\ \hline
\end{tabular}
\label{table:countcorr_10uni}\caption{\small Percentage count
correction results for the uniform noise region of the 10h
field, over the flux density range $2-10$\,mJy, and for signal-to-noise
thresholds of $\rm >1.50\sigma$ to $\rm >4.00\sigma$.}

\begin{tabular}{|c|c|c|c|c|c|c|} \hline
\multicolumn{7}{|c|}{14h field, uniform noise region} \\ \hline
\multicolumn{1}{|c|}{Flux} & 
\multicolumn{6}{c|}{ \% Count Correction} \\ \cline{2-7} 
Density & $\rm >1.50\sigma$  & $\rm >2.00\sigma$ & $\rm >2.50\sigma$ & $\rm
 >3.00\sigma$ &  $\rm >3.50\sigma$ & $\rm >4.00\sigma$ \\ \hline
 $\rm>\phantom{0}2\,mJy$ & $\phantom{0}55.5 \pm \phantom{0}0.6$ & $\phantom{0}83.6 \pm
\phantom{0}1.1$ & $146.4 \pm \phantom{0}2.6$ & $248.4 \pm
\phantom{0}5.8$ & $394.8 \pm 11.6$ & $584.9 \pm \phantom{0}21.0$ \\
 $\rm >\phantom{0}4\,mJy$ & $\phantom{0}34.2 \pm \phantom{0}0.6$ & $\phantom{0}42.3 \pm
\phantom{0}0.9$ & $\phantom{0}55.4 \pm \phantom{0}1.3$ & $\phantom{0}71.9 \pm
\phantom{0}2.0$ & $\phantom{0}96.3 \pm \phantom{0}3.1$ & $126.9 \pm
\phantom{00}4.6$ \\
$\rm >\phantom{0}6\,mJy$ & $\phantom{0}39.0 \pm \phantom{0}1.3$ & $\phantom{0}50.2 \pm
\phantom{0}1.9$ & $\phantom{0}69.0 \pm \phantom{0}3.0$ & $\phantom{0}80.2 \pm
\phantom{0}3.7$ & $\phantom{0}89.5 \pm \phantom{0}4.4$ & $\phantom{0}96.8 \pm
\phantom{00}5.0$ \\
$\rm >\phantom{0}8\,mJy$ & $\phantom{0}45.9 \pm \phantom{0}2.4$ & $\phantom{0}52.8 \pm
\phantom{0}3.0$ & $\phantom{0}69.3 \pm \phantom{0}4.5$ & $\phantom{0}84.3 \pm
\phantom{0}6.0$ & $\phantom{0}92.8 \pm \phantom{0}6.9$ & $\phantom{0}96.5 \pm
\phantom{00}7.3$ \\
$\rm >10\,mJy$ & $\phantom{0}61.1 \pm \phantom{0}5.7$ & $\phantom{0}61.6 \pm \phantom{0}5.8$
& $\phantom{0}71.1 \pm \phantom{0}7.2$ & $\phantom{0}80.2 \pm \phantom{0}8.7$ & $\phantom{0}87.3 \pm
\phantom{0}9.8$ & $\phantom{0}90.8 \pm \phantom{0}10.4$ \\ \hline
\end{tabular}
\label{table:countcorr_14uni}\caption{\small Percentage count
correction results for the uniform noise region of the 14h
field, over the flux density range $2-10$\,mJy, and for signal-to-noise
thresholds of $\rm >1.50\sigma$ to $\rm >4.00\sigma$.}

\begin{tabular}{|c|c|c|c|c|c|c|} \hline
\multicolumn{7}{|c|}{22h field, uniform noise region} \\ \hline
\multicolumn{1}{|c|}{Flux} & 
\multicolumn{6}{c|}{ \% Count Correction} \\ \cline{2-7} 
Density & $\rm >1.50\sigma$  & $\rm >2.00\sigma$ & $\rm >2.50\sigma$ & $\rm
 >3.00\sigma$ &  $\rm >3.50\sigma$ & $\rm >4.00\sigma$ \\ \hline
 $\rm>\phantom{0}2\,mJy$ & $\phantom{0}52.4 \pm \phantom{0}1.9$ & $\phantom{0}82.6 \pm
\phantom{0}3.7$ & $138.7 \pm \phantom{0}8.0$ & $246.5 \pm 18.9$ &
$395.3 \pm 38.4$ & $634.8 \pm \phantom{0}78.1$ \\
 $\rm >\phantom{0}4\,mJy$ & $\phantom{0}40.7 \pm \phantom{0}2.9$ & $\phantom{0}42.9 \pm
 \phantom{0}3.2$ & $\phantom{0}50.3 \pm \phantom{0}4.0$ & $\phantom{0}59.8 \pm
 \phantom{0}5.2$ & $\phantom{0}76.0 \pm \phantom{0}7.4$ & $119.7 \pm \phantom{0}14.7$ \\
$\rm >\phantom{0}6\,mJy$ & $\phantom{0}56.8 \pm \phantom{0}8.6$ & $\phantom{0}56.8 \pm
 \phantom{0}8.6$ & $\phantom{0}56.8 \pm \phantom{0}8.6$ & $\phantom{0}61.0 \pm
 \phantom{0}9.5$ & $\phantom{0}67.6 \pm 11.1$ & $\phantom{0}78.1 \pm \phantom{0}13.8$ \\
$\rm >\phantom{0}8\,mJy$ & $115.4 \pm 32.0$ & $115.4 \pm 32.0$ &
 $115.4 \pm 32.0$ & $115.4 \pm 32.0$ & $115.4 \pm 32.0$ & $125.0 \pm
 \phantom{0}36.1$ \\
$\rm >10\,mJy$ & $100.0 \pm 50.0$ & $100.0 \pm 50.0$ & $100.0 \pm
 50.0$ & $100.0 \pm 50.0$ & $100.0 \pm 50.0$ & $100.0 \pm \phantom{0}50.0$ \\ \hline
\end{tabular}
\label{table:countcorr_22uni}\caption{\small Percentage count
correction results for the uniform noise region of the 22h
field, over the flux density range $2-10$\,mJy, and for signal-to-noise
thresholds of $\rm >1.50\sigma$ to $\rm >4.00\sigma$.}

\begin{tabular}{|c|c|c|c|c|c|c|} \hline
\multicolumn{7}{|c|}{SSA13 deep area field, uniform noise region} \\ \hline
\multicolumn{1}{|c|}{Flux} & 
\multicolumn{6}{c|}{ \% Count Correction} \\ \cline{2-7} 
Density & $\rm >1.50\sigma$  & $\rm >2.00\sigma$ & $\rm >2.50\sigma$ & $\rm
 >3.00\sigma$ &  $\rm >3.50\sigma$ & $\rm >4.00\sigma$ \\ \hline
 $\rm>\phantom{0}2\,mJy$ & $\phantom{0}74.6 \pm \phantom{0}2.5$ &
$\phantom{0}79.0 \pm \phantom{0}2.7$ & $\phantom{0}89.3 \pm
\phantom{0}3.2$ & $102.0 \pm \phantom{0}4.0$ & $125.8 \pm
\phantom{0}5.4$ & $154.4 \pm \phantom{00}7.4$ \\
 $\rm >\phantom{0}4\,mJy$ & $\phantom{0}69.1 \pm \phantom{0}5.0$ &
$\phantom{0}69.4 \pm \phantom{0}5.0$ & $\phantom{0}70.5 \pm
\phantom{0}5.1$ & $\phantom{0}71.7 \pm \phantom{0}5.2$ &
$\phantom{0}72.8 \pm \phantom{0}5.4$ & $\phantom{0}74.0 \pm \phantom{00}5.5$ \\
$\rm >\phantom{0}6\,mJy$ & $\phantom{0}73.3 \pm \phantom{0}9.5$ &
$\phantom{0}73.3 \pm \phantom{0}9.5$ & $\phantom{0}73.3 \pm \phantom{0}9.5$ &
$\phantom{0}73.3 \pm \phantom{0}9.5$ & $\phantom{0}73.3 \pm
\phantom{0}9.5$ & $\phantom{0}74.6 \pm \phantom{00}9.7$ \\
$\rm >\phantom{0}8\,mJy$ & $119.0 \pm 26.0$ & $119.0 \pm 26.0$ &
$119.0 \pm 26.0$ & $119.0 \pm 26.0$ & $119.0 \pm 26.0$ & $119.0 \pm
\phantom{0}26.0$ \\
$\rm >10\,mJy$ & $100.0 \pm 28.9$ & $100.0 \pm 28.9$ & $100.0 \pm
28.9$ & $100.0 \pm 28.9$ & $100.0 \pm 28.9$ & $100.0 \pm \phantom{0}28.9$ \\ \hline
\end{tabular}
\label{table:countcorr_ssa13deepuni}\caption{\small Percentage count
correction results for the uniform noise region of the SSA13 deep area
field, over the flux density range $2-10$\,mJy, and for signal-to-noise
thresholds of $\rm >1.50\sigma$ to $\rm >4.00\sigma$.}

\begin{tabular}{|c|c|c|c|c|c|c|} \hline
\multicolumn{7}{|c|}{SSA13 wide area field, uniform noise region} \\ \hline
\multicolumn{1}{|c|}{Flux} & 
\multicolumn{6}{c|}{ \% Count Correction} \\ \cline{2-7} 
Density & $\rm >1.50\sigma$  & $\rm >2.00\sigma$ & $\rm >2.50\sigma$ & $\rm
 >3.00\sigma$ &  $\rm >3.50\sigma$ & $\rm >4.00\sigma$ \\ \hline
 $\rm>\phantom{0}2\,mJy$ & $\phantom{0}59.7 \pm \phantom{0}0.8$ & $111.6 \pm
\phantom{0}1.9$ & $215.8 \pm \phantom{0}5.2$ & $423.2 \pm 14.2$ &
$841.6 \pm 39.7$ & $>1000$ \\
 $\rm >\phantom{0}4\,mJy$ & $\phantom{0}20.0 \pm \phantom{0}0.3$ & $\phantom{0}30.0 \pm
 \phantom{0}0.6$ & $\phantom{0}51.7 \pm \phantom{0}1.3$ &
 $\phantom{0}96.6 \pm \phantom{0}3.3$ & $189.0 \pm \phantom{0}8.9$ &
 $327.5 \pm \phantom{0}20.4$ \\
$\rm >\phantom{0}6\,mJy$ & $\phantom{0}20.4 \pm \phantom{0}0.5$ &
 $\phantom{0}26.1 \pm \phantom{0}0.7$ & $\phantom{0}35.9 \pm
 \phantom{0}1.2$ & $\phantom{0}54.4 \pm \phantom{0}2.2$ &
 $\phantom{0}91.0 \pm \phantom{0}4.8$ & $138.5 \pm
 \phantom{00}9.1$ \\
$\rm >\phantom{0}8\,mJy$ & $\phantom{0}29.7 \pm \phantom{0}1.4$ & $\phantom{0}34.5 \pm
 \phantom{0}1.7$ & $\phantom{0}41.4 \pm \phantom{0}2.3$ &
 $\phantom{0}49.5 \pm \phantom{0}3.0$ & $\phantom{0}70.6 \pm
 \phantom{0}5.0$ & $\phantom{0}96.5 \pm \phantom{00}8.0$ \\
$\rm >10\,mJy$ & $\phantom{0}50.0 \pm \phantom{0}4.5$ & $\phantom{0}54.8 \pm
 \phantom{0}5.1$ & $\phantom{0}59.4 \pm \phantom{0}5.8$ &
 $\phantom{0}61.8 \pm \phantom{0}6.1$ & $\phantom{0}70.8 \pm \phantom{0}7.5$ &
 $\phantom{0}80.8 \pm \phantom{00}9.1$ \\ \hline
\end{tabular}
\label{table:countcorr_ssa13wuni}\caption{\small Percentage count
correction results for the uniform noise region of the SSA13 wide area
field, over the flux density range $2-10$\,mJy, and for signal-to-noise
thresholds of $\rm >1.50\sigma$ to $\rm >4.00\sigma$.}
\end{table*}

\clearpage
\begin{table*}
\footnotesize
\begin{tabular}{|c|c|c|c|c|c|c|} \hline
\multicolumn{7}{|c|}{SSA17 field, uniform noise region} \\ \hline
\multicolumn{1}{|c|}{Flux} & 
\multicolumn{6}{c|}{ \% Count Correction} \\ \cline{2-7} 
Density & $\rm >1.50\sigma$  & $\rm >2.00\sigma$ & $\rm >2.50\sigma$ & $\rm
 >3.00\sigma$ &  $\rm >3.50\sigma$ & $\rm >4.00\sigma$ \\ \hline
 $\rm>\phantom{0}2\,mJy$ & $\phantom{0}66.5 \pm \phantom{0}1.3$ & $\phantom{0}96.1 \pm
\phantom{0}2.3$ & $148.3 \pm \phantom{0}4.5$ & $226.3 \pm
\phantom{0}8.4$ & $329.7 \pm 14.8$ & $464.7 \pm \phantom{0}24.7$ \\
 $\rm >\phantom{0}4\,mJy$ & $\phantom{0}41.1 \pm \phantom{0}1.4$ & $\phantom{0}46.2 \pm
\phantom{0}1.6$ &  $\phantom{0}56.1 \pm  \phantom{0}2.2$ &
$\phantom{0}69.9 \pm \phantom{0}3.0$ & $\phantom{0}88.7 \pm
\phantom{0}4.3$ & $113.9 \pm \phantom{00}6.2$ \\
$\rm >\phantom{0}6\,mJy$ & $\phantom{0}53.6 \pm \phantom{0}3.2$ & $\phantom{0}57.0 \pm
\phantom{0}3.5$ & $\phantom{0}62.8 \pm \phantom{0}4.1$ &
$\phantom{0}71.0 \pm \phantom{0}4.9$ & $\phantom{0}77.0 \pm
\phantom{0}5.6$ & $\phantom{0}86.0 \pm \phantom{00}6.6$ \\
$\rm >\phantom{0}8\,mJy$ & $\phantom{0}76.3 \pm \phantom{0}7.9$ & $\phantom{0}76.3 \pm
\phantom{0}7.9$ & $\phantom{0}79.8 \pm \phantom{0}8.5$ &
$\phantom{0}83.5 \pm \phantom{0}9.1$ & $\phantom{0}85.5 \pm
\phantom{0}9.4$ & $\phantom{0}87.7 \pm \phantom{00}9.7$ \\
$\rm >10\,mJy$ & $\phantom{0}84.6 \pm 13.5$ & $\phantom{0}84.6 \pm 13.5$ & $\phantom{0}89.2
\pm 14.7$ & $\phantom{0}89.2 \pm 14.7$ & $\phantom{0}91.7 \pm 15.3$ &
$\phantom{0}91.7 \pm \phantom{0}15.3$ \\ \hline
\end{tabular}
\label{table:countcorr_ssa17uni}\caption{\small Percentage count
correction results for the uniform noise region of the SSA17
field, over the flux density range $2-10$\,mJy, and for signal-to-noise
thresholds of $\rm >1.50\sigma$ to $\rm >4.00\sigma$.}

\begin{tabular}{|c|c|c|c|c|c|c|} \hline
\multicolumn{7}{|c|}{SSA22 field, uniform noise region} \\ \hline
\multicolumn{1}{|c|}{Flux} & 
\multicolumn{6}{c|}{ \% Count Correction} \\ \cline{2-7} 
Density & $\rm >1.50\sigma$  & $\rm >2.00\sigma$ & $\rm >2.50\sigma$ & $\rm
 >3.00\sigma$ &  $\rm >3.50\sigma$ & $\rm >4.00\sigma$ \\ \hline
 $\rm>\phantom{0}2\,mJy$ & $\phantom{0}84.3 \pm \phantom{0}1.9$ & $\phantom{0}94.7 \pm
\phantom{0}2.3$ & $107.4 \pm \phantom{0}2.8$ & $126.1 \pm
\phantom{0}3.5$ & $155.9 \pm \phantom{0}4.8$ & $196.2 \pm
\phantom{00}6.8$ \\
 $\rm >\phantom{0}4\,mJy$ & $\phantom{0}65.6 \pm \phantom{0}2.9$ & $\phantom{0}68.6 \pm
\phantom{0}3.1$ & $\phantom{0}71.2 \pm \phantom{0}3.3$ &
$\phantom{0}76.1 \pm \phantom{0}3.6$ & $\phantom{0}80.3 \pm
\phantom{0}3.9$ & $\phantom{0}84.5 \pm \phantom{00}4.2$ \\
$\rm >\phantom{0}6\,mJy$ & $\phantom{0}77.3 \pm \phantom{0}6.2$ & $\phantom{0}77.3 \pm
\phantom{0}6.2$ & $\phantom{0}78.8 \pm \phantom{0}6.4$ &
$\phantom{0}80.4 \pm \phantom{0}6.6$ & $\phantom{0}82.1 \pm
\phantom{0}6.8$ & $\phantom{0}85.0 \pm \phantom{00}7.2$ \\
$\rm >\phantom{0}8\,mJy$ & $\phantom{0}94.9 \pm 12.4$ & $\phantom{0}94.9 \pm 12.4$ & $\phantom{0}94.9
\pm 12.4$ & $\phantom{0}94.9 \pm  12.4$ & $\phantom{0}94.9 \pm 12.4$ & $\phantom{0}98.2 \pm \phantom{0}13.0$ \\
$\rm >10\,mJy$ & $\phantom{0}92.0 \pm 18.4$ & $\phantom{0}92.0 \pm 18.4$ & $\phantom{0}92.0
\pm 18.4$ & $\phantom{0}92.0 \pm 18.4$ & $\phantom{0}92.0 \pm  18.4$ &
$\phantom{0}92.0 \pm \phantom{0}18.4$ \\ \hline
\end{tabular}
\label{table:countcorr_ssa22uni}\caption{\small Percentage count
correction results for the uniform noise region of the SSA22
field, over the flux density range $2-10$\,mJy, and for signal-to-noise
thresholds of $\rm >1.50\sigma$ to $\rm >4.00\sigma$.}

\begin{tabular}{|c|c|c|c|c|c|c|} \hline
\multicolumn{7}{|c|}{Lockman Hole deep field, uniform noise region} \\ \hline
\multicolumn{1}{|c|}{Flux} & 
\multicolumn{6}{c|}{ \% Count Correction} \\ \cline{2-7} 
Density & $\rm >1.50\sigma$  & $\rm >2.00\sigma$ & $\rm >2.50\sigma$ & $\rm
 >3.00\sigma$ &  $\rm >3.50\sigma$ & $\rm >4.00\sigma$ \\ \hline
 $\rm>\phantom{0}2\,mJy$ & $\phantom{0}88.2 \pm \phantom{0}3.3$ & $\phantom{0}98.0 \pm
\phantom{0}3.8$ & $113.1 \pm \phantom{0}4.7$ & $129.5 \pm
\phantom{0}5.8$ & $155.6 \pm \phantom{0}7.6$ & $203.4 \pm \phantom{0}11.4$ \\
 $\rm >\phantom{0}4\,mJy$ & $\phantom{0}63.2 \pm \phantom{0}4.2$ &
$\phantom{0}66.7 \pm \phantom{0}4.5$ & $\phantom{0}72.4 \pm \phantom{0}5.1$ &
$\phantom{0}77.8 \pm \phantom{0}5.7$ & $\phantom{0}82.8 \pm \phantom{0}6.3$ &
$\phantom{0}91.1 \pm \phantom{00}7.3$ \\
$\rm >\phantom{0}6\,mJy$ & $\phantom{0}74.0 \pm \phantom{0}8.7$ &
$\phantom{0}74.0 \pm \phantom{0}8.7$ & $\phantom{0}78.3 \pm
\phantom{0}9.4$ & $\phantom{0}87.1 \pm 11.1$ & $\phantom{0}88.5 \pm
11.3$ & $\phantom{0}88.5 \pm  \phantom{0}11.3$ \\
$\rm >\phantom{0}8\,mJy$ & $\phantom{0}96.6 \pm 17.9$ &
$\phantom{0}96.6 \pm 17.9$ & $\phantom{0}96.6 \pm 17.9$ &
$\phantom{0}96.6 \pm 17.9$ & $\phantom{0}96.6 \pm 17.9$ &
$\phantom{0}96.6 \pm  \phantom{0}17.9$ \\
$\rm >10\,mJy$ & $113.3 \pm 29.3$ & $113.3 \pm 29.3$ & $113.3 \pm
29.3$ & $113.3 \pm 29.3$ & $113.3 \pm 29.3$ & $113.3 \pm \phantom{0}29.3$ \\ \hline
\end{tabular}
\label{table:countcorr_lhdeepuni}\caption{\small Percentage count
correction results for the uniform noise region of the Lockman Hole deep
field, over the flux density range $2-10$\,mJy, and for signal-to-noise
thresholds of $\rm >1.50\sigma$ to $\rm >4.00\sigma$.}

\begin{tabular}{|c|c|c|c|c|c|c|} \hline
\multicolumn{7}{|c|}{Hubble deep field, uniform noise region} \\ \hline
\multicolumn{1}{|c|}{Flux} & 
\multicolumn{6}{c|}{ \% Count Correction} \\ \cline{2-7} 
Density & $\rm >1.50\sigma$  & $\rm >2.00\sigma$ & $\rm >2.50\sigma$ & $\rm
 >3.00\sigma$ &  $\rm >3.50\sigma$ & $\rm >4.00\sigma$ \\ \hline
 $\rm>\phantom{0}2\,mJy$ & $102.5 \pm \phantom{0}4.9$ & $102.5 \pm
\phantom{0}4.9$ & $102.7 \pm \phantom{0}4.9$ & $104.7 \pm
\phantom{0}5.0$ & $108.7 \pm \phantom{0}5.3$ & $118.1 \pm
\phantom{00}6.1$ \\
 $\rm >\phantom{0}4\,mJy$ & $\phantom{0}80.8 \pm \phantom{0}7.4$ & $\phantom{0}80.8 \pm
\phantom{0}7.4$ & $\phantom{0}80.8 \pm \phantom{0}7.4$ & $\phantom{0}80.8 \pm
\phantom{0}7.4$ & $\phantom{0}80.8 \pm \phantom{0}7.4$ & $\phantom{0}80.8 \pm
\phantom{00}7.4$ \\
$\rm >\phantom{0}6\,mJy$ & $\phantom{0}82.7 \pm  11.5$ &
$\phantom{0}82.7 \pm  11.5$ & $\phantom{0}82.7 \pm 11.5$ &
$\phantom{0}82.7 \pm 11.5$ & $\phantom{0}82.7 \pm 11.5$ &
$\phantom{0}82.7 \pm \phantom{0}11.5$ \\
$\rm >\phantom{0}8\,mJy$ & $\phantom{0}81.0 \pm 17.7$ &
$\phantom{0}81.0 \pm  17.7$ & $\phantom{0}81.0 \pm 17.7$ &
$\phantom{0}81.0 \pm 17.7$ & $\phantom{0}81.0 \pm 17.7$ &
$\phantom{0}81.0 \pm \phantom{0}17.7$ \\
$\rm >10\,mJy$ & $\phantom{0}85.7 \pm 32.4$ & $\phantom{0}85.7 \pm
32.4$ & $\phantom{0}85.7 \pm 32.4$ & $\phantom{0}85.7 \pm 32.4$ &
$\phantom{0}85.7 \pm 32.4$ & $\phantom{0}85.7 \pm \phantom{0}32.4$ \\ \hline
\end{tabular}
\label{table:countcorr_hdfuni}\caption{\small Percentage count
correction results for the uniform noise region of the Hubble deep
field, over the flux density range $2-10$\,mJy, and for signal-to-noise
thresholds of $\rm >1.50\sigma$ to $\rm >4.00\sigma$.}
\end{table*}

\renewcommand{\topfraction}{0.1}

\clearpage
\subsection{Output versus input flux density tables}
The following tables give the percentage of sources recovered from the
simulated images, for signal-to-noise thresholds of 
$\rm >1.50\sigma$ to $\rm >4.00\sigma$, according to the following criteria:

\noindent 1) Fainter. The retrieved flux density was fainter than the input
source with which it had been identified ($\rm S_{in}>S_{out}$).\\
\noindent 2) Within error bars. The input flux density lay within the $\rm 1
\sigma_{rms}$ error bars of the retrieved value ($\rm
S_{out}-err_{out} < S_{in} < S_{out}+err_{out}$).\\
\noindent 3) Boosted. The input flux density was less than the lower error
boundary on the output value, but was still within a factor of 2 of
the measured flux density ($\rm S_{out}/2 < S_{in} <
S_{out}-err_{out}$).\\
\noindent 4) Spurious / confused. The fitted flux density to the peak
could not be identified with a source in the input catalogue, located
within 8 arcseconds and a factor 2 in brightness ($\rm S_{in} < S_{out}/2$).\\

The percentage of sources classified as (1), (2), (3) and (4) are
given in Columns 3, 4, 5 and 6 respectively. Those fields 
composed of a deep pencil
beam survey within a wider-area shallower survey have had these two
components treated separately.

\begin{table*}
\footnotesize
\begin{tabular}{|c|c|c|c|c|c|} \hline
\multicolumn{6}{|c|}{ELAIS N2 field, uniform noise region} \\ \hline
\multicolumn{1}{|c|}{} & 
\multicolumn{5}{c|}{Output vs. input flux density statistics} \\ \cline{2-6} 
$\rm S/N$ & \% noise  &\% & \% within 
& \% & \% confused/ \\ 
threshold & only & fainter & $\rm 1\sigma$ errors & boosted & spurious \\ \hline
$\rm >1.50 \sigma$ &  $100.7 \pm  \phantom{0}0.7$ & $\phantom{0}0.4 \pm
\phantom{0}0.1$ & $15.0 \pm \phantom{0}0.1$ & $\phantom{0}4.1 \pm
\phantom{0}0.1$ & $80.6 \pm \phantom{0}0.6$ \\
$\rm >2.00 \sigma$ & $\phantom{0}81.4 \pm \phantom{0}1.0$ &
$\phantom{0}0.6 \pm \phantom{0}0.1$ & $13.8 \pm \phantom{0}0.1$ &
$\phantom{0}7.3 \pm \phantom{0}0.1$ & $78.4 \pm \phantom{0}0.7$ \\
$\rm >2.50 \sigma$ & $\phantom{0}60.0 \pm \phantom{0}0.8$ & $\phantom{0}0.9 \pm
\phantom{0}0.1$ & $16.4  \pm \phantom{0}0.2$ & $12.6 \pm
\phantom{0}0.2$ & $70.1 \pm \phantom{0}0.9$ \\
$\rm >3.00 \sigma$ & $\phantom{0}26.6 \pm \phantom{0}0.5$ &
$\phantom{0}1.6 \pm \phantom{0}0.1$ & $21.8 \pm \phantom{0}0.4$ &
$19.8 \pm \phantom{0}0.4$ & $56.8 \pm  \phantom{0}1.1$ \\
$\rm >3.50 \sigma$ & $\phantom{0}14.8 \pm  \phantom{0}0.4$ &
$\phantom{0}2.1 \pm \phantom{0}0.1$ & $28.6 \pm \phantom{0}0.8$ &
$27.5 \pm \phantom{0}0.7$ & $41.8 \pm \phantom{0}1.1$ \\
$\rm >4.00 \sigma$ &  $\phantom{0}0.0 \pm \phantom{0}0.0$ & $\phantom{0}3.4 \pm \phantom{0}0.1$ & $35.4 \pm
\phantom{0}1.3$ & $35.1 \pm \phantom{0}1.3$ & $26.1 \pm
\phantom{0}0.9$ \\ \hline
\end{tabular}
\label{table:booststats_n2uni}\caption{\small Output versus input flux
density statistics for the uniform noise region of the ELAIS N2
field, for signal-to-noise thresholds in the range $\rm >1.50 \sigma$
to $\rm >4.00 \sigma$. }

\begin{tabular}{|c|c|c|c|c|c|} \hline
\multicolumn{6}{|c|}{ELAIS N2 field, non-uniform noise region} \\ \hline
\multicolumn{1}{|c|}{} & 
\multicolumn{5}{c|}{Output vs. input flux density statistics} \\ \cline{2-6} 
$\rm S/N$ & \% noise  &\% & \% within 
& \% & \% confused/ \\ 
threshold & only & fainter & $\rm 1\sigma$ errors & boosted & spurious \\ \hline
$\rm >1.50 \sigma$ & $\phantom{0}95.8 \pm \phantom{0}1.8$ & $\phantom{0}0.0 \pm
\phantom{0}0.0$ & $\phantom{0}3.4 \pm \phantom{0}0.1$ &
$\phantom{0}0.6 \pm \phantom{0}0.1$ & $96.0 \pm \phantom{0}1.8$ \\
$\rm >2.00 \sigma$ & $\phantom{0}84.8 \pm \phantom{0}2.4$ & $\phantom{0}0.0 \pm
\phantom{0}0.0$ & $\phantom{0}2.6 \pm \phantom{0}0.1$ &
$\phantom{0}1.2 \pm \phantom{0}0.1$ & $96.2 \pm \phantom{0}2.5$ \\
$\rm >2.50 \sigma$ & $\phantom{0}38.5 \pm \phantom{0}1.7$ & $\phantom{0}0.0 \pm
\phantom{0}0.0$ & $\phantom{0}4.0 \pm \phantom{0}0.2$ & $3.5 \pm
\phantom{0}0.2$ & $92.5 \pm \phantom{0}4.1$ \\
$\rm >3.00 \sigma$ & $\phantom{0}70.4 \pm \phantom{0}5.9$ & $\phantom{0}0.0 \pm
\phantom{0}0.0$ & $\phantom{0}7.7 \pm \phantom{0}0.7$ &
$\phantom{0}7.0 \pm \phantom{0}0.6$ & $85.2 \pm \phantom{0}7.2$ \\
$\rm >3.50 \sigma$ & $\phantom{00}0.0 \pm \phantom{0}0.0$ &
$\phantom{0}0.0 \pm \phantom{0}0.0$ & $13.9 \pm \phantom{0}2.3$ &
$25.0 \pm \phantom{0}4.2$ & $61.1 \pm 10.2$ \\
$\rm >4.00 \sigma$ & $\phantom{00}0.0 \pm \phantom{0}0.0$ &
$\phantom{0}0.0 \pm \phantom{0}0.0$ & $\phantom{0}8.3 \pm
\phantom{0}2.4$ & $50.0 \pm 14.4$ & $41.7 \pm 12.0$ \\ \hline
\end{tabular}
\label{table:booststats_n2non}\caption{\small Output versus input flux
density statistics for the non-uniform noise region of the ELAIS N2
field, for signal-to-noise thresholds in the range $\rm >1.50 \sigma$
to $\rm >4.00 \sigma$. }

\begin{tabular}{|c|c|c|c|c|c|} \hline
\multicolumn{6}{|c|}{Lockman Hole wide area field, uniform noise region} \\ \hline
\multicolumn{1}{|c|}{} & 
\multicolumn{5}{c|}{Output vs. input flux density statistics} \\ \cline{2-6} 
$\rm S/N$ & \% noise  &\% & \% within 
& \% & \% confused/ \\ 
threshold & only & fainter & $\rm 1\sigma$ errors & boosted & spurious \\ \hline
$\rm >1.50 \sigma$ & $\phantom{0}89.3 \pm \phantom{0}0.8$ & $\phantom{0}0.3 \pm
\phantom{0}0.1$ & $12.2 \pm \phantom{0}0.1$ & $\phantom{0}2.6 \pm \phantom{0}0.1$
& $84.9 \pm \phantom{0}0.7$ \\
$\rm >2.00 \sigma$ & $\phantom{0}69.8 \pm \phantom{0}1.1$ & $\phantom{0}0.5 \pm
\phantom{0}0.1$ & $11.7 \pm \phantom{0}0.1$ & $\phantom{0}5.3 \pm
\phantom{0}0.1$ & $82.5 \pm \phantom{0}1.0$ \\
$\rm >2.50 \sigma$ & $\phantom{0}64.3 \pm \phantom{0}1.2$ & $\phantom{0}0.7 \pm
\phantom{0}0.1$ & $14.8 \pm \phantom{0}0.3$ & $\phantom{0}9.7 \pm
\phantom{0}0.2$ & $74.7 \pm \phantom{0}1.3$ \\
$\rm >3.00 \sigma$ & $\phantom{0}35.7 \pm \phantom{0}1.0$ & $\phantom{0}1.1 \pm
\phantom{0}0.1$ & $20.5 \pm \phantom{0}0.5$ & $15.0 \pm \phantom{0}0.4$ & $63.3
\pm \phantom{0}1.7$ \\
$\rm >3.50 \sigma$ & $\phantom{00}0.0 \pm \phantom{0}0.0$ &
$\phantom{0}2.4 \pm \phantom{0}0.1$ & $32.7 \pm \phantom{0}1.4$ & $23.3 \pm
\phantom{0}1.0$ & $41.6 \pm \phantom{0}1.7$ \\
$\rm >4.00 \sigma$ & $\phantom{00}0.0 \pm \phantom{0}0.0$ &
$\phantom{0}3.0 \pm \phantom{0}0.2$ & $40.7 \pm \phantom{0}2.3$ &
$32.5 \pm \phantom{0}1.9$ & $23.8 \pm \phantom{0}1.4$ \\ \hline
\end{tabular}
\label{table:booststats_lhuni}\caption{\small Output versus input flux
density statistics for the uniform noise region of the Lockman Hole
wide area
field, for signal-to-noise thresholds in the range $\rm >1.50 \sigma$
to $\rm >4.00 \sigma$. }
\vspace{0.5cm}
\end{table*}

\renewcommand{\topfraction}{0.95}
\newpage
\begin{table*}
\footnotesize
%
\label{table:booststats_lhnon}\caption{\small Output versus input flux
density statistics for the non-uniform noise region of the Lockman Hole
wide area
field, for signal-to-noise thresholds in the range $\rm >1.50 \sigma$
to $\rm >4.00 \sigma$. }

%
\label{table:booststats_lhstrip}\caption{\small Output versus input flux
density statistics for the uniform noise region of the Lockman Hole
deep strip
field, for signal-to-noise thresholds in the range $\rm >1.50 \sigma$
to $\rm >4.00 \sigma$. }

%
\label{table:booststats_03deep}\caption{\small Output versus input flux
density statistics for the uniform noise region of the 03h deep area
field, for signal-to-noise thresholds in the range $\rm >1.50 \sigma$
to $\rm >4.00 \sigma$. }

%
\label{table:booststats_03wuni}\caption{\small Output versus input flux
density statistics for the uniform noise region of the 03h wide area
field, for signal-to-noise thresholds in the range $\rm >1.50 \sigma$
to $\rm >4.00 \sigma$. }
\end{table*}

\newpage
\begin{table*}
\footnotesize
%
\label{table:booststats_10uni}\caption{\small Output versus input flux
density statistics for the uniform noise region of the 10h
field, for signal-to-noise thresholds in the range $\rm >1.50 \sigma$
to $\rm >4.00 \sigma$. }

%
\label{table:booststats_14uni}\caption{\small Output versus input flux
density statistics for the uniform noise region of the 14h
field, for signal-to-noise thresholds in the range $\rm >1.50 \sigma$
to $\rm >4.00 \sigma$. }

%
\label{table:booststats_22uni}\caption{\small Output versus input flux
density statistics for the uniform noise region of the 22h
field, for signal-to-noise thresholds in the range $\rm >1.50 \sigma$
to $\rm >4.00 \sigma$. }

%
\label{table:booststats_ssa13deep}\caption{\small Output versus input flux
density statistics for the uniform noise region of the SSA13 deep area
field, for signal-to-noise thresholds in the range $\rm >1.50 \sigma$
to $\rm >4.00 \sigma$. }
\end{table*}

\newpage
\begin{table*}
\footnotesize
%
\label{table:booststats_ssa13wuni}\caption{\small Output versus input flux
density statistics for the uniform noise region of the SSA13 wide area
field, for signal-to-noise thresholds in the range $\rm >1.50 \sigma$
to $\rm >4.00 \sigma$. }

%
\label{table:booststats_ssa17uni}\caption{\small Output versus input flux
density statistics for the uniform noise region of the SSA17
field, for signal-to-noise thresholds in the range $\rm >1.50 \sigma$
to $\rm >4.00 \sigma$. }

%
\label{table:booststats_ssa22uni}\caption{\small Output versus input flux
density statistics for the uniform noise region of the SSA22
field, for signal-to-noise thresholds in the range $\rm >1.50 \sigma$
to $\rm >4.00 \sigma$.}

%
\label{table:booststats_lhdeep}\caption{\small Output versus input flux
density statistics for the uniform noise region of the Lockman Hole deep
field, for signal-to-noise thresholds in the range $\rm >1.50 \sigma$
to $\rm >4.00 \sigma$. }
\end{table*}

\newpage
\begin{table*}
\footnotesize
%
\label{table:booststats_hdfuni}\caption{\small Output versus input flux
density statistics for the uniform noise region of the Hubble deep field
field, for signal-to-noise thresholds in the range $\rm >1.50 \sigma$
to $\rm >4.00 \sigma$. }
\end{table*}

\end{document}